\documentstyle[twoside,trittsty,psfig,12pt]{report}
\begin{document}
\pagestyle{tritthead}
\markboth{l}{r}

\newcommand{\jhat}{\mbox{$\hat{\jmath}$}}
\newcommand{\del}{\partial}
\newcommand{\diraci}{\left[-i\partial_i\alpha^i+\beta m_e\right]}
\newcommand{\diracj}{\left[-i\partial_j\alpha^j+\beta m_e\right]}
\newcommand{\odel}{\frac{1}{i\partial^+}}
\newcommand{\oddel}{\frac{1}{(i\partial^+)^2}}
\newcommand{\ppsid}{\psi^{\dagger}_+}
\newcommand{\ppsi}{\psi_+}
\newcommand{\xperp}{\vec{x}_{\perp}}
\newcommand{\kp}{k_{\perp}}
\newcommand{\vkp}{\vec{k}_{\perp}}
\newcommand{\nperp}{\vec{n}_{\perp}}
\newcommand{\mperp}{\vec{m}_{\perp}}
\newcommand{\pperp}{\vec{p}_{\perp}}
\newcommand{\afull}{a_{\lambda,p,\pperp}}
\newcommand{\bfull}{b_{s,n,\nperp}}
\newcommand{\dfull}{d_{s,n,\nperp}}
\newcommand{\ux}{\underline{x}}
\newcommand{\uy}{\underline{y}}
\newcommand{\uk}{\underline{k}}
\newcommand{\un}{\underline{n}}
\newcommand{\um}{\underline{m}}
\newcommand{\up}{\underline{p}}
\newcommand{\uq}{\underline{q}}

\newcommand{\adagg}{a^{\dagger}}
\newcommand{\bdagg}{b^{\dagger}}
\newcommand{\ddagg}{d^{\dagger}}

\newcommand{\beq}{\begin{equation}}
\newcommand{\eeq}{\end{equation}}
\newcommand{\beqa}{\begin{eqnarray*}}
\newcommand{\eeqa}{\end{eqnarray*}}


\newcommand{\eeg}{{e\bar{e}\gamma}}
\newcommand{\kperp}{\vec{k}_{\perp}}
\newcommand{\epsvec}{\vec{\epsilon}_{\perp}}
\newcommand{\Rydberg}{Ry\;}
\def\graphpath{}

\begin{titlepage}

\flushright{MPI H-V4-1997}  
\vspace{0.7cm}

\centerline{\Large\bf Quantum Electrodynamics at strong Couplings}
\vspace{0.5cm}

\centerline{Uwe Trittmann and Hans-Christian Pauli}

\centerline{\sl Max-Planck-Institut f\"ur Kernphysik} 

\centerline{D-69029 Heidelberg} 

\centerline{(\today)}
\vspace{1cm}

\centerline{\Large Abstract}
\vspace{0.5cm}

\centerline{\parbox{12.9cm}{
Front form dynamics is not a manifestly rotational invariant formalism.
In particular, the requirement of an invariance under rotations around 
the transverse axes is difficult to fulfill since the corresponding
operators are complicated and involve the interaction.
All approximations in solving Quantum Field Theories using front form dynamics,
as inevitable as 
they are, are destined to destroy full Poincar\'e invariance.
In the present work it is investigated, to which extent rotational 
invariance is restored in the solution of a light-cone quantized field theory. 
The 
positronium spectrum in full (3+1) dimensions is calculated at an unphysically
large coupling of $\alpha=0.3$ to be sensitive for terms breaking rotational
invariance and to accustomize the theory to future QCD applications. 
The numerical accuracy of the formalism is improved to allow for the 
calculation of mass 
eigenvalues for arbitrary components $J_z$ of the total angular momentum. 
We find numerically degenerate eigenvalues as expected from 
rotationally invariant formalisms and the right multiplet structures up to
large principal quantum numbers $n$. The results indicate that rotational 
invariance is unproblematic even in front form dynamics.
Another focus of the work relies on the inclusion of the annihilation 
channel. This enlargement of the model is non-trivial and a consistency check
of the underlying theory of effective interactions. 
The correct numerical eigenvalue shifts, and especially the right hyperfine 
splitting are obtained. Moreover,
the cutoff dependence of the eigenvalues is improved drastically by 
the annihilation channel for the
triplet states. The implications of the applied effective Hamiltonian
approach are discussed in detail.
}}

\end{titlepage}

\pagenumbering{roman}

\tableofcontents



\markboth{Introduction}{Introduction}

\chapter*{Introduction}
\addcontentsline{toc}{chapter}{Introduction}

\pagenumbering{arabic}


After a long line of efforts, Quantum 
Chromodynamics \cite{tHooft,Pollitzer,Gross,Wilson} emerged around 1973
as the theory that correctly describes the interaction between 
{\em quarks} (fermions), considered to be elementary, through the action
of the non-abelian gauge field of {\em gluons} (bosons).
The hierarchy of particles seems to be the following. 
The {\em hadrons}, that is baryons (nucleons) and mesons,
consist of quarks.
The gluons produce the interaction between quarks. They provide the force,
a strong analogue of the van-der-Waals force,
that allows nucleons to bind together to form the atomic nucleus. 
This point of view concerning the nuclear forces clarified the former 
picture that the intermediate boson field is that of the mesons \cite{Yukawa}. 


An open question in modern theoretical physics is the link 
between the high and the low energy domains of Quantum Chromodynamics. 
The description of the high energy region, which was inspired by experiments 
in the late sixties, is well
established. 
Contrary, the low energy region of the theory cannot be attacked
by standard perturbation theory because the coupling depends on the scale and
grows large in this region.
Instead, this region must be described by more phenomenological models
such as the constituent quark model, or chiral perturbation theory.
A major problem in this undertaking to connect the high and the low energy
regimes is the running coupling constant of QCD
which depends very strongly on the four momentum transfer
between the interacting particles. Moreover, it grows large for small 
momentum transfer so that perturbation theory breaks down at a certain
point.
An important task of theoretical high and medium energy physics is 
therefore the 
construction of methods that can produce {\em hadronic} spectra 
and wavefunctions.
The importance of the hadronic properties and their wavefunctions
for physical observables, such as structure functions and decay 
constants, is undeniable. 
Moreover, their calculation would allow for rigorous tests of assumptions
necessary in models of nuclei and hadrons.

The history of methods to calculate quantum mechanical bound states is 
as old as Quantum Mechanics itself \cite{Schroedinger}.
But in hadronic physics the large coupling implies that bound particles form 
highly relativistic systems.
This fact enforces the rigorous application of covariant field theories.
Non-relativistic {\em potential models} \cite{Richardson,Buchmueller,Isgur}
are only applicable for heavy systems, as for instance the 
$J/\psi$ \cite{Aubert,Augustin}
or the $\Upsilon$ \cite{Herb}.
Other covariant methods are {\em lattice gauge theory} \cite{Creutz} 
or the evaluation of the {\em Bethe-Salpeter equation} \cite{BetheSalpeter}.
Within lattice gauge theories only the ground state or the first 
few excited states
are tractable and structure functions are extremely difficult to observe 
directly\cite{Weingarten}.
The application of the Bethe-Salpeter equation\cite{Gromes} 
suffers the typical fate of strongly bound systems:
The wavefunctions are dominated by relativistic momenta and contain 
a considerable number of quark-antiquark pairs.
Therefore, even the vacuum is very complicated. 

The latter point is problematic for all quantum field theories, if a theory
is quantized at equal usual time.
In {\em light-cone quantization}, the results are frame independent, 
as opposed to equal time quantization. There, the Lorentz boosts contain 
the interaction and are therefore complicated dynamical operators, 
whereas in light-cone quantization all Lorentz boosts are 
kinematic \cite[Chap.~2.B]{LeutwylerStern}. 
Consequently, light-cone quantized field theories seem 
to be a promising tool to
understand the physics of relativistic QCD-bound states.
However, neither have all problems 
(Poincar\'e invariance, zero modes, etc.) 
of this Hamiltonian approach been solved, nor is it clear if and how effective
theories \cite{weakWilson,Perry,PauliMIR} which have been proposed to
deal with the problems associated to finite cutoffs, will work.
Some of these deficiencies are due to the neglection of Hamiltonian field
theory in favor of the action-oriented approaches of {\sc Feynman, et.~al}
\cite{Tomonaga,Feynman, Schwinger}.
In any case, it is better to establish that a new method works for known
problems
before applying it to yet unsolved problems.
Bearing this in mind, the basic motivation for the present work is twofold. 
On the one hand, there is the obvious obstacle that light-cone quantized 
theories are rotationally {\em not} manifestly invariant {\em by construction}.
One main part of this thesis is consequently dedicated to the investigation
of the associated question, what this means to the results of such a 
theory: do the {\em results} obey rotational symmetry (as observed in nature)
or not?
Because we have to compare to other calculations and methods, we 
choose a realistic, well-understood, but non-trivial 
problem, namely the QED-bound state {\em par excellence}:
positronium. Our choice has another cause.
The formalism to be applied is ideally suited for this case of two 
particles of same mass. In ordinary approaches, the severe problems
of recoil corrections make it much easier to calculate the hydrogen  
or muonium spectra than to solve for the eigenvalues of 
positronium.   

On the other hand, there is the need for an effective formalism for 
gauge theories: nobody ever has solved rigorously a 
relativistic many-body theory in 
one time and three space dimensions (3+1 dimensional theory). 
It is therefore necessary to use such an effective approach and to investigate
how reasonable its results are. We have done so.  
But 
{\em after} the successfully applying of the effective method, 
we go one step further in asking how the 
excellent results can be understood within a much more general formalism,
originating in the Lagrangian density of a gauge theory. 


\section*{Outline of the present work}
 
This work is structured as follows.
%
%
We start in Chapter \ref{DrAChapterFieldTheories} with an overview of the
development and difficulties of the approach to 
field theories in front form dynamics. 
%
%
In Chapter \ref{ChapterModel}, 
the positronium model used in the present work is introduced.
Starting with a Hamiltonian field theory quantized on the light-cone,
the steps are described that lead to 
the formulation of an {\em effective integral equation} in momentum space.
Although the derivation of the Hamiltonian involved the application of 
the DLCQ formalism, described in the following section,
we stress the fact that we use only the continuum formulation of 
QED$_{(3+1)}$ in front form dynamics in the present work. 
Discretizations occur only in the
context of numerical integrations, which have nothing to do with physical
considerations. Errors due to numerical artifacts, such as effects of 
a finite number of integration points $N$, are discussed. 
Cutoffs determined by physics emerge in two aspects. 
One is the problem of divergencies occuring because
transverse momenta can become infinite. 
Therefore, one must introduce a regulator. 
We analyze
the results of our model concerning their dependencies on this cutoff.
The second is the
restriction of {\em sectors} with different particle content, sometimes
referred to as Tamm-Dancoff approach. 
We will show that (a) a theory truncated in this way is not solvable,
(b) there is a possibility to treat the associated singularity by adjusting
the form of the effective interaction, 
and (c) the treatment of the singularity
can be understood within a general formalism of effective interactions.
The latter point is discussed in Chapter~\ref{DrAChapterEffInt}.

We improved the numerical methods which were used to solve the effective
interaction before \cite{KPW}. 
This leads to an improved convergence of the eigenvalues with the number 
of integration points. However, the basic intention is the need 
of this improvement to attack the generalization of the model, described
in Chapter \ref{DrAChapterJz}.
More effort than in \cite{KPW} is put in the investigation 
of higher excited states to show the range
of applicability of the model. We find the correct multiplet 
structure up to a principal quantum number $n{=}5$. 
Detailed tables allow for the comparison of the results of the present work
with those of equal-time perturbation theory. We emphasize the use of a 
huge coupling constant, $\alpha{=}0.3$, to test our non-perturbative approach: 
it is approximately 40 times larger
than the physical coupling constant of QED. In the calculation
of the hyperfine splitting, we point out that the comparison of our results
with perturbation theory is problematic, because results of the latter depend
noticeably on the order of the calculations at large couplings.   

%
%

As a main point of the thesis,
we generalize the formalism of Chapter \ref{ChapterModel}
to arbitrary $z$-components of the 
total angular momentum, $J_z$.
This is done in Chapter~\ref{DrAChapterJz}.
Before, calculations were performed in the
$J_z{=}0$ sector of the theory only. 
To interpret the results, the theory of the Poincar\'{e} group is studied
in the introductory section of this chapter in 
the context of front form dynamics.
We find that states with the same total angular momentum $J$, but 
different $J_z$, are numerically degenerate. This result is 
expected from physical considerations, but surprising because the associated
operators of transverse rotations are dynamical in front form dynamics
although they do commute with the
light-cone Hamiltonian. As a consequence of our calculations, 
we can classify the 
positronium eigenstates
according to their quantum numbers of $J^2$ and $J_z$, just as in the 
equal-time formalism. 
The wavefunctions are analyzed concerning their symmetry properties. 
Besides, we show that the wavefunctions are {\em not}
rotationally invariant in any sector of $J_z$.

%
%

After the application
of effective interactions and the generalization of 
the theory to arbitrary $J_z$, the next crucial point is 
the introduction of the annihilation channel into the theory 
(Chapter~\ref{DrAChapterAnnihilation}).  
We enlarge the Fock basis by the one-photon state, typical for QED and
absent in QCD. It is shown that the degeneracy of states is maintained. 
This is astonishing, because we find  
a total separation of the dynamic and instantaneous diagrams involved in the
different $J_z$ sectors.
Another consequence of the implementation of the annihilation channel is the
stabilizing of eigenvalues, concerning their dependence on the cutoff 
$\Lambda$. 

%
%

Chapter \ref{DrAChapterEffInt} deals with
the problems constructing effective interactions in front form dynamics.
We show that the prescription to treat the singularity attached to a truncation
of the Fock space is a consequence of the formalism of effective interactions.  
%
%
A summary and a discussion of our results follow.

%
%

All basics and technical details 
are consequently cast into the appendices
to achieve a clear and structured line of arguments 
in the main part of the thesis.
We address the attention of the reader especially 
to Appendix~\ref{AppxQED} where
QED on the light-cone is described, and to Appendix~\ref{AppxNumerics}
where the numerical methods are explained.
The latter should be studied when details of calculations
seem unclear in the main text. 
In Appendix~\ref{AppxRenormalization} the applied renormalization scheme
is described.
The other appendices mainly supply necessary formul\ae.
The source code of the computer program developed in the present work is listed
in Appendix~\ref{AppxProgram}.


\section*{Methodological sketch}

This section describes in short the methods applied in the thesis and
their subtle points.

\begin{itemize}

\item The underlying picture of a composed system in front form dynamics
is that of a state consisting of valence particles carrying its outer quantum
numbers, and of arbitrarily many virtual particles, be it gauge particles or
fermion-antifermion pairs. 
A state is written in Fock-space representation as
\[
|\psi\rangle = \psi_{g}|\psi_{g}\rangle
             + \psi_{q\bar{q}}|\psi_{q\bar{q}}\rangle
             + \psi_{q\bar{q}\gamma}|\psi_{q\bar{q}\gamma}\rangle
             + \psi_{q\bar{q}gg}|\psi_{q\bar{q}gg}\rangle
             + \ldots .
\]
Here, $g$ denotes a gauge boson  (photon or gluon) and $q$
a fermion (electron or quark). In QCD, the first term in this expansion is 
absent because it is not color neutral. 

\item To construct the Hamiltonian, the following steps are performed.
First, starting with the Lagrangian density, the generators of the 
Poincar\'e group are calculated.\footnote{In actual calculations,
one restricts oneself to calculating the momenta $P^{\mu}$, 
because the angular momentum tensor $M^{\mu\nu}$ is irrelevant 
for the mass spectrum. Sometimes $J_z$ is constructed.} 
Next, the dependent fields are expressed in terms of the dynamic fields.

\item 
The independent fields are expanded in plane waves and are discretized by
imposing periodic or anti-periodic boundary conditions along all space-like
directions in a given volume.

\item 
It is important that the expansion into plane waves takes place in 
{\em momentum} space. By this, assuming a reasonable choice of boundary 
conditions, all points in this space are equivalent and the momentum of the 
center of mass is well-defined. 
The enormous improvement of convergence achieved by such a prescription
is seen already in non-relativistic many body theory \cite{Pauli84}.

\item When working in the light-cone gauge, $A_+=0$, the theory is  
consistent in the {\em normal mode} sector only (cf.~next section).
The {\em zero modes} are important for the vacuum structure \cite{DissBvdS},
but most probably not for bound state spectra.

\item 
Having quantized the theory by postulating canonical commutation relations 
for the
creation and destruction operators, the Poincar\'e generators are written
as functionals of these operators. 

\item The reference state, ``the vacuum'', is an eigenstate of the Hamiltonian
with eigenvalue zero.
This follows from the fact, that creation of particles out of the vacuum
is forbidden because of the conservation of longitudinal momentum.

\item Spectrum and wavefunctions are obtained by solving 
the eigenvalue problem
\[
P^+P^- |\psi_n\rangle = M_n^2 |\psi_n\rangle, 
\]
where the $\psi_n$ are eigenfunctions to the mass (squared) eigenvalue
$M_n$. This matrix equation and its solution by diagonalization are the
endpoint of the so-called DLCQ formalism. 
We discuss its difficulties in the next chapter.

\item We can perform the continuum limit of this matrix equation.
This is possible in a 
direct way, as opposed to the non-trivial continuum limit, for example,
of lattice gauge theories. The matrix equation is mapped into an integral 
equation. 

\item The task is now to find an appropriate scheme to solve for the spectrum 
of this (infinite dimensional) operator. 
We use a restriction of the Fock space together with an effective interaction
to account for the effects of the truncated states, and an explicit 
cutoff because of the unrestricted transverse momenta.
To exploit and to justify this approach is the main part of the present work.   

\item The so-obtained {\em effective integral equation} is solved with 
the appropriate method of Gaussian quadratures, which leads in the end
to the diagonalization of a finite dimensional matrix.

\item The dependence of the solutions on the unphysical parameters 
(cutoffs, etc.) has to be investigated to test the significance of the obtained
results. 

\end{itemize}

\markboth{l}{r}

\chapter{\label{DrAChapterFieldTheories}Field theories 
in front form dynamics}

\section{Historical survey}

Light-cone coordinates\protect\footnote{The coordinate vector is  
$x=(x^{+},x^-,x_{\perp})$ with $x^{\pm}=x^0\pm x^3, x_{\perp}=(x_1,x_2)$. 
Cf. Appendix \protect\ref{AppxNotations}.}
were introduced into the Hamiltonian field theory by {\sc Dirac} \cite{Dirac}
in 1949. 
According to his definition an operator is called 
a {\em Hamiltonian} if it propagates a physical system in a fixed 
direction in space-time. 
This direction is subject to certain constraints, but not unique.
{\sc Dirac} listed three different possibilities\protect\footnote{Actually, there
are five such quantization surfaces \protect\cite{LeutwylerStern}.}
in selecting such a ``general time''. The most prominent ones 
are the usual time $x^0$ ({\em instant form dynamics}) 
and the light-cone time $x^+$ ({\em front form dynamics}).
This {\em Hamiltonian} formalism attracted little attention because of
the spectacular achievements (e.g.\ \cite[the anomalous magnetic 
moment of the electron, Eq.~(1.112)]{Schwinger3})
of the {\em action-oriented} work of {\sc Tomonaga}~\cite{Tomonaga},
{\sc Schwinger}~\cite{Schwinger,Schwinger3}, and {\sc Feynman}~\cite{Feynman},
launched at the same time. 

The rejuvenation of the Hamiltonian method in field theory, 
in particular studying front form dynamics, came in several 
steps.
First, {\sc Weinberg}~\cite{Weinberg}, using  
$\phi^3$ theory in the  {\em infinite momentum frame}, discovered that 
the creation and annihilation of particles out of the vacuum 
is forbidden for $p_z\rightarrow\infty$ and that the corresponding 
divergencies are absent in this frame.
After the derivation of the rules for front form perturbation 
theory~\cite{KogutSoper}, it was an important success to 
show the equivalence of this theory with the Feynman rules of
ordinary perturbation theory\cite{ChangRootYan}\cite{Ligterink}. 
The treatment of non-abelian gauge theories and the description of 
exclusive QCD processes are described in
\cite{BLepage,BLepage2}.

To render the theory tractable and to finally implement it on a computer,
the ambitious program of
{\bf D}iscretized {\bf L}ight-{\bf C}one {\bf Q}uantization\cite{BP} 
was proposed by {\sc Pauli} and {\sc Brodsky} in 1985.
The formalism was applied first to the 
Yukawa model in $(1+1)$ dimensions, then to scalar
QED$_{(1+1)}$ \cite{Sawicki}, followed by calculations  
of the $\phi^4_{(1+1)}$ theory \cite{HariVary},
and QCD$_{(1+1)}$\cite{Hornbostel,Heyssler}.
Commencing with a paper by {\sc Eller, Pauli} and {\sc Brodsky}
\cite{EllerPB}, there is a large literature discussing the (massive)
{\em Schwinger model} \cite{McCartor,Elser,Voellinger}. 
   

All of the above examples have been studied in $1{+}1$ dimensions.
Quantum Electrodynamics was the first realistic  gauge
theory in physical space-time to be treated with 
DLCQ \cite{KPW,Tang,Kaluza}, where results were
compared
with experimental data or other theoretical work.
Subsequently, an attempt was made to apply this method to 
QCD$_{(3+1)}$~\cite{DissWoelz}.
Most recently, this was done to a large extent in the 
so-called {\em collinear 
model}~\cite{KalloPauliPinsky,DissBayer}, {\em i.e.}~effectively in lower
dimensions. Here, the method of the 
{\em transverse lattice}~\cite{BvdSBurckardt,BvdSCoralGables} 
seems to open a promising way to proceed, but is still under construction.
Nevertheless,
QED$_{(3+1)}$ is a ``milestone''
for non-perturbative approaches in light-cone quantization 
on the way to understand
full QCD.
QED contains most of the fundamental difficulties found in QCD 
and can be attacked likewise in a truncated Fock basis.
\vspace{-0.1cm}

\section{Problems of front form field theories}
The disadvantages using front form dynamics include a certain 
non-conformity of the
inertial system used, especially the counter-intuitive missing
of a well-defined angular momentum. 
As mentioned in the introduction, the operators of rotations around
the transverse axes are complicated, {\em i.e.}~contain the interaction.
Consequently, although the rotation 
operator ${\cal J}_3$ {\em is} kinematic, the
states of a system cannot be classified with respect to total angular momentum  
$J^2$. 
Of course, this is not necessary from the outset, and calculations have been
performed without 
using equal-time quantum numbers, for example by restricting to a
sector of a definite $z$-component of the total angular momentum $J_z$.
However, the physical results of a calculation should be independent of the 
mathematical method applied. 
{\em A priori} it is not clear in front form dynamics, if the
results of a calculation will be rotationally invariant. 
It seems to be 
impossible to show {\em analytically} that these 
solutions do indeed obey rotational
symmetry. This would require the diagonalization of an operator at least
as complicated as the light-cone Hamiltonian itself.      
The problem, though fundamental for all Hamiltonian formulations 
of field theories, was not of primary interest before, since calculations
were often performed using lower (typically 1+1) dimensional models.
It has been shown \cite{BurkardtLangnau,KalPir} that Lorentz covariance, and in
particular rotational symmetry, is 
explicitly violated, if one evaluates front form perturbation theory
at the one- and two-loop levels. Non-covariant counterterms have to be
constructed to restore covariance. 

Because of its long term use, many
phenomena have been investigated within equal-time formulation.
The number of calculations is understandably much smaller in light-cone
quantization.
Much effort has been made therefore, to reproduce the results of 
{\em instant form dynamics} on the light-cone.
A considerable difficulty in light-cone quantization is the r\^{o}le of the 
so-called {\em zero modes}. 
The zero mode of a function $f(x^-,\xperp)$ in a fixed space direction
$y$ with interval length $L_y$ is defined as 
(cf.~{\em e.g.}~\cite[Eq.~(4)]{KalloPauliPinsky})
\nopagebreak
\[
\langle f(\bar{x})\rangle_0:=\frac{1}{2L_y}\int_{-L_y}^{L_y} dy f(\bar{x},y),
\]
where it is {\em not} integrated over the remaining space directions $\bar{x}$.
As a first step, the zero modes were omitted in actual calculations,
because they are a set of measure zero among the modes of all fields.
Their influence on the spectra of bound systems was considered to be 
negligible.

However, it was found that the convenient {\em light-cone gauge}
$A^+\equiv 0$ is inconsistent with the inclusion of the zero modes into the
formalism. In non-abelian gauge theories, for instance, the
zero mode of the gauge field $A^+$ cannot be gauged away 
\cite{KalloPauliPinsky}.
It was often seen that by working in the light-cone gauge in the 
front form dynamics, 
spurious operators which are singular in the limit $k^+{=}0$ 
are created \cite{Leibbrandt}.
The renormalization of these ultraviolet (UV) and infrared (IR) divergencies
was investigated in perturbation theory by 
{\sc Mustaki et al.}~\cite{MustakiRenormalization}.
Although the most divergent contributions are cancelled in the graphs of 
lowest order, the IR singularity remains.
For example, the photon propagator 
picks up an additional $1/k^+$ singularity. After regularization, it 
gives rise to difficulties only in higher orders, because then one must 
integrate over the photon momentum.
A naive regularization via principal value runs into difficulties
which can be overcome by the 
{\sc Leibbrandt-Mandelstam} prescription \cite{LeibMandelstam}.
This method for Hamiltonian non-abelian 
gauge theories was derived by {\sc Basetto et al.}~\cite{Basetto}.

To put the former results of calculations in 
light-cone quantization on a formally
correct basis, the influence of the zero modes was examined systematically
after pragmatically ignoring them. To name a few,
{\sc Pauli}, {\sc Pinsky} and 
{\sc Kalloniatis}~\cite{KalloPauliPinsky,KalloPauli}, 
{\sc Werner, Heinzl et al.}~\cite{HeinzlWerner}, 
and {\sc McCartor}~\cite{McCartor0}
have made important contributions to this subject.
The problem of zero modes in QED is discussed by {\sc Kalloniatis} and 
{\sc Robertson} in \cite{AlexDave}.
Recently, these more formal examinations have been accompanied by numerical
calculations.
{\sc V\"ollinger}~\cite{Voellinger} found in his investigation of the
massive Schwinger model a vanishing 
influence of the zero modes concerning the
spectrum of this model. However, he considered a vacuum wavefunction that 
is independent of the variable $\theta$, parameterizing the vacuum states. 
In literature \cite{Coleman}, the vacuum is found to be dependent on this
parameter (``$\theta$-vacuum'') which can have an effect on the spectrum. 
As an important result,  
{\sc van de Sande}~\cite{BvdSSmallMass} 
was able to show that DLCQ does give the correct linear 
increase of the squared mass
eigenvalues with the fermion mass in this model.
The work of 
{\sc Elser}~\cite{Elser} displayed a quadratic
rise of this squared mass with the fermion mass. This effect was 
falsely attributed to the zero modes.
The results of {\sc Elser} were numerically unsatisfactory and
the latter interpretation could be excluded in 
Ref.~\cite{Voellinger}, where the method of {\sc van de Sande} was used. 
It was proven
that in the continuum limit the right slope is reached and is
hidden by an extremely slow convergence typical of DLCQ.
The solution to this problem resides in the use of counterterms.


In conclusion, one has strong evidence for a small effect of the zero modes 
on the physics important for the spectra of bound systems. 
An application of the light-cone 
gauge, or equivalently, the consideration of the normal mode sector only,
seems to be justified, at least in the QED case.
Of course, in QCD there are large effects coming from the vacuum structure, 
such as chiral symmetry breaking. The origin of 
these effects has to be investigated separately.

\section{Preceding work}

The investigation of QED within the DLCQ formalism dates back to 
roughly 1988.
In 1990, {\sc Tang} \cite{Tang} set up a {\em matrix equation} for 
QED$_{(3+1)}$ by truncating the Fock space to the sectors
$|e\bar{e}\rangle$ and $|\eeg\rangle$.
In order to solve the associated
eigenvalue problem, he used the diagonalization 
of the discretized Hamiltonian, and variational methods.
With help of the latter, he was only able to produce an {\em upper limit} 
for the mass-squared eigenvalue of the 
triplet ground state of positronium,
three percent larger than the {\sc Bohr} value at
$\alpha=0.6$. But compared to the standard results of 
perturbation theory \cite{Ferrell}, this
upper limit is already below the well-known value
for this triplet term.
To produce significant results by diagonalizing the 
Hamiltonian matrix, 
{\sc Tang} \cite[Ref.~1, p.~46]{Tang}, following his own estimates, 
would have had to include roughly 11 million Fock states 
for the large coupling constant $\alpha=0.3$.
The main cause for this negative conclusion was the slow 
numerical convergence of the method applied. 

With the solution of the Coulomb problem in momentum space~\cite{Woelz},
the method of {\em Coulomb counterterms} was introduced to 
improve numerical convergence.
{\sc Kraut\-g\"art\-ner}~\cite{KPW} applied  this method
to QED$_{(3+1)}$.
He used an {\em effective} interaction, obtained from a 
projection of the $|e\bar{e}\gamma\rangle$-sector onto the 
$|e\bar{e}\rangle$-sector. 
The corresponding effective 
integral equation was solved using Gauss-Legendre quadrature.
His results show excellent convergence and coincide to a high degree
of accuracy with the expected values.
It is worth mentioning that, for the first time, not only the complete
$Bohr$ spectrum is obtained, but also 
relativistic effects, like the hyperfine splitting, are correctly described
within a non-perturbative approach in front form dynamics. 

{\sc Kalu\v{z}a}~\cite{Kaluza} applied the counterterm technique     
to the calculation of the spectrum of the 
light-cone Schr\"odinger equation.
He calculated an eigenvalue for the positronium ground state that 
coincides approximately with the 
{\sc Bohr} value at $\alpha=0.3$.
For the calculation of
relativistic spectra, 
{\sc Kalu\v{z}a} improved the diagonalization
technique to enlarge the  
Fock space feasible with the computer equipment used. 
The convergence of his spectra is rather poor, because he did not use 
counterterms for the singularity of the relativistic problem.
For a comparison with the light-cone Schr\"odinger equation, only 
the ground state is considered, which is certainly too large. 

{\sc W\"olz}~\cite{DissWoelz} subsequently tried to solve the analogous 
problem for QCD. To include typical QCD effects 
(self-coupling of the gluons) he enlarged the Fock basis
implementing the $|q\bar{q}gg\rangle$-sector.
In order to solve for the eigenvalues, he applied a hybrid method by projecting
the new Fock sector unto the other two sectors, considering the so-derived
eigenvalue problem as a matrix equation. 
His results \cite{DissWoelz} 
converge too slowly and suffer strong fluctuations as functions of
the changing Fock space size.
This is due to the fact that the counterterm technique included in the
computer code is not applied in the final calculations
because the necessary two dimensional numerical integration are too time
consuming.

\chapter{The positronium model}
\label{ChapterModel}

When quantizing a field theory on the light-cone,
it is convenient to work in the usual Fock basis. 
Consequently, one thinks of a composite physical system,
{\em e.g.}~a meson or a baryon, as consisting of fermionic valence particles,
bosonic (virtual) gauge particles, and virtual particle-antiparticle pairs.
Considering positronium as the 
physical system of special interest for the present work, 
the wavefunction reads: 
\[
|positronium\rangle = \psi_{\gamma}|\psi_{\gamma}\rangle
 	     + \psi_{e\bar{e}}|\psi_{e\bar{e}}\rangle
	     + \psi_{e\bar{e}\gamma}|\psi_{e\bar{e}\gamma}\rangle
	     + \psi_{e\bar{e}\gamma\gamma}|\psi_{e\bar{e}\gamma\gamma}\rangle
	     + \ldots\, .
\]
The exact wavefunction
is an infinite series and one has to impose simplifiying 
restrictions in actual calculations. The obtained results 
have to be investigated concerning their dependencies on these  
restrictions. 

The Hamiltonian operators ruling the dynamics of a system can be 
derived using light-cone quantization. 
A rough sketch of the procedure 
is given in the introduction and its application to 
QED$_{(3+1)}$ is described in more detail in Appendix \ref{AppxQED}. 
It is important that in this work the {\em continuum} 
formulation of a light-cone
quantized field theory is used. All discretizations come from
numerical integrations, cf.~Appendix \ref{AppxNumerics}.   
The operator 
\[
H_{\rm LC}:=P^{\mu}P_{\mu}
\]
is commonly called the invariant mass (squared) operator. 
For convenience I will refer to it as the
{\em light-cone Hamiltonian}, although in the sense of {\sc Dirac} \cite{Dirac}
only $P^-$ is a Hamiltonian. The light-cone Hamiltonian is obviously 
a Lorentz scalar, and so are
its eigenvalues, which have the dimension of a mass squared.

Our main task is to set up an {\em effective, relativistic} 
Hamiltonian operator, tractable either analytically or 
with the help of a computer, and to solve the eigenvalue equation
\begin{equation}\label{EWP}
H_{\rm LC} |\psi_n\rangle = M_n^2 |\psi_n\rangle,
\end{equation}
whose solutions yield the mass (squared) eigenvalues and the corresponding
eigenfunctions of our positronium model.
The relevant matrix elements of this Hamiltonian are 
tabulated in Appendix \ref{AppxHLC}. 
The full Hamiltonian matrix elements for QED and QCD  
are given in \cite{Tang} and \cite{Kaluza,Schladming}, respectively.

\section{The model}

To construct a manageable physical model for positronium, we
proceed as follows.
One is free to divide the Fock space into two arbitrary subspaces, called the
$P$-{\em space} and the $Q$-{\em space}.  
Restricting the full Fock space to two sectors, being explicit,
one with an electron positron pair and one with an additional 
photon, the associated projection operators 
onto these subspaces are defined to be
\beq\label{Pspace}
\hat{P} := \sum_{n\atop {\rm all\, QN}} 
			|(e\bar{e})_n\rangle \langle (e\bar{e})_n| 
\eeq
and
\beq\label{Qspace}
\hat{Q} := \sum_{n\atop {\rm all\, QN}} 
			|(e\bar{e}\gamma)_n\rangle \langle (e\bar{e}\gamma)_n|. 
\eeq
Within this limited Fock space, the model positronium cannot
decay into photons, contrary to observation. Moreover,
a virtual photon as an intermediate state in an interaction is impossible.
The part of the hyperfine splitting connected to this annihilation 
graph is missing. The inclusion of this graph into the theory 
is the topic of Chapter \ref{DrAChapterAnnihilation}.  
Nonetheless, the vector space is well-defined mathematically,
and we can proceed to solve the eigenvalue equation (\ref{EWP}).
In our restricted Fock-space we have a $2\times2$ block matrix
\beq\label{2x2Matrix}
H_{\rm LC} = \left(
		\begin{array}{cc}
			H_{ PP} & H_{ PQ}\\
			H_{ QP} & H_{ QQ}
		\end{array}
	     \right). 
\eeq
Of course, this truncation of the Fock space violates gauge invariance.
It will be shown in Chapter \ref{DrAChapterEffInt} that one can 
treat the consequences of this violation within a general theory of effective
interactions. 

The attempts to solve for the spectrum of this operator in a matrix 
equation\cite{Tang,Kaluza}
were bound to fail, as described
in Chapter \ref{DrAChapterFieldTheories}, 
because of computer capacity limitations.
We apply a projection method \cite{KPW} (also used in many body 
theory\protect\footnote{It is described as mathematical method in 
\protect\cite{MorseFesh}
and
applied to problems in nuclear theory by {\sc Tamm} \protect\cite{Tamm} and
{\sc Dancoff} \protect\cite{Dancoff}.}) to eliminate  
the $Q$-space at the expense of an in general more complicated
{\em effective} Hamiltonian operating in $P$-space only.  
Formally, one can describe the projection as the solution of a system 
of coupled linear equations. 
The last of the two equations of the eigenvalue problem 
Eq.~(\ref{EWP})
with the Hamiltonian (\ref{2x2Matrix}) reads explicitly
\[
H_{QP}|e\bar{e}\rangle+H_{QQ}|e\bar{e}\gamma\rangle= 
M^2_n|e\bar{e}\gamma\rangle.
\]
We therefore solve for the state
$\hat{Q}|\psi\rangle = |e\bar{e}\gamma\rangle$ and express it with the help of
an inverse Hamiltonian, or resolvent:
\beq\label{resolvent}
G(\omega):=\hat{Q}(\omega - H_{\rm LC})^{-1} \hat{Q},
\eeq
and obtain
\[
\hat{Q}|\psi_n\rangle = G(\omega)H_{\rm LC} 
			\hat{P}|\psi_n\rangle.
\]
The {\em a priori} unknown eigenvalue $M^2_n$ was substituted by $\omega$,
referred to in the remainder of this work as {\em redundant parameter}.
This parameter has to be fixed by an additional condition.

Which interactions can occur in the effective $|e\bar{e}\rangle$-sector?
The only possible graph in $P$-space is shown in Figure (\ref{PSpaceGraphs}).
Because of the projection, one also has iterated graphs, like the one
displayed in Figure (\ref{IteratedGraphs}).
The Q-space seems quite complicated at first glance, but only one graph,
the lower right one
of Figure (\ref{QSpaceGraphs}), survives the so-called
{\em gauge principle} of DLCQ formulated by 
{\sc Tang et al.}~\cite[Ref. 2]{Tang}.
The main idea is to restore gauge invariance
at tree level, which was destroyed by the Fock space truncation. 
The recipe is to
omit all graphs which have intermediate states that are not contained
in the (restricted) Fock space. Instantaneous particles are counted as real 
particles in this procedure.

\begin{figure}[t] 
\begin{minipage}[t]{70mm} \makebox[0mm]{}
\centerline{\psfig{figure=\graphpath seagull.epsi,width=5cm,angle=-90}}
\caption[The seagull graph in $P$-space]
{\label{PSpaceGraphs}The seagull graph in $P$-space.}
\end{minipage}
\hfill 
\begin{minipage}[t]{70mm} \makebox[0mm]{}
\centerline{\psfig{figure=\graphpath dynamic.epsi,width=5cm,angle=-90}}
\caption[The iterated vertex interaction in $P$-space]
{\label{IteratedGraphs}The iterated vertex interaction ($x>x'$) in $P$-space.}
\end{minipage}
\end{figure}

Still not all non-diagonal graphs in Q-space vanish.
To simplify the problem, in \cite{KPW} the remaining seagull graph
was  omitted on an {\em ad hoc} basis. We shall show 
in Chapter \ref{DrAChapterEffInt}, using the 
formalism set up in \cite{PauliMIR}, 
that this
omission is not an assumption, but is a natural consequence of the
projection mechanism. 
We arrive at the {\em nonlinear} equation
\[
H^{\rm eff}_{\rm LC}(\omega) |\psi_n(\omega)\rangle = 
M_n^2(\omega) |\psi_n(\omega)\rangle, 
\]
where the states, the effective Hamiltonian
\begin{equation}\label{Heff}
H^{\rm eff}_{\rm LC}(\omega) := \hat{P}H_{\rm LC}\hat{P} +  
 \hat{P}H_{\rm LC}\hat{Q}(\omega - H_{\rm LC})^{-1} \hat{Q}H_{\rm LC}\hat{P}
\end{equation}
and the mass eigenvalue $M^2_n$ depend on the redundant parameter $\omega$. 
In principle, we also have to satisfy the obvious constraint that
\[
M_n^2(\omega)=\omega.
\]

Although the fixing of the parameter $\omega$ is explained in Chapter 
\ref{DrAChapterEffInt}, 
we introduce a method of calculating an analytic expression
for this parameter. 
This method 
is the way the so-called $\omega^*$-{\em trick} was
introduced before the work of {\sc Pauli} on the Method of 
Iterated Resolvents \cite{PauliMIR}.
The derivation given here should be considered as a {\em plausibility argument} 
rather than as a strict proof.
Nevertheless, it gives some intuition concerning the physics behind the
procedure.

A word on notation seems in 
order\protect\footnote{Cf. Appx.~\protect\ref{AppxNotations}.}.
We label the {\em fermion mass} with $m_f$, the 
{\em longitudinal momentum fraction} with $x$, the {\em transverse momentum} 
with $\vec{k}_{\perp}$, and the {\em helicity} of a particle with $\lambda$. 
The corresponding quantum numbers after an interaction have a
prime $(x',\vkp';\lambda')$.

We have seen that the energy denominator contains the unknown (parameterized)
eigenvalue $\omega$ of the whole eigenvalue problem and the 
sector-Hamiltonian operating in $Q$-space.
The latter consists of a kinetic part $M^2_Q$, {\em i.e.~}the 
free mass (squared)
of the $Q$-space, and an interaction $V_Q$:
\[
H_{LC,Q}=M^2_Q+V_Q.
\]
As pointed out above, this poses a difficult problem: a non-diagonal 
operator has to be inverted {\em and} as a constraint it has to be 
guaranteed that the masses should be equal to the mass parameter. The latter
results in a mathematical fixpoint equation.
To avoid at least the first of these difficulties, one can divide the
interaction into a diagonal part
$\langle V_Q\rangle$ and a non-diagonal part $\delta V_Q$
\[
V_Q=\langle V_Q\rangle+\delta V_Q.
\]
By defining {\em formally}
\beq\label{Tstar} 
T^* := \omega-\langle V_Q\rangle= c \mbox{\sl\bf I}, \quad c\in \mbox{\sl\bf R},
\eeq
I changed the notation of \cite[Eq.~(5.9)]{KPW} to stress two points. Firstly,
$T^*$ is not just a fixed value of $\omega$. This would be inconsistent,
because $\omega$ is a real number, whereas $T^*$ is a function of the
light-cone momenta. Secondly, $T^*$ is a {\em kinetic energy} according to 
Eq.~(\ref{Tstar}), where a potential energy is subtracted from a total energy.
One can expand the resolvent around the diagonal interaction 
$\langle V_Q\rangle$ 
\begin{eqnarray}\label{expansion}
\frac{1}{\omega-H_Q}&=&
\frac{1}{T^*-M^2_Q-\delta V_Q}\\
&=&\frac{1}{T^*-M^2_Q}+ 
\frac{1}{T^*-M^2_Q}
\delta V_Q\frac{1}{T^*-M^2_Q-\delta V_Q},\nonumber
\end{eqnarray}
As a first approximation we consider the first term of the expansion only.
It does not contain any non-diagonal terms and is a simple c-number.
We label
\beq\label{energienenner}
{\cal D}(x,x';T^*):=|x-x'|(T^*-M^2_Q).
\eeq
\begin{figure}
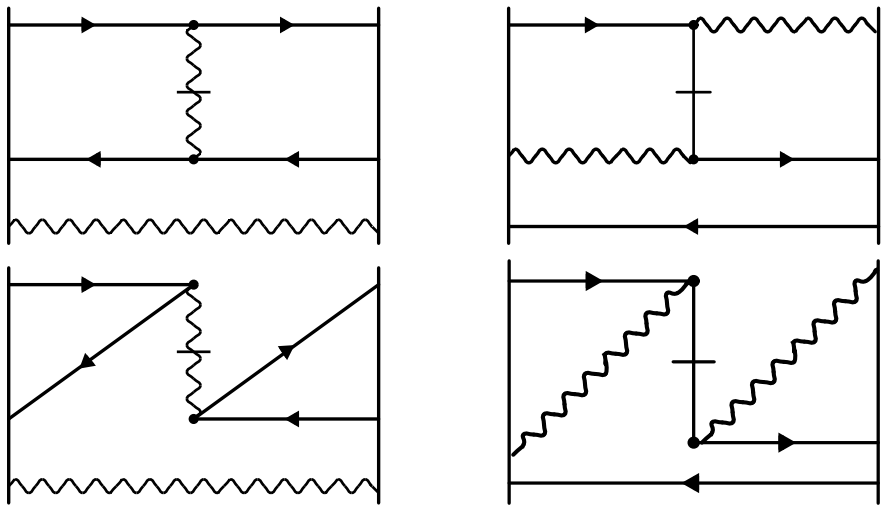

\centerline{
\begin{tabular}{ccc}
\psfig{figure=\graphpath q_seag_a.epsi,width=5cm,angle=-90}
&\hspace{0.8cm} &
\psfig{figure=\graphpath q_seag_c.epsi,width=5cm,angle=-90} \\
\psfig{figure=\graphpath q_seag_b.epsi,width=5cm,angle=-90}
&\hspace{0.8cm} &
\psfig{figure=\graphpath q_seag_d.epsi,width=5cm,angle=-90} \\
\end{tabular}
}
\caption[Graphs of the instantaneous interaction in $Q$-space]
{\label{QSpaceGraphs}Graphs 
of the instantaneous interaction in $Q$-space.}
\end{figure}
This approximation has severe consequences: a collinear singularity occurs.
It is proportional to
\[
\frac{1}{\cal D}\frac{\Delta(x,\kp,x',\kp';T^*)}{|x-x'|}.
\]
We identify now the $|e\bar{e}\rangle$-sector with the $P$-space
and the $|e\bar{e}\gamma\rangle$-sector with the $Q$-space
to calculate the function $\Delta$ by evaluating the two graphs
Fig.\ (\ref{PSpaceGraphs}) and Fig.\ (\ref{IteratedGraphs}) 
according to the rules of front form perturbation theory 
given in Refs.~\cite{BLepage,Schladming},
but using the expression in Eq.~(\ref{energienenner}) 
for the summation over the intermediate states. 
One obtains 
\[
\Delta(x,\kp,x',\kp';T^*)=M^2_{\eeg}-\omega-\frac{(\kperp-\kperp')^2}{x-x'}
-\frac{1}{2}\left(l^-_e-l^-_{\bar{e}}\right),
\]
where
$l_e^{\mu}:=(k_e'-k_e)^{\mu}$ and
$l_{\bar{e}}^{\mu}:=(k_{\bar{e}}-k_{\bar{e}}')^{\mu}$ are the momentum 
transfers of the electron and positron.
One can use this singularity
induced by the truncation of the series (\ref{expansion})
to determine the parameter
$\omega$ in the problem.
One simply demands that this {\em collinear} singularity
vanishes: 
\[
\Delta(x,\kperp,x',\kperp';T^*)=0, \quad\quad \forall x,x',\kperp,\kperp'. 
\]
We are even forced 
to proceed this way to ensure the solubility of the problem. 
One has to fix $\omega$ to the expression
\begin{equation}\label{OmegaStar}
T^*(x,\vec{k}_{\perp};x',\vec{k}'_{\perp})
        =\frac{1}{2}\left( \frac{m_f^2 + 
        \vec{k}_{\perp}}{x(1-x)} + \frac{m_f^2 + \vec{k}'_{\perp}}{x'(1-x')}
        \right),
\end{equation}
which indeed has the form of a kinetic energy.
Although it can be shown that the two kinetic energies of this sum must be
the same, I wrote $T^*$ here in the equivalent suggestive 
form of an {\em average}. 
It cannot be overstated that exactly this form of the expression 
(\ref{OmegaStar}) follows from the structure of the effective theory
\cite{PauliMIR}.

The physical interpretation of this procedure is not easy.
One can imagine that the $Q$-space contains all interactions 
of the higher Fock sectors by effectively summing them.
In this complicated effective interaction, we disregard all non-diagonal 
contributions. But on the other hand, 
this is partly compensated by the special choice 
$\omega\equiv T^*$.
One can therefore state that $T^*$ contains an approximation of the summed
interactions of the higher Fock states.

\section{The effective integral equation}

We can proceed by calculating the matrix elements of the now 
well-defined effective Hamiltonian. It operates only in $P$-space.
To read off the actual definition of the matrix elements, we have to
write down the integral equation in which they are contained.
The continuum version of Eq.\ (\ref{EWP}) is
\begin{eqnarray*}
&&\left(\frac{m_f^2 + \vec{k}^2_{\perp}}{x(1-x)}- M_n^2\right)
\psi_n(x,\vec{k}_{\perp};\lambda_1,\lambda_2)\\
&&+\frac{g^2}{16\pi^3}
\sum_{\lambda'_1,\lambda'_2}\int_D \frac{dx'd^2\vec{k}'_{\perp}}
{\frac{1}{2}\left(l_e^2+l_{\bar{e}}^2\right)}
\frac{\langle x,\vec{k}_{\perp};
\lambda_1,\lambda_2|j(l_e)^{\mu}j(l_{\bar{e}})|x',\vec{k}'_{\perp};
\lambda'_1,\lambda'_2\rangle}
{\sqrt{xx'(1-x)(1-x')}}
 \psi_n(x',\vec{k}'_{\perp};\lambda'_1,\lambda'_2)= 0.\\
\end{eqnarray*}
\beq\label{TDEquation}
\eeq
This form of the {\em effective integral equation} is very useful for a 
comparison with the matrix elements calculated in \cite{KPW}, see also 
Appx.~\ref{AppxCalculation}. However, a main topic of the present work is
the investigation of rotational invariance, or in general 
Poincar\'{e} invariance, of QED on the light-cone. In this context, it is
helpful to write the interaction term of Eq.~(\ref{TDEquation})
a covariant way. The integrand reads now
\beq\label{TDIntEquation}
\frac{g^2}{16\pi^3}
\sum_{\lambda'_1,\lambda'_2}\int_D \frac{dx'd^2\vec{k}'_{\perp}}
{\sqrt{xx'(1-x)(1-x')}}
\frac{j^{\mu}(l_e,\lambda_e)j_{\mu}(l_{\bar{e}},\lambda_{\bar{e}})}
{l_e^{\mu}l_{e,\mu}}
\psi_n(x',\vec{k}'_{\perp};\lambda'_1,\lambda'_2),
\eeq
which makes it obvious that the effective interaction\protect\footnote{Instead 
of providing the arguments of the currents $j^{\mu}$ by bras and kets 
like in Eq.~(\protect\ref{TDEquation}), we wrote them into brackets.}
\[
U_{\rm eff}:=\frac{j^{\mu}(l_e,\lambda_e)j_{\mu}(l_{\bar{e}},\lambda_{\bar{e}})}
{l_e^{\mu}l_{e,\mu}}
\]
is gauge invariant and a Lorentz scalar.
We restrict the integration domain $D$ using the covariant
cutoff of {\sc Brodsky} and {\sc Lepage} \cite{BLepage}:
\beq\label{BLepagecutoff}
\frac{m_f^2+\vkp^2}{x(1-x)}\leq \Lambda^2+4m^2,
\eeq
which allows for states having a kinetic energy below the 
cutoff $\Lambda$.
The matrix elements are given in some detail in
Appendix \ref{AppxCalculation}. The explicit functions are listed in
Appendix \ref{AppxHelicityTables}. 

\begin{figure}
\centerline{
\psfig{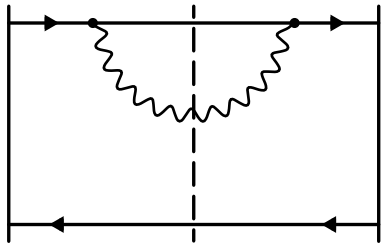}}
\caption[Diverging diagrams in $P$-space: self energy diagram]
{\label{SelfMass}Diverging diagrams in $P$-space: self energy diagram.} 
\end{figure}

\begin{figure}
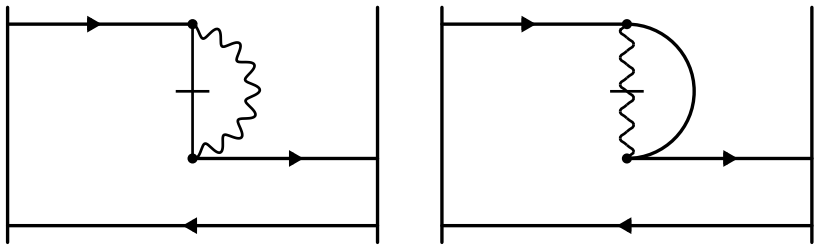

\centerline{
\psfig{figure=\graphpath p_contr_q.epsi,width=5cm,angle=-90}
\hspace{0.5cm}
\psfig{figure=\graphpath p_contr_g.epsi,width=5cm,angle=-90}}
\caption[Diverging diagrams in $P$-space: contraction contributions]
{\label{pcontractions}Diverging 
diagrams in $P$-space: contraction contributions.}
\end{figure}

So far, we have considered only the problems that occur due to the projection
of $Q$-space onto $P$-space. But already in $P$-space we have to face
severe singularities stemming from the graphs of the 
electromagnetic self energy, Fig.\ (\ref{SelfMass}), and the contraction 
graphs, Fig.\ (\ref{pcontractions}).


A renormalization scheme is necessary to remove the divergencies. It is 
described in Appendix \ref{AppxRenormalization} and can be summarized as
follows.
Although each of these graphs is quadratically divergent in the cutoff $\Lambda$,
their sum is only logarithmically divergent. 
The arguments used in the proof 
assume the formulation of the theory in the {\em continuum}. In the
discretized theory the arguments do not necessarily hold.

Before addressing to solve the effective integral equation 
(\ref{TDEquation}),
let us briefly review the steps that lead to its derivation and compare
the method of this work to other attempts to solve for the positronium spectrum.
First, we restricted the Fock space to the sectors $|e\bar{e}\rangle$ and
$|e\bar{e}\gamma\rangle$. By this, we arrived at a positronium model with
one dynamical photon. This constitutes a matrix equation ($2\times 2$ block
Hamiltonian). The advantage of this matrix equation is that all interactions
in the $Q$-space are explicitly taken care of, whereas in our model their
contributions
are occluded by the determination of the redundant parameter $\omega$. 
However, efforts to extract the spectrum and wavefunction from the matrix
equation have failed. In the case of QED,  
{\sc Tang} \cite{Tang} and {\sc Kalu\v za} \cite{Kaluza} produced  
results with no clear
significance. This was due to the convergence problems,
because they did not 
include counterterms for the Coulomb singularity.
The counterterm technique used in the present work is described in 
Appendix~\ref{AppxNumerics}.
Calculating the counterterms, one faces in general 
numerical integrations in two or three dimensions,
which must be performed for each diagonal matrix element. The
numerical effort increases tremendously with the matrix 
dimensions \cite{DissWoelz}.  
We wish to avoid the disadvantages of the matrix equation. 
Therefore we have performed the continuum limit of the DLCQ formalism 
and obtained a coupled 
system of integral equations. The effective integral equation (\ref{TDEquation})
was derived by casting the effects of higher Fock states into 
the $Q$-space by fixing $\omega$ to $T^*$. Then the $Q$-space is projected
onto the $P$-space as a second step. 

If one considers the non-relativistic limit of the
interaction in the effective integral equation (\ref{TDIntEquation}), one
arrives at the so-called {\em light-cone Schr\"odinger equation}.
The main point in this calculation is the relation between
the longitudinal momentum fraction $x$ in light-cone coordinates and the 
equal-time momentum in $z$-direction, $k_z$
\[
x=x(k_z)=\frac{1}{2}\left(1+\frac{k_z}{\sqrt{m^2+\vkp^2+k_z^2}}\right)
\simeq\frac{1}{2}\left(1+\frac{k_z}{m}\right).
\]
The calculation of the non-relativistic limit is then straightforward.
The solution of the light-cone Schr\"odinger equation, 
which has analytically integrable Coulomb counterterms,
was given in \cite[Fig.\ 8]{KPW}, 
shows very good convergence, and yields the correct eigenvalue 
spectrum.
If one fully reduces the effective integral 
equation to the non-relativistic limit,
one arrives at the well-known Coulomb equation in momentum space, 
solved numerically and discussed in Ref.~\cite{Woelz}.  

\section{The positronium mass spectrum} 

The solution to Eq.~(\ref{TDEquation}) was given in \cite{KPW}.
In contrast to the light-cone Schr\"odinger or the Coulomb equation, the 
counterterms for the Coulomb singularity cannot be calculated analytically.
As can be seen in the helicity tables in Appendix 
\ref{AppxHelicityTables}, in the case $J_z{=}0$ essentially two different 
diagonal matrix elements occur: one for parallel, the other for anti-parallel 
helicities.
{\sc Krautg\"artner} \cite{DissKraut} 
was able to integrate out analytically one of two variables contained in 
the continuous part of the counterterm for anti-parallel helicities. 
As a result, he only had to integrate over 
one dimension numerically. However, he did not
succeed in analytically integrating out this variable from both matrix elements 
and consequently had to use the same counterterm for 
both diagonal matrix elements.
The convergence and the spectra he obtained with this method were reasonably
good.
Indeed, the use of identical counterterms in this case ($J_z{=}0$) is well 
justified
since both functions have the same singularity structure and comparable
values. 
We note that this becomes problematic in the case of non-vanishing $J_z$, 
where one has
four distinct diagonal matrix elements, one of which is much smaller than
the others. 
 
\begin{figure}[t]
\begin{minipage}{15.5cm}
\centerline{
\psfig{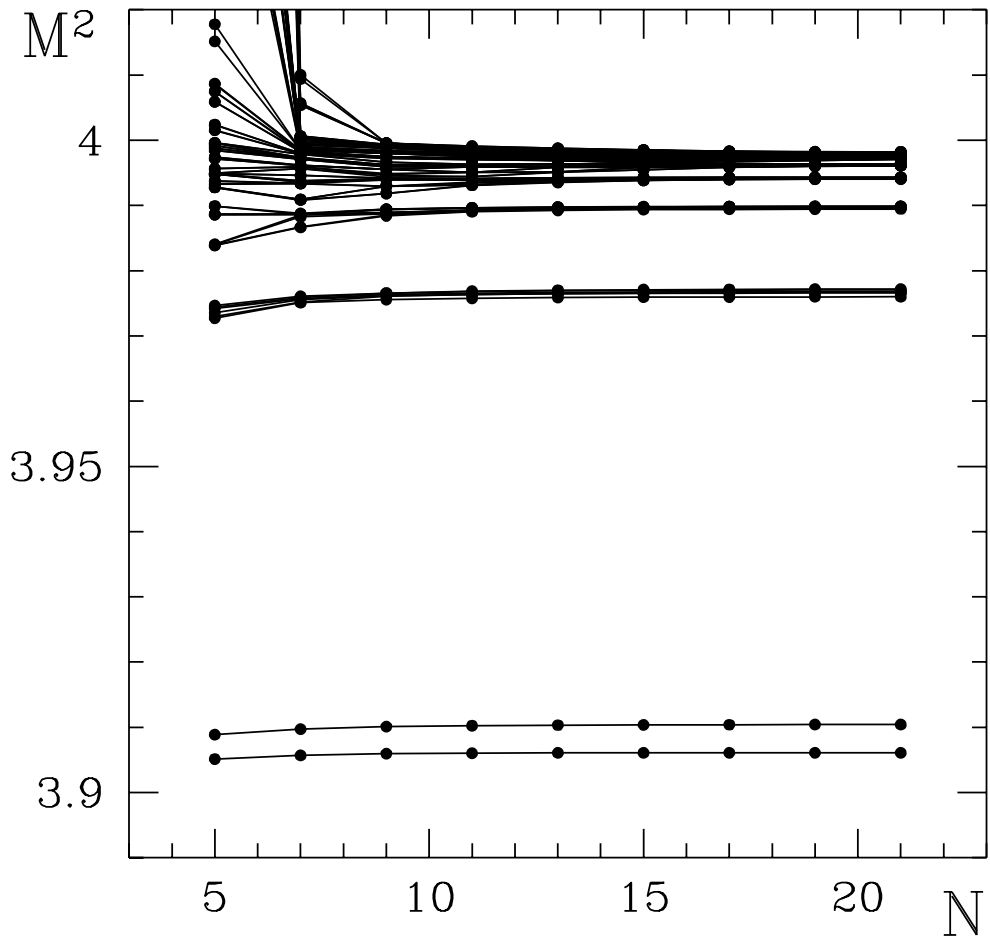}}
\protect\caption[Spectrum in the $J_z{=}0$ sector with numerical counterterms]
{\label{spectrumJ0newCT}The 
spectrum of the effective integral equation for $\alpha=0.3, 
\Lambda=1.0\, m_f, J_z=0$. 
The mass squared eigenvalues $M^2_n$ in units of the electron mass $m^2_f$
are shown as functions of the number of integration points $N\equiv N_1=N_2$.
The calculation was done using the entirely numerically integrated
Coulomb counterterms.
Note the improved convergence of the states with $n{\geq}2$ in comparison with
Fig.~(\protect\ref{spectrumJ0oldCT}). The 100
lowest eigenvalues are displayed.}
\end{minipage}
\end{figure}

\begin{figure}[t]
\begin{minipage}{15.5cm}
\centerline{
\psfig{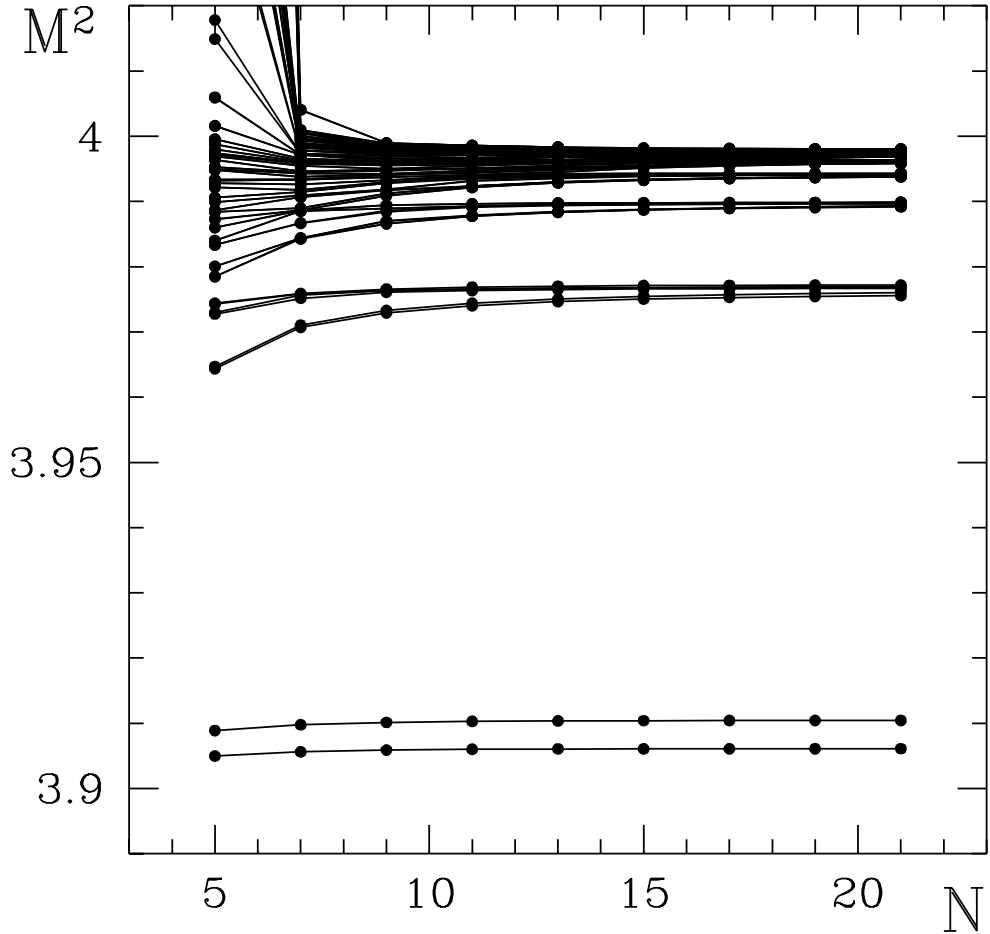}}
\protect\caption[Spectrum in the $J_z{=}0$ 
sector with half-analytical counterterms]
{\protect\label{spectrumJ0oldCT}The 
spectrum of the effective integral equation for $\alpha=0.3, 
\Lambda=1.0\, m_f, J_z=0$. 
The mass squared eigenvalues $M^2_n$ in units of the electron mass $m^2_f$
are shown as functions of the number of integration points $N\equiv N_1=N_2$.
The calculation was done using the half-analytical Coulomb counterterms
of Ref.~\cite{KPW}.
The convergence of states with $n\geq 2$ is not as good as in 
Fig.~(\protect\ref{spectrumJ0newCT}). 
The 100 lowest eigenvalues are displayed.}
\end{minipage}
\end{figure}

We calculate in this chapter the spectrum of the positronium model described
in \cite{KPW} with an improved counterterm technique. This 
means a
rigorous calculation of {\em all four} counterterms corresponding to the
individual diagonal matrix elements. The prize is an entirely numerical
two dimensional integration 
rather than a one dimensional analytic integration over the 
variable\protect\footnote{For the definition of the variables used in the 
calculations on the computer, see Appx.~\protect\ref{AppxNumerics}.}
$\cos\theta$, followed by a numerical integration of the off-shell mass $\mu$.
The effect is an even better 
convergence of the spectra with the number of integration points, as one can
see by comparing 
Fig.~(\ref{spectrumJ0newCT}) with Fig.~(\ref{spectrumJ0oldCT}).
In particular, one notes a better convergence for a principal quantum number 
$n{\ge}2$ and a number of integration points $N{\le}13$.
The lowest states for $n{=}2$ converge much better when
the entirely numerically integrated counterterms are used. This is the 
result of the afore mentioned
small, but distinct difference between the diagonal matrix elements.
If only one counterterm, adjusted to the singularity structure of one
special diagonal matrix element, is used, the Coulomb singularity of the 
other diagonal element 
is over-compensated. As a consequence, the calculated eigenvalue 
is smaller than it
should be. This is exactly the effect observed in the spectra.

We recall the analytic 
results\protect\footnote{The ``state of the art'' theoretical results are given by 
{\sc Gupta et al.}\protect\cite{Gupta}. For the triplets they have
\beqa
E(n{}^3 S_1)&=&-\frac{1}{2}\Rydberg\frac{1}{n^2}
+2\Rydberg\alpha^2\frac{1}{n^3}\left(\frac{1}{12}+\frac{11}{64 n^4}\right)\\
&&+\frac{1}{4\pi}\Rydberg\alpha^3\frac{1}{n^3}\left[-\frac{7}{3}a_n-\frac{109}{15}
+\frac{2}{3}\ln 2+ 6\ln\alpha^{-1}-\frac{16}{3}\ln k_0(n)\right]\\
&&+\frac{1}{6}\Rydberg\alpha^4\frac{1}{n^3}\ln\alpha+\cdots,
\eeqa
with the {\sc Salpeter} terms 
\[
a_{1S}=-2\ln2-3,\quad a_{2S}=-\frac{9}{2},\quad \ldots,
\]
and the {\sc Bethe} logarithms
\[
\ln k_0(1)=2.9841285558\ldots,
\quad
\ln k_0(2)=2.8117698931\ldots,
\quad
\ln k_0(3)=\ldots
\, .
\]
A review of experimental results can be found in 
\protect\cite[Chap.~15]{Kinoshita}.
} 
for the singlet and triplet states
\cite{Ferrell}\cite[p.~10]{DissJungmann} to order ${\cal O}(\alpha^4)$
and write it in the form of {\sc Bethe} and 
{\sc Salpeter} \cite[\S 23]{BetheSalpeterBook}:
\[
E_{n,l}=-\frac{1}{2}\Rydberg
\left[\frac{1}{n^2}-\frac{11}{32}\frac{\alpha^2}{n^4}
+\left(\epsilon_{l,S,J}-\frac{1}{2l+1}\right)\frac{\alpha^2}{n^3}\right]. 
\]
with the principal quantum number $n$ and the Rydberg constant 
$\Rydberg=m_f \alpha^2/2$.
The singlet terms have
\[
\epsilon_{l,S=0,J}=0,
\]
and the triplets 
\[
\epsilon_{l,S=1,J}=\frac{7}{6}\delta_{l0}+\frac{1-\delta_{l0}}{2(2l+1)}\left\{
\begin{array}{lcl}
\frac{3l+4}{(l+1)(2l+3)}& \rm if & J=l+1\\
-\frac{1}{l(l+1)}& \rm if & J=l\\
-\frac{3l-1}{l(2l-1)}& \rm if & J=l-1.\\
\end{array}
\right.
\]

For a comparison of our results to ``experiment'', 
{\em i.e.}~to perturbation theory with a strong coupling constant, 
$\alpha=0.3$, we 
have compiled the positronium mass spectrum in Table (\ref{TableSpectrumJ0}).
Usually one classifies the states according to their quantum numbers
of total angular momentum $J$, orbit angular momentum $L$, and total spin $S$.
This is valid only for rotationally invariant systems, or, in our case,
in the non-relativistic limit. These quantum numbers constitute the 
spectroscopic notation $^{2S+1} L_J$.
We choose a convention for the time reversal operation
\beq\label{Tconvention}
V_{\cal H}|J,J_z\rangle:=(-1)^{(J-J_z)}|J,J_z\rangle
\eeq
to classify the states likewise.
The singlet states are known as {\em parapositronium}, the triplet 
states as {\em orthopositronium}.
We display the non-relativistic notation for the states in 
Table (\ref{TableSpectrumJ0}) to make the comparison to other data easier. 
The eigenvalues are listed in the form of binding coefficients 
$B_n$, defined as
\beq\label{BindingCoeffs}
B_n:=4\, \frac{2-M_n}{\alpha^2}
\eeq
for all eigenvalues.

The finite $N$ error estimates given in the table were obtained 
by comparing the results for 
the maximum number of integration points with those for the next highest number
of points. 
The actual errors are definitely smaller, because the eigenvalues converge 
exponentially with the number of integration points $N$. We will comment
on this in detail, when we have completed our model by introducing the 
annihilation channel in Chapter \ref{DrAChapterAnnihilation}.
A word seems in order on the magnitude of the errors. Surprisingly, the 
largest errors are those of the states with the largest binding coefficients,
in particular of the ground state. The explanation is that we work in 
momentum space. Consequently, the higher excited states, widely spread 
in coordinate space, are condensed in momentum space and therefore 
in the region of many integration points.  

Note the good agreement in Table (\ref{TableSpectrumJ0}), 
including excited states. The singlets $^1S_0$ tend 
to be calculated as too weakly bound, especially for $n{=}1$ and $n{=}2$. 
This effect is reversed for the triplets $^3S_1$.
In principle, the whole 
bound state spectrum is accessible with our method, not only the first few
states.
To support this statement, it is instructive to investigate the properties 
of the higher states. 

The main features of a detailed study of the multiplet structure of 
the spectrum are the following and can be seen from Table 
(\ref{TableSpectrumJ0}):

\begin{itemize}
\item There is a well-defined number of states for each fixed principal quantum 
number $n$. In the case considered here ($J_z{=}0$), there are $4(n-1)+2$
states. This number should be reproduced for as large an $n$ as possible.
It turns out that the multiplets contain the correct number of states 
in our model up to at least
$n=5$.

\item Each state has defined quantum numbers concerning the charge conjugation  
$\pi_{\cal C}$ and ${\cal T}$-parity $\pi_{\cal H}$, 
cf.~Eqs.~(\ref{CSymmetry})~and~(\ref{TSymmetry}).
These quantum numbers can 
be obtained from the non-relativistic notation $^{2S+1}L_J$ by 
\[
\pi_{\cal C}=(-1)^{L+1} \quad \hbox{and} \quad \pi_{\cal H}=(-1)^{J+L+1},
\]
using the convention (\ref{Tconvention}).
It is an important result that only those combinations of  
quantum numbers $\pi_{\cal C}$ and $\pi_{\cal H}$ which
are expected from the theory, occur in each set of eigenvalues for any given $n$
in our calculations.

\item The ordering of the multiplets seems to have minor errors. 
For instance, the
$2^1S_0$ state and the $2^3P_0$ state are permuted, 
cf.~Table~(\ref{TableSpectrumJ0}): the S-state should be 
the lowest according to perturbation theory up to order
${\cal O}(\alpha^4)$. This finite cutoff effect is explained in next paragraph.
\end{itemize}


\begin{table}
\centerline{
\begin{tabular}{|r||c|c|c|}\hline
\rule[-3mm]{0mm}{8mm}$\mbox{Cutoff: }\Lambda$ & $B_s$ & $B_t$ & $C_{hf}$\\\hline\hline 
   1.0\hspace{0.5cm} &  1.04903964 & 1.00046227 &  0.13493713 \\ \hline 
   1.8\hspace{0.5cm} &  1.16373904 & 1.06860934 &  0.26424917 \\ \hline 
   3.6\hspace{0.5cm} &  1.25570148 & 1.10111328 &  0.42941166 \\ \hline 
   5.4\hspace{0.5cm} &  1.29978050 & 1.11163578 &  0.52262422 \\ \hline 
   7.2\hspace{0.5cm} &  1.32941912 & 1.11782782 &  0.58775360 \\ \hline 
   9.0\hspace{0.5cm} &  1.35223982 & 1.12233652 &  0.63862028 \\ \hline 
  10.8\hspace{0.5cm} &  1.37112216 & 1.12596311 &  0.68099735 \\ \hline 
  12.6\hspace{0.5cm} &  1.38744792 & 1.12904455 &  0.71778713 \\ \hline 
  14.4\hspace{0.5cm} &  1.40198469 & 1.13175363 &  0.75064183 \\ \hline 
  16.2\hspace{0.5cm} &  1.41520247 & 1.13419048 &  0.78058886 \\ \hline 
  18.0\hspace{0.5cm} &  1.42740143 & 1.13641774 &  0.80828803 \\ \hline 
\hline ETPT\hspace{0.3cm} & 1.11812500 & 0.99812500 & 0.33333333\\ \hline
${\cal O}(\alpha^6\ln\alpha)$ & \multicolumn{2}{|c|}{ }& 0.23792985\\ \hline
\end{tabular}}
\centerline{
\parbox{10cm}{
\protect\caption[Binding coefficients and hyperfine splitting]
{\protect\label{TableLambda}The 
binding coefficients of the singlet $(B_s)$ and the triplet states 
$(B_t)$ for $\alpha=0.3,$ $N_1=25, N_2=21$ as functions of the cutoff 
$\Lambda$ in electron masses. Additionally the values for 
equal-time perturbation theory up to order ${\cal O}(\alpha^4)$ (ETPT) and up 
to order ${\cal O}(\alpha^6\ln\alpha)$ (cf.~Eq.~[\protect{\ref{Oln}]})
are shown.}}}
\end{table}

\subsection*{Cutoff dependence}

It is stated in Appendix~\ref{AppxRenormalization} that there is, due
to the renormalization scheme used, a logarithmic divergence of the
eigenvalues with the cutoff $\Lambda$. This was already found in 
\cite[Fig.~7]{DissKraut}. In Fig.~(\ref{lambdaJ0n1}) and (\ref{lambdaJ0n2})
the eigenvalues are shown as functions of the cutoff for $n{=}1$ and $n{=}2$,
respectively.
It is obvious that the different eigenstates diverge linearly with the
logarithm of the cutoff, and that the coefficients of these divergencies
are different.
It is reasonable to fit the curves of 
Fig.~(\ref{lambdaJ0n1}) with a polynomial in $\log \Lambda$, because this is
the behavior expected from the renormalization scheme. 
If one omits the points for $\Lambda>20\,m_f$, because there the entirely
numerical counterterm integrations become problematic, one gets good agreement
with the calculated curves if one uses
\begin{eqnarray}\label{LambdaFit}
M^2_{singlet}(\Lambda)&=& 3.90545-0.0350983 \log \Lambda + 0.00745955 
\log^2 \Lambda,
\nonumber\\
M^2_{triplet}(\Lambda)&=& 3.90976-0.0185787 \log \Lambda + 0.00788614 
\log^2 \Lambda.
\end{eqnarray}
The small coefficient of the $\log^2\Lambda$ term verifies the
logarithmic dependence of the eigenvalues on the cutoff.
We will see in Chapter \ref{DrAChapterAnnihilation} that the dependence on 
$\Lambda$ becomes even weaker if one includes the annihilation channel.

One notices several level crossings for $n{=}2$. 
As was stated in the last paragraph, the ordering of the eigenvalues of 
$n{=}2$ for $\Lambda{=}1.0\,m_f$ 
turns out to be wrong for the two lowest states of this multiplet because 
of the crossing. 
That the levels
do indeed cross can be proven by tracing them back to their sectors of 
definite $\cal C$- and $\cal H$-quantum numbers. 
A consequence of these crossings is the fact that the 
order of the eigenvalues is correct in the region between the crossings of 
the $2^1S_0$/$2^1P_0$ and the $2^3P_1$/$2^1P_0$ states, 
$1.5\leq \Lambda \leq 7$. 
A further investigation of these crossings up to $n{=}4$ show, that 
the states with $\pi_{\cal C}{=}+1$ and $\pi_{\cal H}{=}-1$ are those 
that fall off fastest with $\Lambda$ and tend to cross other levels.

An important {\em ratio} of eigenvalues to be compared to results of other
calculations is the hyperfine splitting. 
The hyperfine coefficient
\begin{eqnarray}
C_{hf}&=&\left(M_{triplet}-M_{singlet}\right)/\alpha^4\nonumber\\
&=&\frac{1}{2}\left[\frac{2}{3}+\left(\frac{1}{2}\right)-
\frac{\alpha}{\pi}\left(\ln 2 +\frac{16}{9}\right)
-\frac{5}{12}\alpha^2\ln\alpha +K\alpha^2+K'\alpha^3\right]
\label{Oln}
\end{eqnarray}
was introduced by {\sc Fermi}\cite{Fermi30} in 1930.
{\sc Fermi} calculated 
$C_{hf}=\frac{1}{3}$ for hydrogen-like atoms\footnote{ He investigated 
the cases of Sodium and Caesium, {\sl e.g.}~he naturally did {\em not} consider 
the possible annihilation of a bound particle-antiparticle system.}, which is 
the exact result up to order ${\cal O}(\alpha^4)$. 
The second line of Eq.~(\ref{Oln}) shows the `state of the art' result 
of equal time perturbation
theory \cite[p.~759]{BodwinYennieGregorio}. 
The term in brackets is the contribution from the one-photon annihilation.
This coefficient is listed in Table (\ref{TableLambda}) together with the 
binding coefficients (\ref{BindingCoeffs}). 
They will be compared to the values obtained by including
the annihilation channel in Chapter \ref{DrAChapterAnnihilation}.
The coefficients were calculated with the improved counterterms, as opposed 
to \cite[Table V.]{KPW}. 
Comparing with the values given there, one notices that the singlet falls off
slower with $\Lambda$ for the old counterterms. Consequently the values
for the hyperfine coefficient $C_{hf}$ are smaller there.
The value of this coefficient, using Eq.~(\ref{Oln}) without the annihilation
contribution, is displayed, too. In 
the region of large couplings
considered in this work, also higher orders in the coupling constant are
important.    
Note the remarkable effect: the value $C_{hf}$ is $40\%$ larger up to 
${\cal O}
(\alpha^4)$ than up to ${\cal O}(\alpha^6\ln \alpha)$!

Concluding, one can state that the best values as compared to equal time 
perturbation theory are obtained for $\Lambda\simeq 1.5\, m_f$: the hyperfine 
splitting has the right order of magnitude, the order of the eigenvalues
is correct, and the ground state is at the value of perturbative calculations.
Although one can think of fitting the obtained data to the results
of perturbation theory, I decided to follow the renormalization scheme of 
Appx.~\ref{AppxRenormalization} to show the consistency 
of the positronium model described in this chapter, rather than to produce
results competing with the accuracy of elaborate perturbative calculations.
Moreover, it is not clear in the regime of a large coupling, how reliable
the results of ETPT are, as we have seen from the significant
dependence of the value for
$C_{hf}$ on the order of perturbative calculations.
\vspace{1cm}

\begin{figure}[t]
\begin{minipage}{7.5cm}
\centerline{
\psfig{figure=\graphpath lambda_J0_n1.epsi,width=7.5cm,angle=0}}
\centerline{
\parbox{7.5cm}{
\protect\caption[Cutoff dependence of the ground states ($n{=}1$)]
{\label{lambdaJ0n1}Cutoff 
dependence of the triplet (upper curve) and singlet (lower curve)
ground state, $\alpha=0.3$. The cutoff is given in units of the 
electron mass.}}} 
\end{minipage}
\hfill
\begin{minipage}{7.5cm}
\centerline{
\psfig{figure=\graphpath lambda_J0_n2.epsi,width=7.5cm,angle=0}}
\centerline{
\parbox{7.5cm}{
\caption[Cutoff dependence of the first excited states ($n{=}2$)]
{\label{lambdaJ0n2}Cutoff dependence of the first excited states $(n{=}2)$ 
for $\alpha=0.3$. 
The cutoff is given in units of the electron mass.
Note the level crossings.} 
}}
\end{minipage}
\end{figure}

\begin{table}
\centerline{
\begin{tabular}{|r|c|c|c||c|c|r|c|}\hline
\rule[-3mm]{0mm}{8mm}$n$ & Term &$\pi_{\cal C}$ &$\pi_{\cal H}$ &$B_{ETPT,n}$ & {$B_{theor,n}$} & $\Delta B$\hspace{0.6cm} & $\frac{100\times\Delta B}{B_{exp,n}}$  \\ \hline \hline
 1& $1^1S_0$& $+  1$& $  -1$ & $        1.11812500$ & $        1.04955251$ $\pm$ $        0.00001714$& $0.068572$ & $ 6.13$ \\ \hline 
  2&$1^3S_1$ & $  -1$& $+  1$ & $        0.99812500$ & $        1.00101171$ $\pm$ $        0.00011090$& $-0.002887$ & $ 0.29$ \\ \hline 
\hline
 3& $2^1S_0$& $+  1$& $  -1$ & $        0.26863281$ & $        0.26023681$ $\pm$ $        0.00016870$& $0.008396$ & $ 3.13$ \\ \hline 
  4&$2^3S_1$ & $  -1$& $+  1$ & $        0.25363281$ & $        0.25380363$ $\pm$ $        0.00021696$& $-0.000171$ & $ 0.07$ \\ \hline 
 5& $2^1P_1$& $  -1$& $  -1$ & $        0.25363281$ & $        0.25796923$ $\pm$ $        0.00016055$& $-0.004336$ & $ 1.71$ \\ \hline 
  6&$2^3P_0$ & $+  1$& $+  1$ & $        0.26113281$ & $        0.26706985$ $\pm$ $        0.00015593$& $-0.005937$ & $ 2.27$ \\ \hline 
  7&$2^3P_1$ & $+  1$& $  -1$ & $        0.25550781$ & $        0.25966695$ $\pm$ $        0.00020629$& $-0.004159$ & $ 1.63$ \\ \hline 
  8&$2^3P_2$ & $+  1$& $+  1$ & $        0.25100781$ & $        0.25525791$ $\pm$ $        0.00017678$& $-0.004250$ & $ 1.69$ \\ \hline 
\hline
 9& $3^1S_0$& $+  1$& $  -1$ & $        0.11701389$ & $        0.11520626$ $\pm$ $        0.00031353$& $0.001808$ & $ 1.54$ \\ \hline 
 10&$3^3S_1$ & $  -1$& $+  1$ & $        0.11256944$ & $        0.11344082$ $\pm$ $        0.00036304$& $-0.000871$ & $ 0.77$ \\ \hline 
11& $3^1P_1$& $  -1$& $  -1$ & $        0.11256944$ & $        0.11448968$ $\pm$ $        0.00026966$& $-0.001920$ & $ 1.71$ \\ \hline 
 12&$3^3P_0$ & $+  1$& $+  1$ & $        0.11479167$ & $        0.11713342$ $\pm$ $        0.00027062$& $-0.002342$ & $ 2.04$ \\ \hline 
 13&$3^3P_1$ & $+  1$& $  -1$ & $        0.11312500$ & $        0.11512731$ $\pm$ $        0.00032573$& $-0.002002$ & $ 1.77$ \\ \hline 
 14&$3^3P_2$ & $+  1$& $+  1$ & $        0.11179167$ & $        0.11371654$ $\pm$ $        0.00028065$& $-0.001925$ & $ 1.72$ \\ \hline 
15& $3^1D_2$& $+  1$& $  -1$ & $        0.11168056$ & $        0.11281562$ $\pm$ $        0.00015011$& $-0.001135$ & $ 1.02$ \\ \hline 
 16&$3^3D_1$ & $  -1$& $+  1$ & $        0.11168056$ & $        0.11342693$ $\pm$ $        0.00015480$& $-0.001746$ & $ 1.56$ \\ \hline 
 17&$3^3D_2$ & $  -1$& $  -1$ & $        0.11179167$ & $        0.11297755$ $\pm$ $        0.00016091$& $-0.001186$ & $ 1.06$ \\ \hline 
 18&$3^3D_3$ & $  -1$& $+  1$ & $        0.11136310$ & $        0.11251082$ $\pm$ $        0.00015641$& $-0.001148$ & $ 1.03$ \\ \hline 
\hline
19& $4^1S_0$& $+  1$& $  -1$ & $        0.06507080$ & $        0.06549049$ $\pm$ $        0.00058804$& $-0.000420$ & $ 0.64$ \\ \hline 
 20&$4^3S_1$ & $  -1$& $+  1$ & $        0.06319580$ & $        0.06478625$ $\pm$ $        0.00059833$& $-0.001590$ & $ 2.52$ \\ \hline 
21& $4^1P_1$& $  -1$& $  -1$ & $        0.06319580$ & $        0.06500267$ $\pm$ $        0.00046706$& $-0.001807$ & $ 2.86$ \\ \hline 
 22&$4^3P_0$ & $+  1$& $+  1$ & $        0.06413330$ & $        0.06611514$ $\pm$ $        0.00046966$& $-0.001982$ & $ 3.09$ \\ \hline 
 23&$4^3P_1$ & $+  1$& $  -1$ & $        0.06343018$ & $        0.06533105$ $\pm$ $        0.00048739$& $-0.001901$ & $ 3.00$ \\ \hline 
 24&$4^3P_2$ & $+  1$& $+  1$ & $        0.06286768$ & $        0.06469966$ $\pm$ $        0.00047847$& $-0.001832$ & $ 2.91$ \\ \hline 
25& $4^1D_2$& $+  1$& $  -1$ & $        0.06282080$ & $        0.06396765$ $\pm$ $        0.00029377$& $-0.001147$ & $ 1.83$ \\ \hline 
 26&$4^3D_1$ & $  -1$& $+  1$ & $        0.06282080$ & $        0.06426183$ $\pm$ $        0.00030846$& $-0.001441$ & $ 2.29$ \\ \hline 
 27&$4^3D_2$ & $  -1$& $  -1$ & $        0.06286768$ & $        0.06409883$ $\pm$ $        0.00031866$& $-0.001231$ & $ 1.96$ \\ \hline 
 28&$4^3D_3$ & $  -1$& $+  1$ & $        0.06268687$ & $        0.06386407$ $\pm$ $        0.00030333$& $-0.001177$ & $ 1.88$ \\ \hline 
29& $4^1F_3$& $  -1$& $  -1$ & $        0.06266009$ & $        0.06314142$ $\pm$ $        0.00009618$& $-0.000481$ & $ 0.77$ \\ \hline 
 30&$4^3F_2$ & $+  1$& $+  1$ & $        0.06266009$ & $        0.06329675$ $\pm$ $        0.00011207$& $-0.000637$ & $ 1.02$ \\ \hline 
 31&$4^3F_3$ & $+  1$& $  -1$ & $        0.06267683$ & $        0.06323357$ $\pm$ $        0.00011899$& $-0.000557$ & $ 0.89$ \\ \hline 
 32&$4^3F_4$ & $+  1$& $+  1$ & $        0.06258754$ & $        0.06309753$ $\pm$ $        0.00010310$& $-0.000510$ & $ 0.81$ \\ \hline 
\hline
33& $5^1S_0$& $+  1$& $  -1$ & $        0.04134100$ & $        0.04325281$ $\pm$ $        0.00125908$& $-0.001912$ & $ 4.62$ \\ \hline 
 34&$5^3S_1$ & $  -1$& $+  1$ & $        0.04038100$ & $        0.04291300$ $\pm$ $        0.00127772$& $-0.002532$ & $ 6.27$ \\ \hline 
35& $5^1P_1$& $  -1$& $  -1$ & $        0.04038100$ & $        0.04283966$ $\pm$ $        0.00064062$& $-0.002459$ & $ 6.09$ \\ \hline 
\hline
\end{tabular}}

\protect\caption[Spectrum $J_z{=}0$: perturbation theory and model predictions]
{\protect\label{TableSpectrumJ0}The positronium spectrum for $\alpha=0.3,\Lambda=1.0\, m_f,N_1=N_2=21$.
The $n{}^{2S+1}L_J$ notation, the quantum numbers under charge conjugation, $\cal C$, and $\cal T$-parity, $\cal H$, are shown in the first columns.
The first row of binding coefficients ($B_{ETPT,n}$) comes from equal time perturbation theory calculations up to order ${\cal O}(\alpha^4$). In the following row our results are listed with an 
estimate of finite $N$ errors. Also shown is the difference 
$\Delta B:=B_{ETPT,n}-B_{theor,n}$. The last row contains 
the relative discrepancy to perturbation theory in percent.}

\end{table}



\begin{figure}[ht]
\centerline{
\psfig{figure=\graphpath 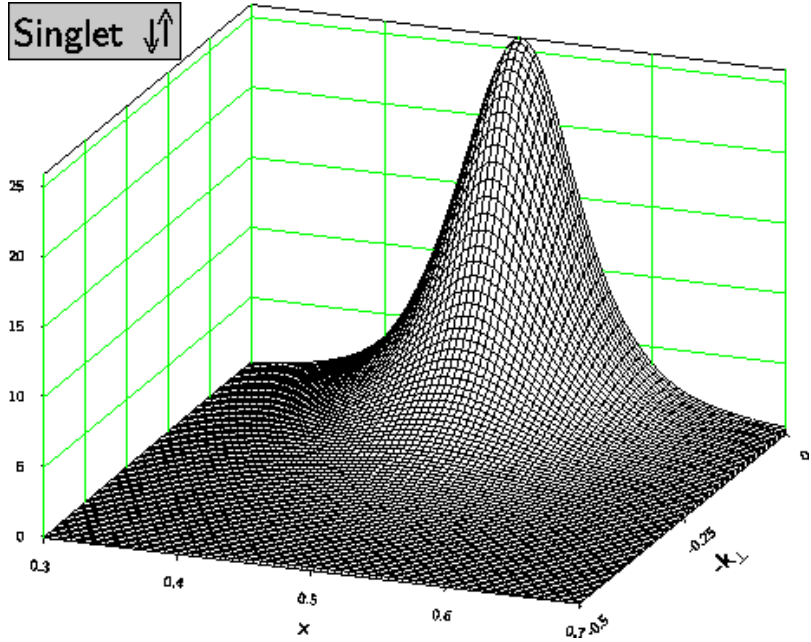,width=11.0cm,angle=0}}
\protect\caption[The $(\uparrow\downarrow)$-component of the $1^1S_0$ 
wavefunction for $J_z{=}0$]
{\protect\label{singletud}The 
singlet wavefunction for anti-parallel spins as a function of 
the longitudinal momentum fraction $x$ and the transverse momentum $\kp$,
omitting the dependence on the angle $\varphi$. The
calculation was done with $\alpha=0.3,\Lambda=1.0\, m_f, J_z=0, 
N_1=41, N_2=11$.}
\end{figure}

\section{Wavefunctions}

A big advantage of the Hamiltonian method applied in the calculations
of the spectrum is the fact that the wavefunctions of positro\-ni\-um are 
obtained in the same calculation as the spectrum.
The wavefunctions 
for the singlet and the triplet components are displayed in this
section.
Similar plots in Ref.~\cite{DissKraut}~seem to indicate 
numerical problems of that work, because they show some
internal structure.
It turns out that this is due merely to mistakes in the graphing package used. 
The smoothing functions are quite sensitive to the data and boundary 
conditions employed\footnote{
For our purpose GLE 3.3 by {\sc C.\ Pugmire} yielded the best results together
with the smoothing function {\tt grid} of PLOTDATA.}.

\begin{figure}
\centerline{
\psfig{figure=\graphpath 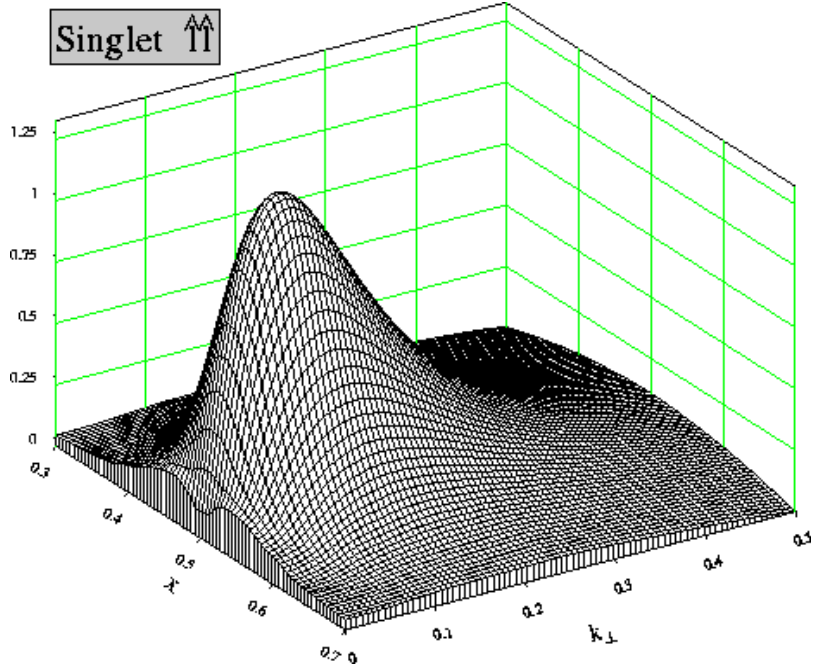,width=11.0cm,angle=0}}
\caption[\protect\label{singletuu}The $(\uparrow\uparrow)$-component 
of the $1^1S_0$ wavefunction for $J_z{=}0$]
{The singlet wavefunction for parallel spins as a function of 
the longitudinal momentum fraction $x$ and the transverse momentum $\kp$,
omitting the dependence on the angle $\varphi$. The
calculation was done with $\alpha=0.3,\Lambda=1.0\, m_f, J_z=0, 
N_1=41, N_2=11$.}
\end{figure}


\begin{figure}
\centerline{
\psfig{figure=\graphpath wf_1j01.epsi,width=11.0cm,angle=0}}
\protect\caption[\protect\label{tripletud}The 
$(\uparrow\downarrow)$-component of the $1^3S_1$ 
wavefunction for $J_z{=}0$]
{The triplet wavefunction for anti-parallel spins as a function of 
the longitudinal momentum fraction $x$ and the transverse momentum $\kp$,
omitting the dependence on the angle $\varphi$. The
calculation was done with $\alpha=0.3,\Lambda=1.0\, m_f, J_z=0, 
N_1=41, N_2=11$.}
\vspace{0.5cm}

\centerline{
\psfig{figure=\graphpath wf_1j00.epsi,width=11.0cm,angle=0}}
\protect\caption[The $(\uparrow\uparrow)$-component of the $1^3S_1$ 
wavefunction for $J_z{=}0$]
{\protect\label{tripletuu}The 
triplet wavefunction for parallel spins as a function of 
the longitudinal momentum fraction $x$ and the transverse momentum $\kp$,
omitting the dependence on the angle $\varphi$. The
calculation was done with $\alpha=0.3,\Lambda=1.0\, m_f, J_z=0, 
N_1=41, N_2=11$.}
\end{figure}

The wavefunctions have two components: one with parallel 
($\uparrow\uparrow$) and the other
with anti-parallel ($\uparrow\downarrow$) helicities.
For the displayed singlet and triplet wavefunctions, the decomposition
in terms of spin components (cf.\ 
Eqs.~[\ref{CSymmetry}]~and~[\ref{TSymmetry}]) read
\[
|\psi_{\rm sing}(\uparrow\downarrow)\rangle=\frac{1}{2}\sum_{i=1}^{N_1}
\sum_{j=1}^{(N+1)/2}\left[\psi \bdagg_1(\uparrow)\ddagg_2(\downarrow)
-\psi^* \bdagg_1(\downarrow)\ddagg_2(\uparrow)
-\psi \bdagg_2(\downarrow)\ddagg_1(\uparrow)
+\psi^* \bdagg_2(\uparrow)\ddagg_1(\downarrow)
\right]|0\rangle,
\]
\[
|\psi_{\rm sing}(\uparrow\uparrow)\rangle=\frac{1}{2}\sum_{i=1}^{N_1}
\sum_{j=1}^{(N+1)/2}\left[\psi \bdagg_1(\uparrow)\ddagg_2(\uparrow)
+\psi^* \bdagg_1(\downarrow)\ddagg_2(\downarrow)
-\psi \bdagg_2(\uparrow)\ddagg_1(\uparrow)
-\psi^* \bdagg_2(\downarrow)\ddagg_1(\downarrow)
\right]|0\rangle,
\]
\[
|\psi_{\rm trip}(\uparrow\downarrow)\rangle=\frac{1}{2}\sum_{i=1}^{N_1}
\sum_{j=1}^{(N+1)/2}\left[\psi \bdagg_1(\uparrow)\ddagg_2(\downarrow)
+\psi^* \bdagg_1(\downarrow)\ddagg_2(\uparrow)
+\psi \bdagg_2(\downarrow)\ddagg_1(\uparrow)
+\psi^* \bdagg_2(\uparrow)\ddagg_1(\downarrow)
\right]|0\rangle,
\]
\[
|\psi_{\rm trip}(\uparrow\uparrow)\rangle=\frac{1}{2}\sum_{i=1}^{N_1}
\sum_{j=1}^{(N+1)/2}\left[\psi \bdagg_1(\uparrow)\ddagg_2(\uparrow)
-\psi^* \bdagg_1(\downarrow)\ddagg_2(\downarrow)
+\psi \bdagg_2(\uparrow)\ddagg_1(\uparrow)
-\psi^* \bdagg_2(\downarrow)\ddagg_1(\downarrow)
\right]|0\rangle,
\]
The wavefunctions are normalized in the polar coordinates
$\mu$ and $ \cos\theta$, such that
\beq\label{Normalization}
\sum_{i=1}^{N_1}\sum_{j=1}^{(N_2+1)/2}\left[\psi(\mu_i,\cos\theta_j;\,
\lambda_1=\lambda_2)+
\psi(\mu_i,\cos\theta_j;\,\lambda_1=-\lambda_2)\right]=1.
\eeq
where  
\beqa
\bdagg_1&=&\bdagg(x,\kp,\varphi)=\bdagg(\mu,\cos\theta,\varphi),\\
\ddagg_2&=&\ddagg(1-x,\kp,\varphi)=\ddagg(\mu,-\cos\theta,\varphi+\pi).
\eeqa
and discretized variables $\mu_i$, $\cos\theta_j$ are used. 
For all details on the
numeric aspects of the calculations see Appx. \ref{AppxNumerics}.

\begin{figure}
\begin{minipage}{15.5cm}
\centerline{
\psfig{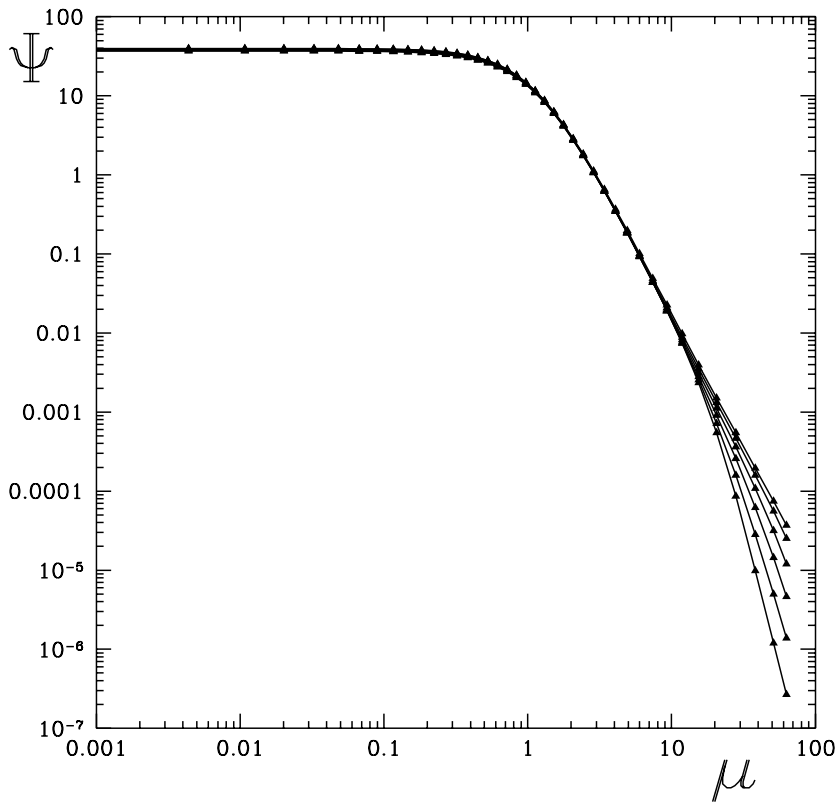}}
\caption[The decrease of the $J_z=0$ singlet ground state]
{\label{fig11}The decrease of the 
$J_z=0$ singlet ground state with anti-parallel helicities 
as function of the off-shell mass variable $\mu$, Eq.~{mu}. 
The parameters used are
$N_1=41,N_2=11,\Lambda=1.0\, m_f, \alpha=0.3$. Plotted are the six different 
decreases corresponding
to the six non-positive values of the discretized angle variable $\cos\theta$. 
One notices the deviations from rotational symmetry for $\mu\ge 10$.}
\end{minipage}
\end{figure}

Figure (\ref{fig11}) shows the decrease of the singlet wavefunction 
with anti-parallel helicities as the off-shell mass $\mu$ increases.
The graph, calculated with entirely numerical counterterms,
is almost the same as that in \cite[Fig.~(11), bottom]{KPW} with 
half-analytical counterterms, since 
a large number of integration points ($N_1=41, N_2=11$) were used 
to calculate the results.
It can be seen that rotational invariance is broken. This component of the 
S-wave state does not depend on any angle and should therefore 
decrease independently of $\cos\theta$.
The broken rotational invariance is no surprise. 
One reason is the fact that the associated operators $F_1$ and $F_2$ of 
rotations around the transverse axes are dynamical, {\em i.e.}~contain the
interaction,  
cf.~Table (\ref{CoesterTable}). The other argument uses the transformation
to equal-time coordinates
\beq\label{rotviol}
dxd^2\vkp=\left|\frac{\partial x}{\partial k_z}\right|=\frac{1}{2E}
\left(1-\frac{k_z^2}{E^2}\right)d^3\vec{k}.
\eeq
One sees directly that the term proportional to $\frac{k_z^2}{E^2}$ breaks
rotational invariance.

It is worth mentioning that the breaking of rotational invariance is 
noticeable only for large cutoffs $\Lambda$. With a cutoff of 
$\Lambda=1.0\, m_{\!f}$, the deviation of curves with different $\cos\theta$
is not visible.
In the next chapter, we investigate the same properties of the 
corresponding wavefunction for $J_z{=}1$.


\chapter{Angular momentum in front form dynamics}
\label{DrAChapterJz}

\section{The Poincar\'e group on the light-cone}

To provide the theoretical background for our numerical calculations, 
we review the properties of angular momenta on the light-cone.
The material presented here was inspired by a section on the same subject
in \cite{Coester}. 

\subsection{Unitary representations of the Poincar\'e group}

The Poincar\'e group has 10 generators: the four-momentum $P^{\mu}$, 
and the generalized angular momenta, constituting an 
anti-symmetric tensor $M^{\mu\nu}=-M^{\nu\mu}$.
They are subject to the commutation relations 
\beqa
\left[P^{\mu},P^{\nu}\right]&=&0,\\
\left[M^{\mu\nu},P^{\rho}\right]&=&
i\left(g^{\mu\nu}P^{\rho}-g^{\rho\mu}P^{\nu}\right),\\
\left[M^{\mu\nu},M^{\rho\sigma}\right]&=&
i\left(g^{\mu\sigma}M^{\nu\rho}-g^{\mu\rho}M^{\nu\sigma}
 -g^{\nu\sigma}M^{\mu\rho}+g^{\nu\rho}M^{\mu\sigma}\right).
\eeqa
The Poincar\'e group has two Casimir operators, {\em i.e.}~two invariant 
functions of the generators. One is the {\em mass} (squared) operator 
\[
M^2:=P^{\mu}P_{\mu},
\]
which cannot be confused with the {\em indexed} tensor of angular momenta
$M^{\mu\nu}$.
The other is the square of the {\em Pauli-Lubanski vector} \cite{Lubanski}
\[
W_{\nu}:=\frac{1}{2}\epsilon_{\rho\sigma\mu\nu}P^{\rho}M^{\sigma\mu}.
\]
The components of this vector obey the commutation relations
\beq\label{WVR}
\left[W^{\mu},W^{\nu}\right]=i\epsilon_{\mu\nu\rho\sigma}W^{\rho}P^{\sigma},
\eeq
and commute with $P^{\mu}$.
The spin operator $\vec{\jmath}\;$ can be defined \cite[Eq.~(6)]{Coester} as
\beq\label{Defspin} 
j_{i}:=u^{\mu}_{i}(P)\frac{W_{\mu}}{M},\quad i=1,2,3.
\eeq
It is therefore a linear function of the Pauli-Lubanski vector. 
The coefficients $u_i$ are three orthonormal, operator valued, space-like basis
vectors, orthogonal to $P^{\mu}$. They have the properties
\beq\label{uproperties}
u^{\mu}_{i}(P)u_{j,\mu}(P)=\delta_{ij}, \quad\quad \mbox{and} \quad\quad
\sum_{i=1}^3 u^{\mu}_{i}(P)u^{\nu}_{i}(P)
=g^{\mu\nu}-\frac{P^{\mu}P^{\nu}}{P^2}.
\eeq
The basis vectors transform under a Lorentz transformation like
\[
U^{\dagger}(\Lambda)u_{i}(P)U(\Lambda)=u_{i}(\Lambda P)\neq\Lambda 
u_{i}(P).
\]
These two basis systems are related by a rotation, since $u_{i}(\Lambda P)$
and $\Lambda u_{i}(P)$ are both orthonormal space-like vectors, orthogonal
to $\Lambda P^{\mu}$. This transformation is referred to as 
a {\em Wigner rotation} by {\sc Coester} \cite{Coester}
\[
{\cal R}_W(\Lambda,P)_{ij}:=u^{\mu}_{i}(\Lambda P)
\Lambda u_{j\mu}(P).
\]
One can show that the Lorentz transformation of the spin operator is 
equivalent to a Wigner rotation of its components
\[
U^{\dagger}(\Lambda)\vec{\jmath}\,(P)U(\Lambda)
={\cal R}_W(\Lambda,P)\vec{\jmath}.
\]
This follows from the definition (\ref{Defspin}) of the spin operator
and holds for every vector $b_{i}(P)=u^{\mu}_{i}(P)V_{\mu}$,
where $V_{\mu}$ is an arbitrary four-vector. 
The spin operator has the correct SU(2) commutation relations, as can be proven
using (\ref{WVR}) and the properties of the basis vectors, 
Eqs.~(\ref{uproperties}).
Lorentz transformations $\Lambda$ have the general property 
\beq\label{lorentz}
\Lambda_{\mu}^{\;\;\,\rho}\;g_{\rho\sigma}\Lambda_{\nu}^{\;\;\,\sigma}=g_{\mu\nu}.
\eeq
If one defines $u_0:=\frac{P}{M}$, one can re-write Eq.~(\ref{uproperties})
into a four dimensional equation
\[
u_{\mu}^{\;\;\,\rho}\;u_{\nu}^{\;\;\,\sigma}\;g_{\rho\sigma}=g_{\mu\nu},
\]
which has the same structure as Eq.~(\ref{lorentz}).
The operator valued matrix $u_{\mu}^{\;\;\nu}$ is thus an $SO(1,3)$ 
representation of 
some Lorentz transformation $B(P)$ which can be interpreted with some care 
as the Lorentz transformation to the rest frame of P: 
$u_{\mu}^{\;\;\nu}(P)P_{\nu}=g_{\mu 0}M$.  
The spin operator $\vec{\jmath}$ can be written, apart from the 
form given in Table (\ref{CoesterTable}), in terms of the basis vectors
\[
j_{i}=\frac{1}{2}u_0^{\;\;\; \mu}(P)u_i^{\;\;\; \nu}(P)
\epsilon_{\mu\nu\rho\sigma}M^{\rho\sigma}.
\]
We are interested in bound state calculations, {\em i.e.}~in composite systems.
One can define the relative momenta $k_n$ of a subsystem by
\[
k_{n,i}:=u_{i}^{\nu}(P)p_{n,\nu}, \quad p^{\mu}_n=\sum_i u^{\mu}_{i}(P)k_{n,i},
\]
where $P=\sum_n p_n$ is the (absolute) momentum of the system.

\subsection{Connection between canonical spin and front form helicity}

So far, everything was independent of the choice of the basis vectors
$u_{\alpha}(P)$, {\em i.e.}~of the form of the Hamiltonian dynamics. 
The implications for {\em instant form} and {\em front form}, 
the two kinds of dynamics of interest for the present work, are compared now. 
The major features of the two types of dynamics are listed in Table 
(\ref{CoesterTable}).

The instant form, {\em i.e.}~canonical, choice of the basis 
$\bar{u}_{\alpha}(P)$
is related to the front form basis $\check{u}_{\alpha}(P)$ 
by the so-called {\em Melosh rotation}
\cite{Melosh}
\[
{\cal R}_M(P)_{ij}:=\bar{u}_{i}(P)\cdot
\check{u}_{j}(P).
\]
Its $SU(2)$ representation is
\[
D^{1/2}[{\cal R}_M(P)]=\frac{M+P^+-i\vec{\sigma}\cdot(\hat{z}\times
\vec{P}_{\perp})}{\sqrt{(M+P^+)^2+P^2_{\perp}}}.
\] 
The essential observation for a relation between instant form spin and
light-cone helicity is that the canonical spin of the $n$-th particle
of a composite system
can be expressed in the front form basis
\[
\bar{s}_{n,i}=\bar{u}_{i}(\bar{k}_{n})\cdot \bar{B}(P)
\frac{W_n}{M_n}={\cal R}_M(P)\bar{u}_{i}(\check{k}_{n})\cdot \check{B}(P)
\frac{W_n}{M_n}.
\]
The translation of the total angular momentum from equal time to light-cone
coordinates is therefore obtained by a Melosh rotation, 
cf.~Table (\ref{CoesterTable}).

\subsection{Field theories on the light-cone}

There are two different ways to construct front form particle dynamics.
One approach starts from the mass and spin operators which have to fulfill 
some constraints on their commutation relations. 
The three Hamiltonians of front form dynamics can then be expressed as 
functions of the kinematic generators and of the mass and spin operators
\beqa
P^-&=&\frac{M^2+P^2_{\perp}}{P^+},\\
F_1&=&\frac{2\left(M \check{\jmath}_2+P_2 \check{\jmath}_3\right)}{P^+}
+\frac{P^-}{P^+}E_1+\frac{2P_1}{P^+}K_z,\\
F_2&=&-\frac{2\left(M \check{\jmath}_1+P_1 \check{\jmath}_3\right)}{P^+}
+\frac{P^-}{P^+}E_2+\frac{2P_2}{P^+}K_z.\\
\eeqa
The dynamical operators commute with each other
\beq\label{VR}
[P^-,\vec{F}_{\perp}]=[F_1,F_2]=0.
\eeq
The other approach is that of Fock space field theories. Here, the fundamental
quantities are the Hamiltonians which are derived from a Lagrangian density.
The spin operators are then formulated in terms of the Hamiltonians 
\beqa
\check{\jmath}_z=\frac{W^+}{P^+}&=&M_z-\frac{\hat{z}\cdot
(\vec{E}_{\perp}\times\vec{P}_{\perp})}{P^+},\\
\check{\vec{\jmath}}_{\perp}&=&
\frac{\vec{W}_{\perp}}{M}-\frac{\vec{P}_{\perp}W^+}{M P^+},
\eeqa
where
\[
\vec{W}_{\perp}=\frac{1}{2}P^+\hat{z}\times\left(\vec{F}-\vec{E}\right)
+\frac{1}{2}\left(P^+-P^-\right)\hat{z}\times\vec{E}_{\perp}-
\left(\hat{z}\times\vec{P}_{\perp}\right)K_z,
\]
and $\hat{z}$ is the unit vector in $z$-direction.
Very important is the fact that Poincar\'e invariance is destroyed, as soon as
truncations of the Fock space or regularizations of Fock sectors are 
implemented. 
In particular, the requirement of an invariance under rotations around 
the transverse axes is difficult to fulfill since the corresponding
operators are complicated and involve the interaction.
Even worse, in a truncated Fock space formalism the full 
Poincar\'e invariance is absent if it is not restored by an additional 
({\em ad hoc}) prescription \cite[p.~11]{Coester}!

It is therefore impossible for our positronium model to be Poincar\'e 
{\em invariant}. I shall rather show its {\em covariance} by looking at the
results, {\em i.e.}~at the physical observables such as the 
invariant mass spectrum.
If full rotational invariance is restored in the solution, the states of 
same total angular momentum $J$ but different $J_z$, become degenerate. 

The most direct way would of course be to construct the operators $F_1$
and $F_2$ explicitly and the to diagonalize the operator of total angular
momentum ${\cal J}^2$. This has not been done up to now. 
Because of the vanishing commutators of the dynamical operators with $P^-$,
Eq.~(\ref{VR}),
it is clear that the diagonalization of the rotation operators will be 
much simpler with an already diagonal Hamiltonian $H_{LC}$: only the 
states with degenerate mass eigenvalues will be coupled by $\vec{F}_{\perp}$.

However, we restrict ourselves in the present work to the 
calculation of the spectrum of the light-cone Hamiltonian and will find 
that we can
classify the eigenstates with regard to $J^2$ even without constructing
or diagonalizing the rotation operators  $\vec{F}_{\perp}$.

\newpage
\vspace*{1.3cm}
\begin{table}[H]
\begin{tabular}{|c|c|c|}
\hline
\rule[-3mm]{0mm}{8mm}&\bf instant form dynamics& \bf front form dynamics\\
\hline
\hline
\parbox{0cm}{\rule[-3mm]{0mm}{26mm}}
\parbox{2cm}{Kinematic operators}
&
\parbox{6.5cm}{
6 operators:
\[
\vec{P},\vec{J}:=(M^{23},M^{31},M^{12})
\]}
&
\parbox{6.5cm}{
7 operators:
\[
\vec{P},K_z:=\frac{1}{2}M^{+-},\vec{E}_{\perp}:=M^{\perp +},J_z
\]}
\\
\hline
\parbox{0cm}{\rule[-3mm]{0mm}{22mm}}
\parbox{2cm}{Dynamic operators}
&
\parbox{6.5cm}{
4 operators:
\[
P^0,\vec{K}
\]}
&
\parbox{6.5cm}{
3 operators:
\[
 P^-, \vec{F}_{\perp}:=M^{\perp -}
\]}
\\
\hline
\parbox{0cm}{\rule[-3mm]{0mm}{16mm}}
\parbox{2cm}{Conditions\\ on basis}
&
\parbox{6.5cm}{
\begin{centering}

$\bar{B}(P)$ are rotationless Lorentz transformations

\end{centering}
}
&
\parbox[th]{6.5cm}{
\begin{centering}

$\check{B}(P)$ are null-plane boosts

\end{centering}
}
\\
\hline
\parbox{2cm}{Basis}
&
\parbox{6.5cm}{
\[
\bar{u}_{i}:=e_{i}+\frac{e_{i}^{\mu}P_{\mu}}
{M-e_0^{\mu}P_{\mu}}\left(\frac{P}{M}+e_0\right)
\]
}&
\parbox{6.5cm}{
\beqa
\check{u}_{\perp}&:=&e_{\perp}+\frac{e_{\perp}^{\mu}P_{\perp\mu}}
{n^{\mu}P_{\mu}}n,\\ 
\check{u}_{3}&:=&\frac{M}{n^{\mu}P_{\mu}}
\left(n-\frac{(n^{\mu}P_{\mu})P}{P^2}\right)
\eeqa
}\\
\hline
\parbox{2cm}{Wigner\\ rotation}
&
\parbox{6.5cm}{
\[
\bar{{\cal R}}_W({\cal R},P)={\cal R}
\]
}&
\parbox{6.5cm}{
\[
\check{\cal R}_W(\check{\Lambda}_{boost},P)_{\alpha\beta}
=\delta_{\alpha\beta}
\]
}\\
\hline
\parbox{2cm}{Angular\\ momentum}
&
\parbox{6.5cm}{
\[
\bar{\vec{\jmath}}=\frac{\vec{W}}{M}-\frac{\vec{P}W_0}{M(M+P^0)}
\]
}&
\parbox{6.5cm}{
\[
\check{\vec{\jmath}}={\cal R}^{-1}_{\rm M}(P)
\bar{\vec{\jmath}}
\]
}\\
\hline
\parbox{2cm}{Spin\\ operator}
&
\parbox{6.5cm}{
\[
\bar{\vec{s}}_i=\bar{\cal R}_{W}(P,p_i)\bar{\vec{\jmath}}_i
\]
}&
\parbox{6.5cm}{
\[
\check{\vec{s}}_i={\cal R}^{-1}_{\rm M}(P)\bar{\vec{s}}_i
\]
}\\
\hline
\hline
Notations&
\multicolumn{2}{|c|}{
\parbox{12cm}{
\[
e_{\sigma}^{\;\;\mu}:=g_{\sigma}^{\;\;\mu}, \quad n^{\mu}:=(1,-\hat{z}),
\quad \check{\Lambda}_{boost}\in \{\vec{E}_{\perp},K_z\}
\]
}}
\\
\hline
\end{tabular}
\caption[Major properties of instant form and front form dynamics]
{Major properties of instant form and front form dynamics.}
\label{CoesterTable}
\end{table}

\newpage

\section{The Hamiltonian matrix with general $J_z$}

The material presented in the previous section shows that the definition
of angular momentum operators in front form dynamics is problematic
since they include the interaction. 
It is therefore a subject of its own merit to study the properties of
angular momentum within a well-defined model on the light-cone.  
Consequently, we investigate the case of a non-vanishing $z$-component 
of the total angular momentum in our positronium model.  

The way to proceed is inferred from the definition of the integral used
to integrate out the angular degree of freedom ($\varphi$) and
substitute it with the discrete quantum number $J_z$
\beq\protect\label{PhiInt}
\langle x,\vec{k}_{\perp};\; J_z,\lambda_1,\lambda_2|\tilde{V}_{\rm eff}|
x',\vec{k}'_{\perp};\; J'_z, \lambda'_1,\lambda'_2\rangle:= \hspace{8cm}
\eeq
\[
\hspace{3cm}
\frac{1}{2\pi}\int_0^{2\pi}d\varphi \; e^{-i L_z \varphi}
\int_0^{2\pi}d\varphi'\; e^{i L_z' \varphi'}
\langle x,\kp,\varphi;\; \lambda_1,\lambda_2| V_{\rm eff}|
x',\kp',\varphi';\; \lambda'_1,\lambda'_2\rangle.
\]
We listed the general matrix elements 
$\langle x,\varphi;s_1,s_2| V_{\rm eff}|x',\varphi';s'_1,s'_2\rangle$ in Table
(\ref{GeneralHelicityTable}).
In general, the functions displayed there contain a dependence on the angles.
Hence, it is not clear how the general $\varphi$ dependence of the 
matrix elements will look like, if one inserts an arbitrary
$L_z=J_z-S_z$ into Eq.~(\ref{PhiInt}). Fortunately, a simple scheme can be 
set up to construct the functions for all $J_z= n, n\in \mbox{\sl\bf Z}$. 
In particular, one can prove that the matrix elements can only depend on 
the difference $\varphi-\varphi'$.
The general function has the shape
\[
F(x,\kp,x',\kp';\lambda_1,\lambda_2)=\frac{1}{2\pi}\int_0^{2\pi}d\varphi 
\int_0^{2\pi}d\varphi' 
\frac{\tilde{F}(x,\vkp,x',\vkp';\lambda_1,\lambda_2)}
{a- 2 k_{\perp}k'_{\perp}\cos(\varphi-\varphi')}
e^{i n (\varphi-\varphi')},
\]
where $n\in\mbox{\sl\bf Z}$ and
\begin{eqnarray*}
a&=&(x-x')^2\frac{m^2_f}{2}\left(\frac{1}{xx'}+\frac{1}{(1-x)(1-x')}\right)
     +k_{\perp}^2+k_{\perp}^{'2}\\
   &&-\frac{1}{2}(x-x')\left[k_{\perp}^{'2}
     \left(\frac{1}{1-x'}-\frac{1}{x'}\right)-k_{\perp}^{2}
     \left(\frac{1}{1-x}-\frac{1}{x}\right)\right]. 
\end{eqnarray*}
It is straightforward to evaluate this expression with the decomposition 
$\exp\{\pm i x\}= \cos x \pm i \sin x$ and the integrals
\beqa
\frac{1}{2\pi}\int_0^{2\pi}d\varphi\; 
\int_0^{2\pi}d\varphi' \,\frac{\cos\{n (\varphi-\varphi')\}}
{a- 2 k_{\perp}k'_{\perp}\cos(\varphi-\varphi')} &=& 
 2\pi(-A)^{-|n|+1}\left(\frac{B}{\kp\kp'}\right)^|n|,\\
\frac{1}{2\pi}\int_0^{2\pi}d\varphi\; 
\int_0^{2\pi}d\varphi' \,\frac{\sin\{n (\varphi-\varphi')\}}
{a- 2 k_{\perp}k'_{\perp}\cos(\varphi-\varphi')} &=& 0.
\eeqa
Here, the definitions
\[
A=\frac{1}{\sqrt{a^2-4k_{\perp}^2 k_{\perp}^{'2}}}\quad\quad \mbox{and} 
\quad\quad
B=\frac{1}{2}\left(1-aA\right),
\]
were used.
Using these relations, one can calculate 
the matrix elements for arbitrary $J_z$. The results
are listed in Table (\ref{HelicityTableJz}).

\section{The positronium spectrum for general $J_z$}

The positronium spectrum is calculated numerically from the matrix
elements of Table (\ref{HelicityTableJz}). Although we implemented a 
more general counterterm technology, described in Chapter 1, which
automatically accounts for the new features of the diagonal 
matrix elements for $J_z\neq0$, we must be even more careful.
In the former calculations ($J_z=0$), we made use 
of the {\em discrete} symmetries
$\cal C$ and $\cal H$, as described in Appendix \ref{AppxNumerics}, \S 2.
These symmetry properties are not explicitly conserved for the more general 
case $J_z\neq 0$. To be careful, we ignore possible point symmetries
in the problem, and solve for the spectrum without symmetrizing the 
Hamiltonian.
We will see in the next section how justified our extra care was.
The numerical effort increases enormously. 
With the unsymmetrized Hamiltonian, 
the dimensions of the matrices to be diagonalized are four times larger. 
Since the number of operations grows with the third power of 
the matrix dimensions
in a standard diagonalization algorithm, the CPU time used is 16 times
longer. 

\begin{figure}
\centerline{
\psfig{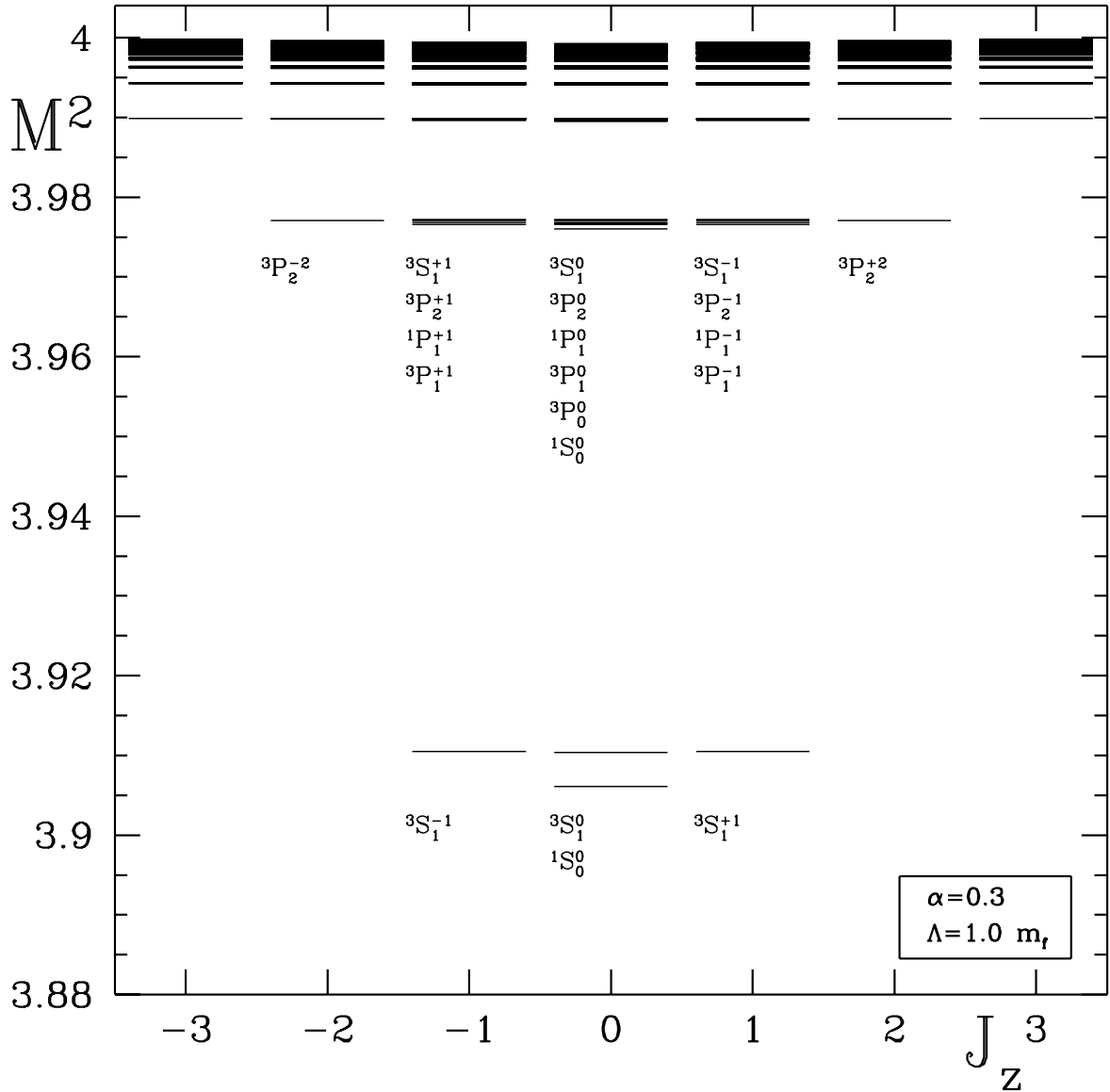}}
\caption[Compiled spectra for $J_z=-3,\ldots,+2,+3$]
{\protect\label{yrast}Compiled 
spectra of positronium with different $J_z=-3,-2,\ldots,+3$.
All spectra have been calculated with $\alpha=0.3, 
\Lambda=1.0\, m_f, N_1=N_2=21$.
The mass squared eigenvalues $M^2_n$ in units of the electron mass $m^2_f$
are shown.  
The notation for the states is $^{3S+1}L^{J_z}_J$.
The resolution of the plot is inadequate for the 
multiplets. Nevertheless, the numerical degeneracy of the three triplet ground
states $^3S^{-1}_1$,$^3S^0_1$, and $^3S^1_1$ becomes very clear.
} 
\end{figure}

We calculated spectra for the seven different values, $J_z=-3,-2,\ldots,+3$.
It is found that the eigenvalues are {\em identical} for $J_z{=}+n$ and 
$J_z{=}-n$ as
can be seen from Fig.\ (\ref{yrast}). 
The individual spectra and the convergence can be seen from Figures
(\ref{spectrumJ1})--(\ref{spectrumJ3}).
In particular, one notices that the singlet ground state is absent 
from all three plots, and that for $J_z{=}2$ and $J_z{=}3$ 
even more states are absent.
The explanation is the impossibility to have states
with $J_z=n$ in multiplets with $J<n$.
The numerical stability ({\em i.e.~}convergence) is very good: in each of the 
figures the lowest eigenvalue is almost independent of the 
number of integration points.
In fact, the eigenvalues converge exponentially with the number of integration
points, as we will show in Chapter \ref{DrAChapterAnnihilation}. 
This is the more surprising, as we adjusted the Coulomb 
counterterm\protect\footnote{{\em i.e.}, the converging function $g(p,p')$ was 
chosen in this manner, cf.\ Appendix \protect\ref{AppxNumerics}.},  
based on the non-relativistic ground state wavefunction, 
and the excited state wavefunctions are quite different.

Fig.~(\ref{yrast}), the summary of the spectra for different $J_z$, is 
central to this chapter. What can we learn from it?
It has two prominent features. 
Firstly, there are multiplets of states with
different $J_z$ that are degenerate. We shall discuss the 
numerical evidence in what follows.
Secondly, there is a limited odd number of degenerate states 
for each eigenvalue.
The interpretation of these facts is obvious. 
The positronium mass spectrum is a physical observable, 
Lorentz invariant 
and therefore independent of the 
mathematical algorithm applied and of the Lorentz frame used. 
Central forces are rotationally invariant, and this should be observed
in the spectrum of an electromagnetic bound system, too. 
Rotational symmetry tells us that there has to be a defined number
of degenerate states for each fixed value of the total angular momentum 
$\vec{J}$. 
Conversely, since this is exactly what we observe here, we can infer the
quantum number $J$ from the number of degenerate states for a fixed 
eigenvalue $M^2_n$. 
Concluding, the $1,3,5,\ldots$ degenerate states constitute the singlets,
triplets, pentuplets, $\ldots$ of a $J=0,1,2,\ldots$ multiplet.

\begin{figure}
\centerline{
\psfig{figure=\graphpath spectrum_J1.epsi,width=10cm,angle=0}}
\protect\caption[Positronium spectrum of the $J_z{=}+1$ sector]
{\protect\label{spectrumJ1}
The spectrum of the effective integral equation for $\alpha=0.3, 
\Lambda=1.0\, m_f, J_z=1$. 
The mass squared eigenvalues $M^2_n$ in units of the electron mass $m^2_f$
are shown as functions of the number of integration points $N\equiv N_1=N_2$.
The calculation was done using the entirely numerically integrated
Coulomb counterterms.
Note that the singlet ground state is absent. For $n{=}1$ only the triplet
$^3S_1$ survives the projection on the $J_z{=}1$ sector. Cf.\ 
Fig.~(\protect\ref{spectrumJ0newCT}).}
\vspace{0.5cm}

\begin{minipage}[t]{75mm} 
\centerline{
\psfig{figure=\graphpath spectrum_J2.epsi,width=7.5cm,angle=0}}
\caption[Positronium spectrum of the $J_z=+2$ sector]
{\protect\label{spectrumJ2}The spectrum of the effective integral equation for $\alpha=0.3, 
\lambda=1.0\, m_f, J_z=2$.
Note the total absence of any $n=1$ state.}
\end{minipage}
\hfill
\begin{minipage}[t]{75mm}
\centerline{
\psfig{figure=\graphpath spectrum_J3.epsi,width=7.5cm,angle=0}}
\caption[Positronium spectrum of the $J_z=+3$ sector]
{\protect\label{spectrumJ3}
The spectrum of the effective integral equation for $\alpha=0.3, 
\Lambda=1.0\, m_f, J_z{=}3$.
Even more states are not compatible with $J_z{=}3$.}
\end{minipage}
\end{figure}


\begin{figure}[ht]
\begin{minipage}[t]{15.5cm} 
\centerline{
\psfig{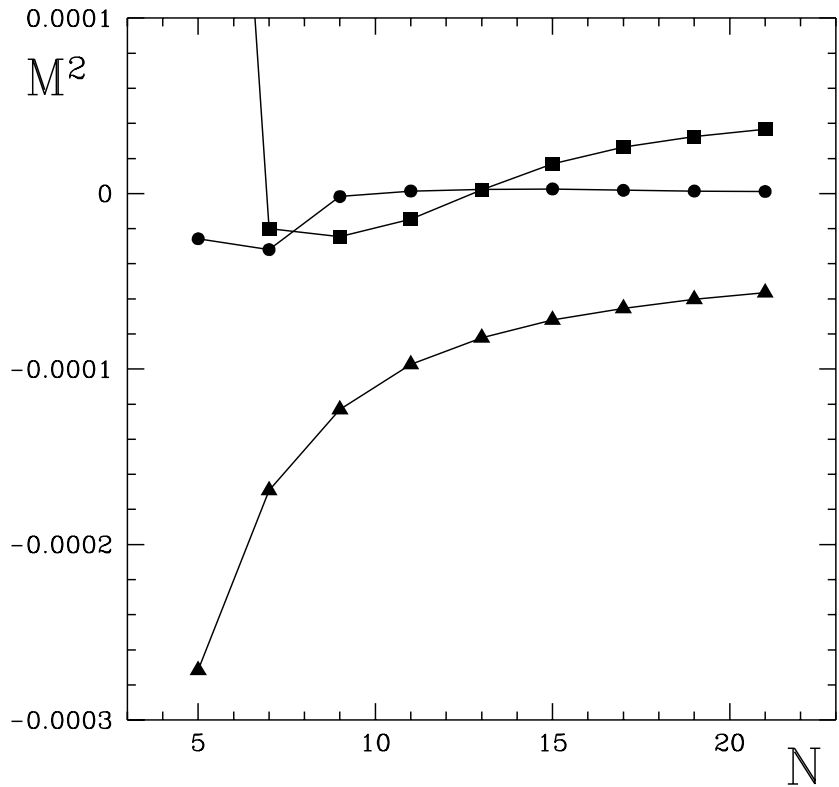}}
\centerline{
\parbox{12cm}{
\caption[Deviation of eigenvalues for $J_z{=}0$ and $J_z{=}1$]
{\protect\label{diff}Deviation of 
corresponding eigenvalues for $J_z{=}0$ and $J_z{=}1$ multiplets
($\alpha=0.3, \Lambda=1.0\, m_f$) with growing number of integration points. 
The graphs show 
$\Delta M^2:= M^2_n(J_z{=}0)-M^2_n(J_z{=}1)$
for the states $1^3S_1\,(\triangle)$, $2^3P_1\,(\Box)$, and $2^1P_1\,(\circ)$.} 
}}
\end{minipage}
\end{figure}
In Table (\ref{TablespecJz}), the spectrum obtained for $J_z{=}1$ is
compiled to compare it with the eigenvalues for $J_z{=}0$. Apart 
from the absence of the states with $J{=}0$, the table displays
an almost complete coincidence of the corresponding states.
Only for the triplet $1^3S_1$, the gap between the two states of different
$J_z$ is
bigger, though the numerical error is actually smaller. 

We have to investigate the significance of the degeneracy with respect to 
the number of integration points and with respect to the cutoff $\Lambda$.
To find out if this discrepancy is merely a numerical artifact, or
a property of the positronium model, consider Figure (\ref{diff}).
Here, the mass (squared) discrepancy between the $J_z{=}0$ and $J_z{=}1$ 
eigenvalues is plotted versus the number of integration points $N$ 
for three different states.
One notices the convergence of $\Delta M^2(1^3S_1)$ with $N$. 
The curve does not converge to zero, as one would want, but
to a value of $\Delta M^2(^3S_1)\simeq -5\times 10^{-5}$.
The mass gap for $2^1P_1$ does, however, go to zero as $N$ grows large.
The mass gap of the other helicity-triplet state $2^3P_1$ has 
the same increase as $\Delta M^2(1^3S_1)$, if one 
disregards the behavior of the corresponding
graph for the untrustworthy values $N{=}5$ and $N{=}7$. 
It converges to 
$4\times 10^{-5}$ as $N$ increases.
We mention that {\sc Kalu\v za} and {\sc Pirner} \cite{KalPir} find
that in light-cone perturbation theory there is a discrepancy between the 
case of $J_z{=}0$ and $J_z{=}1$. This is due to the perturbative
method applied there and even expected from the point of view taken in
the present work. Perturbation theory to any finite order breaks the
symmetries of the theory. Since rotations contain the interaction in front 
form dynamics, the associated symmetry will be broken in a perturbative
approach.

\subsection*{Cutoff dependence}

\begin{figure}
\begin{minipage}[t]{15.5cm} 
\centerline{
\psfig{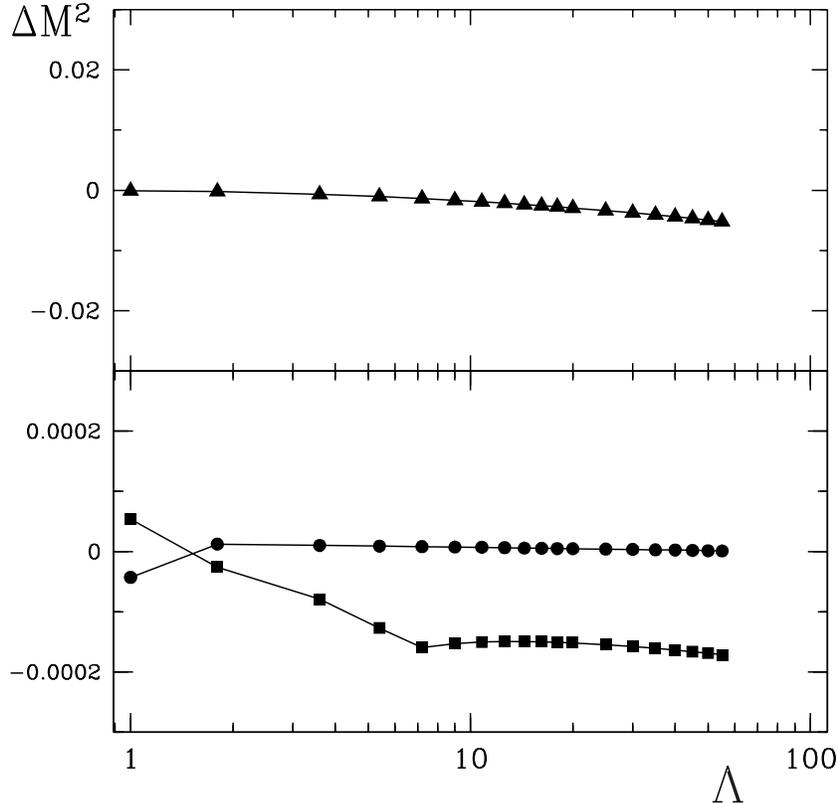}}
\centerline{
\parbox{12cm}{
\caption[Deviation of eigenvalues as function of the cutoff]
{\protect\label{difflambda}Deviation of 
corresponding eigenvalues for $J_z{=}0$ and $J_z{=}1$
($\alpha=0.3$) as functions of the cutoff $\Lambda$. The graphs show 
$\Delta M^2:= M^2_n(J_z{=}0)-M^2_n(J_z{=}1)$
for the states $1^3S_1\,(\triangle)$ [upper box], and 
 $2^3P_1\,(\Box)$, and $2^1P_1\,(\circ)$ [lower box].
Note that the scales of the two boxes differ by a factor of 100!}
}} 
\end{minipage}
\end{figure}

For the cutoff dependence of the eigenvalues for non-vanishing $J_z$ 
the statements
given in the context of Chapter \ref{ChapterModel} hold equally. 
A main result of the present 
work is the documentation of the restoration of rotational symmetry in 
the solution in front form dynamics.
Consequently, the investigation 
of the degeneracy of eigenvalues of same total angular
momentum, but different $J_z$ is crucial.
The discrepancy between corresponding eigenvalues of $J_z{=}0$
and $J_z{=}1$ as functions of the cutoff 
is displayed in Fig.~(\ref{difflambda}).
One notices that there is a slight deviation of the cutoff dependence
of the triplets $1^3S_1$ for different $J_z$. The deviations of the 
other states ($2^3P_1$ and $2^1P_1$) are suppressed by roughly 
a factor of 100.  

This weak dependence of $\Delta M^2_n$ on the cutoff will be suppressed 
even more, if the annihilation channel is included. 


\newpage
\begin{table}[H]
\centerline{
\begin{tabular}{|r|c||c|c|r|}\hline
\rule[-3mm]{0mm}{8mm}$n$ & Term & $B_n(J_z{=}0)$ & $B_n(J_z{=}+1)$ & $\hfill 10^5\;\Delta B_n\hfill$ \\ \hline \hline
 1& $1^1S_0$& 1.049553 $\pm$ 0.000017 & --- & $\hfill\mbox{---}\hfill$ \\ \hline 
  2&$1^3S_1$ & 1.001012 $\pm$ 0.000111 & 1.000376 $\pm$ 0.000071& $ 63.53$ \\ \hline 
\hline
 3& $2^1S_0$& 0.260237 $\pm$ 0.000169 & --- & $\hfill\mbox{---}\hfill$ \\ \hline 
  4&$2^3S_1$ & 0.253804 $\pm$ 0.000217 & 0.253720 $\pm$ 0.000208& $  8.33$ \\ \hline 
  5&$2^1P_1$ & 0.257969 $\pm$ 0.000161 & 0.257982 $\pm$ 0.000166& $ -1.30$ \\ \hline 
 6& $2^3P_0$& 0.267070 $\pm$ 0.000156 & --- & $\hfill\mbox{---}\hfill$ \\ \hline 
  7&$2^3P_1$ & 0.259667 $\pm$ 0.000206 & 0.260075 $\pm$ 0.000159& $-40.80$ \\ \hline 
  8&$2^3P_2$ & 0.255258 $\pm$ 0.000177 & 0.255253 $\pm$ 0.000172& $  0.47$ \\ \hline 
\hline
 9& $3^1S_0$& 0.115206 $\pm$ 0.000314 & --- & $\hfill\mbox{---}\hfill$ \\ \hline 
 10&$3^3S_1$ & 0.113441 $\pm$ 0.000363 & 0.113413 $\pm$ 0.000261& $  2.79$ \\ \hline 
 11&$3^1P_1$ & 0.114490 $\pm$ 0.000270 & 0.114529 $\pm$ 0.000282& $ -3.96$ \\ \hline 
12& $3^3P_0$& 0.117133 $\pm$ 0.000271 & --- & $\hfill\mbox{---}\hfill$ \\ \hline 
 13&$3^3P_1$ & 0.115127 $\pm$ 0.000326 & 0.115116 $\pm$ 0.000273& $  1.13$ \\ \hline 
 14&$3^3P_2$ & 0.113717 $\pm$ 0.000281 & 0.113719 $\pm$ 0.000280& $ -0.26$ \\ \hline 
 15&$3^1D_2$ & 0.112816 $\pm$ 0.000150 & 0.112842 $\pm$ 0.000158& $ -2.66$ \\ \hline 
 16&$3^3D_1$ & 0.113427 $\pm$ 0.000155 & 0.113496 $\pm$ 0.000277& $ -6.90$ \\ \hline 
 17&$3^3D_2$ & 0.112978 $\pm$ 0.000161 & 0.112982 $\pm$ 0.000162& $ -0.43$ \\ \hline 
 18&$3^3D_3$ & 0.112511 $\pm$ 0.000156 & 0.112515 $\pm$ 0.000158& $ -0.41$ \\ \hline 
\hline
19& $4^1S_0$& 0.065490 $\pm$ 0.000588 & --- & $\hfill\mbox{---}\hfill$ \\ \hline 
 20&$4^3S_1$ & 0.064786 $\pm$ 0.000598 & 0.064774 $\pm$ 0.000596& $  1.25$ \\ \hline 
 21&$4^1P_1$ & 0.065003 $\pm$ 0.000467 & 0.065062 $\pm$ 0.000486& $ -5.91$ \\ \hline 
22& $4^3P_0$& 0.066115 $\pm$ 0.000470 & --- & $\hfill\mbox{---}\hfill$ \\ \hline 
 23&$4^3P_1$ & 0.065331 $\pm$ 0.000487 & 0.065282 $\pm$ 0.000475& $  4.88$ \\ \hline 
 24&$4^3P_2$ & 0.064700 $\pm$ 0.000478 & 0.064706 $\pm$ 0.000480& $ -0.65$ \\ \hline 
 25&$4^1D_2$ & 0.063968 $\pm$ 0.000294 & 0.064041 $\pm$ 0.000317& $ -7.32$ \\ \hline 
 26&$4^3D_1$ & 0.064262 $\pm$ 0.000308 & 0.064371 $\pm$ 0.000348& $-10.97$ \\ \hline 
 27&$4^3D_2$ & 0.064099 $\pm$ 0.000319 & 0.064088 $\pm$ 0.000314& $  1.06$ \\ \hline 
 28&$4^3D_3$ & 0.063864 $\pm$ 0.000303 & 0.063875 $\pm$ 0.000305& $ -1.07$ \\ \hline 
 29&$4^1F_3$ & 0.063141 $\pm$ 0.000096 & 0.063106 $\pm$ 0.000103& $  3.51$ \\ \hline 
 30&$4^3F_2$ & 0.063297 $\pm$ 0.000112 & 0.063232 $\pm$ 0.000116& $  6.52$ \\ \hline 
 31&$4^3F_3$ & 0.063234 $\pm$ 0.000119 & 0.063210 $\pm$ 0.000118& $  2.32$ \\ \hline 
 32&$4^3F_4$ & 0.063098 $\pm$ 0.000103 & 0.063422 $\pm$ 0.000158& $-32.44$ \\ \hline 
\hline
33& $5^1S_0$& 0.043253 $\pm$ 0.001259 & --- & $\hfill\mbox{---}\hfill$ \\ \hline 
 34&$5^3S_1$ & 0.042913 $\pm$ 0.001278 & 0.042915 $\pm$ 0.000740& $ -0.19$ \\ \hline 
 35&$5^1P_1$ & 0.042840 $\pm$ 0.000641 & 0.042905 $\pm$ 0.000675& $ -6.53$ \\ \hline 
\hline
\end{tabular}}

\protect\caption[Spectrum $J_z{=}1$: comparison to $J_z{=}0$ ]{\protect\label{TablespecJz}The positronium spectrum for $\alpha=0.3,$ $\Lambda=1.0\, m_f, N_1=N_2=21$.
The non-relativistic notation for the terms and the binding coefficients for $J_z{=}0$ and $J_z{=}+1$ are shown. The numerical errors are estimated from the difference between the values for maximum and next to maximum number of integration points. In fact they are smaller, because 
the eigenvalues converge exponentially with the number 
 of integration points.}

\end{table}

\section{Wavefunctions}

The $J_z\neq 0$ wavefunctions show more structure than
those with $J_z{=}0$, due to the lower symmetry.
In the $J_z{=}0$ case, there are essentially two different components
of the wavefunctions: one for parallel, the other for anti-parallel helicities.
Consequently, there are only two plots shown for the singlet and triplet
wavefunctions in Chapter 1: Figs.~(\ref{singletud}), (2.11)
and Figs.~(2.12), (\ref{tripletuu}), respectively.

This is not a consequence of the symmetrized Hamiltonian. 
If the non-symmetrized $J_z{=}0$ Hamiltonian matrix is diagonalized, 
the same eigenfunctions are found as in the symmetric case, 
but twice as many occur. The only difference is a sign, 
depending on the parity quantum number.

When $J_z{\neq}0$, we encounter four distinct components of 
each of the eigenfunctions, 
corresponding to four different helicity combinations.
We elaborate on this subject by considering the components of the 
triplet wavefunction for $J_z{=}1$, Figs.~(\ref{wftriplet})(a)-(d), 
and 
that of the next higher state, Figs.~(\ref{wfexcited})(a)-(d).
In both cases, the components for anti-parallel helicities are identical,
though displayed differently to show their full shape.
For the triplet, 
the components with parallel helicities have nothing in common: the 
($\uparrow\uparrow$)-component peaks at $x{=}0.5,\, \kp{=}0$ and is 
almost rotationally invariant, whereas the ($\downarrow\downarrow$)-component
vanishes at $\kp{=}0$ and is shaped more like the components with anti-parallel
helicities. Note the extremely differing peak values:
the anti-parallel components 
are suppressed by a factor of 40, compared to the 
($\uparrow\uparrow$)-component, the ($\downarrow\downarrow$)-component 
even by a factor of 1400!
As with the $J_z{=}0$ sector, 
Eq.~(\ref{Normalization}), the normalization is
\[
\sum_{i=1}^{N_1}\sum_{j=1}^{N_2}\sum_{\lambda_1,\lambda_2}
\psi(\mu_i,\cos\theta_j;\; \lambda_1,\lambda_2)=1.
\] 
The missing symmetry of the components is a consequence of the 
missing symmetry of the $J_z\neq0$ sector with respect to $\cal T$-parity. 
This property is found in all wavefunctions of these sectors. We have displayed 
here the wavefunction(s) of the next excited state to show another important
fact: the wavefunction Fig.~(\ref{wfexcited})(a) has a small, but noticeable
deviation from the reflection symmetry with respect to the $x=0.5$ plane.  
This symmetry around $x{=}0.5$ is due to the charge conjugation symmetry 
$\cal C$: if one permutes particle and antiparticle, one substitutes
$x$ with $1-x$. The fact that this property is respected by all wavefunctions
other than Fig.~(\ref{wfexcited})(a) shows, that this symmetry is not broken,
even not in the $J_z{\neq}0$ sector. 
The slight deviations can be
explained as numerical errors, or more likely, as errors due to 
the {\tt grid} function of the graphing package used.
Some examples for higher excitations and larger $J_z$ are given in 
Fig.(\ref{wfdivers}). 

\begin{figure}[t]
\centerline{
\psfig{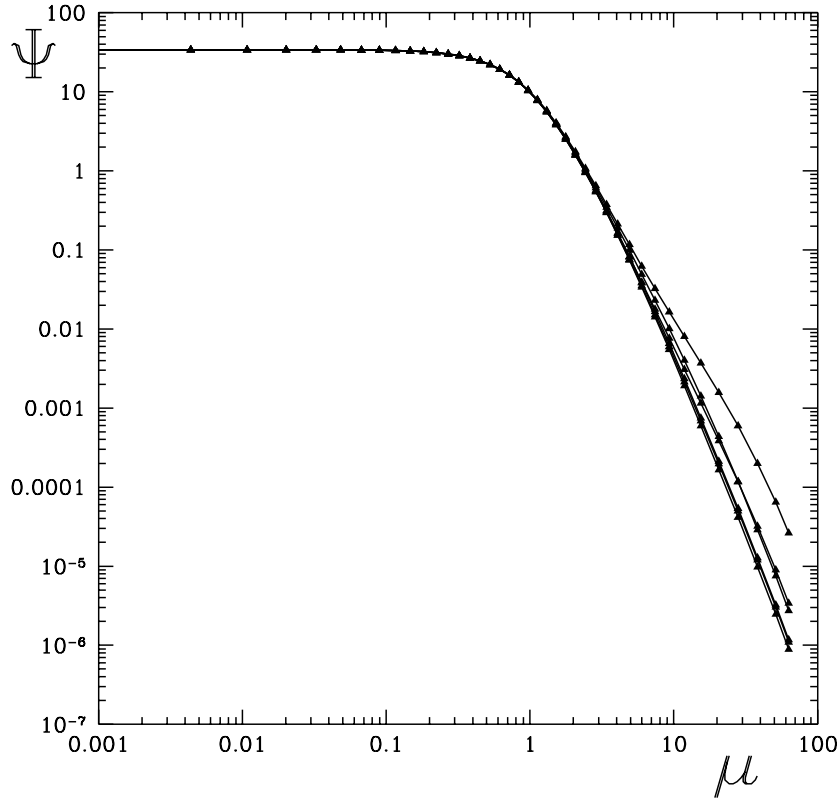}}
\centerline{
\parbox{12cm}{
\caption[Decrease of the $J_z=+1$ triplet ground state wavefunction]
{\protect\label{Jz1fig11}The decrease of the 
$J_z=+1$ triplet ground state wavefunction with parallel helicities 
as a function of the momentum variable $\mu$. The parameters are
$\alpha=0.3$, $\Lambda=20.0\, m_f$, $N_1=41$, $N_2=11$. 
There are six different curves corresponding to six values of the discretized
angle variable, $\theta$: they show the decrease in $\psi$ with increasing 
$\mu$. Notice
the deviations from rotational symmetry for $\mu\ge 10$.}
}}
\end{figure}

The decrease of the triplet wavefunction $1^3S_1(\uparrow\uparrow)$ 
with parallel helicities is plotted in Fig. (\ref{Jz1fig11}).
These curves for different $\cos\theta_j$ have to be compared to those 
of Fig.~(\ref{fig11}). 
In both cases, rotational symmetry is broken, since the decrease 
of the wavefunction is not isotropic but depends on the angle $\theta$.
There are some differences between the sectors $J_z{=}0$ and $J_z{=}1$.
One is the fact that the smallest value of the wavefunction 
for $J_z{=}1$ is roughly $9\times 10^{-6}$, whereas for $J_z{=}0$ 
its approximately three 
times smaller. Another difference is the value of $\mu$ from that  
on the deviations
in the decreases become noticeable. For $J_z{=}0$ this value is at 
$\mu\simeq 10$, contrary to $\mu\simeq 3$ in the $J_z{=}1$ sector. Moreover,
the curves of different $\cos\theta$ seem to be grouped for $J_z{=}1$.
In any case, the important result is the same as in the case of $J_z{=}0$: the
wavefunctions are not rotationally invariant.


\begin{figure}
\centerline{
\psfig{figure=\graphpath wf_0j10.epsi,width=8cm,angle=0}
\psfig{figure=\graphpath wf_0j11.epsi,width=8cm,angle=0}
}
\vspace{2cm}

\centerline{
\psfig{figure=\graphpath wf_0j12.epsi,width=8cm,angle=0}
\psfig{figure=\graphpath wf_0j13.epsi,width=8cm,angle=0}
}
\vspace{2cm}

\caption[The four components of the $1^3S_1$ wavefunction for $J_z=+1$]
{\protect\label{wftriplet}The triplet 
ground state wavefunction for $J_z{=}+1$ as a function of 
the longitudinal momentum fraction $x$ and the transverse momentum $\kp$,
omitting the dependence on the angle $\varphi$. The
calculation was done with $\alpha=0.3,\Lambda=1.0\, m_f, N_1=41, N_2=11$.
Shown are: (a) ($\uparrow\uparrow$)-component, 
(b) ($\uparrow\downarrow$)-component, (c) ($\downarrow\uparrow$)-component,
(d) ($\downarrow\downarrow$)-component.}
\end{figure}


\begin{figure}
\centerline{
\psfig{figure=\graphpath wf_1j10.epsi,width=8cm,angle=0}
\psfig{figure=\graphpath wf_1j11.epsi,width=8cm,angle=0}
}
\vspace{2cm}

\centerline{
\psfig{figure=\graphpath wf_1j12.epsi,width=8cm,angle=0}
\psfig{figure=\graphpath wf_1j13.epsi,width=8cm,angle=0}
}
\vspace{2cm}

\caption[The four components of the $2^3P_1$ wavefunction for $J_z=+1$]
{\protect\label{wfexcited}
The $2^3P_1$ wavefunction for $J_z{=}+1$ as a function of 
the longitudinal momentum fraction $x$ and the transverse momentum $\kp$,
omitting the dependence on the angle $\varphi$. The
calculation was done with $\alpha=0.3,\Lambda=1.0\, m_f, N_1=41, N_2=11$.
Shown are: (a) ($\uparrow\uparrow$)-component, 
(b) ($\uparrow\downarrow$)-component, (c) ($\downarrow\uparrow$)-component,
(d) ($\downarrow\downarrow$)-component.}
\end{figure}


\begin{figure}
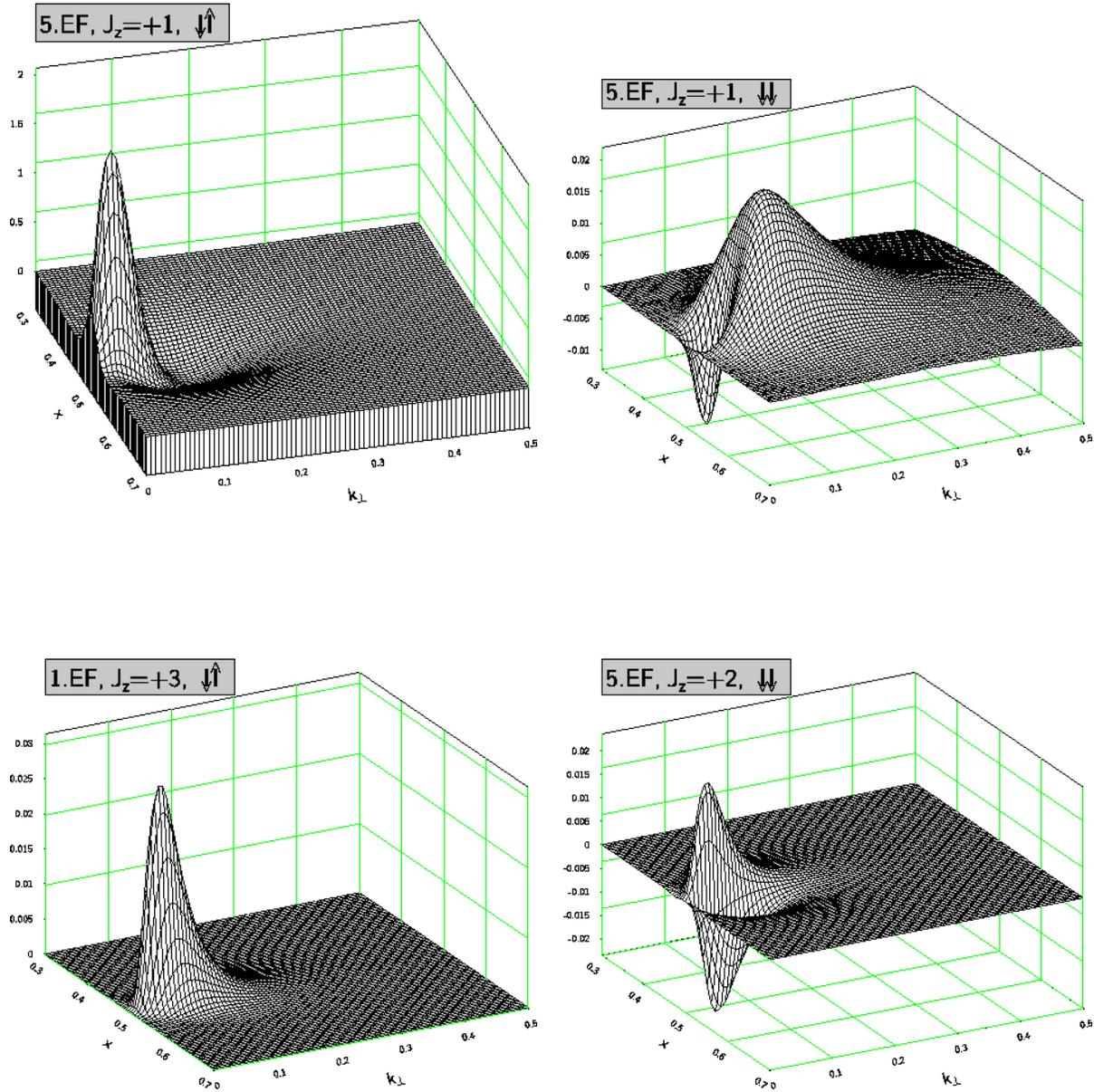

\centerline{
\psfig{figure=\graphpath wf_4j11.epsi,width=8cm,angle=0}
\psfig{figure=\graphpath wf_4j13.epsi,width=8cm,angle=0}
}
\vspace{2cm}

\centerline{
\psfig{figure=\graphpath wf_0j33.epsi,width=8cm,angle=0}
\psfig{figure=\graphpath wf_4j23.epsi,width=8cm,angle=0}
}
\vspace{2cm}

\caption[Components of $2^3S_1$, $2^3S_1$, $3^3D_3$, and $3^1D_2$ for $J_z=+1$]
{\protect\label{wfdivers}Components 
of wavefunctions for larger values of $J_z$ as a function of 
the longitudinal momentum fraction $x$ and the transverse momentum $\kp$,
omitting the dependence on the angle $\varphi$. The
calculation was done with $\alpha=0.3,\Lambda=1.0\, m_f, N_1=41, N_2=11$.
Shown are: (a) ($\uparrow\downarrow$)-component of  $2^3S_1$ , 
(b) ($\downarrow\downarrow$)-component of $2^3S_1$, 
(c) ($\downarrow\uparrow$)-component of $3^3D_3$,
(d) ($\downarrow\downarrow$)-component of $3^1D_2$.}
\end{figure}

\chapter{The annihilation channel}
\label{DrAChapterAnnihilation}

\section{Introduction}


\def\d{$\bullet$}  \def\v{ V }  \def\b{ $\cdot$ } \def\s{ S }   \def\f{ F }
\begin{figure}[t]
\centerline{
\begin {tabular}{|l|r||cc|ccc|cccc|ccccc|}
\hline 
  \rule[-3mm]{0mm}{8mm}  Sector & $n$ & 
     0 & 1 & 2 & 3 & 4 & 5 & 6 & 7 & 8 & 9 &10 &11 &12 &13 
\\ \hline \hline   
    $ |\gamma \rangle$ &  0 & 
    \d & \v &\b &\f &\b &\b &\b &\b &\b &\b &\b &\b &\b &\b 
\\
    $ |e\bar e\rangle $ &  1 & 
    \v &\d &\s &\v &\f &\b &\f &\b &\b &\b &\b &\b &\b &\b 
\\ \hline
    $ |\gamma\ \gamma\rangle$ &  2 & 
    \b & \s &\d &\v &\b &\v &\f &\b &\b &\b &\b &\b &\b &\b 
\\
    $ |e\bar e\, \ \gamma \rangle$ &  3 & 
    \f &\v &\v &\d &\v &\s &\v &\f &\b &\b &\b &\b &\b &\b 
\\
    $ |e\bar e\, e\bar e\rangle $ &  4 & 
    \b&\f &\b &\v &\d &\b &\s &\v &\f &\b &\b &\f &\b &\b 
\\ \hline
    $ |\gamma \ \gamma \ \gamma \rangle$ &  5 & 
    \b& \b &\v &\s &\b &\d &\v &\b &\b &\v &\f &\b &\b &\b 
\\
    $ |e\bar e\, \ \gamma \ \gamma\rangle $ &  6 & 
    \b&\f &\f &\v &\s &\v &\d &\v &\b &\s &\v &\f &\b &\b 
\\
    $ |e\bar e\, e\bar e\, \ \gamma \rangle$ &  7 & 
    \b &\b &\b &\f &\v &\b &\v &\d &\v &\b &\s &\v &\f &\b 
\\
    $ |e\bar e\, e\bar e\, e\bar e\rangle $ &  8 & 
    \b&\b &\b &\b &\f &\b &\b &\v &\d &\b &\b &\s &\v &\f 
\\ \hline
    $ |\gamma \ \gamma \ \gamma \ \gamma \rangle$ & 9 & 
    \b&\b &\b &\b &\b &\v &\s &\b &\b &\d &\v &\b &\b &\b 
\\
     $ |e\bar e\, \ \gamma \ \gamma \ \gamma \rangle$ & 10 & 
    \b&\b &\b &\b &\b &\f &\v &\s &\b &\v &\d &\v &\b &\b 
\\
    $ |e\bar e\, e\bar e\, \ \gamma \ \gamma \rangle$ & 11 & 
    \b&\b &\b &\b &\f &\b &\f &\v &\s &\b &\v &\d &\v &\b 
\\
     $ |e\bar e\, e\bar e\, e\bar e\, \ \gamma \rangle$ & 12 & 
    \b&\b &\b &\b &\b &\b &\b &\f &\v &\b &\b &\v &\d &\v 
\\
     $ |e\bar e\, e\bar e\, e\bar e\, e\bar e\rangle$ & 13 & 
    \b&\b &\b &\b &\b &\b &\b &\b &\f &\b &\b &\b &\v &\d 
\\ \hline
\end {tabular}
}
\caption[The Hamiltonian matrix for QED]{\label{HolyMatrixQED}The 
Hamiltonian matrix for QED. 
The sectors $n$ are numbered starting at zero.
The vertex,
seagull and fork interactions are denoted by V, S, F respectively. Diagonal
matrix elements are symbolized by \d . This table is courtesy of 
{\sc H.-C.~Pauli}.}
\end{figure}

So far, we have considered an effective  Fock space, consisting of two sectors
$|e\bar{e}\rangle$ and $|e\bar{e}\gamma\rangle$. We shall  
justify the absence of any 
higher\protect\footnote{Or rather the substitution of effects of higher 
Fock sectors with the use of {\em effective} matrix elements in the 
remaining sectors.``Higher'' here in the sense of ascending $n$ in Table
(\protect\ref{HolyMatrixQED}).} Fock state in Chapter
\ref{DrAChapterEffInt} from the structure of the applied
formalism of effective interactions. 
The general formalism \cite{PauliMIR} is set up for a 
non-abelian $SU(N)$ gauge theory. Unlike QED, the one boson
state is absent in these theories because of  color neutrality. 
Nevertheless, one has to take care of the
one photon state $|\gamma\rangle$ in QED, and the proceeding to include it 
is as follows.

Firstly, it is important to notice not only the differences between 
the QED Table (\ref{HolyMatrixQED}) and the analogous QCD table in 
\cite[Fig.~1]{PauliMIR}, but also
the similarities. 
Some of the graphs occuring in the QCD case are absent in QED, in particular
those graphs with a three- or four-boson interaction and 
the instantaneous interactions connecting four bosons. 
But, although an additional sector occurs as a first row and a first 
column in the 
QED table,
neither is a change in the higher Fock sectors observed, nor is the ordering 
altered in any way.

In addition to Eqs.~(\ref{Pspace}) and (\ref{Qspace}), the $N$-space 
({\em  i.e.~}the sector 
containing the $|\gamma\rangle$ states)
is added to the system. The corresponding projector is
\[
\hat{N} := \sum_{n\atop {\rm all\;\; QN}} 
|(\gamma)_n\rangle \langle (\gamma)_n|.
\]
The whole procedure of subsequent projections of higher Fock states onto the 
remaining Hilbert space of states can be carried out like before until 
one arrives at a $(2\times 2)$ matrix analogous to (\ref{2x2Matrix}),
but operating in the $N$- and $P$-space rather than in the $P$- and $Q$-space
\beq\label{NPSpace}
H_{\rm LC}\; \psi= \left(
                \begin{array}{cc}
                        H_{NN} & H_{NP}\\
                        H_{PN} & H_{PP}\\
                \end{array}
             \right) 
                \left(
                \begin{array}{c}
                        \psi_{\gamma}\\
                        \psi_{e\bar{e}}\\
                \end{array}
             \right) =\omega
                \left(
                \begin{array}{c}
                        \psi_{\gamma}\\
                        \psi_{e\bar{e}}\\
                \end{array}
             \right). 
\eeq
From Table (\ref{HolyMatrixQED}) one can read off the 
interaction of the one photon
state with all other sectors: the vertex interaction annihilates the photon
into an electron-positron pair and a fork interaction scatters it into the 
sector $|e\bar{e}\gamma\rangle$. The latter interaction is already contained
in the effective interaction Eq.~(\ref{NPSpace}) because of the projection
of the $Q$-space. 

Although we always projected the higher Fock sectors onto the lower 
ones up to now, one is, of course, free to project the (lower) 
$|\gamma\rangle$-sector
onto the (higher) $|e\bar{e}\rangle$-sector,
and obtains 
\[
H_{\rm eff}(\omega)= H_{PP} 
+ H_{PN}\frac{1}{\omega-H_{NN}}H_{NP}
+ H_{PQ}\frac{1}{\omega-H_{QQ}}H_{QP} 
\]

\begin{figure}
\begin{minipage}[t]{7cm}
\centerline{
\psfig{figure=\graphpath annihilation.epsi,width=5cm,angle=-90}}
\caption[The dynamical annihilation graph]
{\label{annihilation}The dynamical annihilation graph.}
\end{minipage} \hfill
\begin{minipage}[t]{7cm}
\centerline{
\psfig{figure=\graphpath anni_seag.epsi,width=5cm,angle=-90}}
\caption[The seagull graph of the annihilation channel]
{\label{anniseag}The seagull graph of the annihilation channel.}
\end{minipage}
\end{figure} 
Of course we do this for convenience; we could just as well solve the 
eigenvalue problem of Eq.~(\ref{NPSpace}).
The projection is depicted in two graphs. One is the {\em dynamic annihilation
graph}, Fig.~(\ref{annihilation}), the other is the corresponding {\em seagull
annihilation graph}, Fig.~(\ref{anniseag}). The latter is a $P$-space graph
and was absent before because of the gauge principle of {\sc Tang}.

\section{Calculation of the matrix elements}

The calculation is straightforward. The matrix elements of the canonical
Hamiltonian are given in Appendix \ref{AppxHLC}.
Up to now, we had to consider merely the first type of vertex interaction
, cf.~Table (\ref{ContractionTable}),
where a photon is irradiated from a fermion. Now the other graph must
be evaluated.
The calculation of the graphs [Fig.~(\ref{annihilation})--(\ref{anniseag})]
is performed in the $(J_z{=}0)$ and $(J_z{=}\pm 1)$ sectors. 
The graph is absent in all
other sectors because of the helicity of the photon: no angular momentum larger
than $J=1$ is possible.
The functions derived depend on the light-cone momenta $(x,\,\vec{k}_{\perp})$.
They are given in Table 
(\ref{HelicityTableAnnihilation}) and have to be added to those of 
Table (\ref{HelicityTableJz}). 
The energy denominator in the one photon sector is simple because the 
photon has zero mass:
\beq\label{Energienenner}
G(\omega) = \frac{1}{\omega - H_{\gamma}}= \frac{1}{T^*},
\eeq
where we used the definition (\ref{OmegaStar}).
Note that this denominator does {\em not} depend on the directions of the 
vectors $\vkp$, $\vkp'$, {\em i.e.}~on the angles $\varphi$, $\varphi'$.

The calculation of the dynamic annihilation graph follows the
steps described in Appendix \ref{AppxCalculation}. We mention 
only some aspects here.
The transversal photon momentum vanishes:
\[
\vec{k}_{\perp\gamma}=\vec{k}_{\perp e}+\vec{k}_{\perp\bar{e}}=0,
\]
and the longitudinal momentum $x_{\gamma}$ is unity.
Consequently, all expressions of the form 
\[
\frac{\vec{k}_{1\perp}}{x_1}\equiv 0
\]
vanish and $1/\sqrt{x_1}\equiv 1$ becomes trivial.
The matrix elements of the vertex interaction $V_{g\rightarrow q\bar{q}}$  
[Table (\ref{VertexTable})], split up into their three components with
different helicity factors, are
\beqa
\langle e\bar{e}|V_A|\gamma\rangle&=&-m\sqrt{\beta}
\left(\frac{1}{x}+\frac{1}{1-x}\right)
\times\delta^{+\lambda_1}_{+\lambda_2}\delta^{+\lambda_1}_{+\lambda_3},\\
\langle e\bar{e}|V_B|\gamma\rangle&=&-\sqrt{\beta}
\frac{\kp}{1-x}e^{+i\lambda_2\varphi}
\times\delta^{+\lambda_1}_{+\lambda_2}\delta^{+\lambda_1}_{-\lambda_3},\\
\langle e\bar{e}|V_C|\gamma\rangle&=&\sqrt{\beta}
\frac{\kp}{x}e^{-i\lambda_2\varphi}
\times\delta^{+\lambda_1}_{-\lambda_2}\delta^{+\lambda_1}_{+\lambda_3}.\\
\eeqa
The complete matrix elements are obtained by symmetry, since
$
\langle e\bar{e}|V_i|\gamma\rangle=\langle \gamma|V_i|e\bar{e}\rangle^*.
$
To substitute the angles,
we multiply the functions 
$\langle e\bar{e}|V_i|\gamma\rangle \;G(\omega)\;
\langle \gamma|V_j|e\bar{e}\rangle$
by 
\beq\label{Factor}
e^{-i(L_z\varphi-L_z'\varphi')} 
\eeq
following Eq.~(\ref{PhiInt}) 
and integrate over $\varphi$ and $\varphi'$.
It turns out that because of the simple energy denominator
(\ref{Energienenner}), the only dependence on the angles comes from the
factor (\ref{Factor}) and is proportional to 
$\cos\{\varphi-\varphi'\}$
or $\sin\{\varphi-\varphi'\}$. As a result, all 
matrix elements of the {\em dynamic} annihilation graph for $J_z{=}0$ 
vanish when integrated over the angles.
Only for $J_z{=}\pm 1$ do some of the matrix elements survive the integration. 

Because of the combination of matrix elements with the factor
(\ref{Factor}), two types of functions emerge for $J_z{=}\pm1$:
one is independent of the angles, the other has a dependence 
proportional to $\exp\{\pm 2i(\varphi-\varphi')\}$.
The latter vanishes after angular integration. The helicity table 
(\ref{SymbolicHelicityTable}) is given to illustrate the helicity dependencies.
It holds for $J_z{=}+1$. The analogous table for $J_z{=}-1$ is obtained 
by the operation 
\[
W_{ij}(J_z{=}\!+\!1;\,\lambda_1,\lambda_2)=-\lambda_1
W_{ij}(J_z{=}\!-\!1;\,-\lambda_1,-\lambda_2).
\]

\begin{table}
\centerline{
\begin{tabular}{|c||c|c|c|c|}\hline
\rule[-3mm]{0mm}{8mm}{\bf final : initial} & $(\lambda_1',\lambda_2')=\uparrow\uparrow$ 
& $(\lambda_1',\lambda_2')=\uparrow\downarrow$ 
& $(\lambda_1',\lambda_2')=\downarrow\uparrow$
& $(\lambda_1',\lambda_2')=\downarrow\downarrow$ \\ \hline\hline
\rule[-3mm]{0mm}{8mm}$(\lambda_1,\lambda_2)=\uparrow\uparrow$ & 
$W_{AA}$   
&$W_{AB}$ & $W_{AC}$ & $0$ \\ \hline
\rule[-3mm]{0mm}{8mm}$(\lambda_1,\lambda_2)=\uparrow\downarrow$ & 
$W_{BA}$ 
& $W_{BB}$ & $W_{BC}$ & $0$ \\ \hline
\rule[-3mm]{0mm}{8mm}$(\lambda_1,\lambda_2)=\downarrow\uparrow$& 
$W_{CA}$ & $W_{CB}$ & $W_{CC}$  & $0$\\ \hline
\rule[-3mm]{0mm}{8mm}$(\lambda_1,\lambda_2)=\downarrow\downarrow$ & $0$ 
& $0$ & $0$ & $0$  \\
\hline
\end{tabular}}
\centerline{
\parbox{14cm}{
\protect
\caption[Symbolic helicity table for the dynamic annihilation graph]
{\protect\label{SymbolicHelicityTable}
Symbolic helicity table for the dynamic annihilation graph.
The functions $W_{ii}$ are identical with the expressions $F_i$ listed in Table 
(\protect\ref{HelicityTableAnnihilation}). Here, terms proportional to 
$\delta_{|J_z|,0}$ are omitted.}}} 
\end{table}

\noindent
The simple kinematics ($x_e+x_{\bar{e}} = 1$) of the
seagull annihilation graph, Fig.(\ref{anniseag}),
result in a constant contribution of this graph to the Hamiltonian 
matrix. 
It is 
\beq
\langle e\bar{e}|S|e\bar{e}\rangle = -2\beta
\delta^{+\lambda_2}_{-\lambda_1}\delta^{+\lambda_2'}_{-\lambda_1'}.
\eeq
Because of its helicity factors, the graph acts 
only between states with 
\beq\label{SpinBed}
S_z = S_z' = 0.
\eeq  
This means that the seagull graph does not contribute when $J_z\neq 0$ because
it has a factor proportional to $(\varphi-\varphi')$ resulting from 
(\ref{SpinBed}). 
A rather surprising consequence is that the dynamic graph is the only 
annihilation channel contribution in the $J_z=\pm 1$ sector, whereas in the 
$J_z=0$ sector the instantaneous graph is the only one. 
This is necessary for rotational invariance: both diagrams must 
yield the same value, though one shows much more structure than the other, 
since degeneracy of the orthopositronium ground state
with respect to $J_z$ 
is found without the annihilation channel and inclusion must not destroy it.

At first glance, there seems to be a manifest breaking of rotational symmetry:
the helicity table (\ref{SymbolicHelicityTable}) separates between states
with $(\lambda_1,\lambda_2)=(\uparrow\uparrow)$ and
$(\lambda_1,\lambda_2)=(\downarrow\downarrow)$. But this is only a consequence
of the integration over the angles: for $J_z=+1$ the $(\downarrow\downarrow)$-
combination
gives no contribution, and likewise does the $(\uparrow\uparrow)$-combination 
for $J_z=-1$.

\begin{figure}[t]
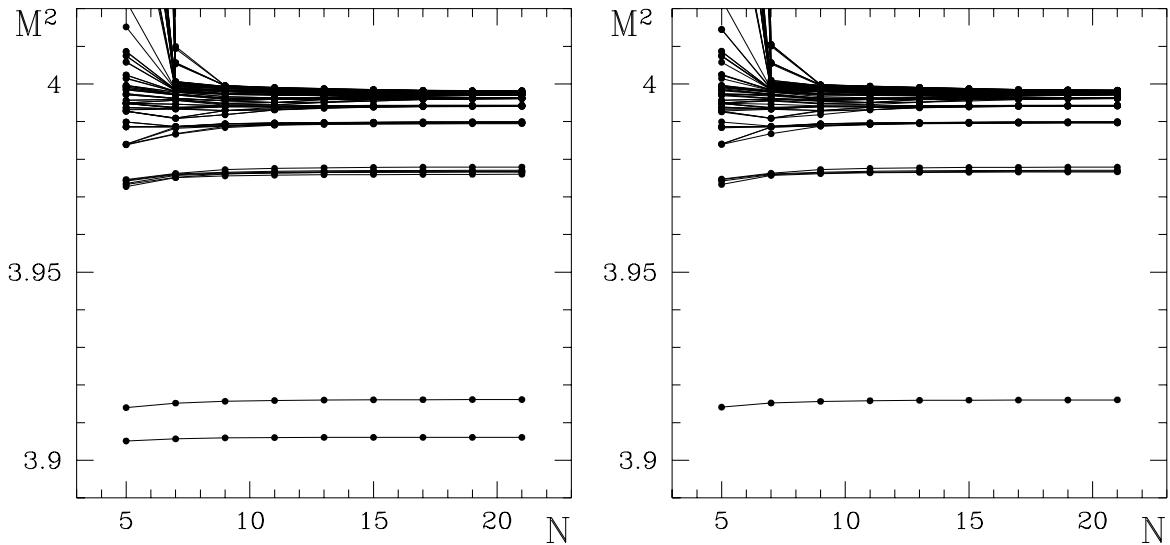

\centerline{
\psfig{figure=\graphpath spec_anni_J0.epsi,width=7.5cm,angle=0}\hfill
\psfig{figure=\graphpath spec_anni_J1.epsi,width=7.5cm,angle=0}}
\protect
\caption[Spectrum including annihilation channel for $J_z{=}0$ and $J_z{=}+1$]
{\protect\label{specanni}The positronium spectrum including the annihilation channel:
(a) $J_z{=}0$ sector, (b) $J_z{=}+1$ sector.  
Parameters of the calculation: $\alpha=0.3, 
\Lambda=1.0\, m_f$. 
The mass squared eigenvalues $M^2_n$ in units of the electron mass $m^2_f$
are shown as functions of the number of integration points $N\equiv N_1=N_2$.
The triplet states, especially $1^3S_1$, are lifted up, the singlet
mass eigenvalues are the same as without annihilation channel. 
Cf.\ 
(\protect\ref{spectrumJ0newCT})
and Table~(\protect\ref{Tablespecanni})
.}
\end{figure}

\section{Spectrum including the annihilation channel}

The spectrum including the annihilation channel shows the 
expected properties: the singlet eigenvalues remain unchanged, only
the triplet states do change at all. 
An essential point in the actual calculations is that one has to use the
same counterterms for the Coulomb singularity as used {\em without}
the annihilation channel. 
This is due to the fact that the one photon annihilation part of the
interaction has no additional singularity 
that needs to be taken care of numerically, because of the simple
energy denominator (\ref{Energienenner}).
We compiled our results in the form of binding coefficients
in Table (\ref{Tablespecanni}). We have used there, contrary to previous
chapters, the bracket convention for the errors to suppress the zeros.

\begin{figure}[t]
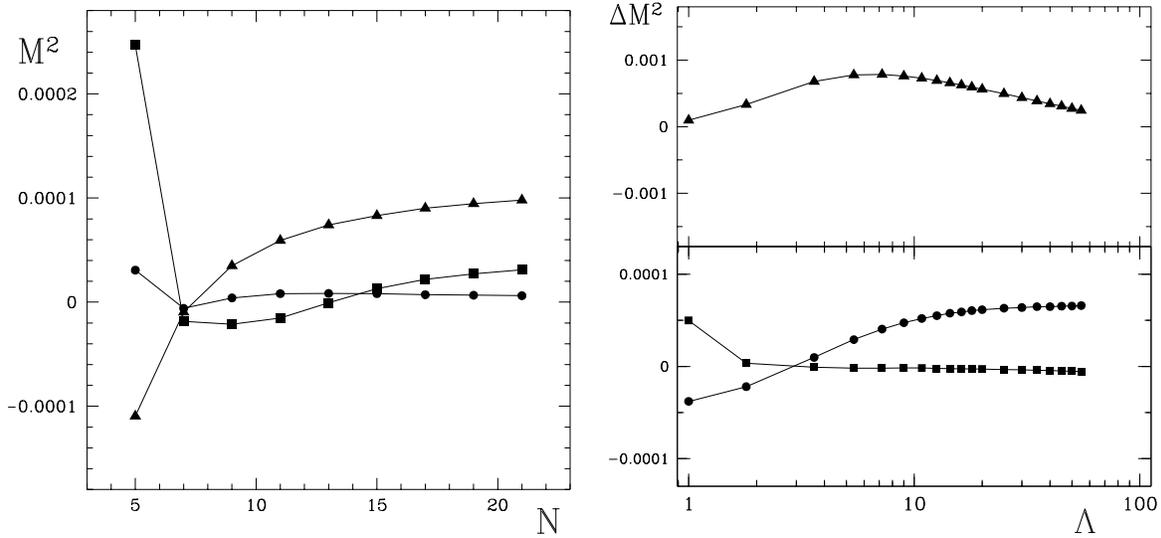

\centerline{
\psfig{figure=\graphpath diff_anni_J1.epsi,width=7.5cm}
\hfill
\psfig{figure=\graphpath diff_lambda_anni.epsi,width=7.3cm}
}
\caption[Deviation of eigenvalues between $J_z{=}0$ and $J_z{=}\pm 1$] 
{\label{diffanni}Deviation of 
corresponding eigenvalues for $J_z=0$ and $J_z=1$ including
the annihilation channel
($\alpha=0.3, \Lambda=1.0\, m_f$): (a) as a function of the number of
integration points $N$, (b)  
as a function of the cutoff $\Lambda$.  
The graphs show 
$\Delta M^2:= M^2_n(J_z=0)-M^2_n(J_z=1)$
for the states $1^3S_1\,(\triangle)$, $2^3P_1\,(\Box)$, and $2^1P_1\,(\circ)$.} 
\end{figure}

One notes a slightly larger breaking of rotational symmetry. 
The triplet ground states
of different $J_z$ are still approximately degenerate, but the discrepancy
is bigger than without the annihilation channel.
The dependence of these discrepancies between corresponding eigenvalues for 
$J_z{=}0$ and $J_z{=}1$ on the number of integration points is shown in 
Fig.~(\ref{diffanni}). The behavior of the curves is similar to those 
of the calculations without annihilation channel, Fig.~(\ref{diff}).   
An additional plot, Fig.~(\ref{diffanni}), shows the dependence 
of the rotational symmetry breaking on the cutoff $\Lambda$. The 
discrepancies of corresponding eigenvalues are almost independent of
the cutoff $\Lambda$. 

To make a comparison of our results to those of perturbation theory easier,
we show in Fig~(\ref{yrastn2}) the eigenvalues for a principal quantum 
number $n{=}2$ for both theories graphically. The structure of the two 
plots is almost the same, 
only the $2^1S_0$ state and the $2^3P_0$ state are exchanged in our results. 
This is due to the 
cutoff dependence of the S-state, which is larger than that of the other 
states (cf.~Fig.~(\ref{lambdaanni})(b) and next paragraph).       
We stress that the results of the perturbative calculations change considerably,
when the
next higher order in $\alpha$ is considered. For example, the mass squared
of the triplet $2^3S_1$ is, according to {\sc Fulton} and  {\sc Martin} 
\cite{Fulton},
$M^2(2^3S_1)=3.9780186070$ up to order $\alpha^3 \Rydberg$, 
which is nearer to our result than the value of 
perturbative calculations up to order ${\cal O}(\alpha^4)$ as displayed
in Fig.~(\ref{yrastn2}). 

\begin{table}[th]
\begin{minipage}{15.5cm}
\centerline{
\begin{tabular}{|r||c|c|c|}\hline
\rule[-3mm]{0mm}{8mm}$\mbox{Cutoff: }\Lambda$ & $B_s$ & $B_t$ & $C_{hf}$\\\hline\hline 
   1.8\hspace{0.5cm} &  1.16373904 & 0.96234775 &  0.55942025 \\ \hline 
   3.6\hspace{0.5cm} &  1.25570163 & 0.96446614 &  0.80898748 \\ \hline 
   5.4\hspace{0.5cm} &  1.29978067 & 0.96482118 &  0.93044303 \\ \hline 
   7.2\hspace{0.5cm} &  1.32941926 & 0.96541695 &  1.01111752 \\ \hline 
   9.0\hspace{0.5cm} &  1.35224000 & 0.96603457 &  1.07279285 \\ \hline 
  10.8\hspace{0.5cm} &  1.37112216 & 0.96661006 &  1.12364471 \\ \hline 
  12.6\hspace{0.5cm} &  1.38744792 & 0.96713137 &  1.16754595 \\ \hline 
  14.4\hspace{0.5cm} &  1.40198469 & 0.96760110 &  1.20662108 \\ \hline 
  16.2\hspace{0.5cm} &  1.41520247 & 0.96802548 &  1.24215830 \\ \hline 
  18.0\hspace{0.5cm} &  1.42740143 & 0.96841025 &  1.27497551 \\ \hline 
\hline ETPT\hspace{0.3cm} & 1.11812500 & 0.90812500 & 0.58333333\\ \hline
${\cal O}(\alpha^6\ln\alpha)$ & \multicolumn{2}{|c|}{ }& 0.48792985\\ \hline
\end{tabular}}
\centerline{
\parbox{10cm}{ 
\caption[Binding coefficients including the annihilation channel]
{\label{TableLambdaAnni}The binding coefficients 
of the singlet ($B_s$), the triplet ($B_t$), 
and the hyperfine coefficient $C_{hf}$ are listed 
for $\alpha=0.3,$ $N_1=41, N_2=11$.}}}
\end{minipage}
\end{table}

\section{Parameter dependence of the spectrum}

The convergence of the eigenvalues with growing number of integration 
points $N$ is the same as the case of no annihilation channel. 
To be explicit, the convergence of the eigenvalues can be shown to be
exponential. One can fit the singlet ground state eigenvalue excellently
with the function
\[
M^2(N)=M^2(21)-\left[M^2(21)-M^2(5)\right]e^{-(N-5)/2.2}.
\]
We did not perform the limit $N\rightarrow\infty$, because the accuracy of the 
results for $N>20$ suffices to compare to other data. 

\begin{figure}[t]
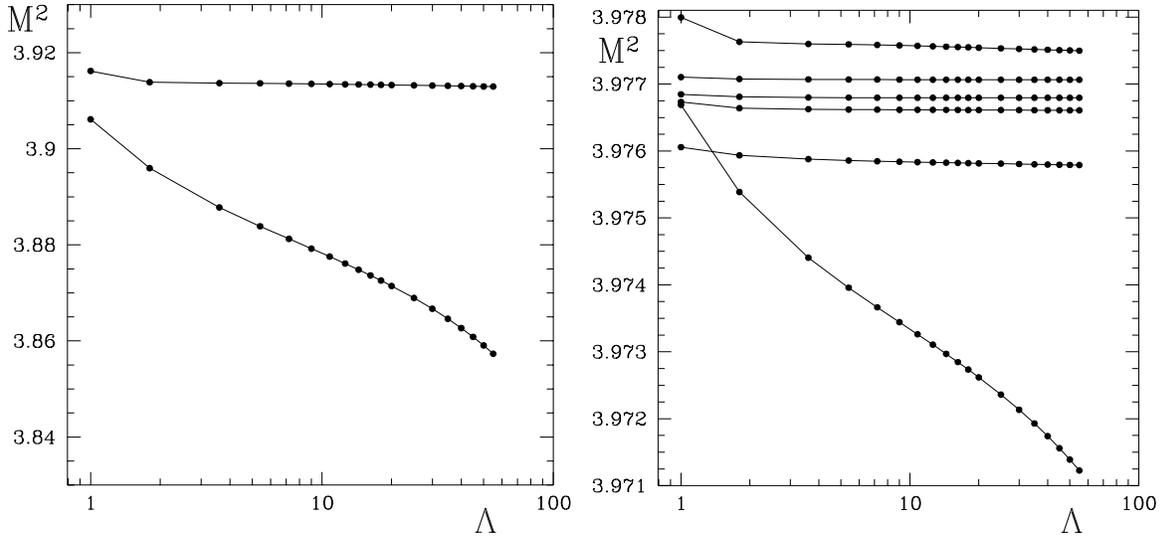

\centerline{
\psfig{figure=\graphpath lambda_J0_anni_n1.epsi,width=7.5cm}\hfill
\psfig{figure=\graphpath lambda_J0_anni_n2.epsi,width=7.5cm}
}
\protect\caption[The spectrum with annihilation channel 
as a function of the cutoff] 
{\protect\label{lambdaanni}The spectrum with 
annihilation channel as a function of the cutoff $\Lambda$: 
(a) ground states ($n{=}1$), (b) first radial excited states ($n{=}2$).
The parameters for the calculation are
$\alpha=0.3$, $J_z=0$, $N_1=25$, $N_2=21$.
One notices a better behavior of the curves with growing $\Lambda$ compared
to Figs.~(\protect\ref{lambdaJ0n1})~and~(\protect\ref{lambdaJ0n2}): 
the decrease of the eigenvalues is smaller,
especially for the ground state triplet $1^3S_1$. Moreover, there are no level
crossings for $n{=}2$, except the crossing of the singlet S-state at 
$\Lambda=1.5$ (not visible in this plot).}
\end{figure}

\begin{figure}[th]
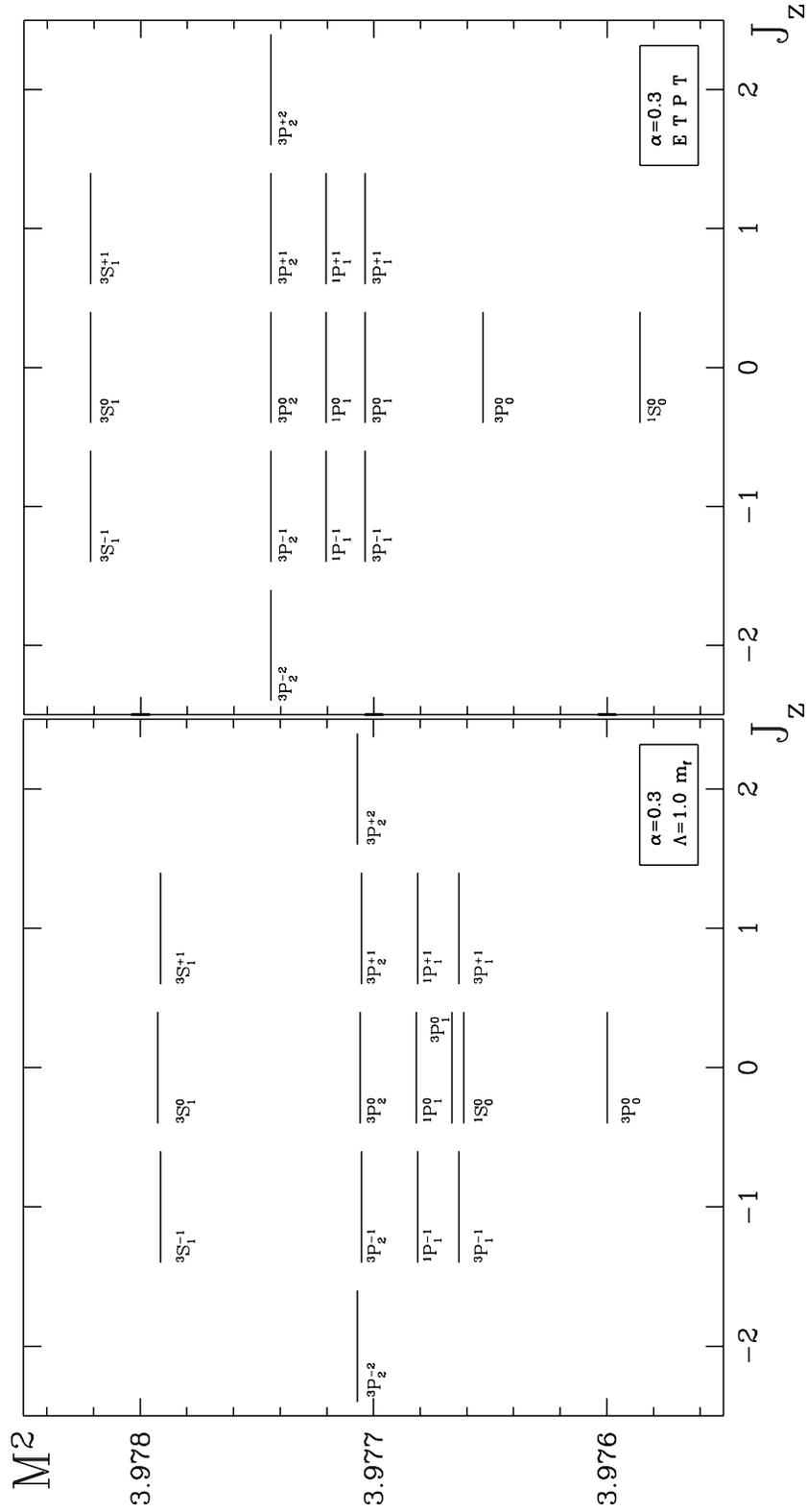

\centerline{
\psfig{figure=\graphpath yrast_n2_E.epsi,width=10.8333cm,angle=90}
\hspace{0.0cm}}
\vspace{-0.2cm}
\centerline{
\psfig{figure=\graphpath yrast_n2.epsi,width=11cm,angle=90}
\hspace{0.175cm}}
\caption[Comparison of multiplets for $n=2$ with perturbation theory] 
{\label{yrastn2}Comparison of 
multiplets for $n{=}2$: (a) results of the present work with
$\alpha=0.3$, $\Lambda= 1.0\,m_f$, $N_1=N_2=21$; 
(b) equal-time perturbation theory (ETPT) up to order ${\cal O}(\alpha^4)$.}
\end{figure}

The cutoff dependence of the positronium spectrum including the 
annihilation channel is comparable to that of the spectrum without it. 
However, a striking difference occurs: the inclusion of the annihilation
channel stabilizes the cutoff dependence of the eigenstates. 
In particular, the triplet ground state in Fig.~(\ref{lambdaanni})(a)
shows only a small dependence on the cutoff, when one compares it to the 
behavior of the same state in Fig.~(\ref{lambdaJ0n1}).
One can fit these curves with a polynomial in $\log \Lambda$. The
singlet ground state eigenvalues are the same as without the annihilation graph
and for the triplet one obtains
\[
M^2_{triplet}(\Lambda)\simeq 3.91392-0.000288079 \log\Lambda
                                    +0.000147268 \log^2\Lambda.
\]
Comparison with Eq.~(\ref{LambdaFit}) shows that the decrease of the triplet 
with $\log\Lambda$ is suppressed by including the annihilation channel by
a factor of 60.
Also the excited states, $n{=}2$, show a different behavior, as compared 
to Fig.~(\ref{lambdaJ0n1}). Here, only one state shows level crossing as
$\Lambda$ grows large. The eigenvalue of the $2^3P_1$ state, however, 
depends only weakly on the cutoff here. 

The values for the binding coefficients $B_n$ and the coefficient of the 
hyperfine splitting $C_{hf}$ are presented in Table~(\ref{TableLambdaAnni}).
The values are correct for 
a cutoff $\Lambda\simeq 2 m_f$ when compared to results of 
perturbation theory up to ${\cal O}(\alpha^4)$. However, 
the effects of higher order correction to perturbative calculations
are significant
for a large coupling such as $\alpha=0.3$. 
The result of perturbation theory to ${\cal O}(\alpha^6\ln \alpha)$
for the coefficient $C_{hf}$ is considerably smaller than that to order 
${\cal O}(\alpha^4)$.  

Concluding, we state that the cutoff dependence of the spectrum 
is improved as compared with the case of missing annihilation
channel. This makes it evident that the annihilation channel is a necessary
part of the theory.

As a last investigation of the properties of our model, we will vary the
coupling constant and interpret the spectrum. A similar procedure
was performed by {\sc Dyks\-hoorn et al.} \cite{Dykshoorn}, who studied 
coupled integral equations for QED-bound states in equal-time quantization
with a variational ansatz. They calculate masses for the 
lowest eigenstates of positronium with and without the annihilation channel
and plot them versus the coupling constant. The prominent feature of their
figures is the occurrence of a critical coupling at which the masses become 
smaller than zero. 
We have performed the analogous calculations within our approach. 
It seems at first glance [Fig.~(\ref{alphaspectrum})(a)] as if the eigenvalues,
after decreasing 
quadratically with the coupling as expected by perturbation theory,
stabilize at a coupling $\alpha\simeq 1.5$.
However, further investigation shows that this 
is merely an effect of the cutoff dependence of the spectrum. The eigenvalues
in  Fig.~(\ref{alphaspectrum})(a) were calculated 
with a cutoff $\Lambda=1.0\; m_f$, which
is too small for an coupling constant of $\alpha=1.0$, since then the Bohr
momentum is of the same order as the cutoff.
Fig.~(\ref{alphaspectrum})(b) shows clearly that there is a critical coupling.
The masses calculated with a cutoff of $\Lambda=20\, m_f$ tend to zero 
at $\alpha\simeq 0.5$. A similar value was found in \cite[Ref.~2]{Dykshoorn}.

\begin{figure}[h]
\centerline{
\psfig{figure=\graphpath yrast_anni.epsi,width=15.5cm,angle=0}}
\caption[Compiled spectra for $J_z=-3,\ldots,+2,+3$ incl. annihilation channel]
{\label{yrastanni}Compiled spectra of 
positronium with different $J_z=-3,-2,\ldots,+3$
including the annihilation channel.
All spectra have been calculated with $\alpha=0.3$, 
$\Lambda=1.0\, m_f$, $N_1=N_2=21$.
The mass squared eigenvalues $M^2_n$ in units of the electron mass $m^2_f$
are shown.  
The notation for the states is $^{3S+1}L^{J_z}_J$.
The resolution of the plot is inadequate for the 
multiplets. Nevertheless, the numerical degeneracy of the three triplet
ground states $^3S^{-1}_1$,$^3S^0_1$, and $^3S^1_1$ becomes very clear.
} 
\end{figure}

\begin{figure}
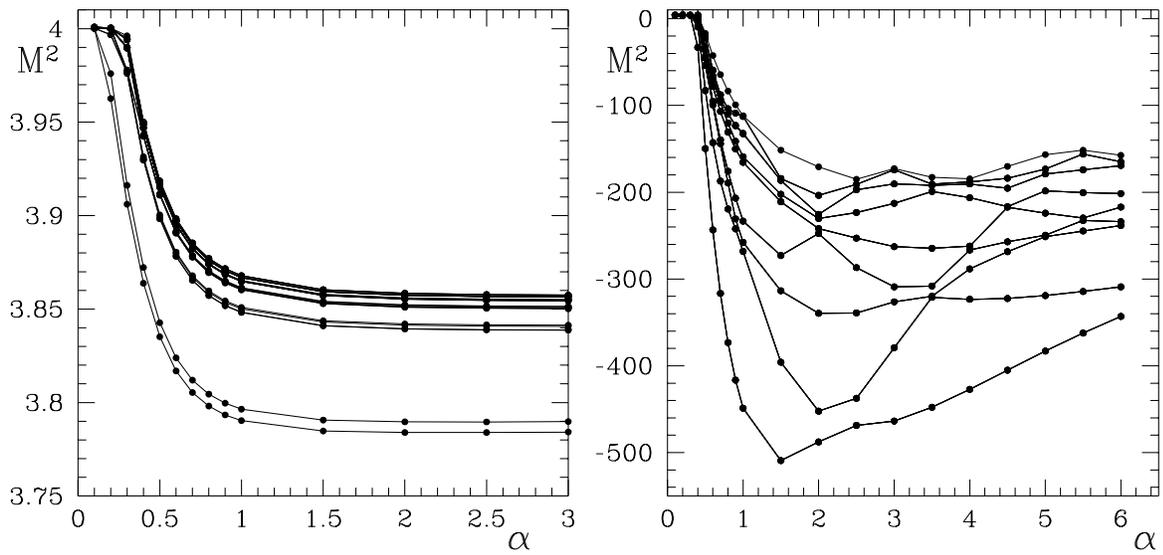

\centerline{
\psfig{figure=\graphpath alpha_1mf.epsi,width=7.5cm}\hfill
\psfig{figure=\graphpath alpha_20mf.epsi,width=7.5cm}
}
\caption[The spectrum as a function of the coupling constant] 
{\label{alphaspectrum}The spectrum as a function of the coupling constant: 
(a) $\Lambda=1.0\,m_f$, (b) $\Lambda=20.0\,m_f$. In (a), the  
eigenvalues seem to converge to a stable value as $\alpha$ grows large.
However, this is just an effect of the cutoff dependence of the spectrum:
for a larger cutoff $\Lambda=20\,m_f$, (b), the eigenvalues become negative
at a critical coupling $\alpha_c\simeq 0.5$.}
\end{figure} 

\newpage
\begin{table}[h]
\centerline{
\begin{tabular}{|r|c||l|l|r|}\hline
\rule[-3mm]{0mm}{8mm}$n$ & Term & $\hfill B_n(J_z{=}0)\hfill $ & $\hfill B_n(J_z{=}+1)\hfill$ & $\hfill \Delta B_n \hfill$ \\ \hline \hline
  1&$1^1S_0$ & 1.049552(17) & $\hfill\mbox{---}\hfill$ & $\hfill\mbox{---}\hfill$ \\ \hline 
  2&$1^3S_1$ & 0.936800(189) & 0.937902(151) & -0.001102 \\ \hline 
\hline
  3&$2^1S_0$ & 0.260237(169) & $\hfill\mbox{---}\hfill$ & $\hfill\mbox{---}\hfill$ \\ \hline 
  4&$2^3S_1$ & 0.255292(184) & 0.255359(179) & -0.000067 \\ \hline 
  5&$2^1P_1$ & 0.257969(160) & 0.258037(168) & -0.000068 \\ \hline 
  6&$2^1P_1$ & 0.267090(160) & $\hfill\mbox{---}\hfill$ & $\hfill\mbox{---}\hfill$ \\ \hline 
  7&$2^3P_1$ & 0.259667(206) & 0.260013(163) & -0.000346 \\ \hline 
  8&$2^3P_2$ & 0.245615(239) & 0.245755(228) & -0.000140 \\ \hline 
\hline
  9&$3^1S_0$ & 0.115206(314) & $\hfill\mbox{---}\hfill$ & $\hfill\mbox{---}\hfill$ \\ \hline 
 10&$3^3S_1$ & 0.113434(156) & 0.113497(179) & -0.000063 \\ \hline 
 11&$3^1P_1$ & 0.114490(269) & 0.114521(283) & -0.000032 \\ \hline 
 12&$3^1P_1$ & 0.117142(272) & $\hfill\mbox{---}\hfill$ & $\hfill\mbox{---}\hfill$ \\ \hline 
 13&$3^3P_1$ & 0.115127(326) & 0.115102(275) & 0.000025 \\ \hline 
 14&$3^3P_2$ & 0.113731(284) & 0.113753(283) & -0.000023 \\ \hline 
 15&$3^1D_2$ & 0.112816(150) & 0.112842(158) & -0.000027 \\ \hline 
 16&$3^3D_1$ & 0.112977(161) & 0.112987(164) & -0.000010 \\ \hline 
 17&$3^3D_2$ & 0.112520(158) & 0.112524(160) & -0.000004 \\ \hline 
 18&$3^3D_3$ & 0.111027(377) & 0.111072(373) & -0.000045 \\ \hline 
\hline
 19&$4^1S_0$ & 0.065490(588) & $\hfill\mbox{---}\hfill$ & $\hfill\mbox{---}\hfill$ \\ \hline 
 20&$4^3S_1$ & 0.064707(480) & 0.064723(481) & -0.000016 \\ \hline 
 21&$4^1P_1$ & 0.065003(467) & 0.065061(486) & -0.000059 \\ \hline 
 22&$4^1P_1$ & 0.066119(470) & $\hfill\mbox{---}\hfill$ & $\hfill\mbox{---}\hfill$ \\ \hline 
 23&$4^3P_1$ & 0.065331(487) & 0.065276(475) & 0.000055 \\ \hline 
 24&$4^3P_2$ & 0.064265(309) & 0.064372(348) & -0.000107 \\ \hline 
 25&$4^1D_2$ & 0.063968(430) & 0.064041(363) & -0.000073 \\ \hline 
 26&$4^3D_1$ & 0.064099(319) & 0.064090(325) & 0.000009 \\ \hline 
 27&$4^3D_2$ & 0.063870(391) & 0.063881(477) & -0.000011 \\ \hline 
 28&$4^3D_3$ & 0.063788(387) & 0.063807(381) & -0.000019 \\ \hline 
 29&$4^1F_3$ & 0.063141(96) & 0.063112(104) & 0.000030 \\ \hline 
 30&$4^3F_2$ & 0.063299(112) & 0.063233(116) & 0.000065 \\ \hline 
 31&$4^3F_3$ & 0.063234(119) & 0.063209(119) & 0.000024 \\ \hline 
 32&$4^3F_4$ & 0.063103(104) & 0.063422(158) & -0.000319 \\ \hline 
\hline
 33&$5^1S_0$ & 0.043253(806) & $\hfill\mbox{---}\hfill$ & $\hfill\mbox{---}\hfill$ \\ \hline 
 34&$5^3S_1$ & 0.043046(682) & 0.042912(739) & 0.000134 \\ \hline 
 35&$5^1P_1$ & 0.043412(1100) & 0.042724(853) & 0.000688 \\ \hline 
\hline
\end{tabular}}

\protect\caption[Spectrum for $J_z{=}0$ and $J_z{=}1$ including  annihilation channel]{\protect\label{Tablespecanni}The positronium spectrum for $\alpha=0.3,$ $\Lambda=1.0\, m_f, N_1=N_2=21$.
The non-relativistic notation for the terms and the eigenvalues for $J_z{=}0$ and $J_z{=}+1$ including the annihilation channel are shown. The discrepancy between the eigenvalues is $\Delta B_n:=B_n(J_z{=}0)-B_n(J_z{=}+1)$. The numerical errors are estimated from the difference between the values for maximum and next to maximum number of integration points. The actual errors are smaller due to 
the exponential convergence of the eigenvalues with $N$.
The $k$ numbers in brackets are the
 errors in the last $k$ digits.}

\end{table}

\newpage
\chapter{\label{DrAChapterEffInt}On the theory of effective 
interactions}

\section{Introduction}

The aim of this chapter is to show that fixing the redundant 
parameter $\omega$ to the expression (\ref{OmegaStar}) is {\em not}
an ad hoc assumption, but a consequence of
the structure of the theory and of the projection method used.
The material presented here 
follows the line of arguments given in \cite{PauliMIR}, 
where the method of iterated resolvents was introduced and applied to 
QCD. The case of QED is discussed in this section.

To understand the following, it is helpful to review the reasoning presented 
in 
Chapter \ref{ChapterModel} and to see things in a more general way.
In the description of the positronium model it was pointed out
that the unrestricted Fock space can be divided 
{\em arbitrarily} into two parts, one called the $P$- the other the $Q$-space.
Since only two Fock sectors ($|e\bar{e}\rangle$, $|e\bar{e}\gamma\rangle$)
were used in the model, it was clear how 
this separation had to be done.

{\sc Tamm} \cite{Tamm} and {\sc Dancoff} \cite{Dancoff} used this method in
a different context in the following way. First, they truncated the Fock space
to two sectors.
Second, they projected one Fock sector onto the other. 
Third, the emerging
energy denominator is {\em modified} to simplify the calculations.
The so-defined procedure fails completely in front form dynamics \cite{KPW}
if the third step is missing. 
A severe (collinear)
singularity occurs that cannot be treated within this approach.
{\sc Tamm} did not recognize this problem, because he considered the 
modification of the energy denominator rather as a simplification than
as an approximation for the effects of the omitted higher Fock states.

What was done in Eq.~(\ref{resolvent}) was to introduce a new,
redundant parameter $\omega$ for the {\em a priori} unknown mass (squared)
eigenvalue $M^2_n$.
This parameter is free and can be fixed on an {\em ad hoc}  
basis to remove the collinear singularity. 
It will be shown that this fixing, Eq.~(\ref{OmegaStar}), is not only 
the natural choice for this parameter,
but that it is a consequence of the structure of the formalism.

\section{The method of iterated resolvents}

The essential cause for the failure of a ``naive'' Tamm-Dancoff approach is
the fact that
the Fock space truncation is performed too early in the formalism.
Because of this, one throws away all possible interactions 
with the omitted sectors. This gives rise to the 
problems described above. 
It is therefore necessary to study the structure of the resolvents
to all orders and to investigate possible approximation schemes consistent
with the solubility of the whole problem.

Of course, one has to construct a finite 
dimensional Hamiltonian from the Lagrangian density in the style of the 
Tamm-Dancoff approach.
But a truncation cannot be the first step. Rather one introduces
unphysical parameters needed to map the Hamiltonian operator onto 
a {\em matrix} in such a way that they remain in the formalism until the end.
One can then investigate rigorously the behavior of the theory 
in the limit when all unphysical parameters are removed.

We consider in this chapter the Hamiltonian within the DLCQ formalism,
{\em i.e.}~with discretized momenta. This maps the Hamiltonian operator
into a matrix with a denumerable number of columns and rows. It may still
be an infinite dimensional matrix.
In a Fock basis, the Hamiltonian is naturally decomposed into sectors of 
different particle content, cf.~Table~(\ref{HolyMatrixQED}). Each sector 
by itself contains an infinite number of states, and is regulated by a 
cutoff $\Lambda$, like in Eq.~(\ref{BLepagecutoff}). The number of Fock sectors,
on the other hand, is limited by the finite 
longitudinal momentum $P^+$, or rather
by the {\em harmonic resolution} $K=\frac{L}{\pi}P^+$.
The finite dimensional eigenvalue problem reads thus:
\beq\protect\label{EVP}
\left(
   \begin{array}{cccc}
      \langle 0|H|0\rangle & \langle 0|H|1\rangle & \cdots &
      \langle 0|H|N\rangle \\
      \langle 1|H|0\rangle & \langle 1|H|1\rangle & \cdots &
      \langle 1|H|N\rangle\\
      \vdots & \vdots & \ddots & \vdots\\ 
      \langle N|H|0\rangle & \langle N|H|1\rangle & \cdots &
      \langle N|H|N\rangle
                \end{array}
\right)
\left(
   \begin{array}{c}
	\langle 0|\psi\rangle\\
	\langle 1|\psi\rangle\\
	\vdots\\
	\langle N|\psi\rangle
   \end{array} 
\right)
=\omega
\left(
   \begin{array}{c}
	\langle 0|\psi\rangle\\
	\langle 1|\psi\rangle\\
	\vdots\\
	\langle N|\psi\rangle\\
   \end{array} 
\right).
\eeq
Like in Chapter \ref{ChapterModel}, one can reduce the dimension of the 
Hamiltonian matrix by projecting the ``highest'' state $\langle N|\psi\rangle$
onto the others. One arrives at an effective Hamiltonian, depending
on the parameter $\omega$, due to the resolvent 
\[
G_n(\omega):=\langle n|\;\omega{-}H\;|n\rangle, \quad 1\leq n\leq N.
\]
In general, one has the eigenvalue problem 
\[
\sum_{j=0}^n \langle i|H_n(\omega)|j\rangle\langle j|\psi(\omega)\rangle
=E(\omega)\langle i|\psi(\omega)\rangle,
\]
where the last state can be expressed by all others
\[
\langle N|\psi(\omega)\rangle=
G_n(\omega)
\sum_{j=0}^{n-1} \langle i|H_n(\omega)|j\rangle\langle j|\psi(\omega)\rangle.
\]
Consequently, the effective Hamiltonian in the $(n{-}1)$-sector becomes
for each matrix element $\langle i|H_{n-1}(\omega)|j\rangle$
\beq\protect\label{recursion}
H_{n-1}(\omega)=H_n(\omega)+H_n(\omega)G_n(\omega)H_n(\omega),
\eeq
which in a sense is a recursion relation.

One has to find a systematic way to express the effective Hamiltonian
after the $i$-th projection in terms of the bare Hamiltonian $H$. 
If one applies the relation Eq.~(\ref{recursion}) repeatedly,
one obtains
\beq\protect\label{recursion2}  
   H _n =  H + \sum_{m=n+1} ^N  H _m G _m  H _m,
\eeq
with the bare Hamiltonian H.
For example, for $n=3$ one gets
\beq\protect\label{H1}
H_1 =  H _3 +  H _3 G _3  H _3 +  H _2 G _2  H _2 .
\eeq 
by inserting $H _2 =  H _3 +  H _3 G _3  H _3$
into $H _1 =  H _2 +  H _2 G _2  H _2 $. 
The general case can be proven by induction.
An illustrative example is given in Section \ref{AnExample}.
 
In Eq.~(\ref{recursion2}), the effective Hamiltonian in Fock sector $n$ 
is expressed in terms of the bare Hamiltonian and scatterings into 
{\em higher} Fock sectors. It is important to notice that no scattering into 
lower Fock sectors occurs. 
From Eq.~(\ref{H1}) one can infer how the general structure of the 
expression for the effective Hamiltonian will be. {\em Chains} of terms
with a different number of resolvents will emerge.
In fact, it turns out to be much better to classify those chains by the number
of its resolvents than, for example, by the order of the coupling constant.  
If all chains with $k$ resolvents are collected in $H^{(k)}$, the effective
Hamiltonian is the sum
\begin{equation}\protect\label{Hn}
    H _n = H ^{(0)} _n + H ^{(1)} _n + H ^{(2)} _n + \ldots \; . 
\end{equation} 
This expansion is finite, because the bare Hamiltonian is a finite matrix.
With this classification scheme, one obtains a recursion relation 
for the general term
by inserting of Eq.~(\ref{Hn}) into Eq.~(\ref{recursion2})
\begin {equation}\protect\label{Hk}
       H ^{(k+1)} _n = \sum_{l>n}  \left( 
       H ^{(0)} _l G _l H ^{(k)} _l
     + H ^{(1)} _l G _l H ^{(k-1)} _l
     + \cdots 
     + H ^{(k-1)} _l G _l H ^{(1)} _l
     + H ^{(k)}   _l G _l H ^{(0)} _l
\ \right).  
\end{equation}
The first terms read 
\begin {eqnarray*}
       H ^{(1)} _n & = & \sum \limits _{l>n} \ \phantom{\Big(}
       H ^{(0)} _l G _l H ^{(0)} _l
\ \phantom{\Big)} \ , \\
       H ^{(2)} _n & = & \sum \limits _{l>n} \ \Big( 
       H ^{(0)} _l G _l H ^{(1)} _l
     + H ^{(1)} _l G _l H ^{(0)} _l
\ \Big) \ .
\end{eqnarray*}
Note the change in the order of the calculation. At first, the  
rows of the Hamiltonian were projected from $N$ to 1 unto one 
another to derive the
expression for the effective Hamiltonian in the lowest sector.
Now, the calculations are reversed in the sense that the effective
Hamiltonian in the lowest sector is evaluated by longer and longer 
chains of resolvents, until one reaches the dimension of the bare Hamiltonian.
The advantage is obvious: one is able to control the limit in which no 
truncation is made at all, {\em i.e.}~$N\rightarrow\infty$.

\section{QED treated with iterated resolvents}

So far, the general method of how to calculate an effective
Hamiltonian out of a finite dimensional Hamiltonian matrix was described.
In particular, no assumptions on the matrix elements of the starting
Hamiltonian (like the one in Eq.~[\ref{EVP}]) were made.
From the structure of the QED Hamiltonian, described in Appendix~\ref{AppxHLC}
and displayed in Table~(\ref{HolyMatrixQED}),
it is obvious that most of the matrix elements will be zero, certainly 
including
all those in which the number of partons created or destroyed by the 
interaction exceeds two. 
This can clearly lead to significant simplifications. The procedure can 
be simplified even more by applying a technique of light-cone 
perturbation theory. There, one takes care of the instantaneous gauge parts
of the Hamiltonian, {\em i.e.}~of the {\em seagulls} and {\em forks}, only at
the end of each calculation by redefining the non-instantaneous propagators
\cite[Appx.~A, Fig.~31]{BLepage}. One adds to the 
propagators
a part containing an instantaneous graph including a $\theta$-function for 
the longitudinal momentum transferred.  
One can apply the same trick here: one only keeps the {\em vertex} 
interaction in the matrix, Table (\ref{HolyMatrixQED}). 
 
\begin{figure}[t]
\begin{minipage}{15.5cm}
\centerline{
\psfig{figure=\graphpath 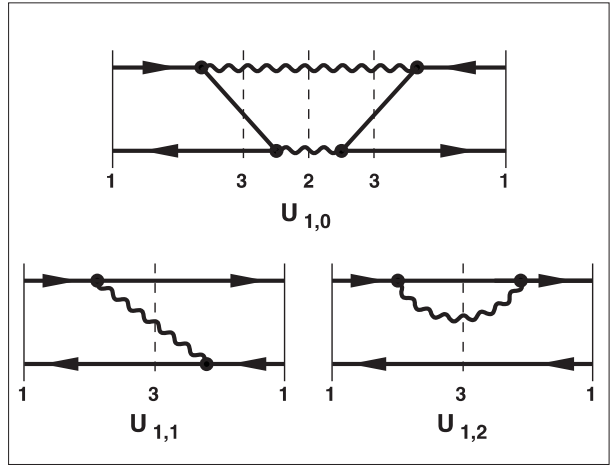,width=8cm,angle=0}}
\caption[The three effective graphs of gauge theories]
{\protect\label{ThreeGraphs}The three 
effective graphs of gauge theories (Courtesy of {\sc H.-C.~Pauli}).}
\end{minipage}
\end{figure}

Since we want to calculate mesonic, {\em i.e.}~fermion-antifermion systems,
the sector $|e\bar{e}\rangle$ plays the r\^{o}le of a ``cornerstone'' sector.  
If one calculates the effective Hamiltonian in this sector one
gets
\beq\protect\label{Hqq}
H_{q\bar{q}}\equiv H_1=T_1+VG_3V+VG_3VG_2VG_3V,
\eeq
where $T_1$ is the kinetic energy in the $|e\bar{e}\rangle$-sector.
This is an expression with at most three resolvents $G_j(\omega)$.
The graphs of these fundamental chains are shown in Figure 
(\ref{ThreeGraphs}): the photon exchange graph, the two-photon annihilation
graph, and the self energy graph.  

It is worth mentioning two things here. Firstly, the resolvents are in general
non-diagonal operators. 
Secondly,   
a single chain can symbolize more than on graph. For example, the 
photon exchange graph and the self energy graph emerge out of the same chain
$VG_3V$.  

The structure of the effective interaction for the 
$|e\bar{e}\rangle$ states seems to be very simple. The question arises what
the effect of the higher Fock states will be.
A first hint can be gained by looking at sectors similar to 
$|e\bar{e}\rangle$, {\em i.e.}~to $|e\bar{e}\gamma\rangle$,
$|e\bar{e}\gamma\gamma\rangle\ldots\;$,
\begin{eqnarray} 
   H_{e\bar{e}\gamma}\equiv H _3 
&=&  T_3 + V G _6 V + V G _6 V  G _5 V G _6 V + V G _4 V, \protect\label{sectors1} \\  
   H_{e\bar{e}\gamma\gamma}\equiv H _5 
&=&   T_6 + V G _ {10} V+ V G _{10} V  G _9 V G _ {10} V 
                 + V G _7 V,\\       
   H_{e\bar{e}\gamma\gamma\gamma}\equiv H _{10} 
&=&  T_{10} + V G _{15} V + V G _{15} V  G _{14} V G _{15}V
      +V G _{11} V .\protect\label{sectorsN}  
\end{eqnarray}
The only difference of these expression 
to Eq.~(\ref{Hqq}) is the absence of the 
last term, the annihilation of a photon into an electron-positron pair.
This is because the first sector does not contain a photon.

These sectors seem to be similar to each other. However, they
are only a small set in the whole Hamiltonian. Their defining property 
is that the photons do not interact with the fermion-antifermion
system. To classify the sectors, this kind of interactions is named
{\em spectator interaction} $\overline{U}_n$. 
The other class of interactions, where the photons do interact, is
called {\em participant interaction} $\widetilde{U}_n$.
Consequently,
the decomposition of the Hamiltonians
read
\begin{equation}  
     H _n =  T  _n + \overline U _n + \widetilde U _n
\ , \quad\quad  n= 3,6,10,15,\dots
\ .  
\end{equation} 

Analogously to Eq.~(\ref{resolvent}), where the 
resolvent was expanded around its
diagonal ({\em i.e.}~free) part, 
the resolvents $\overline G _n$ can be expanded around the part containing 
the spectator interaction $\overline U _n$
\begin{equation} 
     \overline G _n  =   {1 \over  \omega - T _n - \overline U _n }
\ .
\end{equation}
The full resolvent is obtained by the infinite series   
\begin{equation}\protect\label{resolvent2} 
     G _n    =   \overline G _n   +  \overline G _n \,
                        \widetilde U _n \, G _n         
\ 
= \overline G _n
     + \overline G _n \widetilde U _n\, \overline G _n 	
     + \overline G _n \widetilde U _n\, \overline G _n 
                                     \widetilde U _n\, \overline G _n   + \dots
    \ .
\end{equation} 
Note that the unperturbed resolvent contains the interaction!
Moreover, one observes that in this approach, the system does not leave
the sector $n$. 


Let's repeat what we have achieved up to now. 
We divided the interaction in one part that acts only between the electron
and the positron (spectator interaction) and one part where the photons
can interact with the two fermions (participant interaction). 
We expanded the full resolvent $G_n$ as an infinite series around 
the spectator interaction. 
One can show \cite{PauliMIR}, that the spectator interaction allows for  
three fundamental graphs only, Fig.~(\ref{ThreeGraphs}), if one
reduces the bare Hamiltonian by subsequent projections to the 
fermion-antifermion sector. This is the end of the procedure 
for non-abelian gauge theories. In QED, however, one has an additional
sector $|\gamma\rangle$ responsible for 
the annihilation diagram, Fig~(\ref{annihilation}).
The whole spectator interaction resides in the coupling function 
attached to the vertices.

The question arises, if the spectator interactions in the different sectors
are somehow related. Note that the only difference between their diagrams
is the different number of {\em non-interacting} photons.
In fact, this is the crucial point in the formalism.
We will give here a {\em heuristic} explanation, how this can be understood.

Consider the separation of the Fock space 
into two parts, named $P$- and $Q$-space,
like in Chapter~\ref{ChapterModel}. 
According to the discussion of this chapter, the only interaction in 
both spaces is the spectator interaction.
What does the resolvent look like?
It is from Eq.~(\ref{resolvent})  
\beq
G(\omega):=\frac{1}{\omega - H_Q},
\eeq
where $H_Q=T_Q+V_Q$ is the sector-Hamiltonian of the $Q$-space.
The redundant parameter $\omega$ stands for the actual eigenvalue of the
whole eigenvalue problem and contains therefore a kinetic and an interaction
part
\[
\omega=T_{\rm true}+V_{\rm true}.
\] 
The resolvent reads thus
\beq
G(\omega)=\frac{1}{T_{\rm true}+V_{\rm true}-T_Q-V_Q}.
\eeq
With the help of two smallness assumptions on the momenta of the photon 
\begin{equation}\protect\label{smallnessassumption}
      x_{\gamma} \ll 1, \quad\quad \mbox{and} \quad\quad
      \vec k _{\gamma_{\!\perp}} ^2 \ll M^2_{e\bar{e}} . 
\end{equation}
it has been shown in \cite{PauliMIR}, that indeed
\[
\lim_{N\rightarrow \infty}V_{\rm true}-V_Q=0,
\]
which is very plausible from what we have stated before: the interactions in the
different sectors deviate only by a different number of non-interacting 
photons. We stress the importance of the continuum limit at this point.
If one has a finite number of Fock sectors, the argument does not hold.
We obtain finally
\beq\protect\label{finalresult}
G(\omega)=\frac{1}{T_{\rm true}-T_Q}=:\frac{1}{T^*}.
\eeq
By this,
one has shown that the resolvents are diagonal in the
solution if one makes the two smallness assumptions 
(\ref{smallnessassumption}),
which are fully justified within a bound-state formalism.
In fact, one makes the surprising observation that the resolvents are
totally {\em independent} of the redundant parameter $\omega$. 
It is clear that this parameter is merely a mathematical tool
to perform the calculations in a controlled way. 

We recall the result of the plausibility argument in Chapter 
\ref{ChapterModel}, namely the definition of the energy denominator
$\cal D$, Eq.~(\ref{energienenner}), together with the analytic 
expression for $T^*$, Eq.~(\ref{OmegaStar}), in the 
$|e\bar{e}\gamma\rangle$-sector.
Evaluating Eq.~(\ref{finalresult}) in the $|e\bar{e}\gamma\rangle$- and 
$|e\bar{e}\rangle$-sector, one gets by the arguments of the 
general formalism exactly the expression of Chapter \ref{ChapterModel},
Eq.~(\ref{OmegaStar}). 

One might wonder why the correct fixing of the redundant parameter $\omega$
resulting in the diagonal resolvent Eq.~(\ref{finalresult}) could appear
as an approximation of the expansion Eq.~(\ref{expansion}), or even as an 
{\em ad hoc} assumption in \cite[Eq.~(2.7)]{KPW}. 
The answer is that it is very important for the theory,
at which the stage the (necessary) truncations are made. If one truncates
too early in the formalism, it is clear from the work of 
{\sc Tamm} \cite{Tamm} and {\sc Dancoff} \cite{Dancoff}
that other prescriptions, such as {\em ad hoc} assumptions, 
have to guarantee the solubility of the equations.
This is handled better when the truncation happens 
in a controlled way as in the formalism of iterated resolvents \cite{PauliMIR}. 

The general formalism of effective interactions can be understood as a 
summation over all intermediate states in the effective fermion-antifermion
sector. It is, however, no Tamm-Dancoff truncation, as sometimes referred to in 
the literature.
In fact, the Coulomb potential comes out of the formalism correctly only if one
sums over {\em all} photons and fermion-antifermion pairs, {\em i.e}~in the 
limit $N\rightarrow\infty$.

Note that the renormalization problem remains unsolved. After having 
derived the effective interaction within the present formalism, the cutoff
dependence of the results have to be investigated. 
This is still a ``stumbling stone'' for the case of QCD. In QED the problems
are present, but seem to be not as dramatic.

\section{Application to a model Hamiltonian}
\protect\label{AnExample}

\begin{figure}[t]
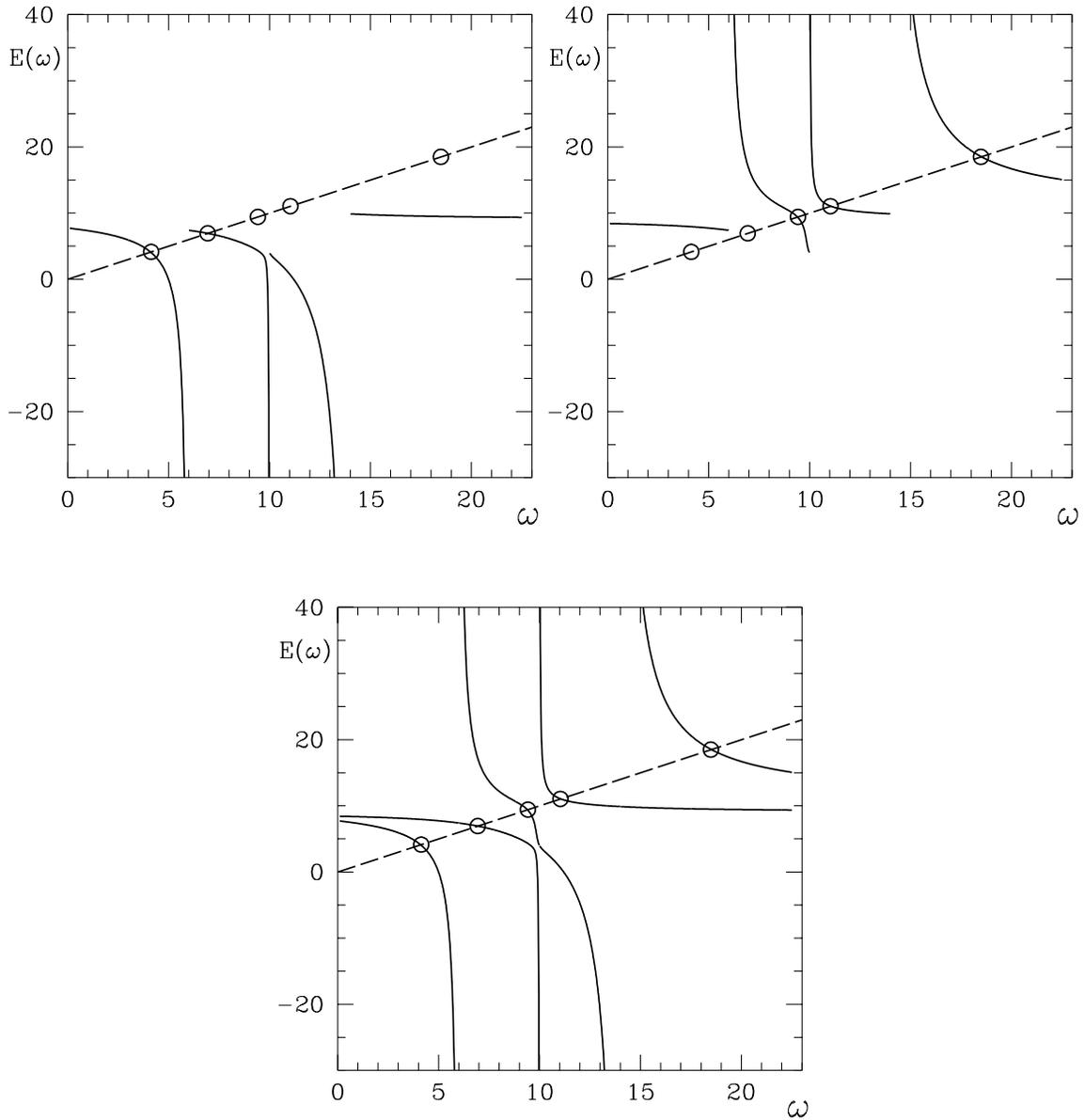

\begin{minipage}{15.5cm}
\centerline{
\psfig{figure=\graphpath effWW_1.epsi,width=7.5cm,angle=0}
\psfig{figure=\graphpath effWW_2.epsi,width=7.5cm,angle=0}}
\vspace*{1cm}
\centerline{
\psfig{figure=\graphpath effWW_0.epsi,width=7.5cm,angle=0}}
\caption[Eigenvalues of the toy Hamiltonian as functions of $\omega$]
{\protect\label{effexample}Eigenvalues of the effective $2\times 2$ dimensional 
toy Hamiltonian as functions of the redundant
parameter $\omega$: (a) the first, (b) the second eigenvalue, (c) the two
plots in one figure. Note the continuous transition from one eigenvalue to the
other.
The five eigenvalues  
at the intersection points with the fixing condition 
$E(\omega)=\omega$ (dashed line) are encircled.}
\end{minipage}
\end{figure}

To give an instructive example of how the method of iterated resolvents works,
we solve the eigenvalue equation
\[
H^{toy}|\psi\rangle=E|\psi\rangle
\]
where the $5\times 5$ dimensional toy Hamiltonian is
\beq\protect\label{toyH}
H^{toy}:=
\left(
   \begin{array}{ccccc}
	D & F & V & S & 0\\
	F & D & V & 0 & F\\
	V & V & D & V & 0\\
	S & 0 & V & D & 0\\
	0 & F & 0 & 0 & D\\
   \end{array}
\right)
\eeq
with $D:=10,V:=4,S:=3,F:=1$. 
The eigenvalues $E_i$ can be calculated by standard methods 
\beq\protect\label{eigenvals}
E_i\in \{4.1386, 6.9387, 9.4239, 11.0368, 18.4908\}.
\eeq	
The space in which this matrix operates 
is divided into two subspaces $|1\rangle$ and $|2\rangle$,
{\em i.e.}~N=2 in Eq.~(\ref{EVP}), so that the diagonal
block matrices are 
\[
\langle 1|H^{toy}|1\rangle =
\left(
   \begin{array}{cc}
	D & F \\
	F & D \\
   \end{array}
\right), \quad\quad
\langle 2|H^{toy}|2\rangle =
\left(
   \begin{array}{ccc}
	D & V & 0\\
	V & D & 0\\
	0 & 0 & D\\
   \end{array}
\right).
\]
The effective Hamiltonian in the $|1\rangle$-sector, according to 
the method of iterated resolvents (cf.~Eq.~[\ref{recursion2}]), is
\[
H^{toy}_{1}(\omega)= H^{toy}_{1}+\sum_{l_1>1}^N 
H^{toy}_{l_1}G_{l_1}(\omega)H^{toy}_{l_1}
=\langle 1|H^{toy}|1\rangle + \langle 1 H^{toy}|2\rangle 
G_{2}(\omega)
\langle 2| H^{toy}|1\rangle,
\]
where the resolvent
\[
G_{2}=\frac{1}{\omega-\langle 2|H^{toy}|2\rangle}=
\left(
   \begin{array}{ccc}
	\omega-D & V & 0\\
	V & \omega-D & 0\\
	0 & 0 & \omega-D\\
   \end{array}
\right)^{-1}
\]
is itself a $3\times 3$
matrix.
The Hamiltonian to diagonalize is a $1\times 1$ block matrix with two columns
and rows. The corresponding two eigenvalues, both functions of $\omega$,
are plotted in Fig.~(\ref{effexample}). Fixing the redundant parameter
$\omega$ by the condition 
\beq\protect\label{line}
E(\omega)=\omega
\eeq
yields exactly the eigenvalues given in Eq.~(\ref{eigenvals}).
Note that the eigenvalue functions for the first and second eigenvalue
have continuous transitions. This is due to the diagonalization algorithm
used which will always consider the lowest eigenvalue to be the first one.
Note also the strange behavior of the curves yielding the second and third
eigenvalue of the whole matrix. One sees that the eigenvalues repel each other
in the numerical program. However, it is clear that there 
should be a crossing of the two curves. 
In fact, we have two functions in the plot, one of which has one and the other 
has two poles. This yields five intersection points with $E(\omega)=\omega$, as 
expected.
Another important observation is the fact that the dependence of the functions 
on $\omega$ is rather weak at the intersection points, which yields numerically
stable eigenvalues.

The result obtained should be compared with the similar plot in 
\cite[Fig.~5]{CoralGables}. There, a $4\times 4$ matrix was solved 
by applying the projection method row by row, and one ended up with
an effective $1\times 1$ Hamiltonian, a function of $\omega$, 
but nonetheless a real number. Here, we have shown 
that the formalism can applied also
to matrices: the effective Hamiltonian is a matrix depending on $\omega$.
But the functions for the different eigenvalues $E_i(\omega)$ combine
in such a way that continuous functions emerge which have $N$ intersection 
points with the line defined by Eq.~(\ref{line}). This  corresponds to the $N$
eigenvalues of the toy matrix (\ref{toyH}), just as in the case discussed in 
\cite{CoralGables}.

\chapter{Summary and Outlook}


In this thesis, the calculation of 
the spectra and the wavefunctions of an electro-magnetically bound,
relativistic fermion-antifermion
system in $(3+1)$ dimensions has been performed.
The problem has been solved for all
components of the total angular momentum, $J_z$, and 
for an arbitrarily large coupling constant.
The latter required the application of non-perturbative 
methods in quantum field theory.
An effective Hamiltonian in the $|e\bar{e}\rangle$-sector has been used. 
The corresponding integral equation is mapped into a matrix eigenvalue
problem and solved numerically.
The spectrum and the wavefunctions of positronium have been calculated in 
the regime of an
unphysically strong coupling constant, $\alpha=0.3$.
A computer code has been created 
that yielded numerically stable results with a minimized demand on computer
equipment. 
The numerical methods described in \cite{KPW} have been tested and improved. 
In particular, the counterterm technology for the Coulomb singularity
has been fully implemented. Since the counterterms may in general involve an
integral which is not analytically solvable, one can  
use only numerical integrations and the numerical effort increases.
As can be seen by comparing the results of this work with 
\cite{KPW}, the convergence is improved noticeably.
The most prominent aspect of the numerical results is that 
the spectrum is found to 
coincide to a very high degree of accuracy with the results of elaborate 
perturbation techniques. 
Previous work in this direction ({\sc Tang}~\cite{Tang}, 
{\sc Kaluza}~\cite{Kaluza})
failed to produce significant results because of numerical problems.

Special care has been taken in the comparison of the eigenvalues of
the positronium model used with results of
perturbation theory. In the sector $J_z{=}0$ the Hamiltonian was split up into
four block matrices with definite quantum numbers of charge conjugation
and $\cal T$-parity. This allows a one-to-one comparison of states with those
resulting from perturbation theory,
since the symmetry quantum numbers can be related to those used in instant form
dynamics.
In addition to the fact that the multiplets of a fixed principal quantum
number $n$ contain the correct number of states up to at least $n=5$, 
we find that the correct combinations of quantum numbers are
included in each of these multiplets.    
A careful study of the higher excited states shows
that the calculations of this work are trustworthy even for these states,
as opposed to, for example, lattice gauge theories.


As was pointed out, rotational invariance is problematic when using
front form dynamics. The operators attached to the transverse rotations
are complicated and contain the interaction explicitly. 
It is therefore important to study the full spectrum of positronium by 
generalizing the model to a non-vanishing $J_z$ component of the total 
angular momentum.  
We find that the matrix elements can be calculated even for arbitrary 
$J_z$. 
The improvement of the counterterm technology was crucial 
in this case. 
It is important that the counterterms are taken care of individually,
because the singularity structures of the diagonal matrix 
elements differ 
considerably from each other.
We stress the fact that this counterterms are due to numerical problems
and have nothing to do with the physics of the positronium model.
We {\em neither} use any special prescription to restore gauge invariance,
as suggested by {\sc Coester} \cite{Coester}, {\em nor}
do we construct non-covariant counterterms for this purpose, 
as necessary in front form perturbation theory \cite{BurkardtLangnau}.
One way to study rotational symmetry in front form dynamics 
requires the diagonalization of operators at least
as complicated as the light-cone Hamiltonian itself. We listed these operators
in the first section of Chapter \ref{DrAChapterJz}. It is clear that this is
a non-trivial task. Almost all efforts up to date are spend to construct
and to diagonalize the light-cone Hamiltonian, and not the other dynamical
operators. Surely, the latter is an interesting project.
However, we were able to construct an effective Hamiltonian and to 
solve for its spectrum in all sectors of different $z$-components of the 
total angular momentum. In this way,
we can learn from the physical observables (mass squared eigenvalues), how 
quantum numbers concerning the complicated operators can be attributed to 
the states.
We point out the importance of the non-perturbative approach used in 
the present work. It has 
been shown in several models (even QED$_{3+1}$) \cite{BurkardtLangnau} 
that perturbation theory in front form dynamics yields results where
rotational symmetry is
explicitly broken. There are no degeneracies of states with 
the same total angular momentum.  
This is clear from our point of view. Exact rotational symmetry is expected
only if it is summed over {\em all} Fock states and {\em all} momenta. 
Because we used an effective theory in the present work, 
where the effects of all Fock states are
included in the effective interaction, 
we observe the desired degeneracies in our approach.
 
The properties of the spectra calculated for different values of $J_z$
are the following. Most remarkably, we find that states of different $J_z$ 
but same total angular momentum $J$ are numerically degenerate.
There are slight deviations from exact degeneracy also for a large number
of integration points. The latter can be explained by the restrictions of the
positronium model used, {\em e.g.}~finite cutoff and others.
These degeneracies make it possible to classify the states according 
to their quantum numbers of total angular momentum {\em a posteriori}
and make 
the diagonalization of the complicated operators of transverse rotations
superfluous. 
Even more, it shows that rotations are unproblematic 
on the light cone,  because rotational symmetry is restored in the solution, 
a previously worrying fact.  
The results show very good agreement with the values obtained by 
equal-time perturbation theory up to order ${\cal O}(\alpha^4)$.


The model was enlarged by including the one photon annihilation channel.
The inclusion is a test for the consistency
of the model. In particular, the conception of an effective 
theory operating with resolvents can 
be falsified if the annihilation channel cannot be implemented without
special assumptions.
Our results show that the implementation of the annihilation channel
is unproblematic in the sense that the same formalism can be applied as 
in the case of the projection of the effective $|e\bar{e}\gamma\rangle$-sector
onto the electron-positron sector.
As an interesting
property of the annihilation channel we find a strict separation
of the instantaneous and the non-instantaneous interaction:
the seagull interaction is present only in the $J_z=0$ sector, whereas the
dynamic graph has non-vanishing matrix elements only for $J_z=\pm 1$.
Our approach passes another test: both graphs 
yield the same contributions to the eigenvalues and consequently 
rotational symmetry is seen also in the
spectrum including the annihilation channel.
Moreover, the inclusion of the annihilation channel 
improves the results for the hyperfine splitting. 

We stress the point that the implementation of the annihilation channel
completes the investigations of how to construct an effective 
interaction for the electromagnetic fermion-antifermion system in the meaning
of the method of iterated resolvents, described in 
Chapter \ref{DrAChapterEffInt}. We have put all effects of higher Fock states
into an effective $|e\bar{e}\gamma\rangle$-sector, as far as the 
{\em spectator interaction} is concerned, {\em i.e.}~the 
interactions in which the photons
are not directly involved. The remainder of the interaction relies
in the coupling function of the vertices. 
A hint to this conclusion is the logarithmic cutoff dependence 
of our results. It is clear that the coupling constant depends on the cutoff
and has to be analyzed.
A future aim, beyond the scope of this work, will be to show that 
the physical results of our model 
become independent of the cutoff, as soon as 
renormalization group techniques are consistently applied.   
This is supported by the fact that we find a stabilizing effect of the
annihilation channel on the dependence of the spectrum on the cutoff:
all eigenvalues show slower variation with growing cutoff $\Lambda$ when
the annihilation graph is added,
in some cases this even prevents level crossings with other states.
We therefore conclude that our model is correct as long as the  
vacuum polarization effects are not considered.
 
The possibility of one boson annihilation is the main difference
between QED and QCD in effectively truncated Fock spaces 
where the dynamic three- or
four-gluon interaction is not possible but resides in the coupling function. 
The one boson sector exists only in QED because of the constraint of 
color neutrality in QCD. 
We showed that this sector can be projected onto the (already) effective  
electron-positron sector.


In Chapter \ref{DrAChapterEffInt} we described the 
method of iterated resolvents which allows for the construction of
effective interactions on the light-cone. We applied this formalism to our
problem. It was 
shown that the assumption is correct
that the effects of the higher
Fock states can be accounted for by the fixing the redundant parameter $\omega$
by a function of the light-cone momenta, Eq.~(\ref{OmegaStar}). 
The fixing was 
necessary to derive an effective integral equation starting from the 
QED$_{(3+1)}$ Hamiltonian and 
is equivalent to the calculation of the effective resolvent in the 
$|e\bar{e}\gamma\rangle$-sector. We showed that this resolvent is indeed
diagonal in the Fock basis used and is therefore a function rather than an 
operator.
This fixing procedure was 
supported before by the numerical precision of the calculations 
using it as an assumption. 

Because of a confusion of technical terms we stress the following in this
context. The formalism applied here is misnamed when referred to 
as a Tamm-Dancoff approach. The Tamm-Dancoff
approach was invented for nuclear forces
in instant form dynamics. Its three main steps become very clear
in the work of {\sc Tamm} \cite[\S 2]{Tamm}. The first step is to
{\em truncate} the Fock space to two sectors ($P$- and $Q$-space). 
Then a projection of one space onto the other is performed. The emerging
energy denominator (or resolvent) is {\em modified} to render the system
solvable as a third step.
What we have done is (a) regulation of the number of Fock sectors,
(b) derivation of an effective Hamiltonian by subsequent projections of
Fock sectors onto another, (c) proof that in the limit of infinitely
many Fock sectors, the resolvents become diagonal for bound states.
In our approach it is clear that a sheer truncation must fail, because
essential parts of the Fock space are missing. 
It remains to treat the so-derived effective Hamiltonian with a 
renormalization group analysis. We have shown that the dependence of our
results is logarithmic (weak) in the cutoff $\Lambda$. The inclusion of the 
annihilation channel stabilizes the eigenvalues. Whether or not this hints
to a weaker dependence of the (running) coupling of QED on the cutoff 
when compared with the coupling of QCD, 
where this channel is absent, is subject to further investigation.


This thesis was motivated by the will to understand the problems of 
non-perturbative methods on the light-cone. 
The hope is clearly to find an adequate description of the low energy
region of QCD and thereby of the QCD bound states, like mesons and hadrons.
Front form dynamics offers very promising 
features, such as a simple vacuum, and more. 
Other methods have been proposed to deal with these problems,
namely approaches which diverge from the constituent quark picture.
The method of {\em similarity transforms} of the bare 
Hamiltonian proposed 
by {\sc Wilson et al.} \cite{weakWilson} and the 
{\em coupling coherence} technique
advocated by {\sc Perry et al.}~seem to be very promising. 
Although the formalism is very elaborate, there are problems in calculating
actual numbers. A model of {\sc Brisudova et al.~}\cite{BrisudovaPerry} aims
at calculating mesons with one heavy constituent quark with these methods. 
However, the authors
are forced to perform a line of approximations, and end up with a 
non-relativistic, rotationally non-invariant Schr\"odinger equation, 
which makes it hard to tell the actual achievements of the 
approach apart from the draw-backs of the approximations made.
An instructive analytical study is that of {\sc Jones et al.~}\cite{Jones},
which makes the 
approach of {\sc Wilson} and {\sc Perry} very clear. The authors
calculate the positronium spectrum up to order ${\cal O}(\alpha^4)$ 
and show that 
the results are independent of the cutoff up to order ${\cal O}(\alpha^5)$.
The triplet ground states are degenerate. This is not surprising 
because the effective interactions are derived for small couplings 
$\alpha\ll 1$ and a non-relativistic limit of the theory is considered.
It is known (Eq.~\ref{rotviol}) that the term that breaks rotational invariance
vanishes in the non-relativistic limit.
A similar method is that of {\sc Wegner} \cite{Wegner}, who uses flow 
equations to renormalize the bare Hamiltonian. Attempts
have been made \cite{Pirner} to apply this formalism to QED and to calculate 
the positronium spectrum.    
A link between the DLCQ approach and the method of {\sc Wilson et al.}~is 
the work of {\sc Ammons} \cite{Ammons}. There, a Tamm-Dancoff truncation
is performed together with an application of the similarity transforms.
These approaches are distinct from the one used in the present work 
in the following sense. First, the Hamiltonian is constructed using 
renormalization theory, then the associated eigenvalue problem is solved.
Up to date, the latter step has been performed using bound state 
perturbation theory or other {\em perturbative} methods, although in principle
a non-perturbative solution seems possible.
Complementary to this formalism, we construct 
{\em first} an effective Hamiltonian to account for the many-body aspects
of the theory \cite{PauliPriv},
solve for the spectra with a non-perturbative method and {\em then} 
have to analyze the results using renormalization techniques to study 
the implications of Quantum Field Theory.


Concluding, the formalism of DLCQ together with the theory of effective 
interactions, as described in Chapter \ref{DrAChapterEffInt}, can be combined
to yield
non-perturbative results for a relativistic bound system 
in very good agreement with other standard methods.
Even in (3+1) dimensions it is possible to construct a consistent and 
solvable model for bound states with an arbitrarily large coupling.


With the results of this work and the computer code at hand, 
one can now proceed to attack QCD bound states in true $(3+1)$
dimensions. 
Because we have reduced the Fock space by effective methods, 
the difference
between QED and QCD relies in the coupling constant, 
or rather coupling function.
The method of iterated resolvents \cite{PauliMIR} allows for the 
calculation of these coupling functions for both QED and QCD, but this is 
tedious work and has not been tackled.
On the other hand, {\sc Merkel et al.} \cite{PauliMerkel} 
used a phenomenological running coupling for QCD, 
inspired by the work of {\sc Richardson} 
\cite{Richardson}, to construct an integral equation, analogous to the one in
the present work, for the bound states of a quark-antiquark system. 
This equation is solved with variational methods. 
The fitted meson masses
show good agreement with experiment.
This phenomenological coupling can now be plugged into the formalism and 
computer code derived in the present work to solve 
for the spectrum of a QCD-bound quark-antiquark system. 
This seems possible, because 
the full effective Hamiltonian
of a gauge theory contains only three essential graphs, which themselves
include a coupling {\em function}.
The computer code created in the present work already 
contains two of these graphs,
and the third one (the two gluon annihilation) is known to be less important. 
Consequently, the first task is to find a reasonable 
phenomenological running coupling constant to construct such a test.
The main problem is to find an appropriate regularization of the 
severe ($p^{-4}$) singularities occuring. 
Because of this, the dependence of the results on the regularization 
scheme used has to be investigated carefully.
Nevertheless, this seems a promising way to proceed to understand 
hadrons as QCD-bound states.

\begin{appendix}

\chapter{Notation}\label{AppxNotations}


The notations used in this work are compiled in the sequel. New technical 
terms are {\em italicized} when first introduced.
The conventions used for discretizing the theory can be found in Appendix 
\ref{AppxQED}.

\bigskip
\noindent
\underline{\bf Coordinates}
\bigskip

\noindent
The {\em light-cone coordinates} are defined by
\[
x^{\pm}:= (x^0\pm x^3).
\]
We use the {\em metric}
\[
g^{\mu\nu}=\left(\begin{array}{cccc}
			 0 & 0 & 0 & 2 \\
			 0 &-1 & 0 & 0 \\
			 0 & 0 &-1 & 0 \\
			 2 & 0 & 0 & 0 
		 \end{array}
	   \right).
\]
The {\em greek} indices run like $(+,1,2,-)$, 
{\em latin} indices denote transversal directions,
{\em e.g.\ } $i=1,2$.
Analogously the {\em momentum coordinates} of a particle are 
$(p^+,\vec{p}_{\perp},p^-)$. Hence, $p^-$ is the {\em light-cone energy}
of a particle.
An underlined variable denotes a merely spatial vector  
$
\ux := (x^-,\vec{x}_{\perp}).
$
\noindent
Some frequently used symbols are:
\bigskip

\centerline{
\begin{tabular}{ll}
$P^+$ & total longitudinal momentum\\
$\vec{P}_{\perp}:= 0$ & total transversal momentum\\
$2L$ & longitudinal box length \\
$2L_{\perp}$ & transversal box length \\
$K:= \frac{L}{\pi}P^+$ & harmonic resolution (integer valued)\\
$\Omega:=2L(2L_{\perp})^2$ & discretization volume \\
all QN & summation over all quantum numbers
\end{tabular}
}
\bigskip

\noindent
The relative coordinates of the i-th particle are
\beq\label{Defx}
x_i := \frac{p^+_i}{P^+},\quad\quad \mbox{and} \quad\quad
\vec{k}_{\perp i} := x_i P^+ -\vec{p}_{\perp i},
\eeq
referred to as the {\em longitudinal momentum fraction} and the (relative)
{\em transversal momentum}, respectively. If $\vec{P}_{\perp}:=0$, 
they have have the properties
\[ 
\sum_i x_i =1, \quad\quad  \mbox{and} \quad\quad \sum_i k_i =0. 
\]

\bigskip
\noindent
\underline{\bf Commutation relations}
\bigskip

\noindent
The commutation relations according to the {\sc Dirac-Bergmann}
\cite{Bergmann,Sunder}
algorithm are
\begin{eqnarray*}
\left\{\psi_{+\alpha}(\ux),\psi^{\dagger}_{+\beta}(\uy)\right\}_{x^+ = x^+_0}
&=&\frac{1}{2}\Lambda_{+\alpha\beta}\delta(x^- - y^-)\delta^{2}(\vec{x}_{\perp}
-\vec{y}_{\perp}),\\
\left[A^i(\ux),\del^+A^j(\uy)\right]_{x^+ = x^+_0}
&=&\frac{i}{2}\delta^{ij}\delta(x^- -y^-)
\delta^{2}(\vec{x}_{\perp}-\vec{y}_{\perp}).
\end{eqnarray*}  
To be consistent with these relations, in the expansions of the fields
the operator-valued coefficients have to obey
\begin{eqnarray*}
\left\{ b_{\lambda,\un}\, ,b^{\dagger}_{\lambda',\um}\right\}=
\left\{ d_{\lambda,\un}\, ,d^{\dagger}_{\lambda',\um}\right\}&=&
\delta_{\lambda,\lambda'}\delta_{n,m}\delta^{2}_{\nperp,\mperp},\\
\left[ a_{\lambda,\up}\, ,a^{\dagger}_{\lambda',\uq}\right]&=&
\delta_{\lambda,\lambda'}\delta_{p,q}\delta^{2}_{\pperp,\vec{q}_{\perp}}.
\end{eqnarray*} 
All other (anti-)commutators vanish.

\bigskip
\noindent
\underline{\bf Coupling constant}
\bigskip

\noindent
We set 
\[
\alpha := \frac{g^2}{4\pi}.
\]
In QED: $g\equiv |e_0|$, $\alpha=1/137.0359895(61)=0.007297353$.
In the discretized theory we use
\[
\beta:=\frac{g^2}{4\pi K L_{\perp}^2}.
\]

\bigskip
\noindent
\underline{\bf Spinors and polarization vectors}
\bigskip

\noindent
The fermion fields are separated into two different helicity eigenstates by 
\[
\psi_{\pm}=\Lambda_{\pm}\psi. 
\]
Here, the projection operators read 
\[
\Lambda_{\pm}:=\frac{1}{2}\gamma^0\gamma^{\pm}
\]
or explicitly
\[
\Lambda_+=\frac{1}{2}\left(\begin{array}{cccc}
			 1 & 0 & 1 & 0 \\
			 0 & 1 & 0 & -1 \\
			 1 & 0 & 1 & 0 \\
			 0 &-1 & 0 & 1 
		 \end{array}
	   \right),\quad
\Lambda_-=\frac{1}{2}\left(\begin{array}{cccc}
			 1 & 0 &-1 & 0 \\
			 0 & 1 & 0 & 1 \\
			-1 & 0 & 1 & 0 \\
			 0 & 1 & 0 & 1 
		 \end{array}
	   \right),
\]
The explicit form of the spinors and polarization vectors used 
is given in Eqs.~(\ref{spinors})~and~(\ref{polvectors}).

\chapter{\label{AppxQED}QED(3+1) on the light-cone}


The Hamiltonian operator of Quantum Electrodynamics in $(3+1)$ dimensions
in front form dynamics is derived.
QED describes the interaction of (electrically) charged
parti\-cles. The interaction is intermediated by the gauge field of the
photons. 
We begin with the {\em Lagrangian density} of QED$_{(3+1)}$
\[
{\cal L}=- \frac{1}{4}F_{\mu\nu}F^{\mu\nu}+
\frac{i}{2}\left[\bar{\psi}\gamma^{\mu}\partial_{\mu}-(\partial_{\mu}
\bar{\psi})\gamma^{\mu}\right]\psi - m_f\bar{\psi}\psi 
- g\bar{\psi}\gamma^{\mu}\psi A_{\mu}.
\]
The {\em field strength tensor} is
\[
F^{\mu\nu}:=\partial^{\mu}A^{\nu}-\partial^{\nu}A^{\mu}.
\]
The equations of motion emerging for the gauge fields are the
{\em Maxwell equations}
\beq\label{MaxwellEqn}
\partial_{\mu}F^{\mu\nu}=j^{\nu}:=g\bar{\psi}\gamma^{\nu}\psi,
\eeq
and for the Fermi fields the {\em Dirac equation}
\beq\label{DiracEqn}
\left(i\gamma^{\mu}D_{\mu}-m_f\right)\psi=0,
\eeq
where
\[
D_{\mu}:=\partial_{\mu}+igA_{\mu}
\]
denotes the {\em covariant derivative}.
We use in this work the {\em light-cone coordinates}
\[
x^{\pm}:=x^0\pm x^3.
\]
$x^+$ plays the role of a time, $x^-$ is a direction of space.
To write scalar products involving Dirac matrices likewise, we define
\[
\gamma^{\pm}:=\gamma^0 \pm \gamma^3.
\]
We use the usual Dirac representation for these matrices.
We work in the so-called {\em light-cone gauge} 
\begin{equation}
A^+ =A^0+A^3\equiv 0,
\end{equation}
which leads to a consistent theory in the normal mode sector.
Deriving the {\em light-cone energy} $P^-$, we shall follow the
line of arguments of {\sc Tang} \cite[Ref.\ 1]{Tang}.
With help of the {\em projection operators}
\[
\Lambda_{\pm}:=\frac{1}{2}\gamma^0\gamma^{\pm}=\frac{1}{4}\gamma^{\mp}
\gamma^{\pm},
\]
we separate the fields into states of different helicity 
\[
\psi_{\pm}=\Lambda_{\pm}\psi.
\]
It turns out that only six out of the twelve fields 
$A^{\mu}$, $\psi_+$, $\psi_+^{\dagger}$, $\psi_-$, 
and $\psi_-^{\dagger}$ occuring in the Lagrangian
density are independent of the others: the conjugate momenta of 
$A^-$, $\psi_-$, and $\psi_-^{\dagger}$ vanish. The Euler-Lagrange equations
in these fields contain no time derivative, and have the rank of constraints.
This is a consequence of the chosen quantization plane. 
The equations of motion for the dependent fields are
\begin{eqnarray}
i\partial^+\psi_-&=&\left(-i\partial_i\alpha^i + \beta m_f + g A_i 
\alpha^i\right)\psi_+\nonumber,\\
i\partial^+\psi^{\dagger}_-&=&-\psi^{\dagger}_+\left(i\stackrel{\leftarrow}
{\partial_i}\alpha^i 
+ \beta m_f + g A_i 
\alpha^i\right)\nonumber,\\
(i\partial^+)^2 A^- &=& 2\partial^+\partial_i A^i + 
4g\psi^{\dagger}_+\psi_+.\label{abhFelder}
\end{eqnarray}
with the usual Dirac matrices $\alpha^i$,$\beta$ and 
$\psi^{\dagger}_+\stackrel{\leftarrow}{\partial_i}:=\partial_i
\psi^{\dagger}_+$.

To quantize correctly, we have to eliminate all dependent degrees of freedom.
This can be achieved 
either\protect\footnote{Cf.~\protect\cite[Chapter 4]{Tang}.} by inverting 
the equations (\ref{abhFelder}) or by applying the 
Dirac-Bergmann algorithm \cite{Bergmann,Dirac2,Sunder}. The latter leads 
for a reasonable Lagrange density necessarily to the right commutation 
relations for the independent fields.

We substitute the dependent fields by functionals of the dynamic 
degrees of freedom
\begin{eqnarray*}
\psi_-&=&\frac{1}{i\partial^+}\left(-i\partial_i\alpha^i + \beta m_f\right)
\psi_+ - \frac{g}{i\partial^+} A^i \alpha^i\psi_+,\\
\psi^{\dagger}_-&=&\frac{-1}{i\partial^+}\psi^{\dagger}_+
\left(i\stackrel{\leftarrow}{\partial_i}\alpha^i 
+ \beta m_f\right) +  \frac{g}{i\partial^+}\psi^{\dagger}_+ A^i \alpha^i,\\
A^-&=&\frac{2}{(i\partial^+)^2}\partial^+\partial_i A^i 
+\frac{4g}{(i\partial^+)^2}\psi^{\dagger}_+\psi_+.
\end{eqnarray*}
The inverse derivatives are defined by Green functions\protect\footnote{These 
definitions are not unique \protect\cite{BP}.}.
From the energy-momentum tensor 
\[
{\cal T}^{\mu\nu}=\sum_r \frac{\delta {\cal{L}}}{\delta(\partial_{\mu}\phi_r)}
\partial^{\mu}\phi_r-g^{\mu\nu}{\cal{L}},
\] 
one obtains the
momenta $P^{\mu}$ by integrating over all space directions
\[
P^{\mu}=\frac{1}{2}
\int_{-\infty}^{\infty} dx^-\int_{-\infty}^{\infty}d^2x_{\perp}
{\cal T}^{+\mu}.
\]
The {\em light-cone energy} $P^-$ has  the following structure
\[
P^- =\frac{1}{2}
\int_{-\infty}^{\infty} dx^-\int_{-\infty}^{\infty}d^2x_{\perp}
{\cal{P}}^-_0+g{\cal{P}}^-_1+g^2 {\cal{P}}^-_2.
\]
The different terms are 
\begin{eqnarray*}
{\cal P}^-_0&=&\del^i A^j\del^i A^j -\del^i A^j\del^j A^i
+\left\{i\del^+ i\del_i A^i\oddel i\del^+ i\del_j A^j\right\}_{sym}\\
&&+2\left\{\ppsid\diraci\odel\diracj\ppsi\right\}_{sym}\\
{\cal{P}}^-_1&=&-2\left\{\ppsid A^i\alpha^i\odel\diraci\ppsi\right\}\\
&&-2\left\{\ppsid \left[i\stackrel{\leftarrow}{\partial_i}
\alpha^i+\beta m_e\right]\odel A^i\alpha^j\ppsi\right\}\\
&&-4\left\{\ppsid\ppsi\oddel i\del^+ i\del_i A^i\right\}_{sym}\\
{\cal{P}}^-_2&=&2\left\{\ppsid A^i\alpha^i\odel A^j\alpha^j\ppsi\right\}  
+4\left\{\ppsid\ppsi\oddel \ppsid\ppsi\right\}_{sym}.
\end{eqnarray*}
Here, the symmetric brackets are defined by
\[
\left\{A\odel B\right\}_{sym}:=\frac{1}{2}\left[A\odel B-\left(\odel A\right) 
B\right],
\]
\[
\left\{A\oddel B\right\}_{sym}:=A\oddel B+\left(\odel A\right)\left(\odel 
B\right)+\left(\oddel A\right)B. 
\]

We expand the system in plane waves at light-cone time $x^+=0$,
and work subsequently in momentum space. The notations used here are those of 
\cite{Schladming}. 
By restricting the system to a box
\beqa
&&-L_{\perp}\leq x^i \leq L_{\perp},\\
&&-L \leq x^- \leq L,
\eeqa 
we discretize the momenta.
We have to impose boundary conditions compatible with the equations of 
motion (\ref{MaxwellEqn},\ref{DiracEqn}).
The expansions of the dynamical fields with the notation
$\ux:=(x^-,\xperp)$ for space and
$\uk:=(k^+,\kp)$ for momentum variables read explicitly
\begin{eqnarray*}
\ppsi(\ux)&=&\frac{1}{\sqrt{\Omega}}
\sum_s\sum_{\un}\left[b_{\un,s}u_+(\lambda)
e^{-i\uk\cdot \ux}+d^{\dagger}_{\un,-s}v_+(\lambda)
e^{i\uk\cdot\ux}\right],\\
A^i(\ux)&=&\frac{1}{\sqrt{\Omega}}
\sum_{\lambda}\sum_{\up}\frac{1}{\sqrt{k^+}}\left[
 a_{\up,\lambda}\epsilon^i(\lambda)e^{-i\uk\cdot\ux} 
+a^{\dagger}_{\up,\lambda}\epsilon^{*i}(\lambda)
e^{i\uk\cdot\ux}\right].
\end{eqnarray*}
Here, $\Omega=2L(2L_{\perp})^2$ symbolizes the volume of the box, 
in which the momentum vectors have to lie.

We are forced to demand periodic boundary conditions in all space directions 
for the gauge fields, since they couple to a bilinear term.
The boundary conditions for the fermion fields are not subject to such a 
constraint.
It is convenient to impose antiperiodic boundary conditions 
merely for the longitudinal fermion field.
The zero mode of this field vanishes by that choice. Of course, this is of
no importance to us, since we consider {\em a priori} 
only the normal mode sector by working in the light-cone gauge.
The summations are performed in the following way.
For fermions the indices are
\begin{eqnarray*}
k^+ &=&\frac{n\pi}{L} \quad n=1,3,5, \ldots\\
k^i &=&\frac{n^i\pi}{L_{\perp}} \quad n^i=0,\pm 1,\pm 2,\ldots
\end{eqnarray*} 
and photons have
\begin{eqnarray*}
k^+ &=&\frac{p\pi}{L} \quad p=2,4,6, \ldots\\
k^i &=&\frac{p^i\pi}{L_{\perp}} \quad p^i=0,\pm 1,\pm 2,\ldots .
\end{eqnarray*} 
The projected spinors are defined according to {\sc Lepage} and 
{\sc Brodsky} \cite{BLepage} to be
$
u_+(\lambda):=\chi(\lambda)
$
and
$
v_+(\lambda):=\chi(-\lambda)
$. The spinors
\begin{eqnarray}\label{spinors}
\chi(\uparrow)=\frac{1}{\sqrt{2}}\left(\begin{array}{c}
1\\0\\1\\0 \end{array}\right), \quad
\chi(\downarrow)=\frac{1}{\sqrt{2}}\left(\begin{array}{c}
0\\1\\0\\-1\end{array}\right)
\end{eqnarray} 
obey the relations
\beqa
\chi^{\dagger}(\lambda)\chi(\lambda')&=&\delta_{\lambda,\lambda'},\\
\sum_{\lambda}\chi_{\alpha}(\lambda)\chi^{\dagger}_{\beta}(\lambda)
&=&\Lambda^+_{\alpha\beta}.
\eeqa
The polarization vectors 
\begin{eqnarray}\label{polvectors}
\epsilon(\uparrow):=\frac{-1}{\sqrt{2}}\left(\begin{array}{c}
1\\i\end{array}\right), \quad\quad \mbox{and} \quad\quad
\epsilon(\downarrow):=\frac{1}{\sqrt{2}}\left(\begin{array}{c}
1\\-i\end{array} \right).
\end{eqnarray}
are orthonormal in the transverse space and complete
\beqa
\epsilon^{*}(\lambda)\epsilon(\lambda')&=&\delta_{\lambda,\lambda'},\\
\sum_{\lambda}\epsilon_i(\lambda)\epsilon^*_j(\lambda)
&=&\delta_{ij}.
\eeqa
The projected spinor $\ppsi(x)$ can easily be generalized to a spinor for the
full free fermion field $\psi(x)$ by substituting $u_+(\lambda)$ and 
$v_+(\lambda)$ by
\beqa
u(\uk,\lambda) &=&\frac{1}{\sqrt{k^+}}\left(k^+ + \beta m_f +
\vec{\alpha}_{\perp}\cdot\vkp\right)\chi(\lambda)\\
v(\uk,\lambda) &=&\frac{1}{\sqrt{k^+}}\left(k^+ + \beta m_f +
\vec{\alpha}_{\perp}\cdot \vkp\right)\chi(-\lambda).
\eeqa
The same is true for the gauge field, if the purely transversal vector
$\vec{\epsilon}_{\perp}(\lambda)$ is substituted by the four vector
\[
\epsilon^{\mu}(\lambda):=\left(\begin{array}{c}
0 \\ 2\frac{\vec{\epsilon}_{\perp}\cdot\vkp}{k^+} \\ 
\vec{\epsilon}_{\perp}(\lambda)
\end{array}\right).
\]
We quantize by postulating the usual commutation relations for the coefficients 
of the Fourier decomposition of the fields. The coefficients become operator
valued and obey
\[
\left\{ b_{s,\uk},b^{\dagger}_{s',\uk'}\right\}=
\left\{ d_{s,\uk},d^{\dagger}_{s',\uk'}\right\}=
\delta_{s,s'}\delta^{(3)}_{\uk,\uk'}
\]
\[
\left[ a_{\uk,\lambda},a^{\dagger}_{\uk',\lambda'}\right]=
\delta_{\lambda,\lambda'}\delta^{(3)}_{\uk,\uk'}
\] 
\[
\left\{ b,b\right\}=
\left\{ d,d\right\}=
\left\{ b,d^{\dagger}\right\}=
\left[ a,b\right]=
\left[ a,d\right]=
\left[ a,\bdagg\right]=
\left[ a,\ddagg\right]=0.
\]
Since we address ourselves in this work to solve an
{\em integral} equation, we additionally list the expansions 
of the fields in the continuum
\beqa
\ppsi(x)&=&\frac{1}{\sqrt{2(2\pi)^3}}
\sum_{\lambda}\int_{0}^{\infty}dk^+\int_{-\infty}^{\infty}d^2\kp
\frac{1}{\sqrt{k^+}}\left[b(\uk,\lambda)u_+(\lambda)e^{-i\uk \cdot \ux}+
d^{\dagger}(\uk,\lambda) u_+(\lambda) e^{+i\uk\cdot \ux}\right]\\
A^{i}(x)&=&\frac{1}{\sqrt{2(2\pi)^3}}
\sum_{\lambda}\int_{0}^{\infty}dk^+\int_{-\infty}^{\infty}d^2\kp
\frac{1}{\sqrt{k^+}}\left[a(\uk,\lambda)\epsilon^i(\lambda)e^{-i\uk \cdot \ux}+
a^{\dagger}(\uk,\lambda)\epsilon^{*i}(\lambda) e^{+i\uk\cdot \ux}\right].
\eeqa

\chapter{Matrix elements of the light-cone Hamiltonian in QED}
\label{AppxHLC}


\noindent
The matrix elements of the QED$_{(3+1)}$ Hamiltonian in 
light-cone quantization have been derived elsewhere \cite{Tang}. 
The light-cone Hamiltonian is defined as
\[
H_{\rm LC}:=P^{\mu}P_{\mu}=P^+P^-.
\]
It is usually divided into the parts
\[
H_{\rm LC}= T+V+S+F+C,
\]
where the {\em kinetic} energy is
\[
T=\sum_q\frac{m_f^2+\vec{k}^2_{\perp q}}{x_q}
\left(\bdagg_q b_q+\ddagg_q d_q\right)
+\sum_q\frac{\vec{k}^2_{\perp q}}{x_q}\adagg_q a_q.
\]
$V$, $S$, $F$ and $C$ are the {\em vertex-}, the {\em seagull-} and
the {\em fork}--interaction, as well as the {\em contractions}
\footnote{
This nomenclature stems from the operator structure of the matrix elements. 
Is the parton number changed by an interaction by  $0,1$ or $2$, the operators
are called seagull,vertex, or fork, respectively. Contractions
occur due to normal ordering of the seagull operators.}.
The Fork interaction is absent in this work because of the truncation of the
Fock space and is not listed.

The creation operators $\bdagg_q$,$\ddagg_q$ and $\adagg_q$ create
plane waves (states) for electrons, positrons and photons, respectively.
These particles are characterized by their quantum numbers
\[
q:=(x,\vkp,\lambda).
\]
A particle with {\em longitudinal momentum fraction} $x$,
{\em transversal momentum} $\vkp$ and {\em helicity} $\lambda$
is annihilated by the 
(annihilation) operators $b_q$,$d_q$,$a_q$, depending on the nature of 
the particle, and represented in the graphs of the matrix elements by
the following symbols

\centerline{
\begin{tabular}{cll}
\parbox{2.5cm}{\vspace{0.0cm}
\psfig{figure=\graphpath symbols_b.epsi,width=2.5cm,angle=-90}}& 
$\bdagg_q=\bdagg(x,\vkp,\lambda)$ & (electron),\\
\parbox{2.5cm}{\vspace{0.0cm}
\psfig{figure=\graphpath symbols_d.epsi,width=2.5cm,angle=-90}}&
$\ddagg_q=\ddagg(x,\vkp,\lambda)$ & (positron),\\
\parbox{2.5cm}{\vspace{0.1cm}
\psfig{figure=\graphpath symbols_a.epsi,width=2.5cm,angle=-90}}& 
$\adagg_q=\adagg(x,\vkp,\lambda)$ & (photon).
\end{tabular}
}
\bigskip

The photons are massless, the fermion mass is $m_f$.
Creation and annihilation operators obey the usual commutation relations.
It is summed over all possible quantum numbers in all expressions. 

The matrix elements are listed in the discretized form where
the normalization volume is
\[
\Omega:= 2L(2L_{\perp})^2,
\]
with the half longitudinal and transversal box lengths $L$ and
$L_{\perp}$, respectively.
The coupling constant in the tables is
$\tilde{g}^2=g^2\frac{2}{P^+\Omega}$, the total longitudinal momentum
$P^+=\frac{\pi}{L}K$, with the {\em harmonic resolution} $K$.   
For all other notations see Appendix \ref{AppxNotations}.
The {\em continuum limit} is obtained by substituting simultaneously
\[
\frac{\tilde{g}^2}{4\pi}\longrightarrow\frac{\alpha}{2\pi^2}
\mbox{\hspace{1cm} and \hspace{1cm}}
\sum_{\rm all\; QN}\longrightarrow \sum_{\lambda=\pm 1}\int^{1}_{0}dx
\int^{\infty}_{-\infty}dk_x\int^{\infty}_{-\infty}dk_y.
\]
\nopagebreak
%
%
%
%
\noindent
\begin{table}[h]
\centerline{
\begin{tabular}{|c|l|}
\hline
\rule[-3mm]{0mm}{8mm}Graph &\hfil Matrix Element \hfil\\
\hline
\hline
\parbox{3cm}{
\vspace{0.3cm}
\psfig{figure=\graphpath contr_g.epsi,width=3cm,angle=-90}
\vspace{0.3cm}
}
&
\parbox{10cm}{ 
\[
C^{(g)}_q(1)=
 \tilde{g}^2 \sum^{\infty}_{x',\vkp'}\left(\frac{1}{(x_1-x')^2}
- \frac{1}{(x_1+x')^2}\right)\\
\]
}
\\\hline 
\parbox{3cm}{
\vspace{0.3cm}
\psfig{figure=\graphpath contr_q.epsi,width=3cm,angle=-90}
\vspace{0.3cm}
}
&
\parbox{10cm}{ 
\[
C^{(q)}_q(1)=
 \tilde{g}^2 \sum^{\infty}_{x',\vkp'}\left(\frac{1}{x'(x_1+x')}
+ \frac{1}{x'(x_1-x')}\right)\\
\]
}
\\\hline 
\parbox{3cm}{
\vspace{0.3cm}
\psfig{figure=\graphpath contr_qg.epsi,width=3cm,angle=-90}
\vspace{0.3cm}
}
&
\parbox{10cm}{ 
\[
C^{(q)}_g(1)=
 -\tilde{g}^2 \sum^{\infty}_{x',\vkp'}\left(\frac{1}{x_1(x'+x_1)}
+ \frac{1}{x_1(x'-x_1)}\right)\\
\]
}
\\\hline 
\hline 
\multicolumn{2}{|c|}{
\parbox{10cm}{ 
\[
C=\sum_{\rm all\; QN}\left[\left(\bdagg_1 b_1 -\ddagg_1 d_1\right)
\left(C^{(g)}_q(1)+C^{(q)}_q(1)\right)+\adagg_1 a_1 C^{(q)}_q(1)\right]
\]
}}
\\\hline
\end{tabular}
}
\caption[Matrix elements of the contractions]
{\label{ContractionTable}
Matrix elements of the contractions. 
For further explanations, see text.}
\end{table}
\pagebreak
%
%
%
%
\begin{table}[t]
\begin{tabular}{|c|c|}
\hline
\rule[-3mm]{0mm}{8mm}Graph & 
\hspace{2cm}Matrix Element \hspace{4.3cm} Helicity\\
\hline
\hline
\psfig{figure=\graphpath ferm_vertex.epsi,width=2.5cm,angle=-90}
&
\parbox{12cm}{ 
\beqa
V_{q\rightarrow qg}(1;2,3)&=&
 +\tilde{g} \frac{m_f}{\sqrt{x_1}}
\left(\frac{1}{x_2} - \frac{1}{x_3}\right)\mbox{\hspace{2.5cm}}
\times \delta^{\lambda_1}_{\lambda_2}\delta^{\lambda_1}_{\lambda_3}\\
&& +\tilde{g} \sqrt{\frac{2}{x_3}}\epsvec(\lambda_3)
\left(\frac{\vec{k}_{\perp 3}}{x_3} - \frac{\vec{k}_{\perp 2}}{x_2}\right)
\mbox{\hspace{1cm}}
\times \delta^{\lambda_1}_{\lambda_2}\delta^{\lambda_1}_{-\lambda_3}\\
&& +\tilde{g} \sqrt{\frac{2}{x_3}}\epsvec(\lambda_3)
\left(\frac{\vec{k}_{\perp 3}}{x_3} - \frac{\vec{k}_{\perp 1}}{x_1}\right)
\mbox{\hspace{1cm}}
\times \delta^{\lambda_1}_{-\lambda_2}\delta^{\lambda_1}_{\lambda_3}\\
\eeqa
}
\\\hline 
\psfig{figure=\graphpath phot_vertex.epsi,width=2.5cm,angle=-90}
&
\parbox{12cm}{ 
\beqa
V_{g\rightarrow q\bar{q}}(1;2,3)&=&
 \tilde{g} \frac{m_f}{\sqrt{x_1}}
\left(\frac{1}{x_2} + \frac{1}{x_3}\right)
\mbox{\hspace{3cm}}
\times \delta^{\lambda_1}_{\lambda_2}\delta^{\lambda_1}_{\lambda_3}\\
&& -\tilde{g} \sqrt{\frac{2}{x_1}}\epsvec(\lambda_1)
\left(\frac{\vec{k}_{\perp 1}}{x_1} - \frac{\vec{k}_{\perp 3}}{x_3}\right)
\mbox{\hspace{1cm}}
\times \delta^{\lambda_1}_{\lambda_2}\delta^{\lambda_1}_{-\lambda_3}\\
&& -\tilde{g} \sqrt{\frac{2}{x_1}}\epsvec(\lambda_1)
\left(\frac{\vec{k}_{\perp 1}}{x_1} - \frac{\vec{k}_{\perp 2}}{x_2}\right)
\mbox{\hspace{1cm}}
\times \delta^{\lambda_1}_{-\lambda_2}\delta^{\lambda_1}_{\lambda_3}\\
\eeqa
}
\\\hline 
\hline 
\multicolumn{2}{|c|} 
{\parbox{15cm}{ 
\beqa
V&=&\sum_{\rm all\; QN}\left(\bdagg_1 b_2 a_3-\ddagg_1 d_2 a_3\right)
V_{q\rightarrow qg}(1;2,3)\\
&&+\sum_{\rm all\; QN}\left(\adagg_3\bdagg_2 b_1-\adagg_3\ddagg_2 d_1\right)
V^*_{q\rightarrow qg}(1;2,3)\\
&&+\sum_{\rm all\; QN}\left[\adagg_1 b_2 d_3
V^*_{g\rightarrow q\bar{q}}(1;2,3)+
\ddagg_3\bdagg_2 a_1 V_{g\rightarrow q\bar{q}}(1;2,3)\right]\\
\eeqa
}}
\\\hline
\end{tabular}
\caption[Matrix elements of the vertex interaction]
{\label{VertexTable}Matrix elements of the vertex interaction. 
For further explanations, see text.}
\end{table}
\vfill

%
%
%
%
\begin{table}
\begin{tabular}{|c|lcl|}
\hline
\rule[-3mm]{0mm}{8mm}Graph & 
\hspace{2cm} Matrix Element & \hspace{1cm} & Helicity \\
\hline
\hline
\parbox{3cm}{
\vspace{0.3cm}
\psfig{figure=\graphpath seag_31.epsi,width=3cm,angle=-90}
\vspace{0.3cm}
}
&
\parbox{7cm}{ 
\[
S^{(s)}_{q\bar{q}\rightarrow q\bar{q}}(1,2;3,4)=
 \tilde{g}^2 \frac{2}{(x_1-x_3)^2}
\]}
&&
$\times \delta^{\lambda_1}_{\lambda_3}\delta^{\lambda_2}_{\lambda_4}$
\\\hline 
\parbox{3cm}{
\vspace{0.3cm}
\psfig{figure=\graphpath seag_32.epsi,width=3cm,angle=-90}
\vspace{0.3cm}
}
&
\parbox{7cm}{ 
\[
S^{(a)}_{q\bar{q}\rightarrow q\bar{q}}(1,2;3,4)=
 \tilde{g}^2 \frac{-2}{(x_1+x_3)^2}
\]}
&&
$\times \delta^{\lambda_1}_{-\lambda_2}\delta^{\lambda_3}_{-\lambda_4}$
\\\hline 
\parbox{3cm}{
\vspace{0.3cm}
\psfig{figure=\graphpath seag_51.epsi,width=3cm,angle=-90}
\vspace{0.3cm}
}
&
\parbox{7cm}{ 
\[
S^{(s)}_{qg\rightarrow qg}(1,2;3,4)=
 \tilde{g}^2 \frac{1}{x_1-x_4}\frac{1}{\sqrt{x_2x_4}}
\]}
&&
$\times \delta^{\lambda_1}_{\lambda_2}\delta^{\lambda_1}_{\lambda_3}
\delta^{\lambda_1}_{\lambda_4}$
\\\hline 
\parbox{3cm}{
\vspace{0.3cm}
\psfig{figure=\graphpath seag_52.epsi,width=3cm,angle=-90}
\vspace{0.3cm}
}
&
\parbox{7cm}{ 
\[
S^{(a)}_{qg\rightarrow qg}(1,2;3,4)=
 \tilde{g}^2 \frac{1}{x_1+x_2}\frac{1}{\sqrt{x_2x_4}}
\]}
&&
$\times \delta^{\lambda_1}_{-\lambda_2}\delta^{\lambda_1}_{\lambda_3}
\delta^{\lambda_1}_{-\lambda_4}$
\\\hline 
\hline 
\multicolumn{4}{|c|} 
{\parbox{15cm}{ 
\beqa
S&=&
\sum_{\rm all\; QN}\bdagg_1\ddagg_2 b_3 d_4
\left[S^{(s)}_{q\bar{q}\rightarrow q\bar{q}}(1,2;3,4)
     +S^{(a)}_{q\bar{q}\rightarrow q\bar{q}}(1,2;3,4)\right]\\
&&+\sum_{\rm all\; QN}\left(\bdagg_1 \adagg_2 b_3 a_4 
+\ddagg_1 \adagg_2 d_3 a_4\right)
\left[S^{(s)}_{qg\rightarrow qg}(1,2;3,4)
     +S^{(a)}_{qg\rightarrow qg}(1,2;3,4)\right]\\
\eeqa
}}
\\\hline
\end{tabular}
\protect\caption[Matrix elements of the seagull interaction]
{\protect\label{SeagullTable}Matrix elements of the seagull interaction used 
or mentioned in the present work. The full table can be found in 
\protect\cite{Schladming}. For further explanations, see text.}
\end{table}

\chapter{Calculation of effective matrix elements}
\label{AppxCalculation}

\begin{figure}
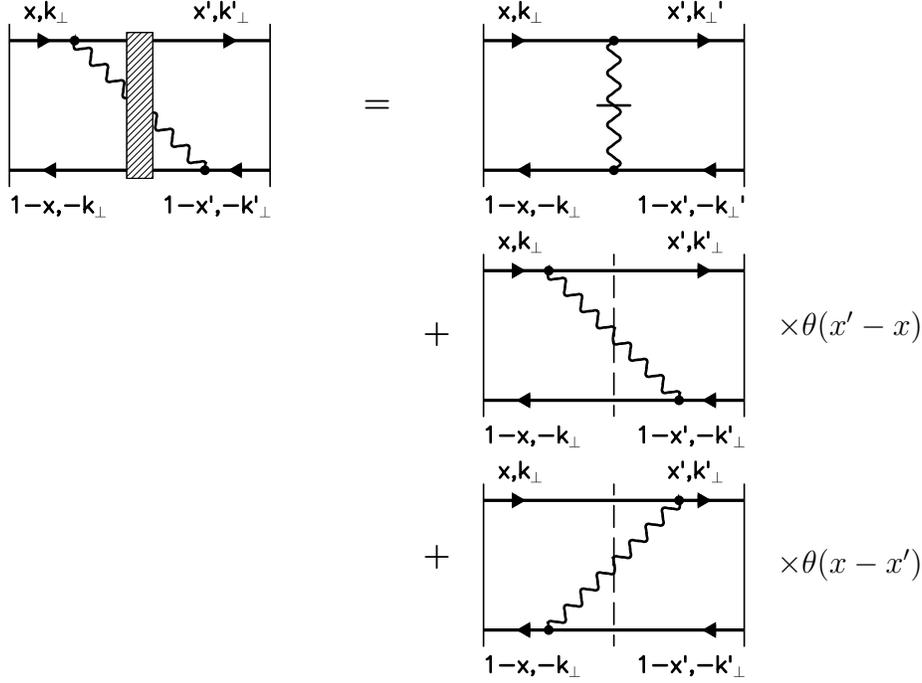

\centerline{
\begin{minipage}{13cm}
\centerline{
\begin{tabular}{ccccccc}
&\psfig{figure=\graphpath dyn_effective.epsi,width=3.5cm,angle=-90}
&\hspace{0.8cm}\raisebox{1.4cm}[-1.4cm]{\bf =}& &
\psfig{figure=\graphpath seagull.epsi,width=3.5cm,angle=-90}& \\
& & &\raisebox{1.4cm}[-1.4cm]{\bf +}&
\psfig{figure=\graphpath dynamic.epsi,width=3.5cm,angle=-90}
&\raisebox{1.5cm}[-1.4cm]{$\times\theta(x'-x)$}\\
& & &\raisebox{1.5cm}[-1.4cm]{\bf +}&
\psfig{figure=\graphpath zeitord_2.epsi,width=3.5cm,angle=-90}
&\raisebox{1.4cm}[-1.4cm]{$\times\theta(x-x')$} 
\end{tabular}
}
\protect\caption[The graphs of the effective interaction]
{\protect\label{Diagramme}The graphs 
of the effective interaction. Shown are the graphs with
an effective photon (a), the instantaneous seagull-graph (b), and 
the dynamic graph (c) and (d) with time ordering (I) and (II), respectively.}
\end{minipage}
}
\end{figure}

The calculation of the matrix elements of the effective interaction
using matrix elements of the canonical Hamiltonian and using currents
is described in short.
We use the derivation of {\sc Krautg\"artner}, {\sc Pauli} and 
{\sc W\"olz} \cite{KPW}.
The interaction within the positronium system can be expressed via currents
\cite[Eq.\ (2.13)]{KPW}
\begin{equation}\label{stromdef}
V_{\rm eff}=\frac{g^2}{\cal{D}}j(l_e)^{\mu}j(l_{\bar{e}})_{\mu}
+\frac{g^2}{\cal{D}}\frac{T^*-\omega}{|x-x'|}j(l_e)^+j(l_{\bar{e}})^+.
\end{equation}
Here the {\em currents} are
\beqa
j(l_e)^{\mu}&=&\bar{u}(k_e')\gamma^{\mu} u(k_e),\\
j(l_{\bar{e}})^{\mu}&=&\bar{u}(k_{\bar{e}}')\gamma^{\mu} u(k_{\bar{e}}),
\eeqa
the {\em momentum transfers} read
\beqa
l_e^{\mu}&=&\left(k_e'-k_e\right)^{\mu},\\
l_{\bar{e}}^{\mu}&=&\left(k_{\bar{e}}-k'_{\bar{e}}\right)^{\mu},
\eeqa
and the {\em energy denominator} is given by
\beq\label{EnergyDenominator}
{\cal D}=|x-x'|(T^*-\omega)-l_e^2,
\eeq
using $l_e^2\equiv\frac{1}{2}\left(l_e^2+l_{\bar{e}}^2\right)$.
In order to derive (\ref{stromdef}), one evaluates the diagrams
shown in Fig.~(\ref{Diagramme}) according to the rules of  
light-cone perturbation theory as formulated by {\sc Lepage} and {\sc Brodsky}
in \cite[Appx.~A]{BLepage}.
In the energy denominator, the average of the energies of the in- 
and outgoing particles is used instead of the sum of the energies of the
incoming particles only.
The second term of Eq.~(\ref{stromdef}) vanishes, if one fixes $\omega=T^*$
as described in Chapter \ref{ChapterModel}.
The definition of the interaction matrix elements is read off 
from the integral equation. We follow \cite[Eq.~(A.5)]{BLepage} and 
\cite[Eq.~(3.19)]{Merkel} by setting

\begin{eqnarray*}
&&\left(\frac{m_f^2 + \vec{k}^2_{\perp}}{x(1-x)}- M_n^2\right)
\psi_n(x,\vec{k}_{\perp};\lambda_1,\lambda_2)\\
&&+\frac{g^2}{16\pi^3}
\sum_{s'_1,s'_2}\int_D \frac{dx'd^2\vec{k}'_{\perp}}
{\frac{1}{2}\left(l_e^2+l_{\bar{e}}^2\right)}
\frac{\langle x,\vec{k}_{\perp};
\lambda_1,\lambda_2|j(l_e)^{\mu}j(l_{\bar{e}})|x',\vec{k}'_{\perp};s'_1,s'_2\rangle}
{\sqrt{xx'(1-x)(1-x')}}
 \psi_n(x,\vec{k}_{\perp};\lambda_1,\lambda_2)= 0.
\end{eqnarray*}

\noindent
For the actual calculation of the effective interaction via the currents
we need the matrix elements of the Dirac spinors:

\begin{table}[h]
\centerline{
\begin{tabular}{|c||c|}
\hline
\parbox{1.5cm}{ \[\cal M\] } & 
\parbox{12cm}{
\[
\frac{1}{\sqrt{k^+k^{'+}}}
\bar{u}(k',\lambda') {\cal M} u(k,\lambda)
\]
}
\\\hline\hline
$\gamma^+$ & \parbox{12cm}{
\[
2\delta^{\lambda}_{\lambda'}
\]}
\\\hline
$\gamma^-$ &
\parbox{12cm}{
\[
\frac{2}{k^+k^{'+}}\left[\left(m^2+
\kp\kp'e^{+i\lambda(\varphi-\varphi')}\right)
\delta^{\lambda}_{\lambda'}
+m\lambda\left(\kp'e^{+i\lambda \varphi'}-
\kp e^{+i\lambda\varphi}\right)
\delta^{\lambda}_{-\lambda'}\right]
\]}
\\\hline
$\gamma_1$ & 
\parbox{12cm}{
\[
\left(\frac{\kp'}{k^{'+}}e^{-i\lambda\varphi'}+
\frac{\kp}{k^+}e^{+i\lambda\varphi}
\right)\delta^{\lambda}_{\lambda'}
-m\lambda\left(\frac{1}{k^{'+}}-\frac{1}{k^+}
\right)\delta^{\lambda}_{-\lambda'}
\]}
\\\hline
$\gamma_2$ &
\parbox{12cm}{
\[ 
i\lambda\left(\frac{\kp'}{k^{'+}}e^{-i\lambda\varphi'}-
\frac{\kp}{k^+}e^{+i\lambda\varphi}
\right)\delta^{\lambda}_{\lambda'}
-im\left(\frac{1}{k^{'+}}-\frac{1}{k^+}
\right)\delta^{\lambda}_{-\lambda'}
\]}\\
\hline
\end{tabular}
}
\caption[Matrix elements of the Dirac spinors]
{Matrix elements of the Dirac spinors.}
\end{table}

\noindent
With the notation
\[
\langle {\cal M}_1{\cal M}_2\rangle:=\frac{
\bar{u}(x',\kp',\lambda'_1){\cal M}_1 u(x,\kp,\lambda_1)
\bar{u}(1-x',-\kp',\lambda'_2){\cal M}_2 u(1-x,-\kp,\lambda_2)}
{\sqrt{xx'(1-x)(1-x')}},
\]
the prescription for the calculation of the matrix elements of the effective 
interaction reads 
\begin{eqnarray*}
\langle x,\vkp;\lambda_1,\lambda_2|j^{\mu}(l_e)j_{\mu}(l_{\bar{e}})|
x',\vkp';\lambda_1',\lambda_2'\rangle&:=&
\frac{\langle x,\vkp;\lambda_1,\lambda_2|j^{\mu}(l_e)j_{\mu}(l_{\bar{e}})|
x',\vkp';\lambda_1',\lambda_2'\rangle}{2\sqrt{xx'(1-x)(1-x')}}\\
&=&\frac{1}{2}\left(\frac{1}{2}
\langle \gamma^+\gamma^-\rangle+\frac{1}{2}\langle \gamma^-\gamma^+\rangle
-\langle \gamma^2_1\rangle-\langle \gamma^2_2\rangle\right).
\end{eqnarray*}
\beq\label{CurrentElements}
\eeq
These functions are listed in Appendix \ref{AppxHelicityTables}.
 
Using canonical Hamiltonian matrix elements, as worked out in detail by
{\sc Tang et al.} \cite{Tang} and listed in Appx.~\ref{AppxHLC},
one can easily obtain the same results as using currents.
As an example, we give the prescription for calculations of effective 
matrix elements originating in the projection of the
$|e\bar{e}\gamma\rangle$-sector onto the $|e\bar{e}\rangle$-sector of the
truncated Fock-space
\beq
\langle (e\bar{e})_i|V_{\rm eff}|(e\bar{e})_j\rangle =\sum_k 
\langle (e\bar{e})_i|V|(e\bar{e}\gamma)_k\rangle\frac{1}
{\langle(e\bar{e}\gamma)_k|\omega-H|(e\bar{e}\gamma)_k\rangle}
\langle (e\bar{e}\gamma)_k|V|(e\bar{e})_j\rangle, 
\eeq
where $V$ is the vertex operator  of a fermion irradiating a photon.
For the time ordering (I), Fig.~(\ref{Diagramme})(c), we obtain
\beqa
\langle(e\bar{e}\gamma)'|V|e\bar{e}\rangle&=&\sqrt{\beta}\frac{1}{\sqrt{(x-x')}}
\left\{m_f\frac{x-x'}{xx'}
\times\delta^{\lambda_1}_{-\lambda_2}\delta^{\lambda_1}_{\lambda_3}\right.\\
&&+\frac{\lambda}{x-x'}\left(-\kp e^{-i\lambda\varphi}
+\frac{x}{x'}\kp' e^{-i\lambda\varphi'}\right)
\times\delta^{\lambda_1}_{\lambda_2}\delta^{\lambda_1}_{\lambda_3}\\
&&+\left.\frac{\lambda}{x-x'}\left(\kp' e^{-i\lambda\varphi'}
-\frac{x'}{x}\kp e^{-i\lambda\varphi}\right)
\times\delta^{\lambda_1}_{\lambda_2}\delta^{\lambda_1}_{-\lambda_3}
\right\}\\
\eeqa
\beqa
\langle (e\bar{e})'|V|e\bar{e}\gamma\rangle&=&-\sqrt{\beta}
\frac{1}{\sqrt{x-x'}}\left\{ m_f\frac{x-x'}{(1-x)(1-x')}
\times\delta^{\lambda_1}_{-\lambda_2}\delta^{\lambda_1}_{\lambda_3}\right.\\
&&+\frac{\lambda}{x-x'}\left(\kp' e^{+i\lambda\varphi'}
-\frac{1-x'}{1-x}\kp e^{+i\lambda\varphi}\right)
\times\delta^{\lambda_1}_{\lambda_2}\delta^{\lambda_1}_{\lambda_3}\\
&&+\left.\frac{\lambda}{x-x'}\left(-\kp e^{+i\lambda\varphi}
+\frac{1-x}{1-x'}\kp' e^{+i\lambda\varphi'}\right)
\times\delta^{\lambda_1}_{\lambda_2}\delta^{\lambda_1}_{-\lambda_3}
\right\}.
\eeqa
The helicity of the irradiated photon is $\lambda$, in all other respects we
use the notation of Appx.~\ref{AppxHLC} for this vertex operator.
The matrix elements for time ordering (II) are evaluated in a straightforward 
way.
They evolve from the functions displayed above by exchanging particles and 
anti-particles\footnote{That is, $(x,\vec{k}_{\perp})
\leftrightarrow(1-x,-\vec{k}_{\perp})$.} 
and a change of the sign because of the $\theta$-function.
We have to multiply these functions by the inverse Hamiltonian,
\[
\frac{1}
{\langle(e\bar{e}\gamma)_k|\omega-H|(e\bar{e}\gamma)_k\rangle},
\]
which is the equivalent of the energy denominator, 
Eq.~(\ref{EnergyDenominator}), in our case.
Together with the much simpler matrix elements of the seagull interaction
Fig.~(\ref{Diagramme})(b), we obtain the functions of 
Table~(\ref{GeneralHelicityTable}) by substituting all photon helicities
by those of fermions.
This is in accordance with \cite{KPW} and our previous treatment via currents
in Eq.~(\ref{CurrentElements}).

\chapter{Renormalization}
\label{AppxRenormalization}




Because of the divergent graphs in our theory, we have to renormalize the 
masses of the electron and positron, respectively.
The singularities are due to the electromagnetic self mass of the electron, 
Fig.~(\ref{SelfMass}), which is a diagram of the iterated vertex interaction, 
and
the contributions of the two contraction graphs, 
Fig.~(\ref{pcontractions})(a+b).
The description of the renormalization scheme to be presented here follows
the line of arguments of {\sc Krautg\"artner et al.} \cite[Appx.\ A]{KPW}.

\section{The electron sector}

As a first step, we consider the electron sector, {\em i.e.}\ we restrict 
the Fock space to contain only the states
\[
|e\rangle=\bdagg_{\lambda}(x{=}1,\vkp{=}0)|0\rangle
\]
and
\[ 
|e\gamma\rangle= \bdagg_{\lambda_1}(x,\vkp)
\adagg_{\lambda_2}(1-x,-\vkp)|0\rangle
\]
which constitute the $P$- and $Q$-space analogously to the case of positronium
described in Chapter \ref{ChapterModel}.
We obtain a $2\times2$ block matrix, but here the $P$-space contains
only one state for fixed helicity. Using the point symmetries $\cal C$ and 
$\cal H$ as described in Appx.~\ref{AppxNumerics}, this reduces to effectively
one single state in each symmetry sector. 
Moreover, the $Q$-sector contains 
only diagonal parts. The seagull graphs are absent due to the properties of the
theory of effective interactions\protect\footnote{See Chapter 
\protect\ref{DrAChapterEffInt} or the plausibility
argument in Chapter \protect\ref{ChapterModel}.}.
Therefore, the inversion in $Q$-space becomes trivial, yielding
\[
G(\omega)=\frac{1}{\langle e\gamma|\omega-H|e\gamma\rangle}=
\frac{1}{\omega-\frac{m^2+\vkp^{'2}}{x}+\frac{\vkp^{'2}}{1-x'}}.
\]    
The contribution of the iterated interaction is therefore
\beqa
\langle e|V_{\rm eff}|e \rangle &=&\sum_{\lambda'_{e},
\lambda'_{\gamma}} 
\int_D dx' d^2\vec{k}^{\; '}_{\perp}
\langle e|V|(e\gamma)'\rangle\frac{1}
{\langle(e\gamma)'|\omega-H|(e\gamma)'\rangle}
\langle (e\gamma)'|V|e\rangle\\
&=&\frac{\alpha}{2\pi^2}\int_D dx' d^2\vec{k}'_{\perp}\frac{1}{1-x'}
\left[\frac{m_f^2(1-x')^2}{x^{'2}}+\frac{\vkp^{'2}}{(1-x')^2}\left(
1+\frac{1}{x^{'2}}\right)\right]\\
&&\times
\frac{1}{\omega-\frac{m^2+\vkp^{'2}}{x}+\frac{\vkp^{'2}}{1-x'}}.
\eeqa
The domain of integration is restricted by imposing the condition 
\[
\frac{m^2+\vkp^{'2}}{x}+\frac{\vkp^{'2}}{1-x'}\le \Lambda^2+m_f^2.
\]
This means that 
the difference of the free invariant masses of the states before
and after an interaction must not exceed a cutoff $\Lambda$, which
plays the r\^{o}le of a regulator.
We use again the fixing of the parameter $\omega$, Eq.(\ref{OmegaStar}),
and arrive at
\beq\label{iteratedIA}
W=\frac{\alpha}{2\pi^2}\int_D dx' d^2\vec{k}'_{\perp}
\left(\frac{2m_f^2}{m_f^2(1-x')^2+\vkp^{'2}}-\frac{1}{x^{'2}}
-\frac{2}{(1-x')^2}\right).
\eeq
The evaluation of the contraction terms is achieved in the 
following way.
We extract from Table (\ref{ContractionTable}) the contributions 
of the {\em instantaneous fermion} and {\em photon} part and
perform the continuum limit
\beqa
C^{(q)}(x,\vec{k}_{\perp},\lambda)
&=&\frac{\alpha}{2\pi^2}\int_0^1 dy \int_{-\infty}^{\infty}d^2\vec{k}'_{\perp}
\frac{1}{2}\left[\frac{1}{y(x-y)}+\frac{1}{y(x-y)}\right],\\
C^{(g)}(x,\vec{k}_{\perp},\lambda)
&=&\frac{\alpha}{2\pi^2}\int_0^1 dy \int_{-\infty}^{\infty}d^2\vec{k}'_{\perp}
\left[\frac{1}{(x-y)^2}-\frac{1}{(x+y)^2}\right].
\eeqa
When we use the {\em principal value} as a regulator for these integrals
\cite{MustakiRenormalization}, we 
obtain\protect\footnote{For the subtleties of these 
integrals cf.~\protect\cite[Appx.\ B]{DissKraut}.}
\beqa
C^{(q)}(x,\vec{k}_{\perp},\lambda)
&=&\frac{\alpha}{2\pi^2}\frac{1}{x}
\int_0^1 dy \int_{-\infty}^{\infty}d^2\vec{k}'_{\perp}\frac{1}{y},
\\
C^{(g)}(x,\vec{k}_{\perp},\lambda)
&=&\frac{\alpha}{2\pi^2}\int_0^{x'} dy 
\int_{-\infty}^{\infty}d^2\vec{k}'_{\perp}\frac{2}{y^2}.
\eeqa
We have to regulate the integral, since both integrands are singular. 
If we use here the same integration domain as for the iterated interaction,
rename the variables and sum the two contraction contributions 
\[
C:=\frac{\alpha}{2\pi^2}\int_D dx\; d^2\vec{k}'_{\perp}
\left[\frac{2}{(1-x')^2}+\frac{1}{1-x'}\right],
\]
we find that the quadratically divergent term 
of the instantaneous photon interaction
is compensated exactly by the analogous expression in the iterated vertex 
interaction, Eq.~(\ref{iteratedIA}). Additionally, the fermion part proportional
to $1/1-x'$ cancels and only a logarithmically divergent
term survives. The self mass 
\beqa
\Delta m^2:=W+C&=&\frac{\alpha}{2\pi^2}\int_D dx d^2\vec{k}'_{\perp}
\left(\frac{2m_f^2}{m_f^2(2-x')^2+\vkp^{'2}}\right)\\
&=&\frac{\alpha}{2\pi}m_f^2\left(3\ln\frac{\Lambda^2+m_f^2}{m_f^2}
-\frac{\Lambda^2}{m_f^2+\Lambda^2}\right),
\eeqa
\beq\label{DeltaM}
\eeq
is therefore only logarithmically divergent, although its three contributions
diverge quadratically with the cutoff $\Lambda$.

\section{The positronium sector}

In the positronium sector as defined in Chapter \ref{ChapterModel}, 
we have the two states  
$|e\bar{e}\rangle$ and $|e\bar{e}\gamma\rangle$.
The calculations of the contractions and the iterated vertex interaction are
performed exactly as in the previous section.
The domain of integration is given by
\[
\left[m_f^2(x-x')^2+(x\vkp'-x'\vkp)^2\right]\frac{1}{xx'(x-x')}\leq \Lambda^2.
\] 
The expression for the electromagnetic self mass of the electron (and positron)
is
\[
\Delta M^2=\frac{\alpha}{2\pi^2}\int_D dx d^2\vec{k}'_{\perp}
\left(\frac{2m_f^2}{m_f^2(x-x')^2+(x\vkp'-x'\vkp)^2}-\frac{1}{xx'}
+\frac{1}{x(x-x')}\right).
\]
We can retrieve an expression similar to Eq.~(\ref{DeltaM}) by substituting 
\[
y:=\frac{x'}{x} \quad {\rm and}\quad \vec{p}_{\perp}:=\vkp'-y\vkp,
\]
where the self masses are related by
\[
\Delta M^2=\frac{1}{x}\Delta m^2.
\]
The renormalization is performed by setting 
\[
m_f\equiv m_e=511\;\mbox{keV},
\]
or in other words, we adjust the 
contraction terms so that their divergent pieces 
cancel exactly those of the iterated interaction, 
{\em i.e.}~$\Delta m^2\equiv0$.

The renormalization is restricted to $P$-space. In principle, the counterterms
will depend on the Fock sector considered. If one takes into account
higher Fock states, these counterterms will in general not be analytically 
calculable. They are essentially non-perturbative.

\chapter{Tables of effective matrix elements}
\label{AppxHelicityTables}

\section{Introduction and Notations}

The dependence of the effective interaction on the helicities of 
in- and out-going particles is displayed in the form of tables.
A part of the compiled functions has been discussed by 
{\sc Krautg\"artner et al.} \cite{KPW}.
The matrix elements of the effective interaction depend on the one hand on the 
momenta of the electron and positron, respectively, and on the other hand  
on their helicities before and after the interaction.
The latter dependencies occur during the calculation of these functions
$M(x,\vkp,\lambda_1,\lambda_2;x',\vkp',\lambda_1',\lambda_2')$ 
as complicated Kronecker deltas. 
A good survey is obtained, however, when displaying 
these functions in the form of a table.

In the remainder of this appendix, the tables are given 
for the general, angle-dependent
effective matrix elements, Table~(\ref{GeneralHelicityTable}), for the
matrix elements of arbitrary $J_z$, after integrating over the angles,
Table~(\ref{HelicityTableJz}), and for the matrix elements of the  
annihilation  graph, Table~(\ref{HelicityTableAnnihilation}).      
The following notation is used for functions of the type
$F(x,\kperp;x',\kperp')$:
\begin{itemize}
\item An asterisk denotes the permutation of particle and anti-particle.
\[
F_3^*(x,\kperp;x',\kperp'):=F_3(1-x,-\kperp;1-x',-\kperp').
\]
\item If the function additionally depends 
on the component of the total angular momentum $J_z=n$,
a tilde symbolizes the operation
\[
\tilde{F}_i(n)=F_i(-n),
\]
\end{itemize}

\pagebreak
\section{General helicity table}\label{phiTabelle}

\begin{table}[h]
\centerline{
\begin{tabular}{|c||c|c|c|c|}\hline
\rule[-3mm]{0mm}{8mm}{\bf final : initial} & $(\lambda_1',\lambda_2')=\uparrow\uparrow$ 
& $(\lambda_1',\lambda_2')=\uparrow\downarrow$ 
& $(\lambda_1',\lambda_2')=\downarrow\uparrow$ &
$(\lambda_1',\lambda_2')=\downarrow\downarrow$ \\ \hline\hline
\rule[-3mm]{0mm}{8mm}$(\lambda_1,\lambda_2)=\uparrow\uparrow$ & $E_1(\vec{k},\vec{k}')$  
& $E_3^*(\vec{k},\vec{k}')$ & $-E_3(\vec{k},\vec{k}')$ & $0$ \\ \hline
\rule[-3mm]{0mm}{8mm}$(\lambda_1,\lambda_2)=\uparrow\downarrow$ & 
$\bar{E}_3^*(\vec{k}',\vec{k})$ & $E_2(\vec{k},\vec{k}')$ & $E_4(\vec{k},\vec{k}')$ 
& $E_3(\vec{k}',\vec{k})$ \\ \hline
\rule[-3mm]{0mm}{8mm}$(\lambda_1,\lambda_2)=\downarrow\uparrow$& $-\bar{E}_3(\vec{k}',\vec{k})$ 
& $E_4(\vec{k},\vec{k}')$ & $E_2(\vec{k},\vec{k}')$  & 
$-E_3^*(\vec{k}',\vec{k})$\\ \hline
\rule[-3mm]{0mm}{8mm}$(\lambda_1,\lambda_2)=\downarrow\downarrow$ & $0$ 
& $\bar{E}_3(\vec{k},\vec{k}')$ & $-\bar{E}_3^*(\vec{k},\vec{k}')$ & 
$E_1(\vec{k},\vec{k}')$\\ \hline
\end{tabular}
}
\protect\caption[General helicity table of the effective interaction]
{\protect\label{GeneralHelicityTable}General helicity 
table of the effective interaction.}
\end{table}
\bigskip

A description of the calculation of the functions displayed in the above 
table was given in some detail in Appendix \ref{AppxCalculation}.
Note that these matrix elements depend in general on the {\em vectors}
$\vkp$ and $\vkp'$.
The functions $E_i(\vec{k},\vec{k}'):=E_i(x,\vkp;x',\vkp')$ read 

\begin{eqnarray*}
E_1(x,\vec{k};x',\vec{k}')&=&
	\frac{\alpha}{2\pi^2}\frac{1}{\cal D}\left[
	m^2_F\left(\frac{1}{xx'}+\frac{1}{(1-x)(1-x')}\right)
	+\frac{k_{\perp}k_{\perp}'}{xx'(1-x)(1-x')}e^{-i(\varphi-\varphi')}
	\right]	\\
E_2(x,\vec{k};x',\vec{k}')&=&
	\frac{\alpha}{2\pi^2}\frac{1}{\cal D}
	\left(m_F^2+k_{\perp}k'_{\perp}e^{i(\varphi-\varphi')}\right)
	\left(\frac{1}{xx'}+\frac{1}{(1-x)(1-x')}\right)\\
	&&+\frac{\alpha}{2\pi^2}\frac{1}{\cal D}\left(\frac{k_{\perp}^2}{x(1-x)}
        +\frac{k_{\perp}^{'2}}{x'(1-x')}\right)\\
E_3(x,\vec{k};x',\vec{k}')&=&
	-\frac{\alpha}{2\pi^2}\frac{1}{\cal D}
	\frac{m_f}{xx'}\left({k'_{\perp}}e^{-i\varphi'}
	-k_{\perp}\frac{1-x'}{1-x}e^{-i\varphi}\right)\\
E_4(x,\vec{k};x',\vec{k}') &=& -\frac{\alpha}{2\pi^2}
	\frac{m_f^2}{\cal D}\frac{(x'-x)^2}{xx'(1-x')(1-x)}.
\end{eqnarray*}
The energy denominator is defined as 
\begin{eqnarray*}
{\cal D}(x,\vec{k}_{\perp};x',\vec{k}_{\perp}')&:=&
	-(x-x')^2\frac{m^2_F}{2}\left(\frac{1}{xx'}+\frac{1}{(1-x)(1-x')}\right)
   	+ 2 k_{\perp}k'_{\perp}\cos(\varphi-\varphi')\\
   &&-\left(k_{\perp}^2+k_{\perp}^{'2}\right)
     +(x-x')\left[\frac{k_{\perp}^{'2}}{2}
     \left(\frac{1}{1-x'}-\frac{1}{x'}\right)-\frac{k_{\perp}^{2}}{2}
     \left(\frac{1}{1-x}-\frac{1}{x}\right)\right].
\end{eqnarray*}

\pagebreak

\section{The helicity table for arbitrary $J_z$}\label{JzTabelle}

\noindent
For an arbitrary $J_z = n$ with $n\in \mbox{\bf\sl Z}$ 
one obtains, following the 
description given in  
Chapter \ref{DrAChapterJz}, the helicity table 
(\ref{HelicityTableJz}).
\bigskip 

\begin{table}[h]
\centerline{
\begin{tabular}{|c||c|c|c|c|}\hline
\rule[-3mm]{0mm}{8mm}{\bf final : initial} & $(\lambda_1',\lambda_2')=\uparrow\uparrow$ 
& $(\lambda_1',\lambda_2')=\uparrow\downarrow$ 
& $(\lambda_1',\lambda_2')=\downarrow\uparrow$ &
$(\lambda_1',\lambda_2')=\downarrow\downarrow$ \\ \hline\hline
\rule[-3mm]{0mm}{8mm}$(\lambda_1,\lambda_2)=\uparrow\uparrow$ & 
$G_1(1,2)$   
& $G_3^*(1,2)$ & $G_3(1,2)$ & $0$ \\ \hline
\rule[-3mm]{0mm}{8mm}$(\lambda_1,\lambda_2)=\uparrow\downarrow$ & 
$G_3^*(2,1)$ 
& $G_2(1,2)$ & $G_4(1,2)$ & $-\tilde{G}_3(2,1)$ \\ \hline
\rule[-3mm]{0mm}{8mm}$(\lambda_1,\lambda_2)=\downarrow\uparrow$& 
$G_3(2,1)$ & $G_4(1,2)$ & $\tilde{G}_2(1,2)$  & 
$-\tilde{G}_3^*(2,1)$\\ \hline
\rule[-3mm]{0mm}{8mm}$(\lambda_1,\lambda_2)=\downarrow\downarrow$ & $0$ 
& $-\tilde{G}_3(1,2)$ & $-\tilde{G}_3^*(1,2)$ & $\tilde{G}_1(1,2)$ \\
\hline
\end{tabular}
}
\protect\caption[Helicity table of the effective interaction for $J_z = \pm n$]
{\protect\label{HelicityTableJz}Helicity table of the effective interaction
for $J_z = \pm n$, $x>x'$.}
\end{table}
\vspace{0.5cm}

\noindent
Here, the functions $G_i(1,2):=G_i(x,\kperp;x',\kperp')$
are given by 
\begin{eqnarray*}
G_1(x,\kp;x',\kp')&=&
      m_f^2\left(\frac{1}{xx'}+\frac{1}{(1-x)(1-x')}\right)Int(|1-n|)\\
      &&+\frac{k_{\perp}k_{\perp}'}{xx'(1-x)(1-x')}Int(|n|)\\
G_2(x,\kp;x',\kp')&=&\left[m_f^2\left(\frac{1}{xx'}+
      \frac{1}{(1-x)(1-x')}\right)+\frac{k_{\perp}^2}{x(1-x)}
      +\frac{k_{\perp}^{'2}}{x'(1-x')}\right]Int(|n|)\\
      &&+k_{\perp}k'_{\perp}\left[\frac{Int(|1-n|)}{xx'}
	+\frac{Int(|1+n|)}{(1-x)(1-x')}\right]\\
G_3(x,\kp;x',\kp')&=&-m_f\frac{1}{xx'}
	\left[k_{\perp}'Int(|1-n|)-k_{\perp}\frac{1-x'}{1-x}Int(|n|)\right]\\
G_4(x,\kp;x',\kp')&=&-m_f^2\frac{(x-x')^2}
        {xx'(1-x')(1-x)}Int(|n|).
\end{eqnarray*}
The function $Int(n)$ is defined as
\[
Int(n) := \frac{\alpha}{\pi}(-A)^{-n+1}
	  \left(\frac{B}{\kp\kp'}\right)^n.
\]
Involved in this expression are the definitions
\begin{eqnarray*}
a&=&(x-x')^2\frac{m^2_f}{2}\left(\frac{1}{xx'}+\frac{1}{(1-x)(1-x')}\right)
     +k_{\perp}^2+k_{\perp}^{'2}\\
   &&-\frac{1}{2}(x-x')\left[k_{\perp}^{'2}
     \left(\frac{1}{1-x'}-\frac{1}{x'}\right)-k_{\perp}^{2}
     \left(\frac{1}{1-x}-\frac{1}{x}\right)\right] 
\end{eqnarray*}
and
\beqa
A&=&\frac{1}{\sqrt{a^2-4k_{\perp}^2 k_{\perp}^{'2}}},\\
B&=&\frac{1}{2}\left(1-aA\right).
\eeqa

\section{Helicity table of the annihilation graph}\label{AnniTabelle}

\begin{table}[h]
\centerline{
\begin{tabular}{|c||c|c|c|c|}\hline
\rule[-3mm]{0mm}{8mm}{\bf final:initial} & $(\lambda'_1,\lambda'_2)=\uparrow\uparrow$ 
& $(\lambda'_1,\lambda'_2)=\uparrow\downarrow$ 
& $(\lambda'_1,\lambda'_2)=\downarrow\uparrow$ &
$(\lambda'_1,\lambda'_2)=\downarrow\downarrow$ \\ \hline\hline
\rule[-3mm]{0mm}{8mm}$(\lambda_1,\lambda_2)=\uparrow\uparrow$ & 
$F_1(1,2)$   
&$F_3(2,1)$ & $F^*_3(2,1)$ & $0$ \\ \hline
\rule[-3mm]{0mm}{8mm}$(\lambda_1,\lambda_2)=\uparrow\downarrow$ & 
$F_3(1,2)$ 
& $F^*_2(1,2)$ & $F_4(2,1)$ &$0$ \\ \hline
\rule[-3mm]{0mm}{8mm}$(\lambda_1,\lambda_2)=\downarrow\uparrow$& 
$F_3^*(1,2)$ & $F_4(1,2)$ & $F_2(1,2)$  & $0$\\ \hline
\rule[-3mm]{0mm}{8mm} $(\lambda_1,\lambda_2)=\downarrow\downarrow$ & $0$ 
& $0$ & $0$ & $0$  \\
\hline
\end{tabular}
}
\protect\caption[Helicity table of the annihilation graph for $J_z\ge 0$]
{\protect\label{HelicityTableAnnihilation}Helicity table 
of the annihilation graph for $J_z\ge 0$.}
\end{table}
\vspace{0.5cm}

\noindent
The functions $F_i(1,2):=F_{i}(x,\kperp;x',\kperp')$ were calculated in Chapter
\ref{DrAChapterAnnihilation}: 
\beqa
F_1(x,\kp;x',\kp') &:=& \frac{\alpha}{\pi}\frac{2m^2}{\omega^*}
	 \left(\frac{1}{x}+\frac{1}{1-x}\right)
	 \left(\frac{1}{x'}+\frac{1}{1-x'}\right)\delta_{|J_z|,1}\\
F_2(x,\kp;x',\kp') &:=&\frac{\alpha}{\pi}\left[\frac{2}{\omega^*}
	 \frac{\kp\kp'}{xx'}
	 \delta_{|J_z|,1}+4\delta_{J_z,0}\right]\\
F_3(x,\kp;x',\kp') &:=& \frac{\alpha}{\pi}\frac{2m}{\omega^*}\lambda_1
	 \left(\frac{1}{x}+\frac{1}{1-x}\right)
	 \frac{\kp'}{1-x'}\delta_{|J_z|,1}\\
F_4(x,\kp;x',\kp') &:=& -\frac{\alpha}{\pi}\left[\frac{2}{\omega^*}
	\frac{\kp\kp'}{x'(1-x)}\delta_{|J_z|,1}-4\delta_{J_z,0}\right].
\eeqa
The table for $J_z=-1$ is obtained by inverting {\em all} helicities.
Note that the table has non-vanishing matrix elements
for $|J_z|\leq 1$ only.
This restriction is due to the angular momentum of the photon.

\chapter{\label{AppxNumerics}Numerics}


To solve the integral equation (\ref{TDEquation}), central to this work, 
we apply several numerical methods, which are described in the sequel.


\section{Transformation of the variables}

The invariant front form wavefunctions coincide with the equal-time
wavefunctions defined in the infinite momentum frame \cite[p.~379]{Karmanov80}. 
Let the relative momentum of the $i$-th particle in a composite
system be $\vec{p}_i$. In the center of mass system holds
\beq\label{cms}
\sum_i\vec{p}_i=0.
\eeq
We want to express the variables $x,\vkp$ in terms of the equal-time variables
$\vec{p}_{\perp}, p_z$. 
With (\ref{cms}) we have
\[
p_1^+=E+p_z,\quad\quad p_2^+=E-p_z,
\]
where $E=\sqrt{m^2+\vec{p}^2}$.
If we write the relative momentum in polar coordinates 
\[
\vec{p}=(\mu\sin\theta\cos\varphi,\mu\sin\theta\sin\varphi,\mu\cos\theta),
\]
it is clear from the definition of the front form momentum fractions 
(\ref{Defx}) that 
\begin{eqnarray*}
x &=& \frac{1}{2}\left(1 + \frac{\mu\cos\theta}{\sqrt{m_f^2+\mu^2}}\right)\\
\vec{k}_{\perp} &=& \vec{\mu}_{\perp} = (\mu\sin\theta\cos\varphi,\mu\sin\theta
\sin\varphi).
\end{eqnarray*}
The inverse transformation is
\begin{eqnarray}\label{Vtransformation}
\mu&:=&\sqrt{\frac{k_{\perp}^2+m^2(2x-1)^2}{1-(2x-1)^2}}\\
\cos\theta&:=&(1-2x)\sqrt{\frac{k_{\perp}^2+m^2}{k_{\perp}^2+m^2(2x-1)^2}}.
\end{eqnarray}
These are exactly the coordinates as used by {\sc Karmanov} \cite{Karmanov81} 
for the deuteron wavefunction 
and by {\sc Sawicki} \cite[Eqs.\ (2.3), (2.4)]{Sawicki}. 
The Jacobian reads
\[
J=\frac{1}{2}\frac{m^2+\mu^2(1-\cos^2\theta)}{\sqrt{m^2+\mu^2}^3}
\mu^2\sin\theta.
\]
For the integration measure holds
\[
\int_0^1dx\;\int_{-\infty}^{+\infty}d^2k_{\perp}=
\int_0^{\infty}d\mu\int_{-1}^1d\cos\theta\int_0^{2\pi}d\varphi\,
\frac{\mu^2}{2}\frac{m^2+\mu^2(1-\cos^2\theta)}
{(m^2+\mu^2)^{3/2}}.
\]
The new variable $\mu$ is referred to as 
as {\em off-shell mass}, because of 
\[
\frac{m_f^2+\vkp^2}{x(1-x)}=4(m^2_f+\mu^2).
\]
The restriction on the momenta (\ref{BLepagecutoff}) translates into
\begin{equation}\label{Lambda2}
4(m^2_f+\mu^2)\le \Lambda^2 +4m^2_f
\end{equation}
and is therefore a restriction of the off-shell mass.

In the next paragraph the integration over the angular variable  $\varphi$
is described. After this, only two variables are left in the game:
$\mu$ and $\cos\theta$. 
The integrations over these coordinates are discretized, {\em i.e.}~the 
integral is mapped onto a sum, with help of the 
Gauss-Legendre algorithm (cf.~{\em e.g.}~\cite{NumRec}).
The domains of the integrations are chosen in such a way, that
\[
\mu \in [0,\frac{\Lambda}{2}],\quad \cos \theta \in [-1,1].
\]
Here, abscissae and corresponding weights are denoted by
\beqa
\mu_i, w_i &\quad& i=1..N_1\\
\cos\theta_j, w_j &\quad& j=1..N_2.
\eeqa
With $\Lambda\rightarrow\infty$ we have to map the interval $[0,\infty]$ 
for numerical integration onto a finite domain.
We follow \cite[p. 60, case $\gamma$]{Woelz} and use as a mapping function
\[
f(\mu)=\frac{1}{1+\mu}.
\]
The mapping of the weights is calculated straightforwardly.

\section{Symmetries}

First, we want to restore the symmetry of the Hamiltonian matrix broken by
the above transformation of the variables. In discrete variables, the 
transformation has the Jacobian   
\beq\label{Jacobian}
a_{ij}:=\mu_i^2\frac{m_f^2+\mu_i^2(1-\cos\theta_j)}{2\sqrt{m^2_f+\mu^2_i}}.
\eeq
From this transformation an unsymmetric matrix evolves. As a consequence,
the numerical effort to diagonalize it increases considerably.
We symmetrize the matrix therefore by substituting for the wavefunction
\[
\Psi(\mu_i,\theta_j) \longrightarrow \Phi(\mu_i,\theta_j) := 
\Psi\sqrt{w_i w_j a_{ij}}.
\]
The so-modified matrix elements of the effective interaction are 
\[
\langle x,\vec{k}_{\perp};\lambda_1,\lambda_2| V_{\rm eff}|
x',\vec{k}'_{\perp};\lambda'_1,\lambda'_2
\rangle\sqrt{w_i w_j w'_k w'_l a_{ij} a'_{kl}}.
\]
Now, we address to the implementation of the point symmetries 
of the theory into the matrix elements.
The Lagrangian density of QED is invariant under the transformations
of time reversal $\cal{T}$, parity $\cal{P}$,
and charge conjugation $\cal{C}$. Opposed to that, the Hamiltonian  
in front form dynamics is invariant under the latter operation only.
The other operations can be used to construct further, in general more
complicated, symmetries.

We consider as a further symmetry of our Hamiltonian matrix the 
charge conjugation,
which is represented mathematically by the unitary operator
 $\hat{U}_{\cal{C}}$. 
Eigenstates to this transformation have the eigenvalues 
$\pi_{\cal{C}}=\pm 1$ and the creation operators of the different particles
transform as follows \cite{Itzykson}
\begin{eqnarray*}
\hat{U}_{\cal{C}}b^{\dagger}_{\lambda}(x,\vec{k}_{\perp})\hat{U}_{\cal{C}}^{-1}
&=&d^{\dagger}_{\lambda}(x,\vec{k}_{\perp}),\\
\hat{U}_{\cal{C}}d^{\dagger}_{\lambda}(x,\vec{k}_{\perp})\hat{U}_{\cal{C}}^{-1}
&=&b^{\dagger}_{\lambda}(x,\vec{k}_{\perp}),\\
\hat{U}_{\cal{C}}a^{\dagger}_{\lambda}(x,\vec{k}_{\perp}\hat{U}_{\cal{C}}^{-1}
&=&-a^{\dagger}_{\lambda}(x,\vec{k}_{\perp}).
\end{eqnarray*}
The helicities are $\lambda=\pm 1$.
We construct out of arbitrary $P$-space wavefunctions with undefined charge
\[
|\psi\rangle=\Psi_{e\bar{e}}(x,\vec{k}_{\perp};\lambda_1,\lambda_2) 
b^{\dagger}_{\lambda_1}(x,\vec{k}_{\perp})
d^{\dagger}_{\lambda_2}(1-x,-\vec{k}_{\perp})| 0 \rangle
\]
eigenstates to the charge conjugation operator
\beq\label{CSymmetry}
|\pi_{\cal{C}}=\pm 1 \rangle = \hspace{12cm}
\eeq
\[
\frac{1}{\sqrt{2}}
\Psi_{e\bar{e}}(x,\vec{k}_{\perp};\lambda_1,\lambda_2) 
\left[b^{\dagger}_{\lambda_1}(x,\vec{k}_{\perp})
d^{\dagger}_{\lambda_2}(1-x,-\vec{k}_{\perp})
\mp b^{\dagger}_{\lambda_2}(1-x,-\vec{k}_{\perp})
d^{\dagger}_{\lambda_1}(x,\vec{k}_{\perp})\right]
| 0 \rangle.
\]  
The remark is in order, that if the quantum
numbers of the creation operators coincide, there is only {\em one} eigenstate,
having the eigenvalue $\pi_{\cal{C}}=-1$.
This explains the different dimensions of the blocks of the Hamiltonian matrix,
which are build of eigenstates to the point symmetries.

As mentioned earlier, further invariants can be constructed from the two
remaining symmetries of the Lagrangian density \cite{Kaluza}.
We use
\[
e^{-i\pi J_z}{\cal PT}.
\]
In Chapter \ref{ChapterModel} of the present
work we address to the special case of $J_z=0$.
In this case, we can use the combination $\cal{H}:=\cal{PT}$ as another
symmetry.
We give again the behaviour of the creation operators under the
transformation via the anti-unitary operator\footnote{It is 
the direct product of the unitary operator of the 
parity transformation and of the anti-unitary operator of the time reversal
transformation $V_{\cal H}:=V_{\cal P}\oplus V_{\cal T}$.} $V_{\cal H}$
\begin{eqnarray*}
\hat{U}_{\cal{H}}b^{\dagger}_s(x,\vec{k}_{\perp})\hat{U}_{\cal{H}}^{-1}
&=&(-1)^{(\lambda-1)/2}d^{\dagger}_{-\lambda}(x,\vec{k}_{\perp}),\\
\hat{U}_{\cal{H}}d^{\dagger}_s(x,\vec{k}_{\perp})\hat{U}_{\cal{H}}^{-1}
&=&(-1)^{(\lambda+1)/2}b^{\dagger}_{-\lambda}(x,\vec{k}_{\perp}),\\
\hat{U}_{\cal{H}}a^{\dagger}_{\lambda}(x,\vec{k}_{\perp})\hat{U}_{\cal{H}}^{-1}
&=&-a^{\dagger}_{-\lambda}(x,\vec{k}_{\perp}).
\end{eqnarray*}
Physically spoken, this operation inverts the helicities and causes a phase
change, dependent on the choice of the spinors \cite{Itzykson}\cite{Nachtmann}.
One can read off the eigenfunctions of the operator $V_{\cal H}$  
\begin{eqnarray}\label{TSymmetry} 
|\pi_{\cal{H}}=\pm 1 \rangle &=& \frac{1}{\sqrt{2}}\left[
\Psi_{e\bar{e}}(x,\vec{k}_{\perp};\lambda_1,\lambda_2) 
b^{\dagger}_{\lambda_1}(x,\vec{k}_{\perp})
d^{\dagger}_{\lambda_2}(1-x,-\vec{k}_{\perp})\right.\\
&&\left.\pm (-1)^{\lambda_1+\lambda_2}\Psi_{e\bar{e}}^*(x,\vec{k}_{\perp};
\lambda_1,\lambda_2) 
b^{\dagger}_{-\lambda_1}(x,\vec{k}_{\perp})
d^{\dagger}_{-\lambda_2}(1-x,-\vec{k}_{\perp})\right]
| 0 \rangle.\nonumber
\end{eqnarray} 

Applying the two point symmetries, the whole Hilbert space is decomposed
into four subspaces with well-defined charge and ``$\cal{T}$-parity'', and
the Hamiltonian matrix becomes block-diagonal. As a consequence, 
only four much smaller matrices need to be diagonalized.
The numerical effort decreases with the third power of the matrix dimensions, 
and a lot of computer time can be saved.
The construction of the simultaneous eigenstates to  $\cal{C}$ and $\cal{H}$
is supplied by the computer routine {\tt GENETIX}, which generates the whole
Hamiltonian matrix by using a subroutine, which calculates the
unsymmetrized matrix elements.

Finally, we use for the implementation of one further symmetry  the fact,
that the Lagrangian density and the Hamiltonian operator of QED are invariant
under rotations in the $x$-$y$-plane in front form dynamics, too. 
Due to this, we can introduce, as another quantum number, the $z$-component of 
the total angular momentum $J_z$, to classify our states.
The angles enter via
\[
\vec{k}_{\perp}=\left(
			\begin{array}{c}
				k_{\perp}\cos\varphi \\
				k_{\perp}\sin\varphi \\
			\end{array}
		\right).
\]
The new effective interaction is obtained by Fourier-transforming the old one
\[
\langle x,k_{\perp},L_z;\lambda_1,\lambda_2|\tilde{V_{\rm eff}}|x',k'_{\perp},L_z';
\lambda'_1,\lambda'_2\rangle:= \hspace{8cm}
\]
\[
\hspace{3cm}
\frac{1}{2\pi}\int_0^{2\pi}d\varphi\; e^{-i L_z \varphi}
\int_0^{2\pi}d\varphi' \;e^{i L_z \varphi'}
\langle x,k_{\perp},\varphi;\lambda_1,\lambda_2| V_{\rm eff}|x',k'_{\perp},
\varphi';\lambda'_1,\lambda'_2\rangle.
\]
But, neither $L_z$ nor $S_z = \lambda_1 +\lambda_2$ 
are good quantum numbers and we set $L_z = J_z-S_z$.
From these calculations we gain the functions of the helicity table 
(\ref{HelicityTableJz}).
This functions are used in the 
computer code in the subroutine {\tt PHYSIX}, which calculates the
unsymmetrized matrix elements.

\section{Coulomb trick}

The underlying structure faced in treating bound states in Quantum 
Electrodynamics is the Coulomb problem.
Obviously, in the non-relativistic limit, the derived effective integral
equation
(\ref{TDEquation}) will be mapped into the  
{\em Coulomb-Schr\"odinger equation}
\protect\footnote{For a proof see 
\protect\cite{DissKraut} or \protect\cite[Appendix C]{Merkel}.}.

In the introduction, we have already made the point that momentum
coordinates are essential, because they render the exact treatment of the
center of mass motion possible.
A major part of the numerical problems in solving Eq.~(\ref{TDEquation})
is corresponding to the solution of Coulomb problems in momentum space.
This problem was examined in detail in \cite{Woelz}. We present the
numerical techniques applied to the treatment of integral equations with
a weakly singular kernel.

In the Coulomb Schr\"odinger equation a kernel is involved with a point
singularity which is analytically integrable. Nevertheless, 
the matrix elements become senseless at that very point.
Therefore, we apply the so-called {\sc Coulomb} trick \cite{Woelz}, which is 
known as the ``Nystr{\o}m method''
in the mathematical literature.

In momentum representation the Schr\"odinger equation of the hydrogen atom
reads
\[
\frac{\vec{p^2}}{2m}\psi(\vec{p})-\frac{\alpha}{2\pi^2}\int d^3\vec{p'}
\frac{\psi(\vec{p'})}{(\vec{p}-\vec{p'})^2} = E\psi(\vec{p}).
\]
For simplicity, we consider only S-waves, {\em i.e.}~rotationally invariant
states. We can integrate out the angular variables and using Bohr units 
we arrive at
\[
p^2\psi(p)+\frac{1}{\pi}\int dp'\;\frac{p'}{p}\ln\left\{
\frac{(p-p')^2}{(p+p')^2}\right\} \psi(p')= E\psi(p).
\]
Next, we translate the integral with help of the 
Gauss-Legendre algorithm in a sum
\[
p_i^2\psi(p_i)+\frac{1}{\pi}\sum_{j=1}^N w_j\frac{p_j}{p_i}\ln\left\{
\frac{(p_i-p_j)^2}{(p_i+p_j)^2}\right\} \psi(p_j)= E\psi(p_i).
\]
This discretized equation has a singularity at $p_i=p_j$. To treat it, we
add and subtract two terms which cancel exactly in the continuum limit.
Both are diagonal, since the divergence occurs on the diagonal only.
The new equation looks like
\begin{eqnarray}\label{CoulombTrickGleichung}
&&p_i^2\psi(p_i)+\frac{1}{\pi}\sum_{j=1}^N w_j\frac{p_j}{p_i}\ln\left\{
\frac{(p_i-p_j)^2}{(p_i+p_j)^2}\right\}
\left[ \psi(p_j)-g(p_i,p_j)\psi(p_i)\right]\\\nonumber
&&+\frac{1}{\pi}\int_{D_1}dp'\frac{p'}{p_i}\ln\left\{
\frac{(p_i-p')^2}{(p_i+p')^2}\right\} \psi(p')= E\psi(p_i).
\end{eqnarray}
One convinces oneself easily, that the second term in the squared bracket,
multiplied by a function $g(p_i,p_j)$ generating better convergence,
nullifies this bracket as $p_i=p_j$. For 
$N\rightarrow\infty$ it is compensated exactly by the 
additional integral expression. 
It must hold
\[
g(p,p)=1.
\]
Since the ground state wavefunction is known, one sets for instance
\[
g(p,p')=\frac{(1+p^2)^2}{(1+p^{'2})^2}.
\]
The integral in (\ref{CoulombTrickGleichung}), the  
{\em continuous part} of the  
Coulomb trick, will in general not be solvable analytically.
It suffices instead, to evaluate it numerically with a much higher 
precision than achieved by doing the sum, {\em i.e.}~the {\em discrete part}.
Here, we obtain the function
\[
f(p,p')=\ln\left\{
\frac{(p_i-p_j)^2}{(p_i+p_j)^2}\right\}
\left[ \psi(p_j)-g(p_i,p_j)\psi(p_i)\right]\\.
\]
It can be continued continuously into $p=p'$ and therefore is 
integrable numerically much easier than the original function. 
There are even possibilities of improving this method \cite{BvdSCoulombTrick}.
The matrix is no longer symmetric. We substitute
\[
\phi(p_i):=\sqrt{w_i}p_i\psi(p_i) 
\]  
and obtain finally
\[
p_i^2\phi(p_i)+\frac{1}{\pi}\sum_{j=1}^N \sqrt{w_i w_j}\ln\left\{
\frac{(p_i-p_j)^2}{(p_i+p_j)^2}\right\}
\phi(p_j)+\mbox{(Coulomb-Counterterms)}= E\phi(p_i).
\]
The result of the manipulations is a considerable improvement of convergence,
documented in Ref.~\cite{KPW}.

\chapter{\label{AppxProgram}Listing of the computer code}



To render the verification of the results of the present work possible and
to demonstrate the implementation of the techniques described throughout the
text {\em in praxi}, we show the listing of the program, written in {\sc C}.


\section{Description of the program}

The program {\tt Mesonix}\footnote{Mnemonic for 
{\em mesonics}=``anything, concerning mesons'', like numerics=``anything, 
concerning the numeric aspects''. 
Here, a meson 
is defined as a fermion-antifermion system. }
is constructed in a modular fashion, {\em i.e.}~one problem corresponds to one 
subroutine. 
The whole program consists of the below parts or subroutines:

\begin{itemize}
\item {\tt meta\_mesonix}:
This part of the program was written to render its application 
more convenient for the user.
An input file for the main program {\tt mesonix} is created. It contains
all parameters, such as cutoff, fermion mass, number of integration points,
etc.
Furthermore, small program tools are supplied, which enable
subsequent calls of {\tt mesonix} to be handled directly. 
The function {\tt vary\_J\_z}, for example, was created
to calculate the spectra for different values of 
$J_z$ at a fixed discretization.
  
\item {\tt mesonix:}
The main program calls the subroutines according to the parameters 
of the input file {\tt input.dat}.

\item {\tt numerix:}
Using the {\sc NAGLib}-Routine {\tt d01bcf} \cite{Naglib}, the 
abscissae and weights for the numerical integrations are obtained using the
Gauss-Legendre algorithm.  

\item {\tt coulomb\_trix:}
Here, the calculation and initialization of the counterterms of the 
Coulomb trick are processed.
It decides whether the results of a former calculation 
are read in from a file, or
if they are (re-)calculated for a different set of parameters.

\item {\tt genetix:}
This routine governs the construction of the {\em symmetrized} 
Hamiltonian matrix. 
By using the charge conjugation symmetry $\cal C$ and the combined
parity  and time reversal symmetry ${\cal H}$, four matrices which 
describe the interaction between the eigenstates 
of these symmetries, are generated.

\item {\tt asymmetrix}
Analogously, this procedure governs the construction of the 
{\em unsymmetrized} Hamiltonian matrix. 
In the case $J_z{\neq}0$ the parity and time reversal operation is not a
symmetry
any more. A special treatment is necessary, which increases the numerical
effort enormously.

\item {\tt physix:}
The matrix elements are calculated by evaluating the function listed in
the helicity tables of Appendix \ref{AppxHelicityTables}. 
The Coulomb counterterms
are introduced when all quantum numbers of the incoming and the outgoing states
are identical.

\item {\tt arithmetix:}
With help of the {\sc NAGLib} routine {\tt f02abf} \cite{Naglib}, the
Hamiltonian matrix is 
diagonalized and the spectrum, as well as the wavefunctions
are calculated and sorted ({\tt m01caf}). It is possible to store 
the wavefunctions. 
For this, they are translated into the original normalization.

\item {\tt publix:}
This routine formats and outputs the results.

\end{itemize}
All global variables are explained in the listing. 
The program is written in standard {\sc C}, but makes no excessive use
of typical {\sc C} structures and should consequently be understandable to 
readers familiar with {\sc Fortran}. 
 
\addcontentsline{toc}{section}{\protect\numberline{H.2}{Listing}}

{\scriptsize

\newpage
\section*{Meta\_mesonix}

\begin{verbatim}
/************************************************************************/
/*                                                                      */
/*  Program    :     M E T A _ M E S O N I X                            */
/*                                                                      */
/*  Class      :     Main program                                       */
/*                                                                      */
/*  Purpose    :     META_MESONIX  calls MESONIX and manipulates the    */
/*                   input file to vary different outer parameters      */
/*                   via INPUT_MES.DAT.                                 */
/*                                                                      */
/************************************************************************/

#include <stdio.h>
#include <math.h>

double alpha;
char default_directory[80]="OUTPUT/";
char special_id[80]="_";
char *name;
FILE *fp;

void create_file(double beta, double lambda, int n1, int n2, int number, 
		 int J_z, int trix)
/* create input file for 'mesonix.c' */
{
        double m;
        int    test,Anni,print_EF;
        char   *name_in = "Input_mes.dat";
        FILE   *fp;
 
        m        = 1.0;   /* fermion mass */
        test       = 1;   /* test variable */
        Anni       = 0;   /* control for incl./excl. of annihilation graph */
        print_EF   = 0;   /* control for writing the eigenfunctions in file */

        if (J_z!=0) test += 4096;   /* asymmetric case */

        fp = fopen(name_in,"w");
        fprintf(fp,"-----------------------------------------------------\n");
        fprintf(fp,"MESONIX_input_file\n");
        fprintf(fp,"-----------------------------------------------------\n\n");
        fprintf(fp,"fermion_mass=\n");
        fprintf(fp,"		%18.12f\n",m);
        fprintf(fp,"cut-off=\n");
        fprintf(fp,"		%18.12f\n",lambda);
        fprintf(fp,"coupling=\n");
        fprintf(fp,"		%18.12f\n",beta);
        fprintf(fp,"test=\n");
        fprintf(fp,"		%3d\n",test);
        fprintf(fp,"J_z=\n");
        fprintf(fp,"		%3d\n",J_z);
        fprintf(fp,"Anni=\n");
        fprintf(fp,"		%3d\n",Anni);
        fprintf(fp,"N1=\n");
        fprintf(fp,"		%3d\n",n1);
        fprintf(fp,"N2=\n");
        fprintf(fp,"		%3d\n",n2);
        fprintf(fp,"number_EV=\n");
        fprintf(fp,"		%3d\n",number);
        fprintf(fp,"trix=\n");
        fprintf(fp,"		%3d\n",trix);
        fprintf(fp,"default_directory=\n");
        fprintf(fp,"		%s\n",default_directory);
        fprintf(fp,"special_id=\n");
        fprintf(fp,"		%s\n",special_id);
        fprintf(fp,"print_EF=\n");
        fprintf(fp,"		%3d\n",print_EF);
        fprintf(fp,"-----------------------------------------------------\n");
        fclose(fp);
}

void vary_coupling(int nn,int N)
/* vary the coupling ALPHA: N1=N2=nn,  */
{	
        int i,j,k;
        double alpha;
        FILE *f_in,*f_out;
        char *name_out = "EVs_couplinx.dat";

        f_out = fopen(name_out,"w");
        fclose(f_out);
        for (i=1; i<=N; ++i)
        {	
                f_in = fopen(name,"w");    /* create output file for MESONIX */
                fclose(f_in);
       	        alpha = 10.0*i/N;
                printf("META_MESONIX:   ALPHA = %8.5f\n\n",alpha);
                create_file(alpha,1.0,nn,nn,10,0,1); 
                system("mesonix.out");
                f_in  = fopen(name,"r");
                f_out = fopen(name_out,"a");
                fscanf(f_in,"%d",&k);
                fprintf("k= %3d\n",k);
                fprintf(f_out,"%18.12f\n",alpha);
                for (j=1; j<=10; ++j) 
                {	
                        fscanf(f_in,"%lf",&alpha);
                        printf("EW = %18.12f\n",alpha);
                        fprintf(f_out,"%18.12f\n",alpha);
                }
                fclose(f_in);
                fclose(f_out); 
        }
}

void vary_N(int begin_step, int end_step, int J_z)
/* vary number of mesh points from <begin_step> to <end_step> */
{	
        int i;

        sprintf(name,"%sEigenValues/EVs%sJ%d.dat",
                default_directory,special_id,J_z);
        printf("%s\n",name);

        fp = fopen(name,"w");        /* create output file for MESONIX */
        fclose(fp);

        for (i=begin_step; i<=end_step; i+=2)
        {	
                printf("\n\nMETA_MESONIX:   N1 = %3d, N2 = %3d\n",i,i);
                create_file(alpha,1.0,i,i,500,J_z,0); 
                system("mesonix.out");
        }
}

void special_case(double lambda, int n1, int n2, int J_z)
/* different N1, N2 */
{	
        int i;

        sprintf(name,"%sEigenValues/EVs%sJ%d.dat",
                default_directory,special_id,J_z);
        printf("%s\n",name);

        fp = fopen(name,"w");        /* create output file for MESONIX */
        fclose(fp);
        printf("\n\nMETA_MESONIX:   N1 = %3d, N2 = %3d\n",n1,n2);
        create_file(alpha,lambda,n1,n2,500,J_z,0); 
        system("mesonix.out");
}

void Coulomb_J_z(int end_step,int max_n) 
/* calculate Coulomb counterterms */
{	
        int i;

        for (i=0; i<=max_n; ++i)
        {	
                printf("\n\nMETA_MESONIX:   J_z = %3d\n",i);
                sprintf(name,"%sEigenValues/EVs%sJ%d.dat",
                	default_directory,special_id,i);
                fp = fopen(name,"w");        /* create output file for MESONIX */
                fclose(fp);

                create_file(alpha,1.0,end_step,end_step,500,i,1); 
                system("mesonix.out");
                if (i>1)
                {
                        printf("\nMETA_MESONIX:   J_z = %3d\n",i);
                        sprintf(name,"%sEigenValues/EVs%sJ%d.dat",
                        default_directory,special_id,-i);
                        fp = fopen(name,"w");        /* create output file for MESONIX */
                        fclose(fp);

                        create_file(alpha,1.0,end_step,end_step,500,-i,1); 
                        system("mesonix.out");
                }
        }
}

void special_Coulomb_J_z(int N1, int N2, int J_z) 
/* calculate Coulomb counterterms for special case: (N1,N2,J_z) */
{	
        printf("\n\nMETA_MESONIX:   J_z = %3d\n",J_z);
        create_file(alpha,1.0,N1,N2,500,J_z,1); 
        system("mesonix.out");
}

void vary_J_z(int begin_N, int end_N, int max_n) 
/* vary J_z from -max_n till +max_n */
{	
        int i;

        for (i=0; i<=max_n; ++i)
        {	
                printf("\n\nMETA_MESONIX:   J_z = %3d\n",i);
                vary_N(begin_N,end_N,i);
                if (i>0)
                {
                        printf("META_MESONIX:   J_z = %3d\n\n",-i);
                        vary_N(begin_N,end_N,-i);
                }
        }
}

double Lambda(int scale, int n)
/* returns Lambda in different scales, n=1..10 */
{
        switch (scale)
        {       
                case 0: return n*1.0*6.0*alpha; break; /* Krautgaertner scale */
                case 1: return n; break;               /* linear scale */
                case 2: return (15.0+5.0*n);           /* linear till 65 scale*/
                        break;                      
        }
        return 1.0;     /* default */
}

void create_environement() 
/* create directories for output */
{
        FILE *fp;

        printf("\nMETA-MESONIX_____________________________________________\n\n");
        fp = fopen("create_dir","w"); /* create file testing if dir. exists */
        fprintf(fp,"if [ ! -d $1 ] \nthen mkdir $1\n");
        fprintf(fp,"echo \"create_dir: Directory '$1' created.\"\n");
        fprintf(fp,"else echo \"create_dir: Directory '$1' exists!\" \nfi");
        fclose(fp);
        name=(char *)malloc(80);
        sprintf(name,"/bin/bash create_dir %s",default_directory);
        system(name);
        sprintf(name,"/bin/bash create_dir %s%s",default_directory,
                                                        "EigenValues/");
        system(name);
        sprintf(name,"/bin/bash create_dir %s%s",default_directory,
                                                        "EigenFunctions/");
        system(name);
        sprintf(name,"/bin/bash create_dir %s%s",default_directory,
                                                        "Coulomb_Trick/");
        system(name);
        system("rm -f create_dir");           /* remove batch file */ 	
        sprintf(name,"%smesonix.log",default_directory);
        printf("Creating '%s'...\n",name);		
        fp = fopen(name,"w");                 /* create LOG file for MESONIX */
        fclose(fp);	
        printf("_________________________________________________________\n\n");
}

void vary_Lambda(int J_z) 
/* vary cutoff Lambda */
{       
        int i,n1,n2;
        double cutoff;

        n1 = 41; 
        n2 = 11;
        for (i=1; i<=5; ++i)
        {       
                cutoff=Lambda(0,i);
                printf("\n\nMETA_MESONIX:   Lambda = %10.8f\n",cutoff);
                sprintf(special_id,"_L%1d_",i-1);
                printf("%s\n",special_id);
                create_file(alpha,cutoff,n1,n2,500,J_z,1); /* Coulomb terms */
                system("mesonix.out");
                special_case(cutoff,n1,n2,J_z);            /* N1!=N2 */
        }
        for (i=6; i<=10; ++i)
        {       
                cutoff=Lambda(0,i);
                printf("\n\nMETA_MESONIX:   Lambda = %10.8f\n",cutoff);
                sprintf(special_id,"_L%1d_",i-1);
                printf("%s\n",special_id);
                create_file(alpha,cutoff,n1,n2,500,J_z,1); /* Coulomb terms */
                system("mesonix.out");
                special_case(cutoff,n1,n2,J_z);            /* N1!=N2 */
        }
        for (i=1; i<=10; ++i)
        {       
                cutoff=Lambda(2,i);
                printf("\n\nMETA_MESONIX:   Lambda = %10.8f\n",cutoff);
                sprintf(special_id,"_LX%1d_",i-1);
                printf("%s\n",special_id);
                create_file(alpha,cutoff,n1,n2,500,J_z,1); /* Coulomb terms */
                system("mesonix.out");
                special_case(cutoff,n1,n2,J_z);            /* N1!=N2 */
        }
}

void Lambda20(int J_z) 
/* calculate for Lambda=20 m_f, N1=41, N2=11 */
{
        int n1=41;
        int n2=11;
        double cutoff=20.0;

        sprintf(special_id,"_l20n4711_");
        printf("%s\n",special_id);
        create_file(alpha,cutoff,n1,n2,500,J_z,1); /* Coulomb terms */
        system("mesonix.out");
        special_case(cutoff,n1,n2,J_z);            /* N1!=N2 */
}

void main()
{
        alpha = 1.0/137.0359895;    /* coupling constant */
        alpha = 0.3;
	
        create_environement();      /* create directories for output */
        Coulomb_J_z(21,3);          /* calc. counterterms */
        vary_N(5,21,0);             /* calc spectra as func. of N, J_z=0,..,3*/
        vary_N(5,21,1);
        vary_N(5,21,2);
        vary_N(5,21,3);
	vary_Lambda(0);             /* calculate spec. for Lambda= 1,..,55 m_f*/
}

\end{verbatim}
\newpage

\section*{Mesonix}

\begin{verbatim}
/************************************************************************/
/*                                                                      */
/*  Program    :     M E S O N I X                                      */
/*                                                                      */
/*  Class      :     Main program                                       */
/*                                                                      */
/*  Purpose    :     MESONIX calculates the eigenstates and -values     */
/*                   of a system consisting of two particles with       */
/*                   same mass, i.e. of mesons.                         */
/*                                                                      */
/************************************************************************/

/************************************************************************/
/*                                                                      */
/*  Structure of the code :     M E S O N I X                           */
/*                             ===============                          */
/*                                                                      */
/*     MAIN PRG              SUBROUTINES            SUB-SUBROUTINES     */
/*    ----------            -------------          -----------------    */
/*                                                                      */
/*       main ---------------> NUMERIX                                  */
/*                |                                                     */
/*                |----------> COULOMB_TRIX                             */
/*                |                                                     */
/*                |----------> GENETIX    ----------->  PHYSIX          */
/*                |                                                     */
/*                |----------> ASYMMETRIX ----------->  PHYSIX          */
/*                |                                                     */
/*                |----------> CALCULATIX                               */
/*                |                                                     */
/*                |----------> PUBLIX                                   */
/*                                                                      */
/*======================================================================*/
/*                                                                      */
/*   Description of the subroutines :                                   */
/*  ----------------------------------                                  */
/*                                                                      */
/*      NUMERIX      calculates the weights and abscissae for the       */
/*                   numerical integration                              */      
/*                                                                      */
/*      COULOMB_TRIX calculates/initializes the Coulomb counter terms   */
/*                                                                      */
/*      GENETIX      generates the Hamilton matrix                      */
/*                                                                      */
/*      PHYSIX       provides the single matrix elements                */
/*                                                                      */
/*      CALCULATIX   diagonalizes the matrix and sorts eigenvalues      */
/*                                                                      */
/*      PUBLIX       controls the output of data                        */
/*                                                                      */     
/*                                                                      */
/*======================================================================*/
/*                                                                      */     
/*  Outer Parameters:                                                   */
/*  -----------------                                                   */
/*                                                                      */     
/*    J_z        = n   ; z component of total angular momentum          */     
/*                                                                      */     
/*    Anni       = 0/1 ; omit/include annihilation graph                */     
/*                                                                      */     
/*    trix       = 0/1 ; read in from file / calculate Coulomb term     */     
/*                                                                      */     
/*    number_EV  = m   ; write m eigenvalues in file 'EVs_Jn.dat'       */     
/*                                                                      */     
/*    print_EF   = 0/1 ; do not write/do write eigenfunctions in file   */
/*                       'EFunc_X.dat'                                  */     
/*                                                                      */     
/*======================================================================*/
/*                                                                      */     
/*                                                                      */
/*   Test Facilities of the Program :                                   */
/*  ----------------------------------                                  */
/*                                                                      */
/* bit    condition       action                                        */
/* ----------------------------------------------------------           */
/*                                                                      */
/* all  test     == 0   nothing will be tested                          */
/*                                                                      */
/* 0  (test &  1)!= 0   NUMERIX will write the abscissae and weights of */
/*                      the numerical integration into the file         */      
/*                      'gauss_vals.dat'                                */
/*                                                                      */
/* 1  (test &  2)!= 0   GENETIX will write a off-diagonal spin table    */
/*                      in "SPIN-TABLE_MES.DAT"                         */
/*                                                                      */
/* 2  (test &  4)!= 0   PUBLIX will write the eigenvalues of the block  */
/*                      matrices in EV_1_MES.DAT, EV_2_MES.DAT,         */
/*                      EV_3_MES.DAT, EV_1_MES.DAT respectively.        */
/*                                                                      */
/*12  (test &4096)!=0   Supress the symmetries, use ASYMMETRIX          */
/*                                                                      */
/************************************************************************/

#include <malloc.h>
#include <stdio.h>
#include <math.h>
#include <unistd.h>
#include <signal.h>

#define PI 3.141592653589793
#define NN1 41                          /* max. # of integr.points (mu)    */
#define NN2 21                          /* max. # of integr.points (theta) */
#define np 1                            /* constant for 'd01alf', etc.     */
#define lp 4000	                        /* constant for 'd01alf', etc.     */
#define lip 2000                        /* constant for 'd01alf', etc.     */
#define N_sym_1 2*(N1*N2)+2*N1          /* dimensions of blocks (symmetrix)*/
#define N_sym_2 2*(N1*N2)+2*N1
#define nagtest 1                       /* use NAGLib for diag if !=0      */
#define old_CT 0                        /* use old(=Krautgaertner) Coulomb */
                                        /* trick if != 0                   */


/* ===================>      external functions       <=================== */

extern long time();
/* time measurements */
 
extern double f02abf_( double *A, int *IA, int *NP, double *R, double *V,
                       int *IV, double *E, int *IFAIL);
/* diagonalization routine from NAGLib*/

extern double Coulomb_trick_function(double *);        
/* Continous Coulomb counterterm, dependend on (mu,theta) */

extern double Coulomb_trick_integrand(double *);
/* Continous Coulomb counterterm, integrated over (theta), dependent on (mu) */

extern double Coulomb_discrete(int jm, int s1, int s2);
/* Discrete Coulomb counterterm, dependend on spins and mu[jm] */

extern double integrand(double *);
/* Krautgartner continous Coulomb counterterm, analytically integrated over */
/* (theta), dependent on (mu) */

extern double old_continous(int jm, int jt);                
/* Krautgartner continous Coulomb counterterm */


/* ===================>     global variables        <=================== */

double m,lambda,ALPHA,p_bohr;   /* mass, cut-off, coupling, Bohr momentum */
double aa,A,B,V_seag;           /* variables used in 'physix' and 'genetix' */ 
double x[3],k[3];               /* old variables (x,k) instead (mu,theta) */
double *mu,*wmu;                /* mu: abscissae and weights for Gauss-Legendre */
double *theta,*wtheta;          /* cos(theta):  "          "     Gauss-Legendre */
double *ma,*r,*v,*e;            /* variables for NAGLIB diagonalization */
double *mat_1,*r_1,*v_1,*e_1;   /* variables for NAGLIB diagonalization */
double *mat_2,*r_2,*v_2,*e_2;   /* variables for NAGLIB diagonalization */
double *mat_3,*r_3,*v_3,*e_3;   /* variables for NAGLIB diagonalization */
double *mat_4,*r_4,*v_4,*e_4;   /* variables for NAGLIB diagonalization */
double mu_global;               /* Coulomb trick:global variable for C_T_function*/
double CT[4][NN1][NN2];         /* matrix for Coulomb data, first index is */
                                /* n = uu,ud,du,dd */
int J_z,test,N1,N2,steps,Anni;  /* OUTER parameters,read in from file */
int trix,number_EV,print_EF;    /* OUTER parameters,read in from file */
int EV_number;                  /* arithmetix: number of eigenfunction */
int IndexMatrix;                /* arithmetix.c */
int IndexMat[4];                /* arithmetix.c */
int jj1,jj2;                    /* global variables for Coulomb trick */
int spins_parallel,spin_1;      /* Coulomb trick: choose function G1 <-> G2 */
int J_z_Coulomb;                /* auxiliary variable for Coulomb trick */
long time1,time2;               /* var. for time measurements */
char *order;                    /* for NAGLib routine m01caf */
char *name_EV;                  /* name of file containing the eigenvalues */
char *name_log;                 /* name of LOG file */
char *default_directory[80];    /* read in by logix, def. dir. for output */
char *special_id[80];           /* read in by logix, special ident. for output*/
char *CT_name[4];               /* names for the Coulomb trick files */
FILE *f_log;                    /* File variable for 'log' file */

/* ===============>   functions for error handling     <===================== */

void nrerror(char error_text[])
/* standard error handler */
{
        fprintf(stderr,"Numerical Recipes run-time error...\n");
        fprintf(stderr,"%s\n",error_text);
        fprintf(stderr,"...now exiting to system...\n");
        exit(1);
}

/* ===============>   functions for (de)allocation     <===================== */

double *dvector(long nl, long nh)
/* allocate a double vector with subscript range v[nl..nh] */
{
        double *v;

        v=(double *)malloc((size_t) ((nh-nl+1+1)*sizeof(double)));
        if (!v) nrerror("allocation failure in dvector()");
        return v-nl+1;
}

void free_dvector(double *v, long nl, long nh)
/* free a double vector allocated with dvector() */
{
        free((char*) (v+nl-1));
}
                
/* ===================>   inclusion of subroutines    <====================== */

#include "numerix.c"
#include "physix.c"
#include "genetix.c"
#include "coulomb_trix.c"
#include "arithmetix.c"
#include "publix.c"
#include "asymmetrix.c"

/* ==============>   functions used for initalization    <=================== */


void init_ma(int dim)
/* initialize matrix for asymmetic NAGLib diagonalization */
{  
        ma = dvector(0,dim*dim - 1);
        e  = dvector(0, dim-1);
        r  = dvector(0, dim-1);	
        v  = dvector(0, dim*dim - 1); 
        IndexMatrix = 0;
}

void init_mat(int dim1, int dim2, int dim3, int dim4) 
/* initialize matrices 1-4 for symmetic NAGLib diagonalization */
{  
        int i;
	
        for (i=0; i<=3; ++i) IndexMat[i] = 0;

        mat_1 = dvector(0,dim1*dim1 - 1);
        e_1   = dvector(0, dim1-1);
        r_1   = dvector(0, dim1-1);	
        v_1   = dvector(0, dim1*dim1 - 1); 

        mat_2 = dvector(0,dim2*dim2 - 1);
        e_2   = dvector(0, dim2-1);
        r_2   = dvector(0, dim2-1);	
        v_2   = dvector(0, dim2*dim2 - 1); 

        if (dim3>0)
        {
                mat_3 = dvector(0,dim3*dim3 - 1);
                e_3   = dvector(0, dim3-1);
                r_3   = dvector(0, dim3-1);	
                v_3   = dvector(0, dim3*dim3 - 1); 

                mat_4 = dvector(0,dim4*dim4 - 1);
                e_4   = dvector(0, dim4-1);
                r_4   = dvector(0, dim4-1);	
                v_4   = dvector(0, dim4*dim4 - 1); 
        }
}

void free_mat(int dim1, int dim2, int dim3, int dim4)
/* deallocate matrices 1-4 used for symmetric NAGLib diagonalization */
{  
        free_dvector(mat_1,0,dim1*dim1-1);
        free_dvector(e_1,0,dim1-1);
        free_dvector(r_1,0,dim1-1);	
        free_dvector(v_1,0,dim1*dim1-1); 

        free_dvector(mat_2,0,dim2*dim2-1);
        free_dvector(e_2,0,dim2-1);
        free_dvector(r_2,0,dim2-1);	
        free_dvector(v_2,0,dim2*dim2-1); 

        if (dim3>0)
        {
                free_dvector(mat_3,0,dim3*dim3-1);
                free_dvector(e_3,0,dim3-1);
                free_dvector(r_3,0,dim3-1);	
                free_dvector(v_3,0,dim3*dim3-1); 

                free_dvector(mat_4,0,dim4*dim4-1);
                free_dvector(e_4,0,dim4-1);
                free_dvector(r_4,0,dim4-1);	
                free_dvector(v_4,0,dim4*dim4-1);
	} 
}

void free_ma(int dim)
/* deallocate matrices used for asymmetric NAGLib diagonalization */
{  
        free_dvector(ma,0,dim*dim-1);
        free_dvector(e,0,dim-1);
        free_dvector(r,0,dim-1);	
        free_dvector(v,0,dim*dim-1); 
}

void read_input()
/* read in inputfile created by 'meta_mesonix.c' */
{
        FILE *fp;
        char *name_IN  = "Input_mes.dat";
        char comment[80];               /* dummy variable to read out comments*/

        fp = fopen(name_IN,"r");        /* read in INPUT_MES.DAT */
        fscanf(fp,"%s",comment);
        fscanf(fp,"%s",comment);
        fscanf(fp,"%s",comment);
        fscanf(fp,"%s",comment);
        fscanf(fp,"%lf",&m);            /* read in <fermion mass> */
        fscanf(fp,"%s",comment);
        fscanf(fp,"%lf",&lambda);       /* read in <cut-off> */
        fscanf(fp,"%s",comment);
        fscanf(fp,"%lf",&ALPHA);        /* read in <coupling> */
        fscanf(fp,"%s",comment);
        fscanf(fp,"%d",&test);          /* read in <test variable> */
        fscanf(fp,"%s",comment);
        fscanf(fp,"%d",&J_z);           /* read in <J_z> */
        fscanf(fp,"%s",comment);
        fscanf(fp,"%d",&Anni);          /* read in control for annih. graph */
        fscanf(fp,"%s",comment);
        fscanf(fp,"%d",&N1);            /* read in # mesh points <N1> */
        fscanf(fp,"%s",comment);
        fscanf(fp,"%d",&N2);            /* read in # mesh points <N2> */
        fscanf(fp,"%s",comment);
        fscanf(fp,"%d",&number_EV);     /* read in # of EVs written in file */
        fscanf(fp,"%s",comment);
        fscanf(fp,"%d",&trix);          /* read in control for Ctrick ab initio*/
        fscanf(fp,"%s",comment);
        fscanf(fp,"%s",default_directory); /* read in <default directory> */
        fscanf(fp,"%s",comment);
        fscanf(fp,"%s",special_id);     /* read in <special identification> */
        fscanf(fp,"%s",comment);
        fscanf(fp,"%s",&print_EF);      /* read in control for writing */
                                        /* eigenfctns in file */
        fclose(fp);
        name_EV  = (char *)malloc(120);
        sprintf(name_EV,"%sEigenValues/EVs%sJ%d.dat",
                default_directory,special_id,J_z); /* create filename for EVs */
}

void logix()
/* write LOG file */
{
        name_log = (char *)malloc(120);  /* name for LOG file */
        sprintf(name_log,"%smesonix.log",default_directory); 			
        f_log = fopen(name_log,"a");
        fprintf(f_log,"\nM E S O N I X                                          LOG file\n");
        fprintf(f_log,"====================================================================\n\n");
        fprintf(f_log,"PARAMETERS:\n");
        fprintf(f_log,"                 m = %18.12f ; fermion mass\n",m);
        fprintf(f_log,"            lambda = %18.12f ; cut-off\n",lambda);
        fprintf(f_log,"             alpha = %18.12f ; coupling\n",ALPHA);
        fprintf(f_log,"\nOPTIONS:\n");
        fprintf(f_log,"              test = %5d ; test variable\n",test);
        fprintf(f_log,"               J_z = %5d ; ang. momentum sector\n",J_z);
        fprintf(f_log,"              Anni = %5d ; annihilation graph\n",Anni);
        fprintf(f_log,"                N1 = %5d ; number of MU values\n",N1);
        fprintf(f_log,"                N2 = %5d ; number of THETA values\n",N2);
        fprintf(f_log,"         number_EV = %5d ; number of eigenvalues\n",number_EV);
        fprintf(f_log,"\n--------------------------------------------------------------------\n\n");
        fclose(f_log);
}

void close_log_file()
/* final comment and close of LOG file */

{
        fprintf(f_log,"\n====================================================================\n\n");
        fclose(f_log);	
}

void allocation()
/* allocate matrices used */ 
{
        init_ma (4*N1*N2);      /* initialize matrix for NAGLIB diag */
        if ((test&16384)!=0) init_mat(N_sym_1,N_sym_2,0,0); 
        else init_mat(N1*N2,N1*N2+N1,N1*N2,N1*N2-N1); 
                                /* initialize matrices 1-4 for NAGLIB diag. */
        mu     = dvector(1,NN1+1); 
        theta  = dvector(1,NN2+1);
        wmu    = dvector(1,NN1+1); 
        wtheta = dvector(1,NN2+1);
}        

void deallocation()
/* deallocate matrices used */ 
{
        if ((test&16384)!=0) free_mat(N_sym_1,N_sym_2,0,0);
        else free_mat(N1*N2,N1*N2+N1,N1*N2,N1*N2-N1); 
        free_ma (4*N1*N2);
}

void  main()
{    	
        int i,j;
        FILE *fp;

        printf("\nMESONIX__________________________________________________\n");
        read_input();		     		/* read in INPUT_MES.DAT     */
        logix();		     		/* write header of LOG file  */
        allocation();                		/* allocate all objects used */
        p_bohr = m*ALPHA/2;          		/* Bohr momentum             */

        if (old_CT==0 ) coulomb_trix(trix,3,N1);     /* init Coulomb trick */
        if (trix==0)  	 	     /* if not C. trick from scratch, then ...*/
        {
            fp = fopen(name_EV,"a"); /* open file for eigenvalues */
            fprintf(fp,"%3d\n",N1);
            fclose(fp);
            f_log = fopen(name_log,"a"); /* open LOG file */        
            fprintf(f_log,"Time for\n");
            fclose(f_log);

            numerix(N1,N2); /* calculate weights for numerical integration */

            f_log = fopen(name_log,"a");  /* open LOG file for subroutines */  
            plot_running_coupling();
            if ((test&(4096+16384)) == 0) /* J_z=0 */
            {
                genetix();     /* set up Hamilton matrix */
                arithmetix(N1*N2,N1*N2+N1,N1*N2,N1*N2-N1); /* diagonalize */
            }
            else /* J_z<>0 */
            {
                if ((test&16384) == 0) /* standard, unsymmetrized Hamiltonian */ 
                {
                        asymmetrix();/* set up Hamiltonian without symmetries */
                        arithmetix(4*N1*N2,0,0,0);
                }
            }
            publix(); /* write eigenvalues in file */
        }
        close_log_file();
        deallocation(); /* deallocate all objects used */
        printf("\n_________________________________________________________\n");
}
\end{verbatim}

\newpage
\section*{Numerix}

\begin{verbatim}
/************************************************************************/
/*                                                                      */
/*  program    :     N U M E R I X                                      */
/*                                                                      */
/*  class      :     subroutine of MESONIX                              */
/*                                                                      */
/*  file name  :     "numerix.c"                                        */
/*                                                                      */
/*  purpose    :     NUMERIX calculates the weights and abscissae       */
/*                   for numerical integration according to the         */
/*                   Gauss-Legendre-algorithm.                          */
/*                                                                      */
/*----------------------------------------------------------------------*/
/*                                                                      */
/*  Subroutines :    d01bcf (NAGLib) for Gauss-Legendre discretization  */
/*                                                                      */
/*  Global variables :    mu[1..NN1]  = abscissa values in mu direction */
/*                        wmu[1..NN1] = weights in mu direction         */
/*                        theta[1..NN2] = abscissa values in cos(theta) */
/*                                        direction                     */
/*                        wmu[1..NN2] = weights in cos(theta) direction */
/*                                                                      */
/************************************************************************/

extern double d01bcf_( int *itype, double *a, double *b, double *c,double *d,
		       int *n, double *wmu, double *mu, int *ifail);
                                 /* NAGLib: Gauss-Legendre discretization */
	
void numerix( int n1, int n2 ) 	       /* n1 = number of mu-values        */
                                       /* n2 = number of theta-values     */ 
                                       /* test == 1 prints values in FILE */ 
/* main function: calculate weights and abscissae */
{
        int i;
        int itype,ifail;         /* for d01bcf */
        double a,b,c,d;          /* for d01bcf */
        double *weight,*abscis;  /* for d01bcf */
        char *name;		 /* name for file in case test=1 */
        FILE *fp;
          
        time(&time1);
        printf("-----------------------------> NUMERIX( N1=%2d, N2 =%2d)\n",
          							     n1,n2);
        f_log = fopen(name_log,"a");                

        /* +++++ Discretize in mu-direction, boundaries [0,lambda/2] ++++ */

        weight = dvector(0,n1-1);      /* initialize fields for d01bcf */
        abscis = dvector(0,n1-1);       
        itype = 0;                     /* parameters for d01bcf */
        a = 1/(1+lambda/2.0/p_bohr);
        b = c = d = 1.0;
	
        d01bcf_(&itype, &a, &b, &c, &d, &n1, weight, abscis, &ifail);
                                       /* Gauss-Legendre discretization */
	
        for (i=1; i<=n1; i++)
        {
                wmu[i] = weight[i-1]/abscis[i-1]/abscis[i-1]*p_bohr;
                mu[i] = (1/abscis[i-1]-1)*p_bohr;		
        }
        free_dvector(weight,0,n1-1);
        free_dvector(abscis,0,n1-1);

        /* ===> Discretize in cos(theta)-direction, boundaries [-1,1] <=== */
 
        weight = dvector(0,n2-1);       /* initialize fields for d01bcf */
        abscis = dvector(0,n2-1);
        itype = 0;                      /* parameters for d01bcf */
        a = -1.0;
        b = c = d = 1.0;

        d01bcf_(&itype, &a, &b, &c, &d, &n2, weight, abscis, &ifail);
                                        /* Gauss-Legendre discretization */
        for (i=1; i<=n2; i++)
        {
                theta[i] = abscis[i-1];
               wtheta[i] = weight[i-1];
        }	
        free_dvector(weight,0,n2-1);
        free_dvector(abscis,0,n2-1);

        /* =========> test==1 : write Gauss-Legendre values in a file <======= */

        if ((test&1) != 0 )
        {
                name = (char *)malloc(120);
                sprintf(name,"%sgauss_values.dat",default_directory); /* filename */
                fp = fopen(name,"w");		
                fprintf(fp,"Gauss-Legendre integration :       MESONIX.C\n");
                fprintf(fp,"----------------------------\n\n");
                fprintf(fp,"               mu-integration                      
                                                        theta-integration\n\n");
                fprintf(fp,"  n         abscissa        weight            
                                                        abscissa            weight\n");
                fprintf(fp,"------------------------------------------------
	                                                ---------------------------\n");
                for (i=1; i<=n1; i++)
                {
                    fprintf(fp,"%3d |    %12.8f   %12.8f    |   %12.8f     
                                     %12.8f \n",i,mu[i],wmu[i],theta[i],wtheta[i]);
                }
                fprintf(fp,"------------------------------------------------
	                                        ---------------------------\n");
                fclose(fp); 
        }	

        /* =======> write time used for calculation in LOG file <========== */

        time(&time2);
        fprintf(f_log,"         ...weights: %d sec\n",time2-time1);
        fclose(f_log);
        return;
}
\end{verbatim}

\newpage
\section*{Coulomb\_trix}
\begin{verbatim}

/************************************************************************/
/*                                                                      */
/*  Program    :     C O U L O M B _ T R I X                            */
/*                                                                      */
/*  Class      :     Subroutine                                         */
/*                                                                      */
/*  Purpose    :     COULOMB_TRIX creates the files for MESONIX         */
/*                   which contain the values for the different         */
/*                   Coulomb tricks, for g e n e r a l  J_z.            */
/*                                                                      */
/*----------------------------------------------------------------------*/
/*                                                                      */
/*  Names of the generated files:                                       */
/*                                                                      */
/*               1)  CTrick_J<n>_uu.dat \                               */
/*               2)  CTrick_J<n>_ud.dat  \_  n = ...-2,-1,0,1,2,...     */
/*               3)  CTrick_J<n>_du.dat  /                              */
/*               4)  CTrick_J<n>_dd.dat /                               */
/*                                                                      */
/*----------------------------------------------------------------------*/
/*  External functions:                                                 */
/*  -------------------                                                 */
/*                                                                      */
/*     d01alf    NAGLib integration of function with singularities      */
/*     d01ajf    NAGLib integration of function without sing.           */
/*                                                                      */
/************************************************************************/


extern double d01alf_( double (*funktion)(double *), double *a, double *b,
                       int *npts, double points[np], double *epsabs,
                       double *epsrel,double *result, double *abserr,
                       double w[lp],int *lm, int iw[lip], int *liw, 
                       int *ifail);
/* integration routine from NAGLib */

extern double d01ajf_( double (*funktion)(double *), double *a, double *b, 
                       double *epsabs, double *epsrel, double *result,
                       double *abserr, double w[lp], int *lm, int iw[lip], 
                       int *liw, int *ifail);
/* integration routine from NAGLib */

double old_continous(int jm, int jt) 
/* old Krautgaertner Coulomb trick */
{
        int ifail,npts,iw[lip],lw,liw;
        double a,b,points[2],w[lp],abserr,epsabs,epsrel,coul_conti;

        a = 0.0;
        b = lambda/2.0;
	
        jj1 = jm;
        jj2 = jt;

        /*   NAGLIB-ROUTINE  */
	
        epsabs = 1e-6; 
        epsrel = 1e-8; 
        npts = 1;
        points[0] = mu[jm]; 				
        liw = lip;
        lw = lp;	
        ifail = -1;	

        d01alf_(integrand,&a,&b,&npts,points,&epsabs,&epsrel,
                &coul_conti,&abserr,w,&lw,iw,&liw,&ifail);
        return coul_conti;
}


double Coulomb_trick_function(double *theta_lokal)
/* continous C. trick counterterm, to be integrated over <theta> */
/* for fixed <mu> */
{	
				/* Parameters = global variables are: 
			                        mu_global
						jj1,jj2
						J_z_Coulomb
						spins_parallel */	
	double result,thetaa;

	thetaa = *theta_lokal;
	result = 0.0;	

	x[1] = (1.0 + mu_global*thetaa/sqrt(m*m + mu_global*mu_global))/2.0;	
	k[1] = mu_global*sqrt(1.0-thetaa*thetaa);
	x[2] = (1.0 + mu[jj1]*theta[jj2]/sqrt(m*m + mu[jj1]*mu[jj1]))/2.0;	
	k[2] = mu[jj1]*sqrt(1.0-theta[jj2]*theta[jj2]);

	if (( fabs(x[1]-x[2]) >= 1e-8 || fabs(k[1]-k[2]) >= 1e-8 )
	       && (fabs(k[1]) >= 1e-8) && (fabs(k[2]) >= 1e-8))
	{
            aa = (x[1]-x[2])*(x[1]-x[2])*m*m/2.0*(1/(1-x[2])/(1-x[1])
                 +1/x[1]/x[2]) + k[1]*k[1] + k[2]*k[2]
                 + (x[1]-x[2])/2*(k[1]*k[1]*(1/(1-x[1])-1/x[1]) -
                                  k[2]*k[2]*(1/(1-x[2])-1/x[2]));

	    if((aa*aa-4*k[1]*k[1]*k[2]*k[2])>0.0)
	    {
                A  = 1.0/sqrt( aa*aa - 4*k[1]*k[1]*k[2]*k[2] );
                B  = (1.0-aa*A)/2.0;

		if (spins_parallel==1) result = G1(2,1,J_z_Coulomb);
		else		       result = G2(2,1,J_z_Coulomb);

		result *= 2.0*mu_global*mu_global*x[1]*(1.0-x[1])/
			  sqrt(m*m + mu_global*mu_global)*
			  (1+mu[jj1]*mu[jj1]*mu[jj1]/p_bohr/p_bohr/p_bohr/8.0)
		          /(1+mu_global*mu_global*mu_global
			  /p_bohr/p_bohr/p_bohr/8.0)/PI;
	    }
	}
	return result;
}

double Coulomb_trick_integrand(double *mu_lokal)
/* C. trick counterterm, already integrated over <theta>, dependent on <mu> */
{
                                  /* Parameters = global variables are
                                                  mu_global
                                                  jj1,jj2
                                                  J_z_Coulomb
                                                  spins_parallel */	
                                                
        int ifail,npts,iw[lip],lw,liw;
        double a,b,points[2],w[lp],abserr,epsabs,epsrel,result_coul;

        mu_global = *mu_lokal;
        a =-1.0;
        b = 1.0;
        epsabs = 1e-12;
        epsrel = 1e-3;
        npts = 1;
        points[0] = theta[jj2];
        liw =lip;
        lw = lp;
        ifail =-1;
        if (fabs(mu_global - mu[jj1]) <= 1e-10)
        {
                d01alf_(Coulomb_trick_function,&a,&b,&npts,points,&epsabs,
                        &epsrel,&result_coul,&abserr,w,&lw,iw,&liw,&ifail);
        }
        else
                d01ajf_(Coulomb_trick_function,&a,&b,&epsabs,&epsrel,
                        &result_coul,&abserr,w,&lw,iw,&liw,&ifail);
        return result_coul;
}	

double Coulomb_discrete(int jm, int s1, int s2) 
/* discrete part of Coulomb trick, called by PHYSIX */
{
        int    km,kt;
        double result,spinfunction;

        result = 0.0;
        for ( km = 1; km <= N1; ++km )
        {
            for ( kt = 1; kt <= N2; ++kt )
            {	
                x[1] = (1.0 + mu[km]*theta[kt]/sqrt(m*m + mu[km]*mu[km]))/2.0;	
                k[1] = mu[km]*sqrt(1.0-theta[kt]*theta[kt]);
	
                if ( fabs(x[1]-x[2]) >= 1e-8 || fabs(k[1]-k[2]) >= 1e-8 ) 
                {
                    aa = (x[1]-x[2])*(x[1]-x[2])*m*m/2.0*(1/(1-x[2])/(1-x[1])
                         +1/x[1]/x[2]) + k[1]*k[1] + k[2]*k[2]
                         + (x[1]-x[2])/2*(k[1]*k[1]*(1/(1-x[1])-1/x[1]) -
       	                                  k[2]*k[2]*(1/(1-x[2])-1/x[2]));
                    A  = 1.0/sqrt( aa*aa - 4*k[1]*k[1]*k[2]*k[2] );
                    B  = (1.0-aa*A)/2.0;

                    if (s1>0)
                    {
                        if (s1==s2) spinfunction = -G1(2,1,J_z);
                        else        spinfunction = -G2(2,1,J_z);
                    }
                    else
                    {
                        if (s1==s2) spinfunction = -G1(2,1,-J_z);
                        else        spinfunction = -G2(2,1,-J_z);
                    }
                    if (old_CT !=0) spinfunction = -G2(2,1,J_z);
    
                    result += 1.0/PI*wtheta[kt]*wmu[km]*spinfunction 
                                *2.0*mu[km]*mu[km]*x[1]*(1.0-x[1])
                                /sqrt(m*m + mu[km]*mu[km])*
                                (1+mu[jm]*mu[jm]*mu[jm]/p_bohr/p_bohr/p_bohr/
                               /(1+mu[km]*mu[km]*mu[km]/p_bohr/p_bohr/p_bohr/8);
                }	
            }
        }
        return result;
}

void install_Coulomb_trick()  
/* read data from 'CTrick_Jn_ss.dat' into CT[] */
{
        int i,j,k,l,lo[2],hi[2],error;
        double value;
        FILE *fp[4];

        error = 0;
        for (i=0; i<=3; ++i)
        { 
                fp[i] = fopen(CT_name[i],"r");
                fscanf(fp[i],"%d",&lo[0]);
                fscanf(fp[i],"%d",&hi[0]);
                if (i>0 && (lo[1]!=lo[0] || hi[1]!=hi[0])) error = 1;
                lo[1]=lo[0];
                hi[1]=hi[0];
        } 
        if (error!=0) 
                printf(">>>>> ERROR(install_Coulomb_trick): lo[i]!=lo[j]\n");
        if (hi[1]<N1) 
                printf(">>>>> ERROR(install_Coulomb_trick): hi[1] < N1\n");

        for (i=lo[1]; i<N1; i+=2) /* ignore first data until actual N1 */
        {
                for (l=0; l<=3; ++l) fscanf(fp[l],"%lf",&k);
                for (j=1; j<=i*i; ++j)
                {
                      for (l=0; l<=3; ++l) fscanf(fp[l],"%lf",&value);
                }
        }
        for (l=0; l<=3; ++l) fscanf(fp[l],"%lf",&k);
        for (i=0; i<=N1-1; ++i)    /* read out needed data */
           for (j=0; j<=N2-1; ++j)
           {
                for (l=0; l<=3; ++l)
                {
                        fscanf(fp[l],"%lf",&value);
                        CT[l][i][j] = value;
                }
        }
        for (l=0; l<=3; ++l) fclose(fp[l]);
}

void special_install_Coulomb_trick()  
/* read data from 'CTrick_Jn_ss.dat' into CT[], special case */
{
        int i,j,k,l,n1[2],n2[2],error;
        double value;
        char *comment[80];
        FILE *fp[4];

        error = 0;
        for (i=0; i<=3; ++i)
        { 
                fp[i] = fopen(CT_name[i],"r");	
                fscanf(fp[i],"%s",comment);
                fscanf(fp[i],"%d",&n1[0]);
                fscanf(fp[i],"%d",&n2[0]);
                if (i>0 && (n1[1]!=n1[0] || n2[1]!=n2[0])) error = 1;
                n1[1]=n1[0];
                n2[1]=n2[0];
        } 
        if (error!=0) 
                printf(">>>>> ERROR(install_Coulomb_trick): lo[i]!=lo[j]\n");
        if (n1[1]<N1) 
                printf(">>>>> ERROR(install_Coulomb_trick): n1[1] < N1\n");

        for (i=0; i<=N1-1; ++i)    /* read out needed data */
           for (j=0; j<=N2-1; ++j)
           {
                for (l=0; l<=3; ++l)
                {
                        fscanf(fp[l],"%lf",&value);
                        CT[l][i][j] = value;
                }
           }
        for (l=0; l<=3; ++l) fclose(fp[l]);
}


void  special_coulomb_trix(int from_scratch)    
/* calculate or initialize the Coulomb counterterms   */
/*   !!!!!!!!!!! special case N1!=N2 !!!!!!!!!!!!!    */
/*----------------------------------------------------*/
/* from_scratch != 0 --> really calculate the terms   */
/*              == 0 --> just install the trick       */
{       
        int i,j,l,NNN,counter;
        int ifail,npts,iw[lip],lw,liw;
        double a,b,points[2],w[lp],abserr,epsabs,epsrel,coul_result;
        FILE *fp[4];

        printf("\n                               (special)\n");
	         
        for (i=0; i<=3; ++i) /* create file names */
        {
            CT_name[i]=(char *)malloc(80);
            sprintf(CT_name[i],"%sCoulomb_Trick/sCTrick%sJ",
                    default_directory,special_id);
            switch (i)
            {
                        case 0:sprintf(CT_name[i],"%s%d_uu.dat",CT_name[i],J_z);
                               break;
                        case 1:sprintf(CT_name[i],"%s%d_ud.dat",CT_name[i],J_z);
                               break;
                        case 2:sprintf(CT_name[i],"%s%d_du.dat",CT_name[i],J_z);
                               break;
                        case 3:sprintf(CT_name[i],"%s%d_dd.dat",CT_name[i],J_z);
                               break;
            }
        }
        if (from_scratch == 0) special_install_Coulomb_trick();
        else 
        {
                for (i=0; i<=3; ++i) /* open file */
                {
                        printf(">>>>> INFO: Opening file '%s'\n",CT_name[i]);
                        fp[i] = fopen(CT_name[i],"w");
                }
                for (i=0; i<=3; ++i) fprintf(fp[i],"Special:\n%2d\n%2d\n",N1,N2);
	    
                numerix(N1,N2);   /* calculate Gauss_Legendre values */

                for (i=1; i<=N1; ++i)
                {	
                    printf("	i = %2d\n",i);
                    for (j=1; j<=N2; ++j)
                    {
                        printf("		j = %2d\n",j);
                        counter = 0;
                        for (l=1; l>=-1; l-=2)
                            for (spins_parallel=1; spins_parallel>=-1;
                                                   spins_parallel-=2)
                            {
                                J_z_Coulomb = l*J_z;
                                x[2] = (1.0+mu[i]*theta[j]/sqrt(m*m+mu[i]*mu[i]))/2.0;
                                k[2] = mu[i]*sqrt(1.0-theta[j]*theta[j]);
                                jj1  = i;
                                jj2  = j;
                                a = 0.0;
                                b = lambda/2.0;
                                epsabs = 1e-6;
                                epsrel = 1e-8; 
                                npts   = 1;
                                liw    = lip;
                                lw     = lp;        
                                ifail  = -1; 
                                points[0] = mu[i];      
                                if (J_z_Coulomb !=0 || l>0) 
                                {
                                    d01alf_(Coulomb_trick_integrand,&a,&b,&npts,
                                            points,&epsabs,&epsrel,&coul_result,
                                            &abserr,w,&lw,iw,&liw,&ifail);
                                    fprintf(fp[counter+spins_parallel*(1-l)/2],
                                            "%18.12f\n",coul_result);
                                    if (J_z_Coulomb==0) 
                                         fprintf(fp[counter+2+spins_parallel],
                                              "%18.12f\n",coul_result);
                                }
                                ++counter;
                           }		 	
                     } /* j loop */
               } /* i loop */
               for (i=0; i<=3; ++i) fclose(fp[i]);    
       } /* END: from_scratch != 0 */
}


void  coulomb_trix(int from_scratch, int lo_N, int hi_N)    
/* calculate or initialize the Coulomb counterterms   */
/*----------------------------------------------------*/
/* from_scratch != 0 --> really calculate the terms   */
/*              == 0 --> just install the trick       */
/* lo_N --> lowest discretization                     */
/* hi_N --> highest discretization                    */
{       
        int i,j,l,n,NNN,counter;
        int ifail,npts,iw[lip],lw,liw;
        double a,b,points[2],w[lp],abserr,epsabs,epsrel,coul_result;
        FILE *fp[4];

        printf("\n-----------------------------> COULOMB_TRIX\n");
	         
        if (N1!=N2) special_coulomb_trix(from_scratch);
        else
        {
        for (i=0; i<=3; ++i) /* create file names */
        {
            CT_name[i]=(char *)malloc(80);
            sprintf(CT_name[i],"%sCoulomb_Trick/CTrick%sJ",
                    default_directory,special_id);
            switch (i)
            {
                        case 0:sprintf(CT_name[i],"%s%d_uu.dat",CT_name[i],J_z);
                               break;
                        case 1:sprintf(CT_name[i],"%s%d_ud.dat",CT_name[i],J_z);
                               break;
                        case 2:sprintf(CT_name[i],"%s%d_du.dat",CT_name[i],J_z);
                               break;
                        case 3:sprintf(CT_name[i],"%s%d_dd.dat",CT_name[i],J_z);
                               break;
            }
        }
        if (from_scratch == 0) install_Coulomb_trick();
        else 
        {
            for (i=0; i<=3; ++i) /* open file */
            {
                printf(">>>>> INFO: Opening file '%s'\n",CT_name[i]);
                fp[i] = fopen(CT_name[i],"w");
            }
            for (i=0; i<=3; ++i) fprintf(fp[i],"%2d\n%2d\n",lo_N,hi_N);
            for (n=lo_N; n<=hi_N; n+=2)
            {
                numerix(n,n);   /* calculate Gauss_Legendre values */

                printf("Counterterms: n/hi_N = %2d/%2d\n",n,hi_N);
                for (i=0; i<=3; ++i) fprintf(fp[i],"%2d\n",n);    
                for (i=1; i<=n; ++i)
                {	
                    printf("                i = %2d\n",i);
                    for (j=1; j<=n; ++j)
                    {
                        printf(" 	               j = %2d\n",j);
                        counter = 0;
                        for (l=1; l>=-1; l-=2)
             	            for (spins_parallel=1; spins_parallel>=-1;spins_parallel-=2)
                            {
                                J_z_Coulomb = l*J_z;
                                x[2] = (1.0+mu[i]*theta[j]/sqrt(m*m+mu[i]*mu[i]))/2.0;
                                k[2] = mu[i]*sqrt(1.0-theta[j]*theta[j]);
                                jj1  = i;
                                jj2  = j;
                                a = 0.0;
                                b = lambda/2.0;
                                epsabs = 1e-6;
                                epsrel = 1e-8; 
                                npts   = 1;
                                liw    = lip;
                                lw     = lp;        
                                ifail  = -1; 
                                points[0] = mu[i];      
                                if (J_z_Coulomb !=0 || l>0) 
                                {
                                    d01alf_(Coulomb_trick_integrand,&a,&b,&npts,
                                            points,&epsabs,&epsrel,&coul_result,
                                            &abserr,w,&lw,iw,&liw,&ifail);
                                    fprintf(fp[counter+spins_parallel*(1-l)/2],
                                            "%18.12f\n",coul_result);
                                    if (J_z_Coulomb==0) 
                                    fprintf(fp[counter+2+spins_parallel],
                                            "%18.12f\n",coul_result);
                                }
                                ++counter;
                            }	 	
                    }
                }
            } 
            for (i=0; i<=3; ++i) fclose(fp[i]);    
        } /* END: from_scratch != 0 */
    } /* END: N1==N2 */
}


double integrand ( double *mu_c )
/* Krautgaertner counterterm, analytically integrated over <theta> */
/* (only used if <old_CT>!=0) */
{
        double e1,e1p,e11,f,g,h,kl,a,b,c,d,x_cc,k_cc,mu_cc;
        double xp1,xp2,xp3,xm1,xm2,xm3,abk;
        double q1,q2,q3,q4,q5;

        mu_cc = *mu_c;

        x_cc = (1.0 + mu[jj1]*theta[jj2]/sqrt(m*m + mu[jj1]*mu[jj1]))/2.0;
        k_cc = mu[jj1]*sqrt(1.0 - theta[jj2]*theta[jj2]);

        e1  = sqrt(m*m + mu[jj1]*mu[jj1]); 
        e1p = sqrt(m*m + mu_cc*mu_cc);
 
        kl = e1p/e1 +e1/e1p;

        a = mu[jj1]*mu[jj1]*mu_cc*mu_cc*(4.0 +theta[jj2]*theta[jj2]*
            (e1p - e1)*(e1p-e1)/e1/e1p/e1/e1p);

        b = -2.0*(mu[jj1]*mu[jj1] + mu_cc*mu_cc)*mu[jj1]*mu_cc*theta[jj2]*kl;

        c =  (mu[jj1]*mu[jj1] + mu_cc*mu_cc)*(mu[jj1]*mu[jj1] + mu_cc*mu_cc)
             -4.0*mu[jj1]*mu[jj1]*mu_cc*mu_cc*(1.0-theta[jj2]*theta[jj2]);

        d = m*m/(1.0 - x_cc)/x_cc/e1p + 2.0*mu_cc*mu_cc/e1p
            + k_cc*k_cc/2.0/e1p/x_cc/(1.0 - x_cc)
            + 0.5/(1.0 - x_cc)/x_cc/e1p*(mu[jj1]*mu[jj1] + mu_cc*mu_cc);

        e11 = 2.0*m*m/(1.0 - x_cc)/x_cc/e1p/e1p*mu_cc*(x_cc - 0.5)
              1.0/x_cc/(1.0 - x_cc)/e1p*
              (-mu[jj1]*mu_cc*theta[jj2]*kl*0.5 + 
              (mu[jj1]*mu[jj1] + mu_cc*mu_cc)*mu_cc/e1p*(x_cc - 0.5));

        f = -2.0*mu_cc*mu_cc/e1p - 0.5*mu_cc*mu_cc/x_cc/(1.0 - x_cc)/
            e1p/e1p/e1p*k_cc*k_cc - 1.0/e1p/e1p/x_cc/(1.0 - x_cc)*
            mu_cc*(x_cc - 0.5)*mu[jj1]*mu_cc*theta[jj2]*kl;

        h = -0.5/x_cc/(1.0 - x_cc)/e1p;

        xp1 = a + b + c;
        xp2 = (mu_cc*mu[jj1]*theta[jj2]*kl - (mu[jj1]*mu[jj1] + mu_cc*mu_cc))*
              (mu_cc*mu[jj1]*theta[jj2]*kl - (mu[jj1]*mu[jj1] + mu_cc*mu_cc));
        xp3 = fabs(mu_cc*mu[jj1]*theta[jj2]*kl-(mu[jj1]*mu[jj1] + mu_cc*mu_cc));

        xm1 = a - b + c;
        xm2 = (mu_cc*mu[jj1]*theta[jj2]*kl + (mu[jj1]*mu[jj1] + mu_cc*mu_cc))*
              (mu_cc*mu[jj1]*theta[jj2]*kl + (mu[jj1]*mu[jj1] + mu_cc*mu_cc));
        xm3 = fabs(mu_cc*mu[jj1]*theta[jj2]*kl + (mu[jj1]*mu[jj1] 
              + mu_cc*mu_cc));

        if (  (2.0*sqrt(a)*xm3 - 2.0*a + b == 0)
            ||(2.0*sqrt(a)*xp3 + 2.0*a + b == 0)) return 0;
        else
        {		
          abk = 1.0/sqrt(a)*log(fabs((2.0*sqrt(a)*xp3 + 2.0*a + b)
           			    /(2.0*sqrt(a)*xm3 - 2.0*a + b)));
          return -1.0/PI*
                 (d*abk*mu_cc*mu_cc + e11*mu_cc*mu_cc/a
                 *(sqrt(xp2) - sqrt(xm2) - 0.5*b*abk)
                 +f*((mu_cc*mu_cc*0.5/a - 0.75*b*mu_cc*mu_cc/a/a)*sqrt(xp2)
                 -(-mu_cc*mu_cc*0.5/a - 0.75*b*mu_cc*mu_cc/a/a)*sqrt(xm2)
                 +(3.0*b*b - 4.0*a*c)*abk/8.0*mu_cc*mu_cc/a/a) 
                 + 2.0 *h*mu_cc*mu_cc)
                 *(1.0+mu[jj1]*mu[jj1]*mu[jj1]/p_bohr/p_bohr/p_bohr/8.0)
                 /(1.0+mu_cc*mu_cc*mu_cc/p_bohr/p_bohr/p_bohr/8.0);
        }
}	
\end{verbatim}

\section*{Genetix}

\begin{verbatim}
/************************************************************************/
/*                                                                      */
/*  program    :     G E N E T I X                                      */
/*                                                                      */
/*  class      :     subroutine of MESONIX                              */
/*                                                                      */
/*  file name  :     "genetix.c"                                        */
/*                                                                      */
/*  purpose    :     GENETIX generates the Hamilton matrix of the       */
/*                   system. It uses the subroutine PHYSIX for          */
/*                   calculation of the single elements.                */
/*                                                                      */
/*----------------------------------------------------------------------*/
/*                                                                      */
/*  External functions:                                                 */
/*  -------------------                                                 */
/*                                                                      */
/*     physix        subroutine of 'mesonix', calculates one matrix     */
/*                   element                                            */
/*                                                                      */
/************************************************************************/


void genetix()
/* create 4 Hamiltonians, symmetries C,H */
{
        int jm,jt,sj1,sj2;                    /* loop variables for rows */
        int im,it,si1,si2;                    /* loop variables for columns */
        int j_parity,i_parity;                /* booleans for loop end */
        double element[17];                   /* matrix elements[1..16] */
        double T,Sqrt2;                       /* kinetic energy, 1/sqrt(2) */
        double factor_j1,factor_j2,factor_j3; /* factors for wavefunctions */
        double factor_i1,factor_i2,factor_i3;
        int jnum_1,jnum_2,jnum_3,jnum_4;      /* actual Hamiltonian row */
        int inum_1,inum_2,inum_3,inum_4;      /* actual Hamiltonian column */

        time(&time1);
        printf("-----------------------------> GENETIX \n");
		
        Sqrt2 = 1/sqrt(2.0);

        jnum_1 = jnum_2 = jnum_3 = jnum_4 = 0;

        sj1 = si1 = 1;    
   
        /* =====================>  rows of Hamiltonian <================= */

        for ( jm=1; jm <= N1; ++jm) /* loop for <mu> */
        {
          for ( jt=1; jt <= (N2+1)/2; ++jt) /* loop for <theta> */
          {	
            printf("jm/N1 =%2d/%2d  jt/N2 =%2d/%2d\n",jm,N1,jt,(N2+1)/2);
            for ( sj2 = -1; sj2 <= 1; sj2 += 2 ) /* loop for <spin> */
            {
              factor_j1 = factor_j2 = factor_j3 = 1;	

              if ( 2*jt-1 != N2 )
              { 	
                ++jnum_1;
                ++jnum_2;
                ++jnum_3;
                ++jnum_4;

                j_parity = 0;         /* not end of loop */
              }
              else
              {
                if (sj1 == sj2)
                {
                       ++jnum_1;
                       ++jnum_2;
                       factor_j1 = factor_j2 = Sqrt2;	
                       j_parity = 1; 	 /* end of loop && s1=s2 */
                }
                else
                {
                       ++jnum_2;
                       ++jnum_3;
                       factor_j2 = factor_j3 = Sqrt2;	
                       j_parity = -1;  /* end of loop && s1!=s2 */	
                }
              }

              /* ===============>  columns of Hamiltonian  <=============== */
	      
              inum_1 = inum_2 = inum_3 = inum_4 = 0;
	
              for ( im=1; im <= N1; ++im) /* loop for <mu> */
              {
                for ( it=1; it <= (N2+1)/2; ++it) /* loop for <theta> */
                {
                  for ( si2 = -1; si2 <= 1; si2 += 2 ) /* loop for <spin> */
                  {
                      factor_i1 = factor_i2 = factor_i3 = 1;
         	
                      if ( (2*it-1) != N2 )
                      { 	
                        ++inum_1;
                        ++inum_2;
                        ++inum_3;
                        ++inum_4;
		
                        i_parity = 0;         /* not end of loop */	     
                      }
                      else
                      {
                        if (si1 == si2)
                        {
                               ++inum_1;
                               ++inum_2;
                               factor_i1 = factor_i2 = Sqrt2;	
                               i_parity = 1;	/* end of loop && s1=s2 */
                        }
                        else
                        {
                               ++inum_2;
                               ++inum_3;
                               factor_i2 = factor_i3 = Sqrt2;	
                               i_parity = -1;	 /* end of loop && s1!=s2 */
                        }
                      }

                    element[1] = physix(jm,jt,sj1,sj2,im,it,si1,si2);
                    element[2] = physix(jm,jt,sj1,sj2,im,it,-si1,-si2);
                    element[3] = physix(jm,jt,sj1,sj2,im,N2+1-it,si2,si1);
                    element[4] = physix(jm,jt,sj1,sj2,im,N2+1-it,-si2,-si1);

                    element[5] = physix(jm,jt,-sj1,-sj2,im,it,si1,si2);
                    element[6] = physix(jm,jt,-sj1,-sj2,im,it,-si1,-si2);
                    element[7] = physix(jm,jt,-sj1,-sj2,im,N2+1-it,si2,si1);
                    element[8] = physix(jm,jt,-sj1,-sj2,im,N2+1-it,-si2,-si1);

                    element[9] = physix(jm,N2+1-jt,sj2,sj1,im,it,si1,si2);
                    element[10] = physix(jm,N2+1-jt,sj2,sj1,im,it,-si1,-si2);
                    element[11] = physix(jm,N2+1-jt,sj2,sj1,
                                         im,N2+1-it,si2,si1);
                    element[12] = physix(jm,N2+1-jt,sj2,sj1,
                                         im,N2+1-it,-si2,-si1);
		    
                    element[13] = physix(jm,N2+1-jt,-sj2,-sj1,im,it,si1,si2);
                    element[14] = physix(jm,N2+1-jt,-sj2,-sj1,
                                         im,it,-si1,-si2);
                    element[15] = physix(jm,N2+1-jt,-sj2,-sj1,
                                         im,N2+1-it,si2,si1);
                    element[16] = physix(jm,N2+1-jt,-sj2,-sj1,
                                         im,N2+1-it,-si2,-si1); 

                    /* ============>  take care of phase <=============	*/

                    if (abs(si1+si2-J_z) == 2) 
                    {
                        element[2] = -element[2];
                        element[3] = -element[3];
                        element[14] = -element[14];
                        element[15] = -element[15];
                    }
                    if (abs(sj1+sj2-J_z) == 2) 
                    {
                        element[5] = -element[5];
                        element[8] = -element[8];
                        element[9] = -element[9];
                        element[12] = -element[12];
                    }
                        if ((abs(si1+si2-sj1-sj2) == 2) || 
                        (abs(sj1+sj2-sj1-sj2) == 3)) 
                    {    
                        element[6] = -element[6];
                        element[7] = -element[7];
                        element[10] = -element[10];
                        element[11] = -element[11];
                    }

                    /* Kinetic energy is diagonal */

                    T = 4*(mu[im]*mu[im] + m*m);

                    if ( (j_parity >= 0) && (i_parity >=0)) 
                    /* CH = ++ */
                    {	 		
                      mat_1[IndexMat[0]] = ALPHA/4*factor_j1*factor_i1* 
                       ( 
                          element[1] + element[2] - element[3] - element[4]
                        + element[5] + element[6] - element[7] - element[8]
                        - element[9] - element[10] + element[11] + element[12]
                        - element[13] - element[14] + element[15] + element[16]
                       );
                      if ( jnum_1 == inum_1 ) mat_1[IndexMat[0]] += T;
                      ++IndexMat[0];
                    }		     
	 		
                    mat_2[IndexMat[1]] = ALPHA/4*factor_j2*factor_i2*
                    /* CH = +- */
                       ( 
                          element[1] - element[2] - element[3] + element[4]
                        - element[5] + element[6] + element[7] - element[8]
                        - element[9] + element[10] + element[11] - element[12]
                        + element[13] - element[14] - element[15] + element[16]
                       );
                    if ( jnum_2 == inum_2 ) mat_2[IndexMat[1]] += T;
                    ++IndexMat[1];
		    
                    if ( (j_parity <= 0) && (i_parity <=0))
                    /* CH = -+ */
                    {	 		
                     mat_3[IndexMat[2]] = ALPHA/4*factor_j3*factor_i3*
                       (
                          element[1] + element[2] + element[3] + element[4]
                        + element[5] + element[6] + element[7] + element[8]
                        + element[9] + element[10] + element[11] + element[12]
                        + element[13] + element[14] + element[15] + element[16]
                       );
                      if ( jnum_3 == inum_3 ) mat_3[IndexMat[2]] += T;
                      ++IndexMat[2];
                    }		    		     
                    if ( (j_parity == 0) && (i_parity ==0))
                    /* CH = -- */
                    {	 		
                     mat_4[IndexMat[3]] = ALPHA/4*
                       ( 
                          element[1] - element[2] + element[3] - element[4]
                        - element[5] + element[6] - element[7] + element[8]
                        + element[9] - element[10] + element[11] - element[12]
                        - element[13] + element[14] - element[15] + element[16]
                       );
                      if ( jnum_4 == inum_4 ) mat_4[IndexMat[3]] += T;
                      ++IndexMat[3];
                    }		     	                    
                  }   /* three column loops: <spin>,<theta>,<mu> */ 
                }
              }       
            }	/* three row loops: <spin>,<theta>,<mu> */
          }      
        }   
        time(&time2);
        fprintf(f_log,"         ...matrix: %d sec\n",time2-time1);
        return;
}

\end{verbatim}
\newpage

\section*{Asymmetrix}

\begin{verbatim}

/************************************************************************/
/*                                                                      */
/*  program    :     A S Y M M E T R I X                                */
/*                                                                      */
/*  class      :     subroutine of MESONIX                              */
/*                                                                      */
/*  file name  :     "asymmetrix.c"                                     */
/*                                                                      */
/*  purpose    :     ASYMMETRIX generates the Hamilton matrix of the    */
/*                   system, but WITHOUT any use of symmetries. It is   */
/*                   as well a test case for the KRAUT code, as a first */
/*                   step to the J_z<>0 case, which has OTHER symmetry  */
/*                   properties. It uses the subroutine PHYSIX for      */
/*                   calculation of the single elements.                */
/*                                                                      */
/*----------------------------------------------------------------------*/
/*                                                                      */
/*  External functions:                                                 */
/*  -------------------                                                 */
/*                                                                      */
/*     physix        subroutine of 'mesonix', calculates one matrix     */
/*                   element                                            */
/*                                                                      */
/************************************************************************/

void asymmetrix()
/* create Hamiltonian, no symmetries */
{
        int jm,jt,sj1,sj2; 	/* loop variables for rows */
        int im,it,si1,si2; 	/* loop variables for columns */
        double T;		/* kinetic energy */
        int j_num,i_num;        /* actual Hamiltonian row/column */

        printf("-----------------------------> ASYMMETRIX\n");
        j_num = 0;    
        IndexMatrix = 0; 

        /* =====================>  rows of Hamiltonian <================= */

        for ( jm=1; jm <= N1; ++jm) /* loop for <mu> */
        {
          for ( jt=1; jt <= N2; ++jt) /* loop for <theta> */
          {	
            printf("jm/N1 =%2d /%2d, jt/N2=%2d /%2d\n",jm,N1,jt,N2);

            for ( sj1 = -1; sj1 <= 1; sj1 += 2 ) /* loop for <spin1> */
            {
              for ( sj2 = -1; sj2 <= 1; sj2 += 2 ) /* loop for <spin2> */
              {
                i_num = 0;
                ++j_num;

        	/* ===============> columns  of Hamiltonian <=============== */

                for ( im=1; im <= N1; ++im) /* loop for <mu> */
                {
                  for ( it=1; it <= N2; ++it) /* loop for <theta> */
                  {
                    for ( si1 = -1; si1 <= 1; si1 += 2 ) /* loop for <spin1> */
                    {
       	              for ( si2 = -1; si2 <= 1; si2 += 2 ) /* loop for <spin2> */
                      {
                         ++i_num;
                         ma[IndexMatrix] = ALPHA*physix(jm,jt,-sj1,-sj2,im,it,-si1,-si2);
                         if ( j_num == i_num ) ma[IndexMatrix] += 4*(mu[im]*mu[im] + m*m);
                         ++IndexMatrix; /* matrix element counter */

                      } /* four column loops: <spin2>,<spin1>,<theta>,<mu> */
                    } 
                  } 
                }
              }	/* four row loops: <spin2>,<spin1>,<theta>,<mu> */
            }
          }      
        }   
        return;
}
\end{verbatim}

\newpage

\section*{Physix}

\begin{verbatim}
/************************************************************************/
/*                                                                      */
/*  program    :     P H Y S I X                                        */
/*                                                                      */
/*  class      :     sub-subroutine of MESONIX, subroutine of GENETIX   */
/*                                                                      */
/*  file name  :     "physix.c"                                         */
/*                                                                      */
/*  purpose    :     PHYSIX calculates one single element of the        */
/*                   Hamilton matrix of the system.                     */
/*                   Takes care of singularities with Coulomb trick.    */
/*                                                                      */
/************************************************************************/

double Int(int n)  
/* the formula for the integral over cos(nx)/(a+b cos(x))*/
{
        double vz;

        vz=1.0;
        if (n%2==0) vz = -1.0;
        return vz*pow(A,1.0-n)*pow(B/k[1]/k[2],1.0*n);
}

double G1(int j, int i, int n)
/* diagonal matrix element, parallel spins */
{ 
        double help;

 
        if ( (k[i]==0) && (k[j]==0) ) help = 0.0;
        else
        {
                help = m*m*(1.0/x[i]/x[j]+1/(1-x[j])/(1-x[i]))*Int(abs(1-n)) 
                       + k[i]*k[j]/x[j]/x[i]/(1-x[j])/(1-x[i])*Int(abs(n));
        }
        return help;
}

double G2(int j, int i, int n)          
/* diagonal matrix element, anti-parallel spins */
{ 
        double help;
 
        help = 1/x[i]/x[j] + 1.0/(1-x[j])/(1-x[i]); 
        help = (m*m*help + k[j]*k[j]/x[j]/(1-x[j])
               + k[i]*k[i]/x[i]/(1-x[i]))*Int(abs(n))+k[i]*k[j]*
               (Int(abs(1-n))/x[i]/x[j]+Int(abs(1+n))/(1-x[i])/(1-x[j]));
        return help;	
}

double G3(int j, int i, int n)
/* off-diagonal matrix element */
{
        double help;

        if ( k[j]==0 ) 	help = 0.0;
        else help = -m/x[i]/x[j]*
                    (k[i]*Int(abs(1-n))-k[j]*(1-x[i])/(1-x[j])*Int(abs(n)));
        return help;
}

double G3_star(int j, int i, int n)
/* off-diagonal matrix element */
{
        double help;

        if ( k[j]==0 ) 	help = 0.0;
        else help = m/(1-x[i])/(1-x[j])*
                    (k[i]*Int(abs(1-n))-k[j]*x[i]/x[j]*Int(abs(n)));
        return help;
}
	
double G4(int j, int i, int n)
/* off-diagonal matrix element */
{
        double help;

        help = -m*m*(x[i]-x[j])*(x[i]-x[j])/(1-x[i])/(1-x[j])
               /x[i]/x[j]*Int(abs(n));
        return help;
}

/* --------------> Functions for dynamical annihilation graph <---------------*/


double F1(int j, int i, int n)
/* dynamical annihilation graph: parallel spins, diagonal */
{ 
        double help,omega;

	help=0.0;
	if (abs(n)==1)
	{
	        omega=((m*m+k[i]*k[i])/x[i]/(1.0-x[i])
	      	      +(m*m+k[j]*k[j])/x[j]/(1.0-x[j]))/2.0;
        	help=2.0*m*m/omega*(1.0/x[i]+1/(1-x[i]))*(1/x[j]+1.0/(1-x[j]));
	}
        return help;
}

double F2(int j, int i, int n)
/* dynamical annihilation graph: antiparallel spins, diagonal, I */
{ 
        double help,omega;

	help = 0.0;
        if (n==0) help=4.0;  /* seagull graph */
	if (abs(n)==1)	     /* dynamic graph */
	{
	        omega=((m*m+k[i]*k[i])/x[i]/(1.0-x[i])
		      +(m*m+k[j]*k[j])/x[j]/(1.0-x[j]))/2.0;
        	help = 2.0/omega*k[i]*k[j]/x[i]/x[j];
	}
        return help;
}

double F2_star(int j, int i, int n)
/* dynamical annihilation graph: antiparallel spins, diagonal, II */
{ 
        double help,omega;

	help = 0.0;
        if (n==0) help=4.0;  /* seagull graph */
	if (abs(n)==1)	     /* dynamic graph */
	{
	        omega=((m*m+k[i]*k[i])/x[i]/(1.0-x[i])
		      +(m*m+k[j]*k[j])/x[j]/(1.0-x[j]))/2.0;
        	help = 2.0/omega*k[i]*k[j]/(1.0-x[i])/(1.0-x[j]);
	}
        return help;
}

double F3(int j, int i, int n)
/* dynamical annihilation graph: spin non-diagonal */
{ 
        double help,omega;

	help=0.0;
	if (abs(n)==1)	    /* dynamic graph */	
	{
	        omega=((m*m+k[i]*k[i])/x[i]/(1.0-x[i])
		      +(m*m+k[j]*k[j])/x[j]/(1.0-x[j]))/2.0;
        	help = 2.0*m/omega*k[i]/(1-x[i])*(1/x[j]+1.0/(1-x[j]));
	}
        return help;
}

double F3_star(int j, int i, int n)
/* dynamical annihilation graph: spin non-diagonal */
{ 
        double help,omega;

	help=0.0;
	if (abs(n)==1)	    /* dynamic graph */	
	{
        	omega=((m*m+k[i]*k[i])/x[i]/(1.0-x[i])
	     	      +(m*m+k[j]*k[j])/x[j]/(1.0-x[j]))/2.0;
        	help = -2.0*m/omega*k[i]/x[i]*(1/x[j]+1.0/(1-x[j]));
	}
        return help;
}

double F4(int j, int i, int n)
/* dynamical annihilation graph: all spins antiparallel, non-diagonal */
{ 
        double help,omega;

	help = 0.0;
        if (n==0) help=4.0;  /* seagull graph */
	if (abs(n)==1) 	    /* dynamic graph */
	{
	        omega=((m*m+k[i]*k[i])/x[i]/(1.0-x[i])+
		       (m*m+k[j]*k[j])/x[j]/(1.0-x[j]))/2.0;
        	help = -2.0/omega*k[i]*k[j]/x[i]/(1-x[j]);
	}
        return help;
}
double annihilation_spintable(int sj1, int sj2, int si1, int si2)
{
        double result;

        /* Evaluation of SPIN-TABLE of dynamic annihilation graph */
        /*                                     J_z == +/-1        */
        /* (row-wise from top to bottom,left to right) */   
        /* (final is row index, initial is column index) */   
	
        result= 0.0;

        if (J_z==-1)                        /* 'transpose' matrix */
        {
                        sj1 = -sj1;
                        sj2 = -sj2;
                        si1 = -si1;
                        si2 = -si2;
        }
        if (((sj1==-1) && (sj2==-1))||((si1==-1) && (si2==-1))) return 0.0;
        else
        {
                     if (sj1 > 0)          /* first and second row */
                     {
                        if (sj2 > 0)       /* first row */
                        {
                           if (si1 > 0)
                           {
                                if (si2 > 0) result = F1(2,1,J_z);
                                else         result = F3(2,1,J_z); 
                           }
                           else              result = F3_star(2,1,J_z);     
                        }
                        else               /* second row */
                        {
                           if (si1 > 0)
                           {
                                if (si2 > 0) result = F3_star(1,2,J_z); /* perm. */ 
                                else         result = F2_star(2,1,J_z);

                           }
                           else              result = F4(2,1,J_z);
                        }
                     }
                     else                  /* third row */
                     {
                        if (si1 > 0)
                        {
                            if (si2 > 0) result = F3_star(1,2,J_z); /* permuted */ 
                            else         result = F4(1,2,J_z);      /* permuted */
                        }
                        else             result = F2(2,1,J_z);
                     }
        }
        return result;
}

double general_spintable(int sj1, int sj2, int si1, int si2)
/* spin table for J_z=n                          */ 
/* --------------------------------------------- */
/* (row-wise from top to bottom,left to right)   */   
/* (final is row index, initial is column index) */   
{
        double result;

        if (sj1 > 0)          /* first and second row */
        {
           if (sj2 > 0)       /* first row */
           {
                if (si1 > 0)
                {
                         if (si2 > 0) result = G1(2,1,J_z);
                         else         result = G3_star(2,1,J_z); 
                }
                else
                {
                         if (si2 >0)  result = G3(2,1,J_z);
                         else         result = 0.0;
                }
           }
           else               /* second row */
           {
                if (si1 > 0)
                {
                         if (si2 > 0) result = G3_star(1,2,J_z); /* permuted */ 
                         else         result = G2(2,1,J_z);
        	}
                else
                {
                         if (si2 >0)  result = G4(2,1,J_z);
      	                 else         result = -G3(1,2,-J_z);  /* permuted */
       	        }
           }
        }
        else                          /* third and forth row */
        {
                   if (sj2 > 0)       /* third row */
                   {
                      if (si1 > 0)
                      {
                         if (si2 > 0) result = G3(1,2,J_z);/* permuted */ 
     	                 else	      result = G4(2,1,J_z);
                      }
                      else
                      {
                         if (si2 > 0) result = G2(2,1,-J_z);
                         else	      result = -G3_star(1,2,-J_z);/* permuted */
                      }
                   }
                   else               /* forth row */
                   {
                      if (si1 > 0)
                      {
                         if (si2 > 0) result = 0.0;
                         else         result = -G3(2,1,-J_z); 
                      }
                      else
                      {
                         if (si2 >0)  result = -G3_star(2,1,-J_z); 
                         else         result = G1(2,1,-J_z);
                      }
                   }
        }
        if (Anni!=0) result += annihilation_spintable(sj1,sj2,si1,si2);
        return result;
}


int spin2number(int s1, int s2) 
/* transforms (uu,ud,du,dd) --> (0,1,2,3) */
{
        return 2-s1-s2-(1-s2)/2;
}


double physix(int jm,int jt,int sj1,int sj2,int im,int it,int si1,int si2)
/* main routine: calculates one Hamiltonian matrix element */	
/*---------------------------------------------------------*/
/* parameters:  (j=final, i=initial)                       */
/*	        jm,jt   = <mu>,<theta>                     */
/*              sj1,sj2 = <spin electron>,<spin positron>  */
{ 
        double result,jacobian,epsilon;

        V_seag = -2; /* seagull graph */
        epsilon = 1e-8;

        /* ========>  transformation: (mu,theta)-->(x,k)  <============ */
        /*                                    j --> 2                   */
        /*                                    i --> 1                   */ 

        x[2] = (1.0 + mu[jm]*theta[jt]/sqrt(m*m + mu[jm]*mu[jm]))/2.0;
        k[2] = mu[jm]*sqrt(1.0-theta[jt]*theta[jt]);

        /* ======================>  Coulomb trick  <=================== */

        if ((jm==im) && (jt==it) && (sj1==si1) && (sj2==si2))
        {	        			
                result = 0.0; /* default */
                result = Coulomb_discrete(jm,sj1,sj2); /* discrete counterterm */
                if (old_CT!=0) result += old_continous(jm,jt);
                else result += CT[spin2number(sj1,sj2)][jm-1][jt-1];
                                                       /* continuous counterterm*/
        }
       				
        /* =====================>  No Coulomb trick  <================== */

        else 
        {   
                x[1] = (1.0 + mu[im]*theta[it]/sqrt(m*m + mu[im]*mu[im]))/2.0;
                k[1] = mu[im]*sqrt(1.0-theta[it]*theta[it]);

                if ( jt==it && jm==im ) return 0.0;

                /* Note: aa = -a(Kraut), 09/11/95  */
                /* -----                           */
                /* -a,A,B as defined in KPW, etc.  */

                aa = (x[1]-x[2])*(x[1]-x[2])*m*m/2.0*(1/(1-x[2])/(1-x[1])
                     +1/x[1]/x[2]) + k[1]*k[1] + k[2]*k[2]
                     + (x[1]-x[2])/2*(k[1]*k[1]*(1/(1-x[1])-1/x[1]) -
                                      k[2]*k[2]*(1/(1-x[2])-1/x[2]));
                A  = 1.0/sqrt( aa*aa - 4*k[1]*k[1]*k[2]*k[2] );
                B  = (1.0-aa*A)/2.0;
	
                jacobian = sqrt(wmu[jm]*wmu[im]*wtheta[jt]*wtheta[it])
                           *mu[jm]*mu[im]*sqrt( 4*x[1]*(1-x[1])*x[2]*(1-x[2])
                           /sqrt(m*m + mu[im]*mu[im])/sqrt(m*m + mu[jm]*mu[jm]));
                result = jacobian/PI*general_spintable(sj1,sj2,si1,si2);
        }
        return result;
}
\end{verbatim}

\section*{Arithmetix}

\begin{verbatim}
/************************************************************************/
/*                                                                      */
/*  program    :     A R I T H M E T I X                                */
/*                                                                      */
/*  class      :     subroutine of MESONIX                              */
/*                                                                      */
/*  file name  :     "arithmetix.c"                                     */
/*                                                                      */
/*  purpose    :     ARITHMETIX diagonalizes the four block matrizes    */
/*                   of the Hamiltonian and sorts the Eigenvalues from  */
/*                   small to big ones.                                 */
/*                   The eigenfunctions are written in a file, if       */
/*                   <print_EF>!=0                                      */
/*                                                                      */
/*----------------------------------------------------------------------*/
/*                                                                      */
/*  External functions:                                                 */
/*  -------------------                                                 */
/*                                                                      */
/*     m01caf        NAGLib routine, sorts vector                       */
/*                                                                      */
/************************************************************************/

extern double m01caf_( double *r, int *m1, int *m2, char *order, 
        	       int *IFAIL);
/* sorting routine from NAGLib */

double normalisation(int i, int j) 
/* the factor to resubstitute Phi --> Psi */
{
        double aij;
                         
        aij = mu[i]*mu[i]*(m*m+mu[i]*mu[i]*(1.0-theta[j]))
              /2.0/sqrt(m*m+mu[i]*mu[i]);
        aij = 1.0/sqrt(wtheta[j]*wmu[i]*aij);
        if (j<(N2+1)/2) aij *= 1.0/sqrt(2.0);
        return aij;
}

double asym_normalisation(int i, int j) 
/* the factor to resubstitute Phi --> Psi */
{
        double aij;
                         
        aij = mu[i]*mu[i]*(m*m+mu[i]*mu[i]*(1.0-theta[j]))
              /2.0/sqrt(m*m+mu[i]*mu[i]);
        aij = 1.0/sqrt(wtheta[j]*wmu[i]*aij);
        aij *= 1.0/sqrt(2.0);
        return aij;
}

void store_asy_eigenfunctions()
/* write eigenfunctions of asymmetric Hamiltonian in file 'EFunc_asy_<l>.dat' */
{
        int i,j,l,n;
        char *name_EF[4];
        FILE *fp[4];

        for (l=0; l<=3; l++) name_EF[l]=(char *)malloc(120);
        for (n=0; n<=5; n++) /* <n>th eigenvalue */
        {
            for (l=0; l<=3; l++)
            {
                sprintf(name_EF[l],"%sEigenFunctions/EF%1d_J%1dasy_%1d.dat",
                                                    default_directory,n,J_z,l);	
                printf("%s\n",name_EF[l]);
                fp[l]=fopen(name_EF[l],"w");
            }
            for (i=1; i<=N1; i++)
               	for (j=1; j<=N2; j++) 
                {
                  for (l=0; l<=3; l++)
                  {
                      x[1] = (1.0 + mu[i]*theta[j]/sqrt(m*m + mu[i]*mu[i]))/2.0;
                      k[1] = mu[i]*sqrt(1.0-theta[j]*theta[j]);
                      fprintf(fp[l],"%18.12f  %18.12f  %18.12f\n",x[1],k[1],
                                        v[n*4*N1*N2+(i-1)*N2*4+4*(j-1)+l]*
                                        asym_normalisation(i,j));
                  }
                } 
            for (l=0; l<=3; l++) fclose(fp[l]);
        } /* loop n */
}


void store_eigenfunctions()
/* write eigenfunctions in files 'EFunc_<n>.dat' */
{
        int i,j;
        char name_EF[120];
        FILE *fp1;

        /* CH = ++ */
        sprintf(name_EF,"%sEigenFunctions/EFunc_1.dat",default_directory);
        fp1=fopen(name_EF,"w");
        for (i=1; i<=N1; i++)
                for (j=1; j<=(N2+1)/2; j++) 
                {
                        fprintf(fp1,"%18.12f\n",v_1[(i-1)*N2+2*j-2]*normalisation(i,j));
                        if (j!=(N2+1)/2) fprintf(fp1,"%18.12f\n",v_1[(i-1)*N2+2*j-1]*normalisation(i,j));
                        else fprintf(fp1,"\n"); /* dummy line */
                } 
                fclose(fp1);

        /* CH = +- */
        sprintf(name_EF,"%sEigenFunctions/EFunc_2.dat",default_directory);
        fp1=fopen(name_EF,"w");
        for (i=1; i<=N1; i++)
                for (j=1; j<=(N2+1)/2; j++) 
                {
                    fprintf(fp1,"%18.12f\n",v_2[(i-1)*(N2+1)+2*j-2]*normalisation(i,j));
                    fprintf(fp1,"%18.12f\n",v_2[(i-1)*(N2+1)+2*j-1]*normalisation(i,j));
                } 
        fclose(fp1);

        /* CH = -+ */
        sprintf(name_EF,"%sEigenFunctions/EFunc_3.dat",default_directory);
        fp1=fopen(name_EF,"w");
        for (i=1; i<=N1; i++)
                for (j=1; j<=(N2+1)/2; j++) 
                {
                    fprintf(fp1,"%18.12f\n",v_3[(i-1)*N2+2*j-2]*normalisation(i,j));
                    if (j!=(N2+1)/2)
                       fprintf(fp1,"%18.12f\n",v_3[(i-1)*N2+2*j-1]*normalisation(i,j));
        	    else fprintf(fp1,"\n"); /* dummy line */
                } 
        fclose(fp1);

        /* CH = -- */
        sprintf(name_EF,"%sEigenFunctions/EFunc_4.dat",default_directory);
        fp1=fopen(name_EF,"w");
        for (i=1; i<=N1; i++)
        {
                for (j=1; j<(N2+1)/2; j++) 
                {
                    fprintf(fp1,"%18.12f\n",v_4[(i-1)*(N2-1)+2*j-2]*normalisation(i,j));
                    fprintf(fp1,"%18.12f\n",v_4[(i-1)*(N2-1)+2*j-1]*normalisation(i,j));
                }
                fprintf(fp1,"\n"); /* dummy line */
                fprintf(fp1,"\n"); /* dummy line */
        }		
        fclose(fp1);

        /* ================> write (x,k) in file 'x_k.dat' <======== */

        sprintf(name_EF,"%sEigenFunctions/x_k.dat",default_directory);
        fp1 = fopen(name_EF,"w");       
        for (i=1; i<=N1; ++i) 
                for (j=1; j<=(N2+1)/2; ++j) 
                {
                      x[1] = (1.0 + mu[i]*theta[j]/sqrt(m*m + mu[i]*mu[i]))/2.0;
                      k[1] = mu[i]*sqrt(1.0-theta[j]*theta[j]);
                      fprintf(fp1,"%22.16f      %22.16f\n",x[1],k[1]);
                      fprintf(fp1,"%22.16f      %22.16f\n",x[1],k[1]);
                }
        fclose(fp1);
 
        /* mu values */
        sprintf(name_EF,"%sEigenFunctions/abscissae.dat",default_directory);
        fp1 = fopen(name_EF,"w");       
        for (i=1; i<=N1; i++) fprintf(fp1,"%22.16f\n",mu[i]/p_bohr);
        fclose(fp1);
}

void arithmetix(int dim_1, int dim_2, int dim_3, int dim_4)
/* main function: diagonalize Hamiltonians */
{
        int i;					/* loop variable */
        int IA_nag,N_nag,IV_nag,IFAIL_nag;	/* variables for NAGLib */
        int m1,m2;                              /* variables for m01caf */
        FILE *ffp;

        time(&time1);
        printf("\n-----------------------------> ARITHMETIX\n");
	      
        if ((test&(4096+16384)) == 0) 
        /* J_z=0 case, 4 bloc matrices */
        {
                /* ==========> diagonalize the 4 bloc matrices <============ */

                IA_nag = IV_nag = N_nag = dim_1;IFAIL_nag=-1;
                f02abf_(mat_1,&IA_nag,&N_nag,r_1,v_1,&IV_nag,e_1, &IFAIL_nag);

                IA_nag = IV_nag = N_nag = dim_2;IFAIL_nag=-1;
                f02abf_(mat_2,&IA_nag,&N_nag,r_2,v_2,&IV_nag,e_2, &IFAIL_nag);

                IA_nag = IV_nag = N_nag = dim_3;IFAIL_nag=-1;
                f02abf_(mat_3,&IA_nag,&N_nag,r_3,v_3,&IV_nag,e_3, &IFAIL_nag);

                IA_nag = IV_nag = N_nag = dim_4;IFAIL_nag=-1;
                f02abf_(mat_4,&IA_nag,&N_nag,r_4,v_4,&IV_nag,e_4, &IFAIL_nag);

                ffp=fopen("EVs_1.dat","w");
                fprintf(ffp,"C=+,H=+\n");
                for (i=1; i<=dim_1; ++i) fprintf(ffp,"%18.12f\n",r_1[i-1]);
                fclose(ffp);
                ffp=fopen("EVs_2.dat","w");
                fprintf(ffp,"C=+,H=-\n");
                for (i=1; i<=dim_2; ++i) fprintf(ffp,"%18.12f\n",r_2[i-1]);
                fclose(ffp);
                ffp=fopen("EVs_3.dat","w");
                fprintf(ffp,"C=-,H=+\n");
                for (i=1; i<=dim_3; ++i) fprintf(ffp,"%18.12f\n",r_3[i-1]);
                fclose(ffp);
                ffp=fopen("EVs_4.dat","w");
                fprintf(ffp,"C=-,H=-\n");
                for (i=1; i<=dim_4; ++i) fprintf(ffp,"%18.12f\n",r_4[i-1]);
                fclose(ffp);

                for (i=1; i<=dim_1+dim_2+dim_3+dim_4; ++i) 
                /* put all EVs together */
                {
                        if (i<=dim_1) r[i-1] = r_1[i-1];
                        else
                        {
                                if (i<=dim_1+dim_2) r[i-1] = r_2[i-1-dim_1];
                                else
                                {
                                        if (i<=dim_1+dim_2+dim_3) 
                                           r[i-1] = r_3[i-1-dim_1-dim_2];
                                        else r[i-1] = r_4[i-1-dim_1-dim_2-dim_3];
                                }
                        }
                }	
                if (print_EF > 0) store_eigenfunctions();

        } /* end of (test&4096==0) */
        else
        {
            if ((test&16384)==0) /* asymmetric Hamiltonian (1 bloc) */
            {
                printf("                                (asymmetric case)\n");
                IA_nag = IV_nag = N_nag = dim_1; IFAIL_nag = -1;
                f02abf_(ma,&IA_nag,&N_nag,r,v,&IV_nag,e, &IFAIL_nag);
                if (print_EF > 0) store_asy_eigenfunctions();
            }		
            else /* use C-symmetric Hamiltonian (2 blocs) */
            {
                printf("                                (symmetric case)\n");

                IA_nag = IV_nag = N_nag = dim_1; IFAIL_nag = -1;
                f02abf_(mat_1,&IA_nag,&N_nag,r_1,v_1,&IV_nag,e_1,&IFAIL_nag);

                IA_nag = IV_nag = N_nag = dim_2; IFAIL_nag = -1;
                f02abf_(mat_2,&IA_nag,&N_nag,r_2,v_2,&IV_nag,e_2,&IFAIL_nag);

                /* ============> write all eigenvalues in <r[i]> <========== */

                for (i=0; i<=dim_1-1; ++i) 
                {
                    r[i] = r_1[i];
                    printf("H_EW[%2d] = %12.8f\n",i,r_1[i-1]);
                }
                for (i=0; i<=dim_2-1; ++i)
                {
                    r[dim_1+i+1] = r_2[i];
                    printf("H_EW[%2d] = %12.8f\n",i,r_2[i-1]);
                }
            }
        }
        order = "A"; m1 = 1; m2 = dim_1+dim_2+dim_3+dim_4;
        m01caf_(r,&m1,&m2,order,&IFAIL_nag);     /* sort eigenvalues */

        time(&time2);
        fprintf(f_log,"         ...diagonalization: %d sec\n",time2-time1);
        return;
}
\end{verbatim}

\newpage
\section*{Publix}

\begin{verbatim}
/************************************************************************/
/*                                                                      */
/*  program    :     P U B L I X                                        */
/*                                                                      */
/*  class      :     subroutine of MESONIX                              */
/*                                                                      */
/*  file name  :     "publix.c"                                         */
/*                                                                      */
/*  purpose    :     PUBLIX is the output routine of MESONIX.           */
/*                                                                      */
/************************************************************************/

void publix()
/* write eigenvalues in file 'EVs_J<n>.dat' */
{
        int i;                          /* loop variable */
        double bs,bt,chf;               /* binding coefficients */
        FILE *fev;                      /* pointer onto file */

        printf("-----------------------------> PUBLIX\n");
        if ((test & 4) != 0) /* write EVs of 4 blocs in different files */
        {
                printf(">>>>> INFO : 2nd test bit set ON => eigenvalues in 'EV_<n>.dat'\n"); 
		
                fev = fopen("EV_1.dat","w");
                fprintf(fev,"Sector: CH = ++\n---------------\n\n",i,r_1[i]);
                for (i=0; i<=N1*N2-1; ++i) fprintf(fev,"%18.12f\n",r_1[i]);
                fclose(fev);

                fev = fopen("EV_2.dat","w");
                fprintf(fev,"Sector: CH = +-\n---------------\n\n",i,r_1[i]);
                for (i=0; i<=N1*N2+N1-1; ++i) fprintf(fev,"%18.12f\n",r_2[i]);
                fclose(fev);

                fev = fopen("EV_3.dat","w");
                fprintf(fev,"Sector: CH = -+\n---------------\n\n",i,r_1[i]);
                for (i=0; i<=N1*N2-1; ++i) fprintf(fev,"%18.12f\n",r_3[i]);
                fclose(fev);

                fev = fopen("EV_4.dat","w");
                fprintf(fev,"Sector: CH = --\n---------------\n\n",i,r_1[i]);
                for (i=0; i<=N1*N2-N1-1; ++i) fprintf(fev,"%18.12f\n",r_4[i]);
                fclose(fev);
        }
        fev = fopen(name_EV,"a"); /* open file for eigenvalues */
        i=-1;
        while ((++i<4*N1*N2)&&(i<number_EV)) fprintf(fev,"%18.12f\n",r[i]);
        for (i=4*N1*N2+1; i<=number_EV; ++i) fprintf(fev,"\n"); 
        /* fill with empty lines up to <number_EV> */ 
        fclose(fev);

        /* ==============>  calculate binding coefficients  <============== */

        bs  = 4.0/ALPHA/ALPHA*(2.0-sqrt(r[0])); 
        bt  = 4.0/ALPHA/ALPHA*(2.0-sqrt(r[1])); 
        chf = (sqrt(r[1])-sqrt(r[0]))/ALPHA/ALPHA/ALPHA/ALPHA;
        fprintf(f_log,"\n       B_s  = %12.8f\n",bs);
        fprintf(f_log,"       B_t  = %12.8f\n",bt);
        fprintf(f_log,"       C_hf = %12.8f\n",chf);
        return;
}
\end{verbatim}

}

\end{appendix}


\begin{thebibliography}{999}

\bibitem{tHooft} {\sc G.\ t'Hooft},``{\sl Renormalization of Massless Yang-Mills
		Fields}'', Nucl.\ Phys.\ {\bf B33} (1971) 173.
\bibitem{Pollitzer} {\sc H.D.\ Politzer}, ``{\sl Reliable Perturbative Results 
		for Strong Interactions? }'',
		Phys.\ Rev.\ Lett.\ {\bf 30} (1973) 1346;
		``{\sl Asymptotic Freedom: An Approach to Strong 
		Interactions}'',
		Phys.\ Rep.\ {\bf C14} (1974) 129.
\bibitem{Gross} {\sc D.J.\ Gross, F.\ Wilczek}, ``{\sl Ultraviolet Behavior of
		Non-Abelian Gauge Theories}'',
		Phys.\ Rev.\ Lett.\ {\bf 30} (1973) 1343.
\bibitem{Wilson} {\sc K.G.\ Wilson}, ``{\sl Confinement of Quarks}'',
		Phys.\ Rev.\ {\bf D10} (1974) 2445.
\bibitem{Yukawa} {\sc H.\ Yukawa}, 
	``{\sl On the Interaction of Elementary Particles}'',
	Proc.\ Phys.\ Math.\ Soc.\ Jap.\ {\bf 17} (1935) 48--57.
\bibitem{Schroedinger} {\sc E. Schr\"odinger}, ``{\sl Quantisierung als 
	        Eigenwertproblem (Erste Mitteilung.)}'', 
		Ann.Phys.(Leipzig) {\bf 79} (1926) 361-376;  
		Ann.Phys.(Leipzig) {\bf 79} (1926) 489-527;
		Ann.Phys.(Leipzig) {\bf 80} (1926) 437-490;
		Ann.Phys.(Leipzig) {\bf 81} (1926) 109-139;
		Ann.Phys.(Leipzig) {\bf 79} (1926) 734-756. 
\bibitem{Richardson} {\sc J.\ L.\ Richardson}, ``{\sl The heavy Quark 
	Potential and the $\Upsilon$, $J/\psi$ Systems}'', 
        Phys.\ Lett.\ {\bf 82B} (1979) 272.
\bibitem{Buchmueller} {\sc W.Buchm\"uller, S.-H.\ H.\ Tye}, ``{\sl Quarkonia
	and Quantum Chromodynamics}'', 
	Phys.\ Rev.\ {\bf D24} (1981) 132.
\bibitem{Isgur} {\sc S.\ Gottfrey, N.\ Isgur}, 
	``{\sl Mesons in a relativized quark 
	model with chromodynamics}'', 
	Phys.\ Rev.\ {\bf D32} (1985) 189-231.
\bibitem{Aubert} {\sc J.J.\ Aubert et al.},
	``{\sl Experimental Observation of a Heavy Particle J }'', 
	Phys.\ Rev.\ Lett.\  {\bf 33} (1974) 1404--1406.
\bibitem{Augustin} {\sc J.E.\ Augustin et al.},
	``{\sl Discovery of a Narrow Resonance in $e^+e^-$ Annihilation}'' 
	Phys.\ Rev.\ Lett.\  {\bf 33} (1974) 1406--1408.
\bibitem{Herb} {\sc S.W.\ Herb et al.},
	``{\sl Observation of a Dimuon Resonance at 9.5 GeV in 400-GeV 
	Proton-Nucleus Collisions}'',
	Phys.\ Rev.\ Lett.\  {\bf 39} (1977) 252--255.
\bibitem{Creutz} {\sc M.\ Creutz}, ``{\sl Quarks, Gluons and Lattices}'',
		Cambridge University Press, Cambridge 1983.
\bibitem{BetheSalpeter} {\sc E.E.\ Salpeter, H.A.\ Bethe}, ``{\sl A 
	Relativistic Equation for Bound-State Problems}'', 
	Phys.\ Rev.\ {\bf 84} (1951) 1232; {\sc E.E.\ Salpeter},
	``{\sl Mass Corrections to the Fine Structure of Hydrogen-Like Atoms}'',
	Phys.\ Rev.\ {\bf 87} (1952) 328--343;
	Phys.\ Rev.\ {\bf 84} (1951) 1226.
\bibitem{Weingarten} {\sc D.\ Weingarten}, 
	Nucl. Phys.\ (Proc.\ Supp.) {\bf B34} (1994) 29;
        {\sc C.T.H.\ Davies, A.J.\ Lindsey, K.\ Hornbostel, G.P.\ Lepage, 
	J.\ Shigemitsu, J.\ Sloan}, 
	``{\sl $B_c$ und $\Upsilon$ Spectra from Lattice NRQCD}'',
	{\tt hep-lat/9510052}, October 1995.
\bibitem{Gromes} {\sc D.\ Gromes}, 
 	``{\sl Construction of Bethe-Salpeter Wave Functions and 
	Applications in QCD}'',
	Z.\ Phys.\ {\bf C57} (1993) 631--638; 
	``{\sl More on Bethe Salpeter 
	Wave Functions for Quark-Antiquark Systems}'', 
        Preprint HD-THEP-93-15, 1993. 
\bibitem{LeutwylerStern} {\sc H.\ Leutwyler, J.\ Stern}, ``{\sl Relativistic 
		Dynamics on a Null Plane}'', Ann.~Phys.~(NY) {\bf 112}
		(1978) 94. 
\bibitem{weakWilson} {\sc K.G.\ Wilson, T.S.\ Walhout, A.\ Harindranth, 
	W.-M. Zhang, R.J.\ Perry},
	``{\sl Nonperturbative QCD: A weak-coupling treatment on the 
	light front}'',
	Phys.\ Rev.\ {\bf D49} (1994) 6720.
\bibitem{Perry} {\sc R.J.\ Perry},
	``{\sl Hamiltonian Light-Front Theory and Quantum Chromodynamics}'',
	in {\sl Proceedings of Hadrons 94}, V.~Herscovitz, C.\ Vasconcellos,
	Eds., World Scientific, Singapore 1995.
\bibitem{PauliMIR} {\sc H.-C.\ Pauli},
	 ``{\sl Solving Gauge field Theory by 
	Discretized Light-Cone Quantization}'', 
	submitted to Phys.~Rev.~{\bf D} (1996).
\bibitem{Tomonaga} {\sc S.\ Tomonaga}, ``{\sl On a Relativistically Invariant 
	Formulation of the Quantum Theory of Wave Fields}'',
        Prog. \ Theoret.\ Phys.\ {\bf 1} (1946) 27--42; 
        {\sc Z.\ Koba, T.\ Tati, S.\ Tomonaga}, ``{\sl  On a 
	Relativistically Invariant 
	Formulation of the Quantum Theory of Wave Fields. II.	
	--- Case of Interacting Electromagnetic and Electron Fields ---}'',
        Prog. \ Theoret.\ Phys.\ {\bf 2} (1947) 101--116; 
	{\sc Z.\ Koba, T.\ Tati, S.\ Tomonaga}, ``{\sl  On a 
	Relativistically Invariant Formulation of the Quantum Theory of 
	Wave Fields. III.
	--- Case of Interacting Electromagnetic and Electron Fields ---}'',
        Prog. \ Theoret.\ Phys.\ {\bf 2} (1947) 198--208;
	{\sc S.\ Kanesawa, S.\ Tomonaga}, ``{\sl  On a 
	Relativistically Invariant
	Formulation of the Quantum Theory of Wave Fields. VI.
	--- Case of Interacting Electromagnetic and Meson Fields ---}'',
        Prog. \ Theoret.\ Phys.\ {\bf 3} (1948) 1--13; 
	{\sc S.\ Kanesawa, S.\ Tomonaga}, ``{\sl   
	On a Relativistically Invariant
	Formulation of the Quantum Theory of Wave Fields. V.
	--- Case of Interacting Electromagnetic and Meson Fields ---}'',
        Prog. \ Theoret.\ Phys.\ {\bf 3} (1948) 101--113; 
	{\sc S.\ Tomonaga}, Phys.\ Rev.\ {\bf 74} (1948) 224. 
\bibitem{Feynman} {\sc R.\ P.\ Feynman}, ``{\sl Space-Time Approach to 
        Non-Relativistic Quantum Mechanics}'',
        Rev.\ Mod.\ Phys.\ {\bf 20} (1948) 367--387; 
	``{\sl A Relativistic Cut-Off for 
        Classical Electrodynamics}'',
        Phys.\ Rev.\ {\bf 74} (1948) 739--746; 
        ``{\sl A Relativistic Cut-Off for 
        Quantum Electrodynamics}'',
        Phys.\ Rev.\ {\bf 74} (1948) 1430--1438;
	``{\sl The Theory of Positrons}'',
        Phys.\ Rev.\ {\bf 76} (1949) 749--759; 
	``{\sl Space-Time Approach to 
        Quantum Electrodynamics}'',
        Phys.\ Rev.\ {\bf 76} (1949) 769.
\bibitem{Schwinger} {\sc J.\  Schwinger}, ``{\sl Quantum Electrodynamics. I.
        A Covariant Formulation}'',
        Phys.\ Rev.\ {\bf 74} (1948) 1439; 
	``{\sl Quantum Electrodynamics. II.
        Vacuum Polarization and Self-Energy}'',
        Phys.\ Rev.\ {\bf 75} (1948) 651--679. 
\bibitem{KPW} {\sc M. Krautg\"artner, H.-C.~Pauli, F. W\"olz}, 
	``{\sl Positronium and Heavy Quarkonia as Testing Case for DLCQ}'', 
	Phys.\ Rev.\ {\bf D45} (1992) 3755. 
\bibitem{Pauli84} {\sc H.-C.\ Pauli}, 
	``{\sl On a Numerical Exact Solution to the Many-Body Problem in One
	Dimension}'',
	Z.\ Phys.\ {\bf A319} (1984) 303-314. 
\bibitem{DissBvdS} {\sc B.~van de Sande},
	``{\sl Renormalization and the Zero-Mode in Light-Front field Theory}'',
	Dissertation, Ohio State University, 1994;
	``{\sl Renormalization of Tamm-Dancoff Integral Equations}'',
	{\tt hep-th/9210147};
	{\sc B.~van de\ Sande, S.S.\ Pinsky, et al.}
	``{\sl Spontaneous Symmetry Breaking of $\phi^4$ Theory in Light-Front
	Field Theory}'',
	Phys.\ Rev.\ {\bf D48} (1993) 816-821;
	Phys.\ Rev.\ {\bf D49} (1994) 2001-2013;
	Phys.\ Rev.\ {\bf D51} (1994) 726-733.
\bibitem{Dirac} {\sc P.A.M. Dirac}, 
	``{\sl Forms of Relativistic Dynamics}'',
	Rev.\ Mod.\ Phys.\ {\bf 21} (1949) 392.
\bibitem{Schwinger3} {\sc J.\  Schwinger}, ``{\sl Quantum Electrodynamics. III.
        The Electromagnetic Properties of the Electron --- 
        Radiative Corrections to Scattering}'',
        Phys.\ Rev.\ {\bf 75} (1948) 790--817. 
\bibitem{Weinberg} {\sc S.\ Weinberg}, ``{\sl Dynamics at Infinite Momentum}'',
		Phys.\ Rev.\ {\bf 150} (1966) 1313.
\bibitem{KogutSoper} {\sc J.B.\ Kogut, D.E.\ Soper}, ``{\sl Quantum 
		Electrodynamics in the Infinite-Momentum Frame}'',
		Phys.\ Rev.\ {\bf D1} (1970) 2901.
\bibitem{ChangRootYan} {\sc S.J.\ Chang, R.G.\ Root, T.M.\ Yan},
	``{\sl Quantum field theories in the infinite momentum frame. 
	I--III.}'',  
	Phys.\ Rev.\ {\bf D7} (1973) 1133;1147;1760.
\bibitem{Ligterink} {\sc N.E.\ Ligterink},
		``{Light-front Hamiltonian field theory: 
		covariance and renormalization}'', Dissertation, University
		of Amsterdam, 1996; 
		{\sc N.E.\ Ligterink, B.L.G.~Bakker},
		``{Equivalence of light-front and covariant field theory}'',
		Phys.\ Rev.\ {\bf D52} (1995) 5954.
\bibitem{BLepage} {\sc G.P.\ Lepage, S.J.\ Brodsky}, ``{\sl Exclusive 
		processes in perturbative Chromodynamics}'',
		Phys.\ Rev.\ {\bf D22} (1980) 2157.
\bibitem{BLepage2} {S.J.\ Brodsky, \sc G.P.\ Lepage}, in ``{\sl Perturbative
		Quantum  Chromodynamics}'', A.H.\ Mueller, Ed.,
		World Scientific, Singapore, 1989.
\bibitem{BP} {\sc H.-C.\ Pauli, S.J.\ Brodsky}, 
	``{\sl Solving field theory in one space and one time dimension}'',
	Phys.\ Rev.\ {\bf D32} (1985) 1993; 
	``{\sl Discretized light-cone quantization: Solution to a 
	field theory in one space and one time dimension}'',
	Phys.\ Rev.\ {\bf D32} (1985) 2001. 
\bibitem{Sawicki} {\sc M.~Sawicki},
		``{\sl Solution of the light-cone equation for the relativistic
		bound state}'',  
		Phys.\ Rev.\ {\bf D32} (1985) 2666;
		``{\sl Eigensolutions of the light-cone equation for a
		scalar field model}'', 
		Phys.\ Rev.\ {\bf D33} (1986) 1103.
\bibitem{HariVary} {\sc A.\ Harindranath, J.P.\ Vary},
	``{\sl Solving $\phi^4_2$ by discretized light front quantization}'',
        Phys.\ Rev.\ {\bf D36} (1987) 3666.
\bibitem{Hornbostel} {\sc K.\ Hornbostel, S.J.\ Brodsky, H.-C. Pauli},
	``{\sl Light-Cone-Quantized  QCD in 1+1 Dimensions}'', 
	Phys.\ Rev.\ {\bf D45} (1990) 3814;
	{\sc K.\ Hornbostel}, Ph.D. thesis, SLAC Report 333.
\bibitem{Heyssler} {\sc M.\ Heyssler}, ``{\sl Numerische L\"osungsverfahren zur
	QCD$_{(1+1)}$ im Rahmen der DLCQ-Methode}'', Diplomarbeit,
	Heidelberg 1994.
\bibitem{EllerPB} {\sc S.\ Eller, H.-C.\ Pauli, S.J.\ Brodsky}, ``{\sl 
	DLCQ: The massless and the massive Schwinger Model}'', Phys.\ Rev.\ 
	{\bf 35} (1987) 1493.
\bibitem{McCartor} {\sc G.~McCartor}, 
	``{\sl Schwinger Model in the light cone representation}'', 
	Z.\ Phys.\ {\bf C64} (1994) 349--354;
	``{\sl Light cone gauge Schwinger Model}'', 
	Z.\ Phys.\ {\bf C52} (1991) 611--626.
\bibitem{Elser} {\sc S.\ Elser}, ``{\sl Das Spektrum der QED$_{(1+1)}$ im 
	Rahmen der DLCQ-Methode}'', Diplomarbeit,
	Heidelberg 1994. 
\bibitem{Voellinger} {\sc M.\ V\"ollinger}, ``{\sl \"Uber die 
	Lichtkegel-Nullmoden in der QED$_{(1+1)}$}'', Diplomarbeit,
	Heidelberg 1996. 
\bibitem{Tang} {\sc A. Tang}, ``{\sl DLCQ: Application to Quantum 
		Electrodynamics}'', Dissertation, SLAC-Report-351, Juni 1990;
	{\sc A.~Tang, S.J.~Brodsky, H.-C.~Pauli},
	``{\sl Discretized light-cone quantization: Formalism for quantum 
	electrodynamics}'', 
	Phys.\ Rev.\  {\bf D44} (1991) 1842.
\bibitem{Kaluza} {\sc M.~Kaluza, H.-C.~Pauli}, 
	``{\sl Discretized light-cone quantization: $e^+e^-(\gamma)$ Model for
	Positronium}'', 
	Phys.\ Rev.\ {\bf D45} (1992) 2968;
        {\sc M. Kaluza}, ``{\sl Facets of solving 3+1 gauge 
	theory with DLCQ}'', Dissertation, Heidelberg 1991.
\bibitem{DissWoelz} {\sc F.~W\"olz}, ``{\sl \"Uber das Spektrum der Normalmoden 
	in der QCD und die Theorie der effektiven Wechselwirkung nach der 
	Tamm-Dancoff Methode}'', Dissertation (unpublished), Heidelberg 1995. 
\bibitem{KalloPauliPinsky} {\sc A.C.\ Kalloniatis, H.-C.\ Pauli, S.S.\ Pinsky},
	``{\sl Towards solving QCD: The transverse zero modes in light-cone 
	quantization}'',
	Phys.\ Rev.\ {\bf D52} (1995) 1176-1189.
\bibitem{DissBayer} {\sc R.M.~Bayer}, ``{\sl Berechnung von Spektren
	in einem kollinearen Modell der Quantenchromodynamik}'', 
	Dissertation(unpublished), Heidelberg 1996.	
\bibitem{BvdSBurckardt} {\sc B.~van de\ Sande, M.\ Burckardt},
	``{\sl Tube model for light-front QCD}'',
	Phys.\ Rev.\ {\bf D53} (1996) 4628--4637.
\bibitem{BvdSCoralGables} {\sc B.~van de\ Sande, S.\ Dalley},
	``{\sl The transverse lattice in 2+1 dimensions}'' in
        {\sc B.N. Kursunoglu, S.\ Mintz, and A.\ Perlmutter}
        (Eds.), ``{\sl Neutrino Mass, Monopole Condensation, 
        Dark Matter, and Gravitational Waves}'', 
        Plenum Publishing Corp., New York, to be published in 1996. 
\bibitem{BurkardtLangnau} {\sc M.\ Burkardt, A.\ Langnau}, ``{\sl Rotational
	invariance in light-cone quantization}'', 
	Phys.\ Rev.\ {\bf D44} (1991) 3857--3867.
\bibitem{KalPir} {\sc M.~Kaluza, H.-J.~Pirner}: ``{\sl Hyperfine 
	splitting in the light-cone Tamm-Dancoff equation of QED and QCD}'', 
	Phys.\ Rev.\ {\bf D47} (1993) 1620-1628. 
\bibitem{Leibbrandt}{\sc G.\ Leibbrandt},
	``{\sl Introduction to non-covariant gauges}'',
        Rev.\ Mod.\ Phys.\ {\bf 59} (1987) 1067.
\bibitem{MustakiRenormalization} {\sc D.\ Mustaki, S.S.\ Pinsky, 
	J.\ Shigemitsu, K.G.\ Wilson},
	``{\sl Perturbative renormalization of null-plane QED}'',
	Phys.\ Rev.\ {\bf D43} (1991) 3411-3427.
\bibitem{LeibMandelstam} {\sc S.\ Mandelstam},
	``{\sl Light-cone superspace and ultraviolet finiteness of 
	the N=4 model}'',
        Nucl.\ Phys.\ {\bf B213} (1983) 149;
	{\sc G.\ Leibbrandt},
	``{\sl Light-cone gauge in Yang-Mills theory}'',
        Phys.\ Rev.\ {\bf D29} (1984) 1699.
\bibitem{Basetto} {\sc A.\ Basetto, M. Dalbolsco, I.\ Lazzizzera, R.\ Soldati},
	``{\sl Yang-Mills theories in the light-cone gauge}''
        Phys.\ Rev.\ {\bf D31} (1985) 2012.
\bibitem{KalloPauli} {\sc A.C.\ Kalloniatis, H.-C.\ Pauli},
	``{\sl Bosonic zero modes and gauge theory in discrete light-cone 
	quantization}'',
	Z.\ Phys.\ {\bf C60} (1993) 255-264;
	``{\sl Beyond the Light-Cone Gauge in Discrete Light-Cone 
	Quantization of Quantum Electrodynamics}'',
	Z.\ Phys.\ {\bf C63} (1994) 161.
\bibitem {HeinzlWerner} 
   	{\sc T.~Heinzl, S.~Krusche, S.~Simb\"urger, and E.~Werner}, 
	``{\sl Nonperturbative light cone quantum field theory beyond the 
	tree level}'',
   	Z.\ Phys.\ {\bf C56}, (1992) 415. 
\bibitem{McCartor0} {\sc G.~McCartor}, 
	``{\sl Bosonic Zero Modes in Discretized Light Cone Field Theory}'',
	Z.\ Phys.\ {\bf C53} (1992) 679--686.
\bibitem{AlexDave} {\sc A.C.\ Kalloniatis, D.\ Robertson},
	``{\sl On the Discretized Light-Cone Quantization of Electrodynamics}'',
	Phys.\ Rev.\ {\bf D50} (1994) 5262.
\bibitem{Coleman} {\sc S.\ Coleman, R.~Jackiw, L.~Susskind}, 
	Ann.~Phys.~{\bf 93} (1975) 267;
	{\sc S.\ Coleman}, 
	Ann.~Phys.~{\bf 101} (1976) 239.
\bibitem{BvdSSmallMass} {\sc B.~van de Sande},
	``{\sl Convergence of Discretized Light-Cone Quantization in the 
	Small Mass Limit}'',
	Preprint MPIH-V18-1996, {\tt hep-ph/9605409}.
\bibitem{Ferrell} {\sc J.\ Pirenne}, Arch.\ Sci.\ phys.\ nat.\ {\bf 29} (1947)
	121, 207, 265; 
	{\sc V.\ Berestetski, L.D.\ Landau},
	``{\sl Interaction between an Electron and a Positron}'', 
	translated from the Russian original, 
	JETP(USSR) {\bf 19} (1949) 673--679;
	{\sc V.\ Berestetski}, 
	``{\sl The spectrum of Positronium}'', translated from the 
	Russian original, 
	JETP(USSR) {\bf 19} (1949) 1130-1135; 
	{\sc R.A.\ Ferrell}, ``{\sl The Positronium Fine Structure}'',
	Phys.\ Rev.\ {\bf 84} (1951) 858 (Letter to the editor); 
	{\sc R.A.\ Ferrell}, Ph.D. thesis,
	Princeton University, 1952.
\bibitem{Woelz} {\sc F.~W\"olz}, ``{\sl Numerische L\"osung des Coulomb-Problems
	in der Impulsdarstellung}'', Diplomarbeit (unpublished), 
	Heidelberg 1990.	
\bibitem{Schladming} {\sc S.J.~Brodsky, H.-C.~Pauli}, ``{\sl Light-Cone 
                Quantization of Quantum Chromodynamics}'' in 
                {\sc H. Mitter, H. Gausterer} (Eds.) : ``{\sl Recent Aspects
                of Quantum Fields}'', Proceedings of the 
                XXX. Int. Universit\"atswochen f\"ur Kernphysik in {Schladming} 
                1991, Springer-Verlag.
\bibitem{MorseFesh} {\sc P.M. Morse, H. Feshbach}, ``{\sl Methods of 
		Theoretical Physics}'' (2 Volumes) McGraw-Hill, New York 1953.
\bibitem{Tamm} {\sc I.J.\ Tamm},
	``{\sl Relativistic Interaction of Elementary Particles}'',	
	J.~Phys.\ (USSR) {\bf 9} (1945) 449.
\bibitem{Dancoff} {\sc S.M.\ Dancoff}, 
	``{\sl Non-Adiabatic Meson Theory of Nuclear Forces}'',	
	Phys.~Rev. {\bf 78} (1950) 382.
\bibitem{Gupta}{\sc S.N.\ Gupta, W.W.\ Repko, C.J.\ Suchyta},
	``{\sl Muonium and positronium potentials}''  
	Phys.\ Rev.\ {\bf D40} (1989) 4100--4104. 
\bibitem{Kinoshita} {\sc T.\ Kinoshita}(Ed.),	
	``{\sl Quantum Electrodynamics}'', World Scientific, Singapore 1990.
\bibitem{DissJungmann} {\sc K.-P.\ Jungmann}, ``{\sl Kontinuierliches 
	Multimoden-Laserspektrometer zur Zweiphotonenspektroskopie am
	Positronium-Atom}'', 
	Dissertation (unpublished), Heidelberg 1985.	
\bibitem{BetheSalpeterBook} {\sc H.A.\ Bethe, E.E.\ Salpeter}, 
	``{\sl Quantum Mechanics of One and Two-Electron Atoms}'',
	A Plenum/Rosetta Edition, Plenum Publishing Corporation, 
	New York, 1957, 1977.
\bibitem{DissKraut} {\sc M. Krautg\"artner}, ``{\sl Anwendung der DLCQ auf die
		QED}'', Dissertation (unpublished), Heidelberg 1992.
\bibitem{Fermi30} {\sc E.\ Fermi},
	``{\sl \"Uber die magnetischen Momente der Atomkerne}'',
	Z.\ Phys.\ {\bf 60} (1930) 320--333.
\bibitem{BodwinYennieGregorio}{\sc G.T.\ Bodwin, D.R.\ Yennie, M.A.\ Gregorio}, 
 	``{\sl Recoil Effects in the hyperfine structure of QED bound states}'',
	Rev.\ Mod.\ Phys.\ {\bf 57} (1985) 723--782. 
\bibitem{Coester} {\sc F.\ Coester}, ``{\sl Null-Plane Dynamics of Particles 
	and Fields}'',
        Prog.\ Part.\ Nucl.\ Phys.\ {\bf 29} (1992) 1--32.
\bibitem{Lubanski} {\sc K.\ Lubanski}, ``{\sl Sur la th\'eorie des particules
	\'el\'ementaires de spin quelconque}'',
        Physics\ {\bf 9} (1942) 310--338.
\bibitem{Melosh} {\sc H.J.\ Melosh}, 
	``{\sl Quarks: Currents and Constituents}'',
        Phys.\ Rev.\ {\bf D9} (1974) 1095--1112.
\bibitem{Fulton} {\sc T.~Fulton, P.C.~Martin},
	``{\sl Two-Body System in Quantum Electrodynamics. Energy Levels of 
	Positronium}'', 
	Phys.~Rev.~{\bf 95} (1954) 811--822.
\bibitem{Dykshoorn} {\sc W.D.~Dykshoorn, R.~Koniuk}
	``{\sl Ultrarelativistic bound states in spinor and scalar QED}'', 
	Phys.~Rev.~{\bf A41} (1990) 60--63;
	``{\sl Spin effects in highly relativistic systems}'', 
	Phys.~Rev.~{\bf A41} (1990) 64--67.
\bibitem{CoralGables} {\sc H.-C.\  Pauli} in 
                {\sc B.N. Kursunoglu, S.\ Mintz, and A.\ Perlmutter}
                (Eds.), ``{\sl Neutrino Mass, Monopole Condensation, 
                Dark Matter, and Gravitational Waves}'', 
                Plenum Publishing Corp., New York, to be published in 1996.  
\bibitem{BrisudovaPerry} {\sc M.\ Brisudov\'a, R.J.\ Perry},
	``{\sl Initial bound state studies in light-front QCD}'',
	Phys.\ Rev.\ {\bf D54} (1996) 1831;
	``{\sl Note on restoring manifest rotational symmetry in
	hyperfine and fine structure in light-front QED}'',
	{\tt hep-th/9605363}, 1996.
\bibitem{Jones} {\sc B.D.\ Jones, R.J.\ Perry, S.D.\ G\/{l}azek},
	``{\sl Nonperturbative QED: An analytical treatment on the
	 light front}'',
	{\tt hep-th/9605231}, submitted to Phys.\ Rev.\ {\bf D}.
\bibitem{Wegner} {\sc F.\ Wegner},
	``{\sl Flow-equations for Hamiltonians}'', 
	Ann.~Phys.~{\bf 3} (1994) 77--91.
\bibitem{Pirner} {\sc H.J.\ Pirner}, private communication.
\bibitem{Ammons} {\sc E.\ Ammons},
	``{\sl Bloch-Wilson Renormalization of Tamm-Dancoff Light-Front QED}'', 
	Phys.~Rev.~{\bf D54} (1996) 5153--5162.
\bibitem{PauliPriv} {\sc H.-C.\ Pauli}, private communication.
\bibitem{PauliMerkel} {\sc H.-C.~Pauli, J.\ Merkel},
	 ``{\sl On the Masses of the Physical Mesons:
	Solving the Effective QCD-Hamiltonian by DLCQ}'', 
	submitted to Phys.~Rev.~{\bf D} (1996).
\bibitem{Bergmann} {\sc J.L.\ Anderson, P.G.\ Bergmann}, ``{\sl Constraints in 
		Covariant Field Theories}'', Phys.\ Rev.\ {\bf 83} (1951) 1018.
\bibitem{Sunder} {\sc K. Sundermayer}, ``{\sl Constrained Dynamics}'', 
		Lecture Notes
		in Physics {\bf 169}, Springer, Berlin 1982.
\bibitem{Dirac2} {\sc P.A.M. Dirac}, ``{\sl Lectures on Quantum Mechanics}'',
		Belfer Graduate School of Science Monograph Series, 
		New York 1964.
\bibitem{Merkel} {\sc J.\ Merkel}, ``{\sl \"Uber das Massenspektrum der 
	physikalischen Mesonen: Ein effek\-tiver QCD-Hamiltonian mit laufender 
	Kopplungskonstante}'', Diplomarbeit, Heidelberg 1994.
\bibitem{Karmanov80} {\sc V.A.\ Karmanov},
		``{\sl Light-front wave function of a relativistic composite
		system in an explicitly solvable model}'', 
		Nucl.~Phys.~{\bf B166} (1980) 378--398.
\bibitem{Karmanov81} {\sc V.A.\ Karmanov},
		``{\sl Relativistic deuteron wave function on the light 
		front}'', 
		Nucl.\ Phys.\ {\bf A362} (1981) 331--348.
\bibitem{NumRec} {\sc W.H. Press, B.P. Flannary, S.A. Teukolsky,
		W.T. Vetterling}, ``{\sl Numerical Recipes}'',
		Cambridge Unviversity Press, 1989.
\bibitem{Itzykson} {\sc C. Itzykson, J.B. Zuber}, 
		``{\sl Quantum Field Theory}'', McGraw Hill, 1980.
\bibitem{Nachtmann} {\sc O. Nachtmann}, ``{\sl Konzepte und Ph\"anomene der
		Elementarteilchenphysik}'', Vieweg, Braunschweig 1986. 
\bibitem{BvdSCoulombTrick} {\sc B. van de Sande}, 
		``{\sl On the Coulomb Singularity}'',
		unpublished notes.
\bibitem{Naglib} ``{\sl The NAG Fortran Library Manual, Mark 16}'',
	The Numerical Algorithms Group Limited, 1993.
\end{thebibliography}
\end{document}